\pdfoutput=1
\documentclass[12pt,a4paper]{article}

\usepackage{ifthen} 
\newboolean{pdflatex}
\setboolean{pdflatex}{true} 

\newboolean{articletitles}
\setboolean{articletitles}{true} 

\newboolean{uprightparticles}
\setboolean{uprightparticles}{false} 

\def\paperasciititle{Unleashing the full power of LHCb to probe Stealth New Physics}
\def\papertitle{Unleashing the full power of LHCb to probe Stealth New Physics}

\usepackage[top=1in, bottom=1.25in, left=1in, right=1in]{geometry}

\usepackage{microtype}
\usepackage{lineno}
\usepackage{xspace}
\usepackage{caption}

\usepackage{comment}

\usepackage{graphicx}
\usepackage{color}
\usepackage{colortbl}
\graphicspath{{./figs/}}

\usepackage{amsmath}
\usepackage{amssymb}
\usepackage{amsfonts}
\usepackage{upgreek}

\newcommand*\patchAmsMathEnvironmentForLineno[1]{%
\expandafter\let\csname old#1\expandafter\endcsname\csname #1\endcsname
\expandafter\let\csname oldend#1\expandafter\endcsname\csname
end#1\endcsname
 \renewenvironment{#1}%
   {\linenomath\csname old#1\endcsname}%
   {\csname oldend#1\endcsname\endlinenomath}%
}
\newcommand*\patchBothAmsMathEnvironmentsForLineno[1]{%
  \patchAmsMathEnvironmentForLineno{#1}%
  \patchAmsMathEnvironmentForLineno{#1*}%
}
\AtBeginDocument{%
\patchBothAmsMathEnvironmentsForLineno{equation}%
\patchBothAmsMathEnvironmentsForLineno{align}%
\patchBothAmsMathEnvironmentsForLineno{flalign}%
\patchBothAmsMathEnvironmentsForLineno{alignat}%
\patchBothAmsMathEnvironmentsForLineno{gather}%
\patchBothAmsMathEnvironmentsForLineno{multline}%
\patchBothAmsMathEnvironmentsForLineno{eqnarray}%
}

\usepackage{hyperxmp}

\usepackage[pdftex,
            pdftitle={\paperasciititle},
            hidelinks]{hyperref}

\usepackage{enumitem}

\usepackage[bottom,flushmargin,hang,multiple,stable]{footmisc}

\usepackage[all]{hypcap}

\usepackage{xspace} 
\usepackage{upgreek}

\def\lhcb   {\mbox{LHCb}\xspace}

\def\velo   {VELO\xspace}

\def\MagUp {\mbox{\em Mag\kern -0.05em Up}\xspace}

\def\hltone {HLT1\xspace}
\def\hlttwo {HLT2\xspace}

\ifthenelse{\boolean{uprightparticles}}%
{

 \def\Pmu         {\ensuremath{\upmu}\xspace}

 \def\Ppi         {\ensuremath{\uppi}\xspace}

 \def\Ptau        {\ensuremath{\uptau}\xspace}

 \def\Ppsi        {\ensuremath{\uppsi}\xspace}

 \def\PDelta      {\ensuremath{\Delta}\xspace}                 
 \def\PXi         {\ensuremath{\Xi}\xspace}                 
 \def\PLambda     {\ensuremath{\Lambda}\xspace}                 
 \def\PSigma      {\ensuremath{\Sigma}\xspace}                 
 \def\POmega      {\ensuremath{\Omega}\xspace}                 
 \def\PUpsilon    {\ensuremath{\Upsilon}\xspace}

 \def\PB      {\ensuremath{\mathrm{B}}\xspace}                 
                  
 \def\PD      {\ensuremath{\mathrm{D}}\xspace}

 \def\PJ      {\ensuremath{\mathrm{J}}\xspace}                 
 \def\PK      {\ensuremath{\mathrm{K}}\xspace}

 \def\Pe      {\ensuremath{\mathrm{e}}\xspace}

 \def\Pi      {\ensuremath{\mathrm{i}}\xspace}

 \def\Ps      {\ensuremath{\mathrm{s}}\xspace}

 \def\thebaroffset{0.0em}
}
{

 \def\Pmu         {\ensuremath{\mu}\xspace}

 \def\Ppi         {\ensuremath{\pi}\xspace}

 \def\Ptau        {\ensuremath{\tau}\xspace}

 \def\Ppsi        {\ensuremath{\psi}\xspace}                 
                  
 \mathchardef\PDelta="7101
 \mathchardef\PXi="7104
 \mathchardef\PLambda="7103
 \mathchardef\PSigma="7106
 \mathchardef\POmega="710A
 \mathchardef\PUpsilon="7107
                  
 \def\PB      {\ensuremath{B}\xspace}                 
                  
 \def\PD      {\ensuremath{D}\xspace}

 \def\PJ      {\ensuremath{J}\xspace}                 
 \def\PK      {\ensuremath{K}\xspace}

 \def\Pe      {\ensuremath{e}\xspace}

 \def\Pi      {\ensuremath{i}\xspace}

 \def\Ps      {\ensuremath{s}\xspace}

 \def\thebaroffset{0.18em}
}
\newcommand{\offsetoverline}[2][\thebaroffset]{\kern #1\overline{\kern -#1 #2}}%

\makeatletter
\ifcase \@ptsize \relax
  \newcommand{\miniscule}{\@setfontsize\miniscule{4}{5}}
\or
  \newcommand{\miniscule}{\@setfontsize\miniscule{5}{6}}
\or
  \newcommand{\miniscule}{\@setfontsize\miniscule{5}{6}}
\fi
\makeatother

\DeclareRobustCommand{\optbar}[1]{\shortstack{{\miniscule (\rule[.5ex]{1.25em}{.18mm})}
  \\ [-.7ex] $#1$}}

\def\epem       {{\ensuremath{\Pe^+\Pe^-}}\xspace}

\def\mup        {{\ensuremath{\Pmu^+}}\xspace}
\def\mun        {{\ensuremath{\Pmu^-}}\xspace} 

\def\taup       {{\ensuremath{\Ptau^+}}\xspace}
\def\taum       {{\ensuremath{\Ptau^-}}\xspace}

\def\squark    {{\ensuremath{\Ps}}\xspace}

\def\pion   {{\ensuremath{\Ppi}}\xspace}
\def\piz    {{\ensuremath{\pion^0}}\xspace}
\def\pip    {{\ensuremath{\pion^+}}\xspace}
\def\pim    {{\ensuremath{\pion^-}}\xspace}

\def\kaon    {{\ensuremath{\PK}}\xspace}

\def\KorKbar {\kern \thebaroffset\optbar{\kern -\thebaroffset \PK}{}\xspace}

\def\Kp      {{\ensuremath{\kaon^+}}\xspace}
\def\Km      {{\ensuremath{\kaon^-}}\xspace}

\def\D       {{\ensuremath{\PD}}\xspace}

\def\DorDbar {\kern \thebaroffset\optbar{\kern -\thebaroffset \PD}\xspace}

\def\Dp      {{\ensuremath{\D^+}}\xspace}
\def\Dm      {{\ensuremath{\D^-}}\xspace}

\def\DpDm    {\ensuremath{\Dp {\kern -0.16em \Dm}}\xspace}

\def\B       {{\ensuremath{\PB}}\xspace}

\def\BorBbar {\kern \thebaroffset\optbar{\kern -\thebaroffset \PB}\xspace}

\def\Bd      {{\ensuremath{\B^0}}\xspace}

\def\BdorBdbar {\kern \thebaroffset\optbar{\kern -\thebaroffset \Bd}\xspace}
\def\Bu      {{\ensuremath{\B^+}}\xspace}

\def\Bp      {{\ensuremath{\Bu}}\xspace}

\def\Bs      {{\ensuremath{\B^0_\squark}}\xspace}

\def\BsorBsbar {\kern \thebaroffset\optbar{\kern -\thebaroffset \Bs}\xspace}

\def\jpsi     {{\ensuremath{{\PJ\mskip -3mu/\mskip -2mu\Ppsi}}}\xspace}

\def\Y#1S{\ensuremath{\PUpsilon{(#1S)}}\xspace}

\def\LorLbar     {\kern \thebaroffset\optbar{\kern -\thebaroffset \PLambda}\xspace}

\def\to                 {\ensuremath{\rightarrow}\xspace}

\def\order   {{\ensuremath{\mathcal{O}}}\xspace}

\def\met {{\ensuremath{E^{\rm miss}_T}}\xspace}

\def\AT#1     {\ensuremath{A_{\mathrm{T}}^{#1}}\xspace}           

\def\C#1      {\ensuremath{\mathcal{C}_{#1}}\xspace}                       
\def\Cp#1     {\ensuremath{\mathcal{C}_{#1}^{'}}\xspace}                    
\def\Ceff#1   {\ensuremath{\mathcal{C}_{#1}^{\mathrm{(eff)}}}\xspace}        
\def\Cpeff#1  {\ensuremath{\mathcal{C}_{#1}^{'\mathrm{(eff)}}}\xspace}       
\def\Ope#1    {\ensuremath{\mathcal{O}_{#1}}\xspace}                       
\def\Opep#1   {\ensuremath{\mathcal{O}_{#1}^{'}}\xspace}                    

\newcommand{\aunit}[1]{\ensuremath{\text{\,#1}}}       
\newcommand{\unit}[1]{\aunit{#1}\xspace}                   

\newcommand{\tev}{\aunit{Te\kern -0.1em V}\xspace}
\newcommand{\gev}{\aunit{Ge\kern -0.1em V}\xspace}
\newcommand{\mev}{\aunit{Me\kern -0.1em V}\xspace}
\newcommand{\kev}{\aunit{ke\kern -0.1em V}\xspace}
\newcommand{\ev}{\aunit{e\kern -0.1em V}\xspace}
 
\newcommand{\mevc}{\ensuremath{\aunit{Me\kern -0.1em V\!/}c}\xspace}
\newcommand{\gevc}{\ensuremath{\aunit{Ge\kern -0.1em V\!/}c}\xspace}
\newcommand{\mevcc}{\ensuremath{\aunit{Me\kern -0.1em V\!/}c^2}\xspace}
\newcommand{\gevcc}{\ensuremath{\aunit{Ge\kern -0.1em V\!/}c^2}\xspace}
\newcommand{\tevcc}{\ensuremath{\aunit{Te\kern -0.1em V\!/}c^2}\xspace}

\def\cm   {\aunit{cm}\xspace}

\def\mm   {\aunit{mm}\xspace}

\def\fb   {\ensuremath{\aunit{fb}}\xspace}
\def\invfb   {\ensuremath{\fb^{-1}}\xspace}

\def\sec  {\ensuremath{\aunit{s}}\xspace}

\def\mhz  {\ensuremath{\aunit{MHz}}\xspace}

\def\order{{\ensuremath{\mathcal{O}}}\xspace}

\def\gsim{{~\raise.15em\hbox{$>$}\kern-.85em
          \lower.35em\hbox{$\sim$}~}\xspace}
\def\lsim{{~\raise.15em\hbox{$<$}\kern-.85em
          \lower.35em\hbox{$\sim$}~}\xspace}

\def\pt         {\ensuremath{p_{\mathrm{T}}}\xspace}

\def\pythia     {\mbox{\textsc{Pythia}}\xspace}

\def\tell1  {TELL1\xspace}
\def\ukl1   {UKL1\xspace}

\usepackage{cite}
\usepackage{mciteplus}

\usepackage{longtable}
\usepackage{subfigure}
\makeatletter\g@addto@macro\bfseries{\boldmath}\makeatother

\begin{document}

\renewcommand{\thefootnote}{\fnsymbol{footnote}}
\setcounter{footnote}{1}
\begin{titlepage}

\vspace*{-3.5cm}

\noindent

\vspace*{4.0cm}

{\normalfont\bfseries\boldmath\huge
\begin{center}
  \papertitle
\end{center}
}

\vspace*{2.0cm}

\begin{center}
\textbf{Editors:} M. Borsato\ref{hei}, X. Cid Vidal\ref{usc}, Y. Tsai\ref{mar}$^,$\ref{notredame}, C. V\'{a}zquez Sierra\ref{cern}, J. Zurita\ref{ific}

\vspace{1cm}
\textbf{Authors:} 
G.~Alonso-\'{A}lvarez\ref{mcgill}, A.~Boyarsky\ref{lorentz}, A.~Brea~Rodr\'iguez\ref{usc}, D.~Buarque~Franzosi\ref{chalmers}$^,$\ref{goth}, G.~Cacciapaglia\ref{lyon}$^,$\ref{lyon2}, A.~Casais~Vidal\ref{usc}, M.~Du\ref{nanjing}, G.~Elor\ref{seattle},
M.~Escudero\ref{tum}, G.~Ferretti\ref{chalmers}, T.~Flacke \ref{IBS}, P.~Foldenauer\ref{ippp},
  J.~Hajer\ref{louvain}$^,$\ref{basel},  L.~Henry\ref{cern}, P.~Ilten\ref{cin}, J.~Kamenik\ref{jsi}$^,$\ref{ljubljana}, B.~Kishor~Jashal\ref{ific}, S.~Knapen\ref{cern}, F.~L.~Redi\ref{epfl}, M.~Low\ref{fnal}, Z.~Liu\ref{nanjing}$^,$\ref{beijing}$^,$\ref{CAS}, A.~Oyanguren~Campos\ref{ific}, E.~Polycarpo\ref{ufrj}, M.~Ramos\ref{granada}$^,$\ref{minho}, M.~Ramos~Pernas\ref{war}, E.~Salvioni\ref{cern},  M.~S.~Rangel\ref{ufrj}, 
R.~Sch\"afer\ref{itphei}, 
L.~Sestini\ref{pad}, Y.~Soreq \ref{technion},  V.~Q.~Tran\ref{nanjing}, I.~Timiryasov\ref{epfl}, M.~van~Veghel\ref{gron}, S.~Westhoff\ref{itphei}, M.~Williams\ref{mit}, J.~Zupan\ref{cin}

\vspace{1cm}

\end{center}

{\footnotesize
\begin{enumerate}[label = {$^{\arabic*}$}, labelsep = 1pt]
\item Physikalisches Institut, Ruprecht-Karls-Universit\"{a}t Heidelberg, Heidelberg, Germany \label{hei}
\item Instituto Galego de F{\' i}sica de Altas Enerx{\' i}as (IGFAE), Universidade de Santiago de Compostela, Santiago de Compostela, Spain \label{usc}
\item Maryland Center for Fundamental Physics, Department of Physics, University of Maryland, College Park, MD 20742-4111, United States \label{mar}
\item Department of Physics, University of Notre Dame, South Bend, IN 46556, USA \label{notredame}
\item European Organization for Nuclear Research (CERN), Geneva, Switzerland \label{cern}
\item Instituto de F\'isica Corpuscular (CSIC-UV), Valencia, Spain  \label{ific}
\item McGill University Department of Physics \& McGill Space Institute,
3600 Rue University, Montr\'eal, QC, H3A 2T8, Canada\label{mcgill}
\item Intituut-Lorentz, Leiden University, 2333 CA Leiden, the Netherlands\label{lorentz}
\item Institute for Theoretical Physics, Heidelberg University, 69120 Heidelberg, Germany \label{itphei}
\item Department of Physics, Chalmers University of Technology, Fysikg{\r{a}}rden, 41296 G{\" o}teborg, Sweden \label{chalmers}
\item Physics Department, University of Gothenburg, 41296 G\"oteborg, Sweden \label{goth}
\item University of Lyon, Universit{\' e} Claude Bernard Lyon 1, F-69001 Lyon, France \label{lyon}
\item Institut de Physique des 2 Infinis (IP2I) de Lyon, CNRS/UMR5822, F-69622 Villeurbanne, France \label{lyon2}
\item Department of Physics, Nanjing University, Nanjing 210093, China \label{nanjing} 
 \item Department of Physics, University of Washington, Seattle, WA 98195, USA 
 \label{seattle}
\item Physik-Department, Technische Universit{\" a}t, M{\" u}nchen, James-Franck-Stra{\ss}e, 85748 Garching, Germany \label{tum}

\item Center for Theoretical Physics of the Universe, Institute for Basic Science (IBS), Daejeon 34126, Korea
 \label{IBS}
 \item Institute for Particle Physics Phenomenology,
Durham University, Durham DH1 3LE, United Kingdom \label{ippp}
\item Centre for Cosmology, Particle Physics and Phenomenology,
Universit{\' e} catholique de Louvain, Louvain-la-Neuve B-1348, Belgium  \label{louvain}
\item Department of Physics, Universit{\" a}t Basel, Klingelbergstraße 82, CH-4056 Basel, Switzerland \label{basel}
\item Department of Physics, University of Cincinnati, Cincinnati, Ohio 45221, USA \label{cin}
\item Jo\v{z}ef Stefan Institute, Jamova 39, 1000 Ljubljana, Slovenia \label{jsi}
\item 
Faculty of Mathematics and Physics, University of Ljubljana, Jadranska 19, 1000 Ljubljana, Slovenia 
\label{ljubljana}
\item Institute of Physics, Ecole Polytechnique F{\' e}d{\' e}rale de Lausanne (EPFL), Lausanne, Switzerland \label{epfl}
\item Theoretical Physics Department, Fermilab, P.O. Box 500, Batavia, IL 60510 \label{fnal}
\item Center for High Energy Physics, Peking University, Beijing 100871, China \label{beijing}
\item CAS Center for Excellence in Particle Physics, Beijing 100049, China \label{CAS}
\item Universidade Federal do Rio de Janeiro (UFRJ), Rio de Janeiro, Brazil \label{ufrj}
\item CAFPE and Departamento de F\'isica Te\'orica y del Cosmos, Universidad de Granada,
Campus de Fuentenueva, E–18071 Granada, Spain \label{granada}
\item Laborat\'orio de Instrumenta\c{c}ao e Física Experimental de Partículas, Departamento de Física da Universidade do Minho, Campus de Gualtar, 4710-057 Braga, Portugal \label{minho}
 \item Department of Physics, University of Warwick, Coventry, United Kingdom \label{war}
 \item Istituto Nazionale di Fisica Nucleare (INFN), Padova Division, Padova, Italy \label{pad}
\item Physics Department, Technion – Israel Institute of Technology, Haifa 3200003, Israel \label{technion}
\item Van Swinderen Institute, University of Groningen, Groningen, Netherlands \label{gron}
\item Laboratory for Nuclear Science, Massachusetts Institute of Technology, Cambridge, MA 02139, U.S.A. \label{mit}
\end{enumerate}
\vspace{\fill}
}

\begin{abstract}
  \noindent
  In this paper, we describe the potential of the LHCb experiment to detect Stealth physics. This refers to dynamics beyond the Standard Model that would elude searches that focus on energetic objects or precision measurements of known processes. Stealth signatures include long-lived particles and light resonances that are produced very rarely or together with overwhelming backgrounds. We will discuss why LHCb is equipped to discover this kind of physics at the Large Hadron Collider and provide examples of well-motivated theoretical models that can be probed with great detail at the experiment.
\end{abstract}

\vspace*{2.0cm}
\vspace{\fill}

\pagestyle{empty}

\end{titlepage}

\pagestyle{empty}

\newpage
\setcounter{page}{2}
\mbox{~}

\renewcommand{\thefootnote}{\arabic{footnote}}
\setcounter{footnote}{0}

\tableofcontents
\cleardoublepage

\pagestyle{plain}
\setcounter{page}{1}
\pagenumbering{arabic}

\section{Introduction}
\label{sec:Introduction}
While particle colliders have not yet been able to find physics beyond the Standard Model (BSM), there is a general consensus within the particle physics community that BSM physics must exist, given the limitations of the Standard Model (SM) \cite{Allanach:2016yth}. Therefore, this lack of discovery might be due to three reasons: a) simply not being possible to produce BSM physics at colliders; b) BSM physics lying beyond the energy frontier, and being only accessible by building more powerful accelerators or; c) BSM physics being within our reach but produced in “Stealth” mode, {\it{i.e.}}, very rarely and/or with significant backgrounds, hence not having been detected so far. We advocate here for option c), which is comprised of what we have dubbed \textbf{Stealth physics}. 
With this description, Stealth physics would involve signatures including rarely-produced particles, soft objects, or decays from new light resonances. A prominent example is that of long-lived particles (LLPs) \cite{Alimena:2019zri}, which are BSM particles that travel a macroscopic distance\footnote{For LHCb this would mean roughly a mm in the laboratory frame.}. Given how rare they are, these types of signatures are ideally suited to be studied at the precision frontier, {\it{i.e.}}, by accumulating large amounts of data from particle colliders, and using precise detectors that can overcome the difficulties that these signatures impose. Moreover, novel theoretical approaches solving some key problems of the SM naturally include Stealth signatures.

Out of the existing experimental efforts to find BSM physics, the Large Hadron Collider (LHC) is one of the most ambitious. The LHC collides protons at unprecedented luminosity, producing some of the largest recorded collision datasets in history, that allow the exploration of the precision frontier. The results of these collisions are registered by four main detectors located at different collision points. In this paper we will focus in one of them, the LHCb detector, which has been taking data since 2009 and is still expected to continue doing so for the next 10 to 15 years. 
The LHCb detector is a single-arm spectrometer, with a design complementary to the rest of the LHC experiments, and very well instrumented in the forward region. These features allow the LHCb detector to tackle Stealth signatures with surgical precision. 

This paper aims at providing a useful guide both for:
\begin{itemize}
    \item Theorists who want to study whether (and to what extent) their favorite model could be probed at LHCb.
    \item Experimentalists at LHCb who are looking for new exciting signatures with a sound theoretical motivation. 
\end{itemize}

With those two goals in mind, the paper is divided in three main sections. Section \ref{sec:experimental} is conceived as a dictionary for anyone to check the main considerations one should make to study Stealth physics at LHCb. Next, Sec.~\ref{sec:review} summarizes the main results achieved so far by LHCb in the area of Stealth physics. Section~\ref{sec:theory} contains a brief review of interesting theoretical models for which LHCb can have great discovery potential. Finally, conclusions are drawn in Section~\ref{sec:conclusions}.


\section{Experimental perspective}
\label{sec:experimental}

The LHCb experiment at CERN was originally designed to study the decays of heavy flavour particles~\cite{Alves:2008zz}, that is, particles containing a $b$ or a $c$ quark. These particles are characteristically light, long-lived and produced in the forward direction. They have masses few times larger than that of the proton and they typically fly several \mm (up to few \cm) in the detector before decaying. Consequently, the detector is designed to select forward displaced decays with soft (low transverse momentum) final states exploiting an outstanding Vertex Locator (VELO) and a fast and precise trigger. 
Unique particle identification capabilities are provided by muon stations, a calorimeter system and Cherenkov light detectors. Furthermore, excellent invariant mass resolution is key to reduce backgrounds due to partially reconstructed decays or random combinations of tracks. This is provided by a precise spectrometer composed of several tracking stations upstream and downstream of a 4\,Tm dipole magnet. 
With all of these features combined, the experiment has become an outstanding resource to search for new particles, especially in decays involving soft and/or displaced final states. The ranges of masses and lifetimes targeted by these searches are similar to those typical of heavy flavour particles, which are very difficult to probe at other LHC experiments like ATLAS and CMS, and where the LHCb detector has unique and complementary capabilities.

The capabilities of LHCb will be significantly better in LHC Run~3, thanks to upgrades of the LHCb vertex \cite{LHCb-TDR-013} and track \cite{LHCb-TDR-015} reconstruction systems, as well as a new trigger system completely based on software \cite{LHCb-TDR-016}.

\subsection{The LHCb detector: Stealth considerations}

In this section we present general considerations that concern the status of the detector and its upgrades, the trigger and the particle reconstruction and identification.

The unique detector design of LHCb provides advantages and disadvantages with respect to other LHC experiments in terms of Stealth physics. One key disadvantage of LHCb is reduced data rates compared to that of the ATLAS and CMS detectors:

\begin{itemize}
\item \textit{Luminosity}: To ensure the ability of the LHCb VELO to resolve displaced vertices from heavy flavour decays, the luminosity at the LHCb collision point was reduced via {\it{luminosity levelling}}, to an avaerage of about one proton-proton collision per bunch crossing ($\lesssim 5\,\%$ of the nominal LHC luminosity)\cite{Follin:2014nva}. While this does decrease the total collected data, it also provides a cleaner environment with significantly less pile-up. Reduced pile-up is particularly important for any type of Stealth physics emerging from central exclusive production, such as some axion-like particle searches.

\item \textit{Acceptance}: LHCb instruments 4\% of the solid angle in the pseudo-rapidity range $2<\eta<5$. The LHCb acceptance for fully reconstructing the two-body decay of a $100\gevcc$ resonance is roughly $10\,\%$, and $1\,\%$ for a $1\tevcc$ resonance. The detector is able to reconstruct charged stable particles with transverse momentum (\pt) above\footnote{compared to $\sim 1$ \gevc for, e.g., CMS\cite{Chatrchyan:1704291}.} $\sim 80$ MeV/c, if these are roughly produced in $0<z<500$ mm and $0<\rho<30$ mm, $z$ and $\rho$ being cylindrical coordinates. 

Due to the non isotropic production angle of heavy quarks, 40 \% of the heavy quark production cross-section can be resolved within the LHCb forward acceptance. But, this also means the signal acceptance for LHCb is smaller than for a more traditional full coverage detector. Additionally, the limited geometrical acceptance reduces the ability to use missing transverse momentum to search for invisible signatures. However, the VELO does have full coverage from $-5 < \eta < 5$, albeit without momentum determination, which can be used to search for large rapidity gaps in events. 
\end{itemize}

The overall combined factors of {\it{luminosity levelling}} and acceptance result in LHCb collecting the equivalent of $1\,\%$ of the nominal LHC luminosity when searching for new resonances above the electroweak scale. However, this reduction is counteracted by a number of LHCb advantages, which include the following:

\begin{itemize}
\item \textit{Particle identification}: LHCb has a comprehensive set of detectors for particle identification, including Ring Imaging Cherenkov (RICH) detectors, as explained in Section \ref{sec:PID}. 

\item \textit{Small displacements}: The VELO used during Run~1 and Run~2 has a lifetime resolution of $\approx 50~\text{fs}$, an impact parameter resolution of\footnote{again, to provide a significant comparison, a ``soft'' track with \pt of $\sim1$ \gevc would have an impact parameter resolution of $\sim 100~\mu\text{m}$ in CMS\cite{Chatrchyan:1704291}.} $13$ to $20\,\mu\text{m}$, and a secondary vertex precision of $0.01$ to $0.05\mm$ in the $xy$ plane\cite{LHCb:2001aa, LHCbCollaboration:2013bkh, LHCb-DP-2014-002}. Similar resolutions are expected after the Phase-I upgrade, as it will be seen in Section \ref{lhcb-future}. Not only do these resolutions allow for displaced searches to be performed for new particles with lifetimes on the order of the $D$-mesons or larger, but also the ability to discriminate against backgrounds using the flight vector determined from the secondary vertex. This is particularly effective when constructing variables such as the {\it{corrected mass}}, introduced in Sec.~\ref{sec:tau}. Similarly, requiring the momentum and flight vector to align can suppress large backgrounds for pointing signals.

\item \textit{Narrow}: The Run~1 and Run~2 momentum resolution is $\approx 0.5\,\%$ for charged particles with momenta of $5\gevc$ and $1\,\%$ at $100\gevc$. This excellent single track momentum resolution translates to a mass resolution of $0.4\,\%$ for muon pair production~\cite{Aaij:2017rft}. As many Stealth signatures have narrow natural widths, this mass resolution significantly enhances searches when performing bump hunts on continuum backgrounds.

\item \textit{Trigger}: The LHCb trigger, described in Section \ref{sec:trigger}, is very flexible and efficient in several key topologies for Stealth physics, which the trigger of other LHC experiments cannot select. For instance, during Run~2, every muon pair candidate where each of the muons satisfies $\pt(\mu) > 0.5\gevc$ and $p(\mu) > 10\gevc$ was recorded, across all masses~\cite{Aaij:2017rft}. 
\end{itemize}


\subsubsection{LHCb past, present and future detector status}\label{lhcb-future}

\begin{figure}
    \centering
    \includegraphics[width=\textwidth]{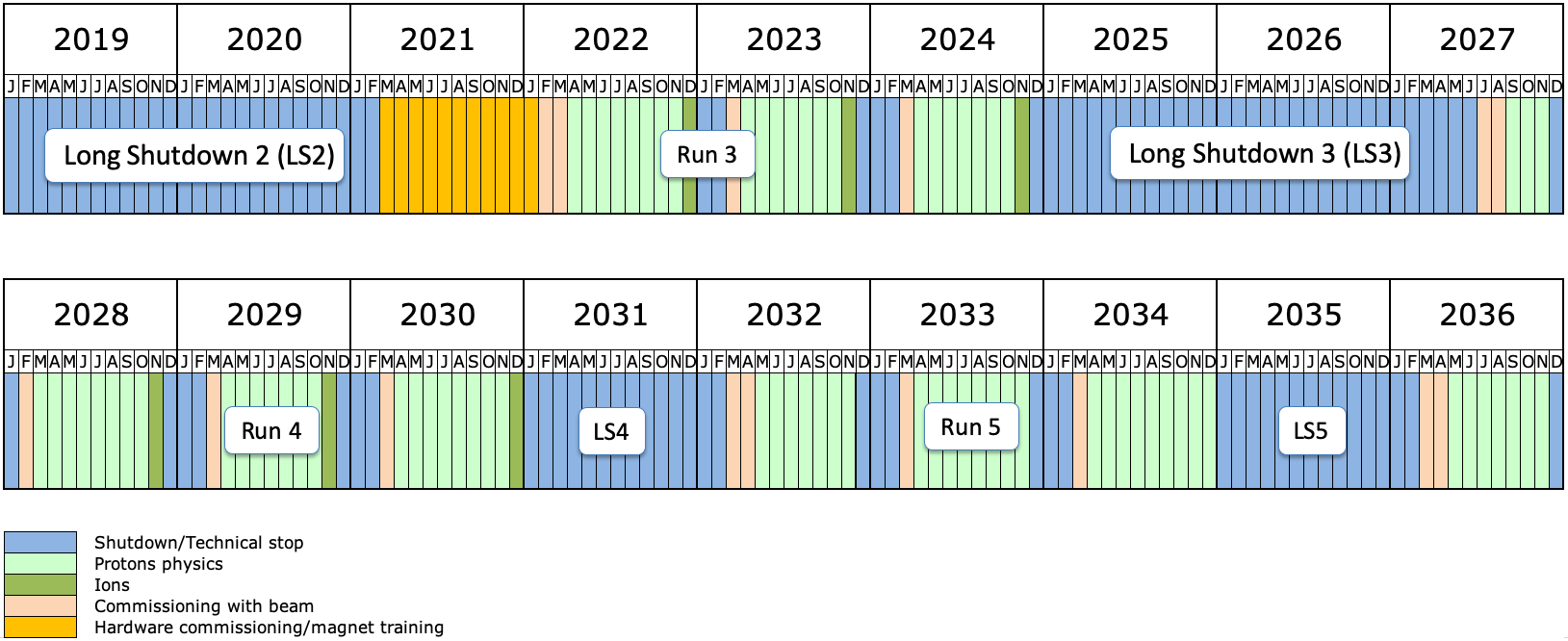}
    \caption{Longer term LHC schedule as of September 2020. Figure from Ref.~\cite{schedule}.}
    \label{fig:lhc_runs}
\end{figure}

During the LHC Run~1 and Run~2, the LHCb detector has proven to be the ideal environment to study heavy flavour particles. The schedule for LHC Runs, as well as the foreseen long shutdowns, are all collected in Fig.~\ref{fig:lhc_runs}. On top of heavy flavor, the physics program of the detector has been expanded well beyond its design, transforming LHCb in a forward general-purpose detector at the LHC.

The LHCb collaboration has a long-term plan of upgrades, with a schedule that is different from that of other LHC experiments.
Its upgrades are divided in two phases~\cite{LHCb-TDR-012, LHCb-PII-Physics}. 

\begin{figure}
    \centering
    \includegraphics[width=.8\textwidth]{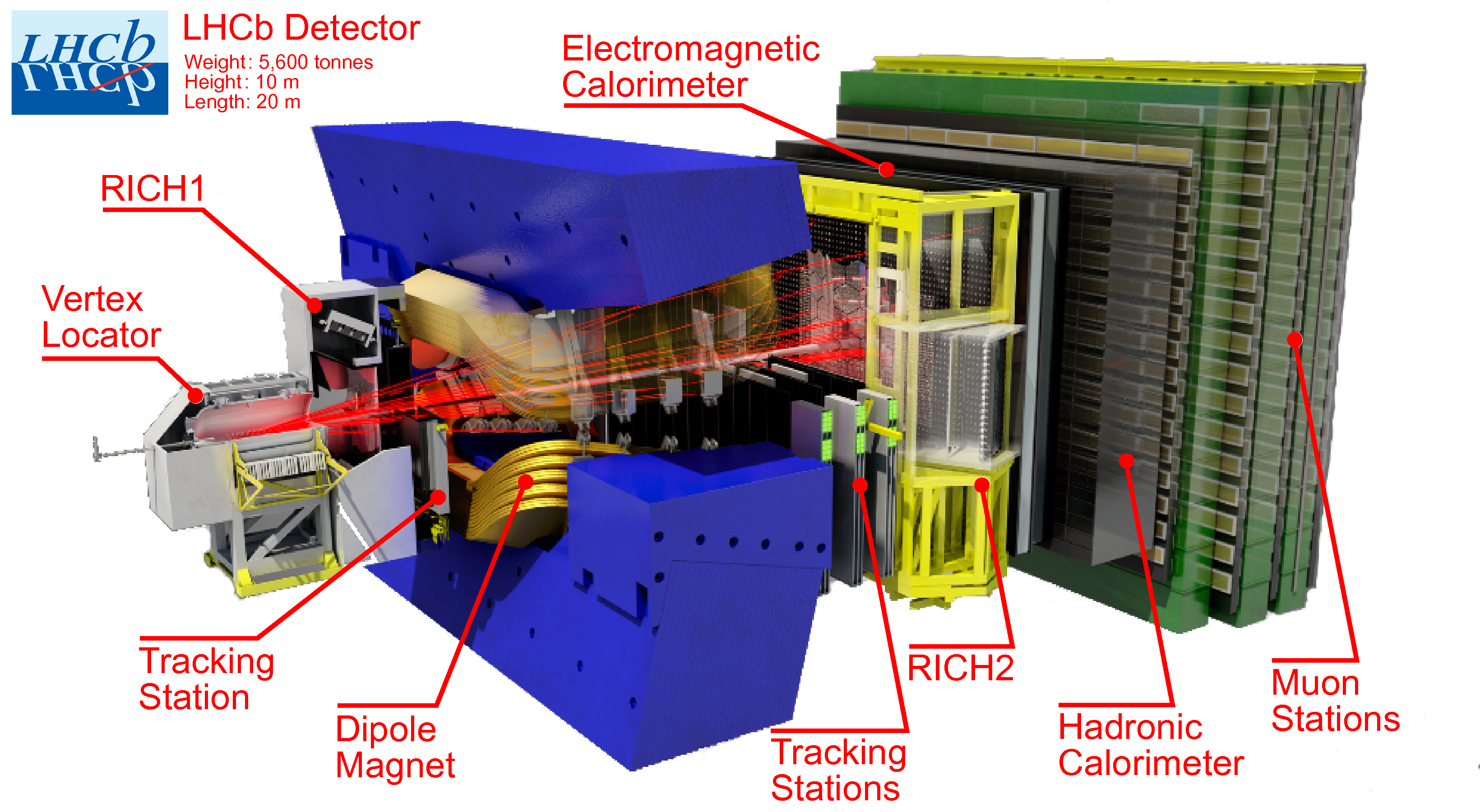}
    \caption{3D sketch of the LHCb detector. Figure from Ref.~\cite{Koppenburg:2059944}.}
    \label{fig:detector}
\end{figure}

The Phase-I LHCb upgrade is being installed during the LHC Long Shutdown 2 (LS2) and will allow the LHCb detector to collect up to 23 (50) \invfb before the end of Run~3 (4) \cite{Nakada:1100545, CERN-LHCC-2011-001, LHCb-TDR-012, LHCb-TDR-013, LHCb-TDR-014, LHCb-TDR-015,LHCb-TDR-016}. 
This upgraded detector is designed to run the experiment at higher luminosity compared to the previous detector and improve the flexibility of the trigger system. To better understand the different components of the detector, a 3D sketch of it is shown in Fig.~\ref{fig:detector}.
The previous trigger system was based on high transverse momentum signatures in the calorimeters or in the muon system. This allowed to fully read-out events at a 1.1\,MHz rate~\cite{LHCb-TDR-010}.
In order to keep the same read-out rate in the upgrade, these transverse momentum requirements would have to be tightened further, which would largely cancel the gains due to the higher luminosity.
Moreover, the sub-detectors were originally designed to operate for five years at an instantaneous luminosity of $2\times10^{32}$\,cm$^{-2}$, but at the end of Run~2, they had already reached an exposure of seven years at more than $3\times10^{32}$\,cm$^{-2}$. 
As a result, during the upgrade of the LHCb detector, many sub-detectors are being replaced, improving the Stealth physics reach of LHCb.
As discussed in Sec.~\ref{sec:trigger}, one of the largest effects will come from the trigger which will move to a software-only scheme.
It will be able to fully process 30 MHz of inelastic event rate, collecting data from all sub-detectors which will be needed for a full event reconstruction~\cite{LHCb-TDR-016,Aaij:2019zbu}. 
Data will then be buffered to disk for online calibration and alignment.

Another important sub-detector for Stealth physics is the VELO~\cite{LHCb-TDR-013}. 
The upgraded VELO is based on pixel technologies rather than silicon strips as the previous VELO, with a smaller sensor pitch size and located closer to the interaction point (just 5.1\mm from the beam line). It is expected to have better primary vertex and impact parameter resolution than the current VELO, even for Run~3 challenging conditions (higher particle multiplicity and radiation damage)~\cite{LHCb-TDR-013}. Thanks to an improved spatial resolution close to the interaction point, the upgraded VELO will allow to perform a more precise fit of the vertex and thus to reduce the combinatorial background and resolve decays with shorter lifetimes.

The Tracker Turicensis (TT), located upstream from the magnet, will be completely replaced by a four-layered silicon strip detector, the Upstream Tracker (UT)~\cite{LHCb-TDR-015}. The new detector will be able to withstand at least 50 \invfb of integrated luminosity, while having an increased granularity and a 40\mhz read-out frequency.

The Phase-II LHCb upgrade was originally proposed to capitalise on the opportunities that the High Luminosity LHC (HL-LHC) will offer~\cite{LHCb-PII-EoI,LHCb-PII-Physics}. 
This second upgrade is scheduled to be installed during the Long Shutdown 4 (LS4), for the start of Run~5, at the beginning of 2031. 
Specifically, if the LHCb detector in the Phase-I upgrade configuration is capable of operating in a high luminosity scenario but does not require it, a Phase-II LHCb upgrade will allow the detector to operate at a value of instantaneous luminosity of $1-2\times10^{34}\cm^{-2}\textrm{s}^{-1}$ and collect at least $300\invfb$. 
Among the many sub-detectors that will be introduced, one of particular potential for the Stealth physics programme are tracking stations installed on the internal surfaces of the dipole magnet. This upgrade will allow LHCb to track particles with momenta well below 1\gevc. 
A promising first proposal is to construct this sub-detector based on scintillating fibres with a readout of Silicon Photo-Multipliers (SiPMs).
This will allow the readout electronics to be placed outside of the acceptance in a region of lower magnetic field and lower neutron flux. 

The particle identification (PID) capabilities of LHCb rely mostly on information recorded by the RICH sub-detector.
Currently, the RICH provides no information for kaon and proton candidates below 10 \gevc.
The proposed TORCH (Time Of internally Reflected CHerenkov light) detector, to be installed as part of Phase-II upgrade, would measure the time-of-flight (ToF) of charged particles~\cite{Bhasin:2729016}. 
This would  effectively allow a $3\,\sigma$ separation between kaons and protons all the way down to 2\gevc.
TORCH would be installed immediately upstream of the RICH~2 substations, with planes of 1\cm-thick quartz that would cover a total area of $5 \times 6\,\mathrm{m}^2$. 
The modes that involve protons in the final state are those that would benefit the most from the installation of this sub-detector.
Nevertheless, the fact that $K - \pi$ separation in the \gevc momentum range is also enhanced, will directly benefit the searches for Stealth signal candidates.
The expected gains are in the range of 10 to 30\,\% in PID efficiency~\cite{Bhasin:2729016}.

Another expected detector development related to Stealth Physics is the upgrade of the muon system. 
Extra shielding has been introduced during LS2 to cope with the increased luminosity but this will not be sufficient for the future and therefore some modifications will be required during Phase-II. 
The current proposal is to replace the hadron calorimeter (HCAL) with up to 1.7\,m of iron~\cite{LHCb-PII-Physics}.
This corresponds to an addition four interaction lengths compared to the current setup. This change, which might seem drastic, is possible since the HCAL information is mostly used in the hardware trigger, which will be obsolete starting from Run~3.

Overall, the LHCb experiment is in a unique position to extend its role in the current Stealth physics scenario.
In particular, the removal of the hardware level trigger during the Phase-I upgrade and the installation of tracking stations on the surface of the magnet during the Phase-II upgrade will prove great tools for the future physics searches.

\subsubsection{Trigger}
\label{sec:trigger}

The structure of the LHCb trigger has been in continuous evolution since its design for Run~1 data taking~\cite{LHCb-TDR-10}.
Important features of the system are a high flexibility and the capability of changing its configuration to adapt it to new physics goals.
The Run~1~\cite{Aaij:2012me,Albrecht:2013fba} and Run~2~\cite{Aaij:2018jht} triggers were implemented in two levels. 

First, the L0 trigger, implemented in hardware, reduced the visible bunch crossing rate to 1\mhz by selecting events with at least one particle with high transverse momentum or energy, using the muon chambers and the calorimeter system. 
Events with high occupancy\footnote{this referring to having a large fraction of the sub-detectors actually firing.} and complexity were rejected with an additional requirement on SPD hit multiplicity, this being one of the components of the calorimeter. 
The transverse momentum and energy thresholds varied over time and were different among the several signatures.
The lowest thresholds were used for muons (1.4-2.8\gevc), while the highest were used for hadron signatures (3.5-3.7\gevc).
    
Then, the high level trigger (HLT) was designed in software, with high flexibility.
It was subdivided in two different levels:
\begin{itemize}
    \item HLT1: This level comprised a partial event reconstruction and comprised a set of inclusive selections based on topological information of signal candidates. 
    \item HLT2: The last stage of the trigger was composed by a mixture of exclusive and inclusive selections running after a full event reconstruction. Selections also included particle identification.
    \end{itemize}

Due to the rigidity of the L0 trigger, only the thresholds of the selections could be changed across the years, but not the set of physical quantities used.
However, the HLT has undergone many modifications during its lifetime.
For instance, significant advances in reconstruction techniques allowed for full reconstruction in the HLT, including particle flow and jet reconstruction. This flexible trigger during Run~1 and Run~2 already allowed for very inclusive triggers. Apart from these, the major modification for Run~2 data taking has been the inclusion of real-time calibration and alignment of the detector~\cite{Dujany:2015lxd}.
Special trigger selections at HLT1 fed dedicated algorithms in order to align and calibrate the different sub-detectors. These procedures were performed in a matter of minutes, so they interfered minimally with regular data taking. 
A dedicated buffer was placed between the HLT1 and HLT2 stages, able to accumulate up to two weeks of regular data taking before running the HLT2 selections.
This buffer served as a safety mechanism to check and modify the calibration and alignment and to delay the HLT2 execution in case of problems with the detector.
Therefore, it was possible to get the same performance at the HLT2 stage as for any offline selection.

The interest on storing large samples of beauty and charm hadrons for data analysis pushed towards the design of a data stream based on exclusive selections where only the information from the signal candidates is persisted.
For the Run~2 of the LHCb experiment, the {\it{Tesla}} application~\cite{LHCb-DP-2016-001} was introduced in the online LHCb infrastructure for that purpose.
This new application generated a new data stream at HLT2, profiting from the real time alignment and calibration achieved after HLT1 and from the full event reconstruction.
The new data stream, known as the {\it{Turbo}} stream~\cite{Benson:2015yzo}, allowed to reduce the event size by one order of magnitude with respect to the default data stream and thus increase the rate at which events can be written. 
This new approach of data taking becomes particularly efficient for studies where the analysis strategy is clear and the backgrounds are well-known. 
These two factors become crucial since the information from the underlying event is dropped after data processing, making it impossible to compute some quantities offline. This strategy has been essential in the unprecedented precision of results obtained, for instance, on charm $CP$-violation using LHCb Run~2 data \cite{LHCb-PAPER-2019-006}, or for dark photon searches, which fall in the category of Stealth physics (see Secs.~\ref{sec:review} or~\ref{sec:dp} for details). 

For the Phase-I Upgrade of the LHCb experiment~\cite{LHCb-TDR-017, LHCb-TDR-018}, the high increase of instantaneous luminosity, from \(4\times10^{32}\cm^{-2}\sec^{-1}\) to \(2\times10^{33}\cm^{-2}\sec^{-1}\), together with the aspiration to widen the physics program of the experiment, forced a full re-design of the trigger.
One of the biggest limiting factors of the trigger schemes existing so far was the L0 stage.
The requirements on the transverse momentum and transverse energy of the particle candidates present at this trigger level became inefficient for many searches at LHCb.
In the Phase-I Upgrade, the L0 will be removed, leading to a trigger composed of two software-based triggers similar to those present in the high-level trigger of Run~2~\cite{LHCb-TDR-016}.
This will reduce the thresholds on the transverse momentum to roughly \(0.5\gevc\), imposed by timing constraints on the reconstruction, and increasing the LHCb capabilities in the low-\pt region.
A simplified event reconstruction with basic muon identification will be performed in the first stage, whilst full event reconstruction, including real-time alignment and calibration, together with full particle identification, will be available in the second stage.
The overall performance of the trigger will therefore become even more similar to that achieved offline.

Due to the processing necessities of LHCb in terms of physics reach and also due to the limited bandwidth of the experiment, the second stage of the upgraded LHCb trigger is expected to be composed mainly by exclusive selections based on the Turbo stream~\cite{LHCb-TDR-018}, described above.
At least \(60\,\%\) of the bandwidth will be reserved for this stream, and the remaining \(40\,\%\) will consist of inclusive and exclusive selections where the information from the whole event is kept.

During the last years, there was an increasing effort to design new techniques in order to boost the reconstruction and selection processes at HLT1.
The low event size at LHCb, together with the reduced relative cost with respect to a CPU implementation, suggested the use of Graphics Processing Units (GPUs) at HLT1~\cite{LHCb-TDR-021,Aaij:2019zbu,Aaij:2021mzf}. Implementing a full HLT1 trigger on GPUs will allow to run the selection with no thresholds in the reconstruction sequence, increasing the efficiency to trigger on soft signatures. 

The current \hltone strategy is partly tuned towards the fast reconstruction of displaced vertices due to heavy-flavour decays.
This entails that most of its discriminatory power, and	time budget, are ascribed to the \velo, which is the detector most suited for this task.
However, many Stealth signatures could still be located in detectors situated downstream, due to long-lived particles in the final state.
As a result, these channels rely on the	rest of	the event to trigger the \hltone in order to have a chance to be studied further in \hlttwo, which relies on much more diverse signatures.
Consequently, the search for new physics in these channels pays a statistical price due to the absence of a dedicated \hltone algorithm.\footnote{The exact reduction factor depends on the production mechanism of the channel.}
The \lhcb collaboration, in synergy with increased interest in SM decays containing long-lived particles, is considering the possibility to run part of the \hlttwo reconstruction algorithms in the \hltone.
This would possibly allow for dedicated, high-efficiency trigger selections on high-\pt long-lived particles, and thus significantly increase the lifetime reach for LLPs with the \lhcb detector.

In the LHCb Phase-I Upgrade data-taking scheme, only a small portion of	the data is recorded without any dedicated \hlttwo selection, most of which are exclusive triggers. 
This makes it crucial, in order	to make the most of the	delivered luminosity, to consider as soon as possible interesting signatures and write trigger selections for them.
Whereas	some Stealth searches could be performed with no dedicated \hltone selection, for instance if those searches include a high-\pt muon or a displaced \jpsi, those same searches would still need a dedicated \hlttwo selection during data-taking.
It is for instance possible technically, but not yet implemented, to write \hlttwo selections for modes containing a long-lived particle decaying after the UT.	
Such line would	allow to greatly increase the lifetime reach of	searches for BSM long-lived particles.
Another	possibility is to look for charged long-lived particles that decay inside of the VELO. Triggering on such decays would be done by associating tracks to a VELO segment, such as what would be done for the $\Kp\to\pip\pim\pip$ decay, which has a similar topology.


\subsubsection{Offline reconstruction, with a focus on jets}
\label{sec:off_and_jets}

\begin{figure}
    \centering
    \includegraphics[width=0.6\textwidth]{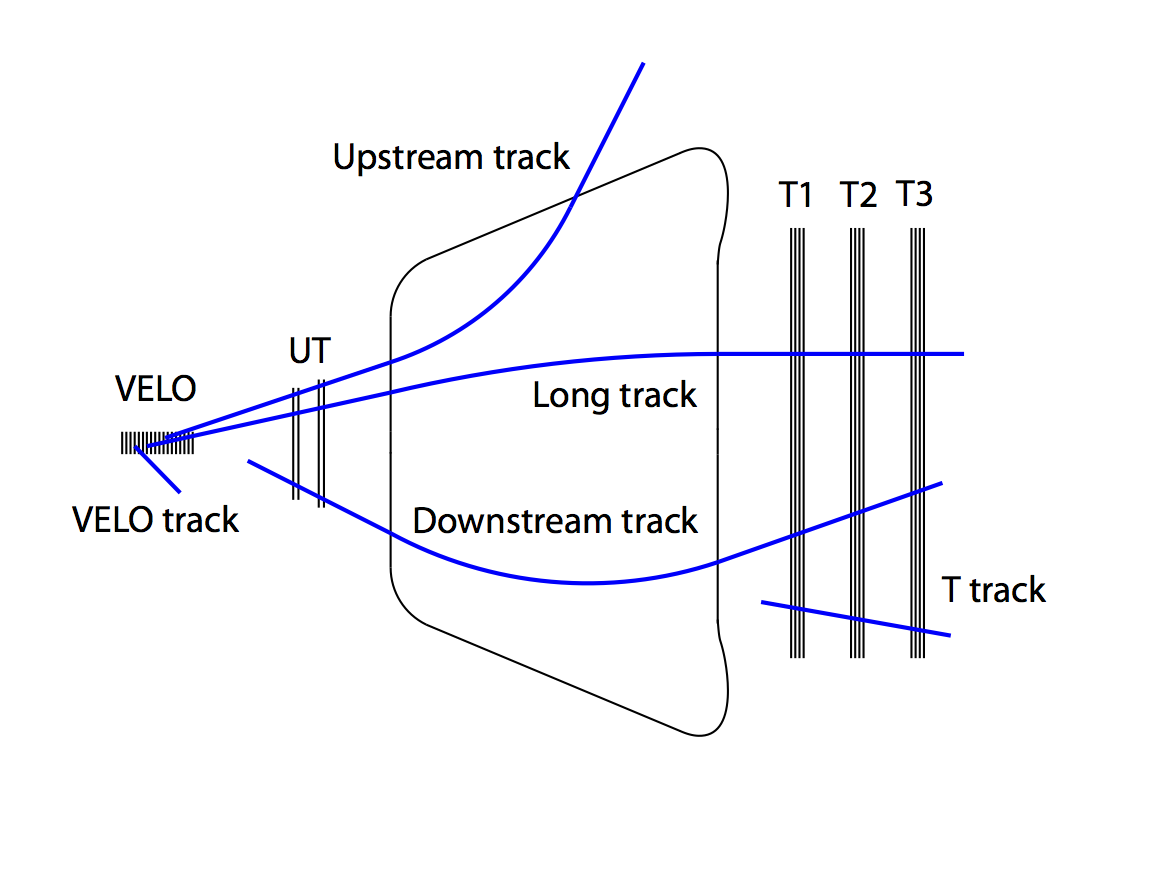}
    \caption{Scheme of the main track types existing at the upgraded LHCb \cite{LHCb-TDR-015}.}
    \label{fig:tracking}
\end{figure}

Tracks in LHCb are classified according to the sub-detectors used in their reconstruction. They belong to five main categories\footnote{note this refers to the upgraded version of the detector, as introduced above.} (as shown in Fig.~\ref{fig:tracking}):
\begin{itemize}
    \item \textit{Long} tracks that are reconstructed using hits from the VELO and the SciFi trackers, and possibly hits from the UT. The scintillating fiber tracker (SciFi) is the new tracking detector located right after the magnet (the T--stations as presented in in Fig.~\ref{fig:tracking}), to be installed during Phase-I LHCb Upgrade and to replace the current post-VELO silicon tracking system, together with the UT. They have the best momentum resolution, typically of the order of 0.1\,\%;
    \item \textit{Downstream} tracks that are reconstructed using information from the UT and SciFi trackers only. Their momentum resolution is slightly worse than that of \textit{Long} tracks;
    \item \textit{Upstream} tracks that are reconstructed using information from the VELO and UT trackers only. They often correspond to low-momentum charged particles that are swept away by the magnet. Close to no momentum information is available for those tracks;
    \item \textit{VELO} tracks that are reconstructed using only hits in the VELO. No momentum information is available for these tracks;
    \item \textit{T}-tracks that are reconstructed using only hits in the SciFi tracker. Due to the residual magnetic field in the SciFi region, a typical momentum resolution of 20\,\% can be achieved on these tracks.
\end{itemize}

Trigger decisions rely solely on \textit{Long} and \textit{Downstream} tracks, and only the former type is used in the first level of the trigger.
As a result, analyses using any of the three other types of tracks needs to access full events and apply offline reconstruction techniques.

A part of the LHCb data flow is stored in the ``full stream'', where all the event information is stored so that further reconstruction can be performed offline.
Due to the much larger average event size, these streams have a limited acquisition rate.
Accessing all the underlying information of the events however allows to run some algorithms that are too costly for online execution, as well as allow for some flexibility to design innovative new physics searches.
For instance, track extrapolation through the LHCb magnetic field can be run from scratch, with the aim of increasing the vertex reconstruction performance (inside and after the magnet region) with the aid of a Runge-Kutta extrapolator\cite{Bos:1070314}.
The study of decays in flight of charged particles also benefits from this, allowing to look further into kinks of the track trajectories, which would otherwise be associated with scattering.

Jet reconstruction at LHCb uses the anti-kT algorithm~\cite{Cacciari:2008gp} as implemented Ref.~\cite{Cacciari:2011ma}.
The inputs used in the jet reconstruction are built using a particle-flow strategy. Charged particles and calorimeter information are selected in order to build unique charged or neutral particle candidates with a minimum overlap between them.
In Run~2 data, well reconstructed jets had a minimum $\pt$ of 15\gevc within the range $2.2 < \eta < 4.2$~\cite{Aaij:2013lla}.

Jet energy correction is determined from simulation and applied to recover the jet energy at particle level. Jet quality requirements can be applied to reduce fake jets with efficiencies close to unity.

Jets can also be heavy-flavour tagged, i.e. identified as containing a $b-$ or $c-$hadron.
In Run~2, the jet heavy-flavour tagging was based on the presence of secondary vertices within the jet cone~\cite{Aaij:2015yqa}. Its performance was reproduced in simulation at about the $10\,\%$ level for heavy-flavour jets, and at the $30\,\%$ level for light-parton jets. 
A typical heavy-flavour tagging efficiency for identifying a $b$ ($c$) jet was measured to be about $65\,\%$($25\,\%$) with a fake rate of $0.3\,\%$.

In addition to the standard jet definition, jets originating from a hypothetical long-lived particle can also be reconstructed by selecting displaced vertices and/or tracks as inputs~\cite{Aaij:2014nma}.
This strategy can also be used if a specific particle is required to be in the jet in order to study hadronisation phenomena (see Ref.~\cite{Aaij:2017fak}).

In Run 3 the jet reconstruction is expected to be challenging, given the higher multiplicity of pile-up interactions. Neutral particles reconstructed in the calorimeters cannot be associated to the primary vertex, resulting in a degradation of the jet energy resolution in events with many interactions. For this reason advanced techniques for the neutral recovery that take into account the pile-up should be employed. Algorithms like the Pileup Mitigation With Machine Learning (PUMML) \cite{Komiske:2017ubm} can be used to effectively subtract the pile-up contamination. In particular this algorithm takes as input the energy distribution of charged leading vertex particles, and gives as outputs the energy distribution of neutral particles coming from leading vertex alone.

Novel jet clustering techniques can also be used to face the Run 3 environment. For example clustering algorithms based on Graph Neural Networks are considered promising \cite{GNN,GNN2}. These algorithms can be trained on specific channels ($i.e.$ a di-jet resonance from new physics) and it has been demonstrated that they can give a better performance with respect to classical algorithms in high pile-up conditions.

Improvements on the heavy flavour tagging are also expected in Run 3. As explained above, during Run 1-2, LHCb has based its tagging capability on the reconstruction of displaced vertices, in order to keep the light flavour contamination low. However, in many cases, this approach may not give the best performance in terms of significance for new physics channels. For this reason we foresee the development on new tagging algorithms \textit{\`{a} la carte}, that can be tuned depending on the target search. As an example, these can be done by employing more general vertexing algorithms with adjustable parameters. Moreover deep learning algorithms that exploit the jet sub-structure can be used for the jet tagging, as done at the CMS experiment \cite{CMSTagging}. The advantage of this last category of algorithms is that the displaced vertex reconstruction is not explicitly required, potentially improving the efficiency.


\subsubsection{Particle identification}
\label{sec:PID}

PID at the LHCb experiment is based on the combined information from the electromagnetic and hadronic calorimeters, the muon chambers, two Cherenkov-light based RICH detectors and the momentum measurement from the spectrometer.
The decay products of long-lived particles might be harder to identify if the decay occurs downstream of some subdetectors.
The LHCb PID system is optimised for the momentum range of heavy meson decay products. For example, only muons with a momentum in excess of 3~GeV can be correctly identified in the muon chambers, while electrons with transverse momenta larger than 20~GeV saturate the electromagnetic calorimeter readout and are harder to identify. The identification of charged kaons, which is provided mostly by the RICH system is optimal in the momentum range between 10 and 40~GeV (kaon efficiency of 95\% for pion mis-identification probability below 1\%). On the other hand, the RICH system can provide excellent identification of electrons with very soft momenta, as pions of the same momentum have significantly lower velocity $\beta$ and do not emit any Cherenkov light (the Cherenkov momentum threshold for pions is $2.6\gev$ in RICH1 and $4.4\gev$ in RICH2~\cite{LHCb-TDR-014}). 

Photons reconstructed in the calorimeter are distinguished from neutral pions by the shape of their energy deposit. Neutral pions with transverse momenta in excess of 2~GeV result in a merged energy deposit cluster in the calorimeter and are hard to distinguish from photon clusters (roughly 45\% of neutral pions as misidentified if one requires a photon identification efficiency of 95\%~\cite{LHCb-DP-2014-002}). Despite the low radiation length of the LHCb trackers, a fraction of photons convert to di-electrons in the detector material. These photons can be reconstructed with excellent momentum resolution and information on the photon trajectory which can be used to identify their displaced production vertex. However, detecting photons through their conversion is a factor 30 to 100 less efficient than via the calorimeter~\cite{LHCb-DP-2014-002}. 


\subsection{Final state considerations}

This section reviews the main experimental challenges involved in studying Stealth physics at LHCb. It covers some of the most popular experimental signatures, such as $\mup\mun$ or jets, as well as some for which fewer studies have been done, such as final states with invisible particles or $\taup\taum$. 


\subsubsection{Beauty hadron decays to invisible}

Given it has no full coverage, \lhcb cannot in principle measure \met, which severely compromises direct searches for invisible particles. An exception to that arises in displaced decays of particles to invisible plus reconstructible objects. In that case, if the reconstructible objects allow the determination of the position of the decay vertex, it is possible to measure a proxy \met variable for displaced semi-invisible decays: the missing \pt with respect to the direction of flight of the parent particle, $\pt^{\rm miss}$. An example of this is the search for dark matter (DM) particles in $b$ hadrons, $b \to {{\rm DM} + \rm X_{\rm SM}}$. The $b$ hadron line of flight determined with the production and decay vertices is misaligned with the visible momentum $p(X_{\rm SM})$ This technique has been used widely at \lhcb to select $b$ decays involving neutrinos, and is explained in more detail in Sec.~\ref{sec:tau}. 

Moreover, whenever an assumption is made about the mass of the parent particle, it becomes possible to relate the maximum $\pt^{\rm miss}$ observed in the decay with the mass of the missing particle. This is useful to distinguish, {\it{e.g.}}, DM produced in $b$ decays from semileptonic $b$ decays involving neutrinos.

An additional handle to help with the reconstruction of the missing particle in displaced decays is the reconstruction of the heavy, usually prompt, parent of the displaced particle. Examples are the prompt $D^{\ast +}\to D^0\pip$ or $B_{s2}^{\ast 0}\to \Bp \Km$ decays. In this regard, \lhcb has just published a search for the lepton flavor violating decay $\Bp \to K^+ \mu^- \tau^+$ where the kinematics of the $\tau$ lepton are obtained through the reconstruction of the excited $B^{\ast 0}_{s2}$ meson, in the $B^{\ast 0}_{s2}\to \Bp \Km$ decay~\cite{Aaij:2020mqb}. This or other excited $b-$hadron decays could be crucial to measure the kinematics of decays involving DM, although at the cost of reduced production rates.

The main background to search for DM in $b$ hadron decays arises from other $b$ decays where some of the final state particles are not reconstructed. As mentioned above, semileptonic $b$ decays are a perfect example of this. Other events that can produce background involve cases in which some of the $b$ hadron decay particles either a) are outside of the geometric acceptance of the detector (pseudorapidities roughly between 2 and 5) or b) they are neutral and are missed by the calorimeters or; c) are charged but their momentum is too small and they get swept out of the acceptance by the magnet. While a part of these backgrounds is mostly irreducible, and can only be suppressed through the techniques described above, powerful isolation techniques can also provide discrimination. This would involve, {\it{e.g.}}, the use of tracks without hits at all the tracking layers or the reinterpretation of the data from the calorimeters.

Semileptonic $b$ decays involving a neutrino provide a good benchmark for determining the expected trigger efficiencies for searches of this kind. The \lhcb trigger was designed to fire also on partially reconstructed decays of $b$ hadrons. Therefore, high trigger efficiencies are expected in this type of channels, provided at least two charged tracks that do not originate from the primary vertex form a good displaced vertex.
Specific examples of trigger efficiencies at LHCb so far for this kind of channels are given by Ref.~\cite{Likhomanenko:2015aba}, which obtains HLT efficiencies for several semileptonic $b$ decays of around 50\,\%. An additional benchmark comes from Ref.~\cite{Aaij:2017deq}, measuring an efficiency during Run~1 of $\sim50$\% for $B\to \tau +X$ decays, with $\tau$ decaying in turn to at least 3 charged pions and the whole decay chain passing some additional selection cuts. 
As for the future trigger, given its structure will be close to that of Runs 1 and 2 (but with the big advantage of having no hardware level trigger, as discussed in Sec.~\ref{sec:trigger}), the expected efficiencies will be better. In general, final states including charged tracks with impact parameters well above 0.1 $mm$ and \pt above 1 \gevc guarantee high trigger efficiencies. For instance, Ref.~\cite{LHCb-PUB-2017-006} shows \hltone efficiencies close to $\sim50$\% to several semileptonic $B$ decays loosely selected in simulation and Ref.~\cite{LHCb-PUB-2019-013} quotes \hlttwo efficiencies in the $\sim10-50$\% range for different decays where one of the final state particles is not reconstructed. It should be highlighted that developing more exclusive selections, incorporating the $\pt^{\rm miss}$ proxies or isolation criteria, could increase the \hlttwo efficiencies to values closer to 100\,\%.


\subsubsection{Displaced hadronic decays}
\label{exp:hadronicdecays}

The LHCb experiment has published a search (using Run~1 data) for long-lived particles decaying hadronically~\cite{Aaij:2017mic}, where the signature consists of a single displaced vertex decaying into two heavy quarks which later hadronise into jets. The di-jet invariant mass range between 25 and 50\gevcc was explored. In this search, displaced vertices are reconstructed using tracks selected with a minimum impact parameter requirement, and then required to be highly displaced (the vertex displacement is required to be more than 0.4\mm in the transverse plane). A minimum track multiplicity and an invariant mass higher than the typical $B$-meson masses, is also imposed. 
A more efficient and inclusive selection strategy would be to use displaced tracks to reconstruct a displaced jet with a cone radius of ${\Delta}R<0.5$, employing the anti-$k_t$ clustering algorithm. This was tested during the 2018 data-taking period as a proof-of-concept, and it is described as follows: after the displaced jet is reconstructed, loose minimum displacement and minimum \pt requirements, $\pt>10$-$15\gevc$, are imposed. Then, a vertex reconstruction algorithm is ran using tracks associated to the jet, reconstructing in this way all displaced vertices inside the jet cone. Finally, a certain number of both displaced vertices and children particles of the displaced vertices are required: these numbers are chosen depending on the mass of the jet and its flavour. 
With this technique, a number of displaced jets can be required as well to reduce QCD background as much as possible, especially for very low lifetimes. However, even if the proof-of-concept showed promising results, for Run~3 it would be needed to improve the selection requirements by performing dedicated optimisation studies. 

On top of this, for long-lived particle masses between 5 and 10\gevcc, the two jets in the final state would be identified as a single one, hence new tools need to be developed to distinguish these merged jets from QCD background jets as well as to disentangle the substructure of these objects~\cite{Marzani:2017mva, Larkoski:2014wba} (more details can be found in \ref{exp:jets}). Finally, additional improvements to jet reconstruction algorithms are considered, {\it{i.e}}., the use of machine learning (ML) tasks \cite{Komiske:2017ubm} to identify the higher pile-up contributions expected during Run~3 and beyond.

The LHCb collaboration has also published a search for long-lived particles decaying semileptonically (one high \pt muon, well isolated, and associated to a displaced vertex consisting of two heavy quarks) \cite{Aaij:2016xmb}. Prospects for both searches described above, for Run~3 and beyond can be found in~\cite{LHCb-CONF-2018-006}.
A search of long-lived particles decaying to lighter quarks that are fully reconstructed as charged hadrons rather than jets, has been recently proposed \cite{CidVidal:2019urm}. In principle, analyses involving hadronic decays in LHCb make use of the so-called {\it{topological triggers}}, standard \hlttwo triggers designed for $B$-meson decays rely on the identification of a displaced vertex formed by at least two tracks~\cite{Williams:2011aza, Likhomanenko:2015aba}. Unfortunately, they include requirements on the invariant mass and the flight distance of the multi-body object which are tailored for $B$-mesons and can be inefficient for long-lived particles with larger mass and lifetime. This analysis can benefit from a dedicated trigger selection such as an inclusive selection for multiple displaced vertices in the event, decaying hadronically. High \pt tracks identified as kaons or pions can be required to form a good displaced vertex while their combined momentum can be required to point back to the PV.  
The amount of remaining hadronic background needs to be studied. If needed, it can be reduced by requiring more than one displaced vertex with these characteristics in the decay, a feature found in models such as~\cite{Pierce:2017taw}. For these models, another possibility would be to develop a trigger selection similar to the {\it{i.e. topological triggers}}, but optimised for short lifetimes, {i.e.} 2 ps. 
On the other hand, if the stealth particles come from the decay of a SM Higgs such as in~\cite{CidVidal:2019urm}, a tighter \pt requirement can be used to reduce QCD background. 
However, a significant amount of background would be still present: contamination from $K_S^0$ and $D^0$ decays (produced in the PV or from a $B$-meson decay), as well as from $\Lambda$-baryon decays into a $pK$ or $p\pi$, where the proton is mis-identified as a hadron by the RICH system. 
%


\subsubsection{Pairs of muons}\label{sec:LHCbmuon}

Searches for new particles decaying into a pair of oppositely-charged leptons are a crucial part of the LHCb physics programme, leading to many publications in the past years~\cite{Aaij:2016qsm, Aaij:2015tna, Aaij:2013lla,Aaij:2017rft,Aaij:2018xpt, Aaij:2019bvg}.
Previous di-lepton searches at LHCb are based on pairs of muons rather than electrons or tau leptons, for two main reasons: LHCb is very good at identifying muons with very low \pt and can do so at the first trigger level, and low-energy muons momenta are precisely measured by the LHCb spectrometer. 

The muon hardware trigger used during the LHC Run~1 and Run~2 required either one muon with $\pt(\mu) \gtrsim 1.8\gevc$ or a muon pair with $\pt(\mu^+)\pt(\mu^-) \gtrsim (1.5\gevc)^2$ \cite{Aaij:2018jht}. These requirements became an important bottleneck for light muon pairs and will be removed in the upcoming LHCb upgrade, as seen in Sec.~\ref{sec:trigger}.
The kinematic selection at the software stage of the trigger depends on the event topology considered. The background rate is larger for the inclusive prompt di-muon signature due to the Drell-Yan continuum, semi-leptonic heavy-flavour hadron decays, and misidentified pions. Yet, in Run~2 LHCb 
recorded prompt muon pairs with $\pt(\mu)>0.5\gevc$, $p(\mu)>10\gevc$, $\pt(\mu^+\mu^-)>1\gevc$ and $\pt(\mu^+)\pt(\mu^-)>(1.0\gevc)^2$, while no requirement in the minimum di-muon invariant mass is imposed~\cite{Aaij:2019bvg}.
Thresholds this low were reached by precisely identifying and reconstructing muon candidates in real time, thanks to the techniques pioneered by LHCb, such as the {\it{Tesla}} application.
Indeed, online analysis allows to drastically reduce the mis-identified background component and, as we have seen, to discard most of the lower-level information, leading to a dramatic reduction of data rate to be recorded.

Note that, contrary to other experiments, the muon system at LHCb has no magnetic field, so that the momentum of muons is measured as for any other charged track, i.e., through the trackers located upstream of the detector. More specifically, the momenta of muons produced in the VELO and leaving hits in all tracking stations (reconstructed as {\it{Long tracks}}, as explained in Sec.~\ref{sec:off_and_jets}) are measured with a resolution of about 5 per mille for $p(\mu)< 20\gevc$ (rising to about 8 per mille for $p(\mu)\sim 100 \gevc$)~\cite{LHCb-DP-2014-002}. Despite the much higher multiplicity, slightly better momentum resolution is expected for the tracking system of the upgraded LHCb detector~\cite{LHCb-DP-2014-001}.

Note that, depending on the signature that is being reconstructed, the dimuon mass resolution can be improved by imposing that muons come from the same vertex, or by mass constraints on known narrow resonances. For example, in the supplementary material from Ref.~\cite{Aaij:2016qsm}, it is shown that $\sigma_m/m$ of $m(\mu\mu)$ from $B^+\to K^+\chi(\mu^+\mu^-)$ is at the level of 1 per mille in the region of large $m(\mu\mu)$. This is thanks to a kinematic constraint that forces $m(K^+\chi(\mu^+\mu^-))$ to match the known $B^+$ mass.

Muon tracks that leave no hits in the VELO but interact with the tracking stations before and after the magnet ({\it{Downstream tracks}}, see again Sec.~\ref{sec:off_and_jets}) have a resolution about two times worse and are much harder to trigger and reconstruct online. Currently, they are not used in published LHCb di-lepton searches, but, as introduced above, there are plans for the future, especially as the interest on long-lived particles increases.

Concerning the backgrounds, these depend pretty much on the properties of particle decay that we one wants to reconstruct, such as invariant mass or lifetime. For prompt decays, Drell-Yan production becomes an irreducible background, usually very hard to distinguish from signal (other than through the invariant mass or any additional product produced accompanying the signal). For displaced decays, heavy QCD, such as $b\bar{b}$ production, is very abundant, since $B$ mesons decay semi-leptonically very often to muons. Other than these, hadrons misidentified as muons can be relevant, specially if low momentum muons ($p\lesssim 10$ \gev) are not excluded from the kinematic selection. These considerations are obviously dependent on the mass of the decaying mother, but generally hold in the regime of more interest at LHCb, $\lesssim 20 $ \gevcc. 


\subsubsection{Pairs of electrons}

The momentum measurement of electron tracks is affected by significant losses of energy before the magnet due to the emission of bremsstrahlung radiation. These result in a dielectron mass resolution that is significantly worse than that of dimuons.
Furthermore, in Run~1 and Run~2, selections on final states with di-electrons and di-photons were limited by stringent requirements on deposits in the electromagnetic calorimeter (ECAL), at the hardware first trigger level. 
An inclusive di-electron software-level trigger was implemented during Run~2, which used electromagnetic clusters identified at the L0 trigger stage, but now linked against HLT1 tracks.
Triggers similar to the inclusive di-muon selections used to search for low masses resonances, with \pt thresholds as low as $1~\gev$, were introduced. Additional inclusive displaced triggers, as well as exclusive $\pi^0 \to e^+ e^- \gamma$ and $\eta \to e^+ e^- \gamma$ triggers were also implemented. While no analyses have yet been published using these selections, initial studies indicate the efficacy of these triggers. 
In addition, searches for di-photons with conversions benefit from the selections already available for displaced tracks, which are designed to select di-electron tracks from conversions. 

For Run~3, with the removal of the L0 level, a direct selection based on information from the tracking system is possible, both at \hltone and \hlttwo, allowing to select on more discriminate information like displacement, vertex quality and others, in comparison to just ECAL information. In addition, at \hltone level, full reconstruction of clusters in the ECAL should be possible. The information obtained from the full ECAL reconstruction can in turn be matched with the information from the tracking systems, opening up the identification of electrons within the set of available charged tracks at \hltone, which was only possible during Run~2 using the more rudimentary L0 electromagnetic clusters. 
In addition, the use of partial track-reconstruction for di-electrons can considerably decrease material-interaction related inefficiencies, which can be implemented at both \hltone and \hlttwo. 

The aforementioned additions in the trigger system of LHCb open up the selection of very soft (di-)electrons and converted (di-)photons that are predicted BSM scenarios like dark photons, axions, and dark showers. 
With the full clustering of the ECAL, {\it{bremsstrahlung}} recovery will also be available at the \hltone, and will help increase momentum resolution and improve sensitivity in resonance searches. For exclusive final states with photons and electrons, this can significantly reduce trigger rates.


\subsubsection{Pairs of taus}
\label{sec:tau}
LHCb has performed several analyses of decays involving tau leptons in the final state. Most of them consider only a single $\tau$ produced from the semileptonic decay of a $B$-meson, where the $\tau\to\pi\pi\pi\nu$ decay mode is considered in most of the cases~\cite{Aaij:2017deq}. 
Fewer analyses use pairs of $\tau$ leptons. In the search for $B_s^0\to\tau\tau$ decays both $\tau$ leptons are reconstructed hadronically~\cite{Aaij:2017xqt} while in the $Z^0\to\tau\tau$ analysis both leptonic and hadronic $\tau$ decays were used~\cite{Aaij:2624023}. 

The $\tau$ leptons can be reconstructed either in their hadronic three-prong decay $\tau\to\pi\pi\pi\nu$ or the leptonic decay $\tau\to{l}\nu\nu$. 
The leptonic $\tau$ decay has a larger branching fraction but does not allow to reconstruct the $\tau$ decay vertex due to the presence of two neutrinos in the final state. 
For hadronic $\tau$ decays, combinatorial background can be reduced by estimating the $\tau$ invariant mass using the so-called {\it{corrected mass}}, $\sqrt{m^2(\pi\pi\pi) + \pt^2(\pi\pi\pi)} + \pt(\pi\pi\pi)$, where the \pt is computed with respect to the $\tau$ direction of flight\footnote{so this quantity necessarily requires knowing the $\tau$ decay vertex.}. This technique is widely used in LHCb analyses involving neutrinos and can only be used if the $\tau$ direction of flight is known. 
In addition, the hadronic three-prong decay mode consists of three reconstructible tracks in the final state in comparison with the single-track leptonic decay mode, which leads to a lower selection efficiency.  

Pairs of $\tau$ leptons can be formed using either modes, following the strategy in Ref.~\cite{Aaij:2624023}, namely:

\begin{enumerate}
    \item Purely hadronic decays: $\tau_1\to\pi\pi\pi\nu$ and $\tau_2\to\pi\pi\pi\nu$,
    \item Purely leptonic decay with different flavours: $\tau_1\to{l_1}\nu\nu$ and $\tau_2\to{l_2}\nu\nu$,
    \item Both hadronic and leptonic decays: $\tau_1\to\pi\pi\pi\nu$ and $\tau_2\to{l}\nu\nu$.
\end{enumerate}

A large QCD component is expected to pollute the $\tau\tau$ system reconstructed with the first method, however, since both $\tau$ leptons are reconstructed, an isolation technique can be used to reduce it. 
With the second method, requiring leptons with different flavour allows to efficiently reduce background contributions from SM di-leptons. 
However, depending on the $\tau\tau$ invariant mass and lifetime, the optimal reconstruction strategy in terms of signal efficiency and background cancellation can be a combination of these three methods. 

There is also the possibility that either both $\tau$ leptons are produced from a displaced decay vertex, or that only one of them is produced from a prompt decay vertex and associated with an invisible particle, which then flies and decays into some other objects plus the second $\tau$ lepton. For the displaced $\tau$ lepton(s), additional requirements on the displacement of this object can be imposed, potentially having a sizable impact in the selection: the more displaced, the less prompt QCD background will be present. However, the {\it{corrected mass}} technique for hadronic $\tau$ decays can not be applied, since the direction of flight of the $\tau$ lepton would not be known anymore.


\subsubsection{Pairs of photons}

LHCb can reconstruct the energy of photons when they reach the ECAL. Their identification is discussed in Sec.~\ref{sec:PID}. The position of the photon energy cluster in the calorimeter provides the photon direction if it is assumed to have originated from the $pp$ interaction point. The momentum measurement is more accurate if photons convert in the material detector to an \epem pair which are reconstructed in the tracker system. Furthermore, converted photons have a well defined trajectory, which is useful to detect those originating from displaced objects. However, only a fraction of roughly 5\,\% of the detected photons are reconstructed as conversions \cite{LHCb-DP-2014-002}.

\begin{figure}[h]
    \centering
    \includegraphics[width=0.7\textwidth]{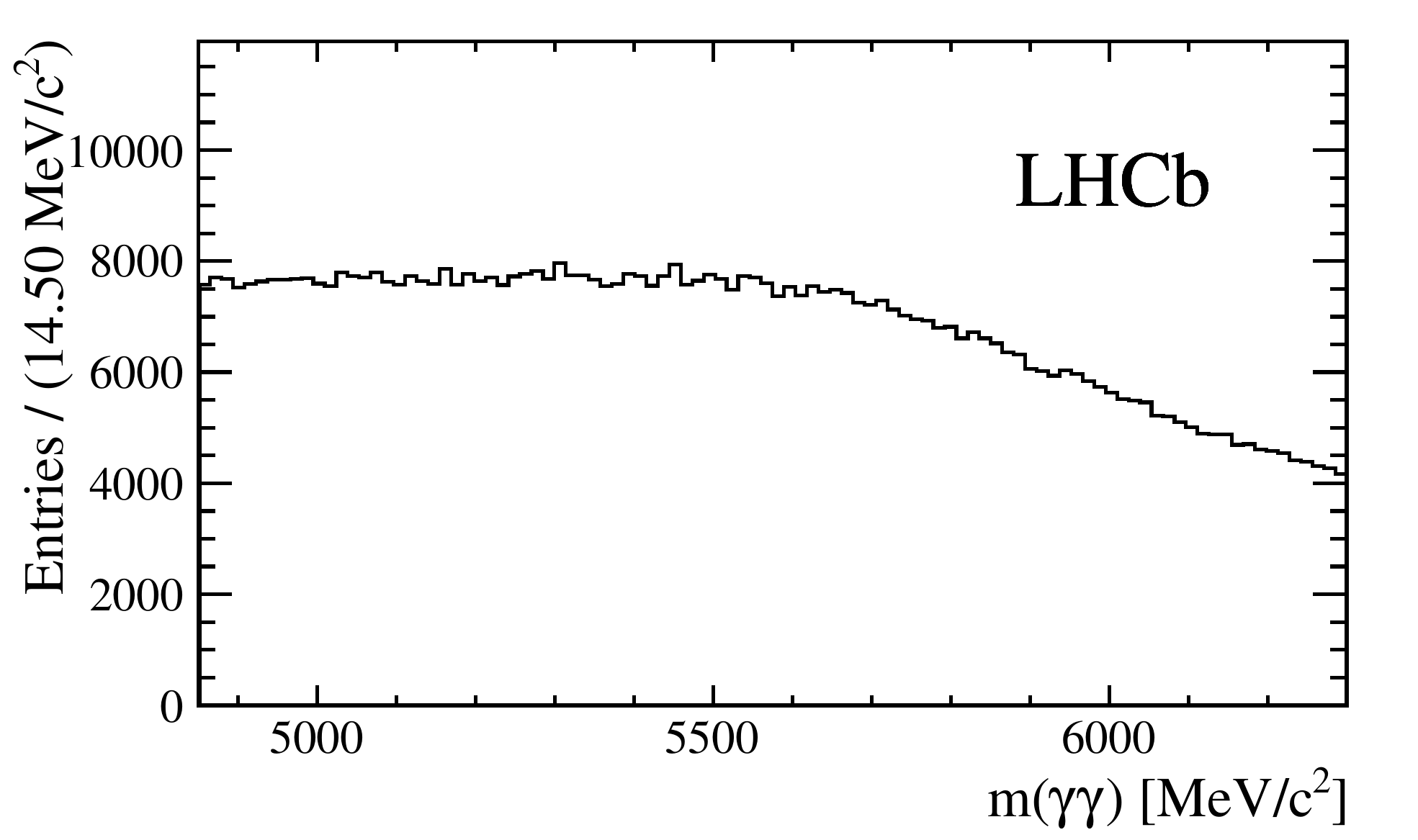}
    \caption{Figure taken from \cite{Benson:2314368}. Mass profile of unconverted photon selection based on 80 pb$^{-1}$ of 2015 LHCb data}
    \label{fig:B2gg_Unconv}
\end{figure}

Dedicated di-photon trigger selections for Run~2 were first introduced in 2015 for the search of the rare decay $B_s^0 \to \gamma \gamma$.  The reconstruction of unconverted photons at \hltone was made possible by the introduction of combinations of ECAL clusters coming from L0, given the fact that no dedicated ECAL reconstruction was present at \hltone level. The shape and size of these clusters is fixed to fit an array of 2 x 2 cells. Due to this topology and to saturation of the ECAL electronics, these objects do not allow for effective cuts for  $E_T$ values above 10\gev. This is because these ECAL objects saturate over this threshold. Figure \ref{fig:B2gg_Unconv} shows the invariant mass spectrum of unconverted di-photons selected by this trigger sequence. The composition of the background present in this plot mainly consists of \piz mesons and photons that do not proceed from heavy quark processes. 
For Run~3, a dedicated reconstruction sequence will allow for more efficient requirements on high $E_T$ photons. 

Displaced di-photon searches that include converted photons benefit from existing displaced track trigger selections. Moreover, during 2018 data taking, a trigger to select both converted and unconverted di-photons based on a Neural Network was introduced. This is a novel technique which was never used before for this kind of multi-body decays at LHCb~\cite{CidCasaisPhotonTrigger}. %
A similar trigger strategy could be adopted in Run~3 by introducing selection features to account for different conditions such as higher multiplicity and pile-up. 


\subsubsection{Prompt jets (includes b/c and light)}
\label{exp:jets}

Many new physics models predict new particles that couple with quark and gluons, allowing them to decay into a pair of jets. For this reason, prompt di-jet searches have been extensively pursued at hadron colliders.
LHCb has already demonstrated its capability in reconstructing di-jet resonances, by measuring the $Z \to b \bar{b}$ cross section \cite{LHCb-PAPER-2017-024}, and in principle it could push down the lower mass limit for this kind of searches at the LHC.

At LHCb, di-jet searches have been performed exclusively without any requirements on the jet substructure, selecting either all possible jet flavours, or identifying $b$-hadrons inside the jets ($b$-tagging).
The requirement of one or two $b$-tagged jets in the pair could enhance the sensitivity on several new physics signal. The $b$-tagging could also help to reduce the overwhelming QCD background in generic di-jet searches.
ATLAS and CMS explored the di-jet invariant mass region above 450 \gevcc by requiring two resolved jets \cite{Sirunyan:2016iap, Khachatryan:2016ecr, Aaboud:2017yvp, Aaboud:2018fzt, Sirunyan:2018xlo} and down to 325 \gevcc by requiring $b$-tagging \cite{Aaboud:2018tqo, Sirunyan:2018pas}. The lower bound of their searches is mainly limited by the trigger thresholds on the jet transverse momentum. Moreover, by requiring a significant initial and final state radiation at trigger level, CMS pushed down the mass range in the 50-350 \gevcc region in the search for $b\bar{b}$ resonances \cite{Sirunyan:2018ikr}. However the latter method could put constraints on the production mechanism of the resonance.
The requirement of a $c$-hadron signature inside the jets ($c$-tagging) can be motivated by several models, but it has not been experimentally exploited yet in di-jet resonances searches, apart from the direct search of the SM Higgs decay into a pair of $c$-quarks by ATLAS, CMS and LHCb \cite{Sirunyan:2019qia, Aaboud:2018fhh, LHCb-CONF-2016-005}. 

The LHCb detector could be used to search for such resonances, in a complementary phase space with respect to ATLAS and CMS. Thanks to the low pile-up events and its low thresholds trigger system, LHCb could push down the lower mass bound of the di-jet searches.
Moreover, thanks to its excellent vertex reconstruction system, LHCb can efficiently identify $b$- and $c$-jets with a low mis-identification probability for light jets ($i.e.$ jets generated from $u$, $d$ or $s$ quark or gluons).
This is mainly done by reconstructing Secondary Vertices (SV), compatible with $b$- or $c$-hadron decays, inside the jet cone.

The jet reconstruction at LHCb has been introduced in Section \ref{sec:off_and_jets}. LHCb can reconstruct jets with $\pt > 20$ \gevc with an efficiency greater than 90$\%$, and with a \pt resolution between 10$\%$ and 15$\%$.
In the Run~2 data taking a non-prescaled trigger selection that selected pair of jets with $\pt > 17$ \gevc and a reconstructed SV in each jet has been implemented.  
LHCb can therefore search for heavy flavour tagged di-jet resonances starting from an invariant mass of about 40 \gevcc, lower than the 325 \gevcc limit of ATLAS and CMS, obtained by requiring two resolved jets, and lower than the 50 \gevcc limit obtained by CMS requiring extra activity other than the jets.  
At LHCb, the efficiency of reconstructing a SV inside a $b$-jet is of about 65$\%$ while it is of about 25$\%$ for a $c$-jet, with a light jet misidentification in the order of 0.3$\%$.
A further discrimination of jet flavours is achieved by studying the observables related to the SV, like the SV mass and the number of tracks. These are usually used as inputs to multivariate discriminators, and the outputs of these algorithms can be used to reject the backgrounds but preserving most of the signals, improving the significance of the search.
As example, as demonstrated in \cite{LHCb-CONF-2016-006}, it is possible to remove 90$\%$ of $b \bar{b}$ events around the SM Higgs mass, while retaining 62$\%$ of $H \rightarrow c \bar{c}$ events.
Novel techniques should be studied to improve the tagging efficiency, in particular for the identification of the $c$-jets. Techniques that exploits the whole jet substructure and not just the SV properties, could be developed for such purpose, in a similar way of what done by ATLAS and CMS \cite{Asquith:2018igt}.

One of the main experimental challenges of the di-jet searches is the description of the QCD background. It is widely known that Monte Carlo simulations are not able to properly reproduce the di-jet mass distribution observed in data.
For low mass searches at LHCb the di-jet mass distribution is distorted by acceptance and trigger effects, that make the description of the region between 40 and 80 \gevcc even more challenging.
Several data-driven techniques can be used to determine the QCD di-jet mass shape. Control regions where the signal is not expected can be defined ($i.e.$ by asking a different tagging decision for the two jets as in \cite{Sirunyan:2018ikr}), and the QCD model obtained in the control region can be translated in the signal one by the means of transverse functions. Alternatively, a sliding-window fit technique could be tested in the LHCb environment \cite{Aaboud:2018fzt}.

At masses below 40 \gevcc the new particles may produce two collimated jets that merge together forming a fat jet \cite{Plehn:2009rk}. Experimentally fat jets can be reconstructed by using a large radius parameter in the clustering algorithm, typically between 1 and 2. Identification algorithms that exploits the fat jet substructure could be employed to distinguish between $bb$, $cc$ and backgrounds, like light flavour and single $b$- and $c$-jets. At LHCb such techniques have not been explored yet. Given the trigger with low \pt thresholds, the fat jet reconstruction at LHCb could be the key to access the mass region between 10 and 40 \gevcc in the search for di-jet resonances.


\subsubsection{More convoluted/blue sky ideas}

Beyond the more {\it standard} final states presented so far, a number of other possibilities to do Stealth physics at LHCb could be considered. For these, no specific studies exist so far, so we give more qualitative and speculative ideas. They concern not only the actual detector, but also future upgrades, for which some of the characteristics are not known in detail yet. Note that, if enough interest exists from the theoretical community, detailed studies could be done to be more quantitative about the actual LHCb reach to these final states. Examples of this type of speculative ideas are given next:
\begin{itemize}
    \item  The RICH detectors have the capability of detecting any charged long-lived particle that traverses them. This could also include fractionally charged objects. Note that the number of Cherenkov photons produced is proportional to $Q^2$, where $Q$ is the electric charge. This means, for instance, that a fractionally charged object with $Q=e/3$ would produce a fraction of $1/9$ photons with respect to a standard particle with $Q=e$, posing a detecting challenge. Also, the efficiency of the rest of the tracking layers would be compromised by this smaller charge, and the bending effect of the magnet would be different too. 
    
    \item New Charged Massive Stable Particles (CMSP) would behave as a very slow ionising particles passing through the RICH stations. All SM quasi-stable charged particles (the heaviest being the proton) emit Cherenkov light in RICH1 (RICH2) if their momentum is larger than 18 (30)\gev. A CMSP with the same momentum but higher mass could be identified by the absence of detected Cherenkov light. This detection principle was already used at LHCb in a proof-of-concept search \cite{LHCb-PAPER-2015-002}. 
    \item A new set of magnet stations \cite{LHCb-PII-EoI} has been proposed to improve the reconstruction of low-momentum particles, with the goal of improving the reconstruction efficiency of various decays relevant for flavour physics. However, these stations could also be useful to help detecting disappearing tracks. That signature usually corresponds to a BSM charged particle that decays to a BSM neutral undetectable particle with a similar mass, accompanied by a SM particle too soft to be detected. However, the magnet stations could be able to detect this SM particle, which would significantly help to reduce the SM backgrounds.
    
    \item One last possibility to consider would be that of soft unclustered energy patterns (SUEPs~\cite{Knapen:2016hky}) as predicted in some models of Neutral Naturalness (see Sec.~\ref{sec:neutralnatura}  where the calorimeter occupancy could be used as source of discrimination. 
    
\end{itemize}


\section{Review of Stealth results at LHCb}
\label{sec:review}

One of the great strengths of the LHCb physics programme is its complementarity to the general purpose detectors at the LHC. Because LHCb is a dedicated $b$-physics experiment, its primary focus is not Stealth physics. However, many of its core $b$-physics measurements are able to indirectly probe Stealth models. Anomalies in a comprehensive set of $b \to s \ell^+ \ell^-$ measurements~\cite{Aaij:2017vbb,Aaij:2014ora,Aaij:2014tfa,Aaij:2020ruw,Aaij:2019wad,Aaij:2021vac}, including both angular coefficients and the ratio between semileptonic decay channels, may indicate possible tensions with lepton universality. Some models which may be able describe these anomalies could also result in Stealth physics signatures. LHCb also provides a host of supporting measurements to Stealth and DM searches, including measurements of $D$-meson production at low proton momentum fraction values~\cite{Aaij:2013mga,Aaij:2015bpa}, relevant for neutrino telescopes~\cite{Gauld:2015yia}, and the cross-section measurement of antiprotons in proton-helium collisions~\cite{Aaij:2018svt}, critical for cosmic ray spectrometer analysis~\cite{Giesen:2015ufa}. 

LHCb has also performed world leading direct searches for a variety of Stealth signals, making the best of its unique features. This includes both long-lived particles and light new resonances. The most important among these searches are summarized next.

\subsection{Jets from Hidden Valleys and MSSM}

A large class of models from hidden valley scenarios can produce Stealth signatures of long-lived particles. The hidden valley sector can exist at a low mass scale, but only weakly couple with the SM. This coupling may arise from new massive particles beyond the reach of current experimental searches, or via the Higgs boson, considering the current lower limit on invisible Higgs boson decays is $\approx 20\,\%$. Consequently, LHCb has performed a number of searches for long-lived hidden valley particles produced from SM Higgs boson decays~\cite{Aaij:2014nma, Aaij:2017mic}. The signature is $H \to \pi_V[\to q \bar{q}] \pi_V$ where one hidden valley pion, $\pi_V$, decays within the LHCb fiducial volume. Here, $\pi_V$ is a proxy for the lightest state of the hidden sector. Requiring a single hidden valley pion, rather than both, increases the signal acceptance by a factor of $\approx 5$ or more, without significantly increasing background.

\begin{figure}
    \centering
    \includegraphics[width=0.6\columnwidth]{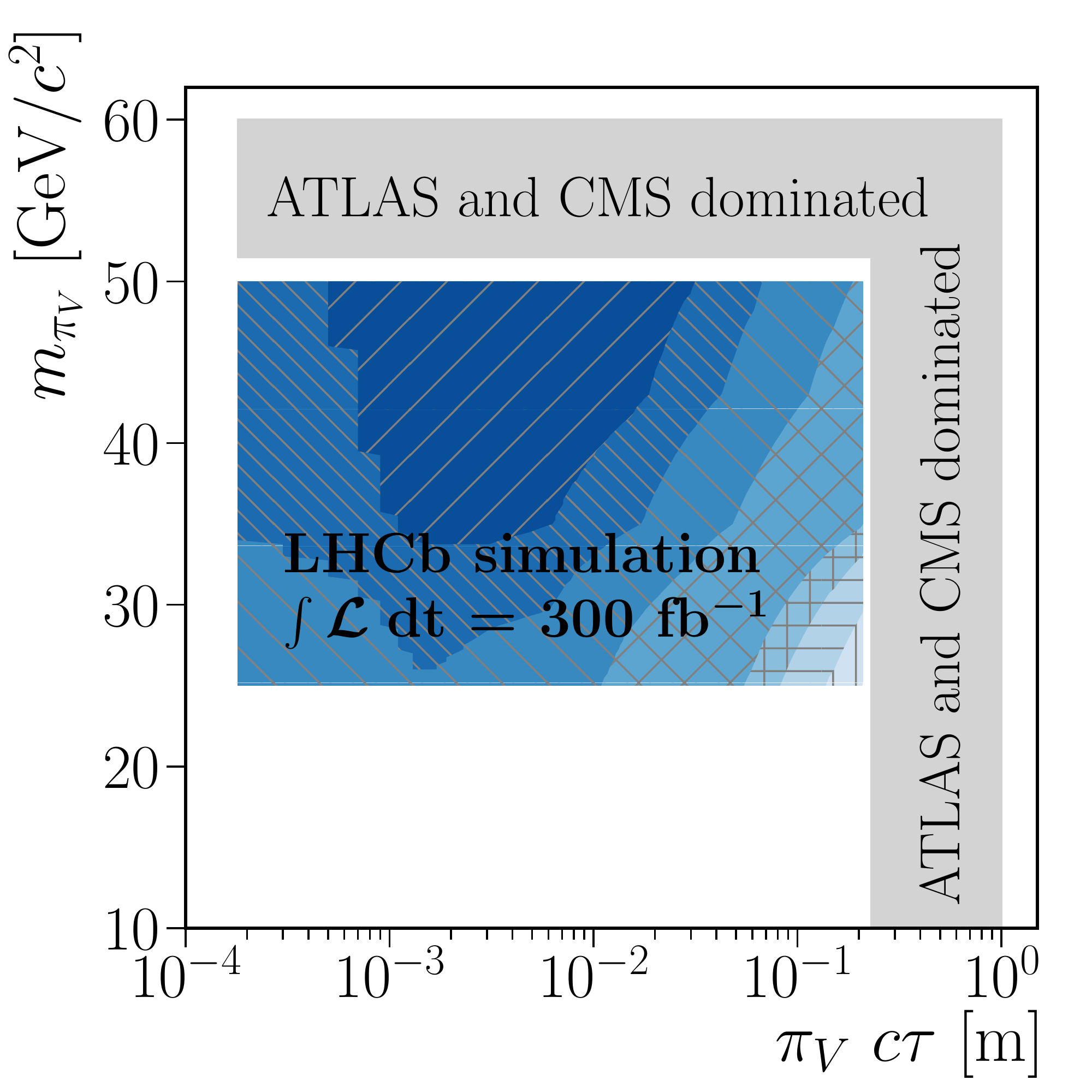} \begin{minipage}{0.28\textwidth}
    \vspace{-10cm}
    \includegraphics[width=\textwidth]{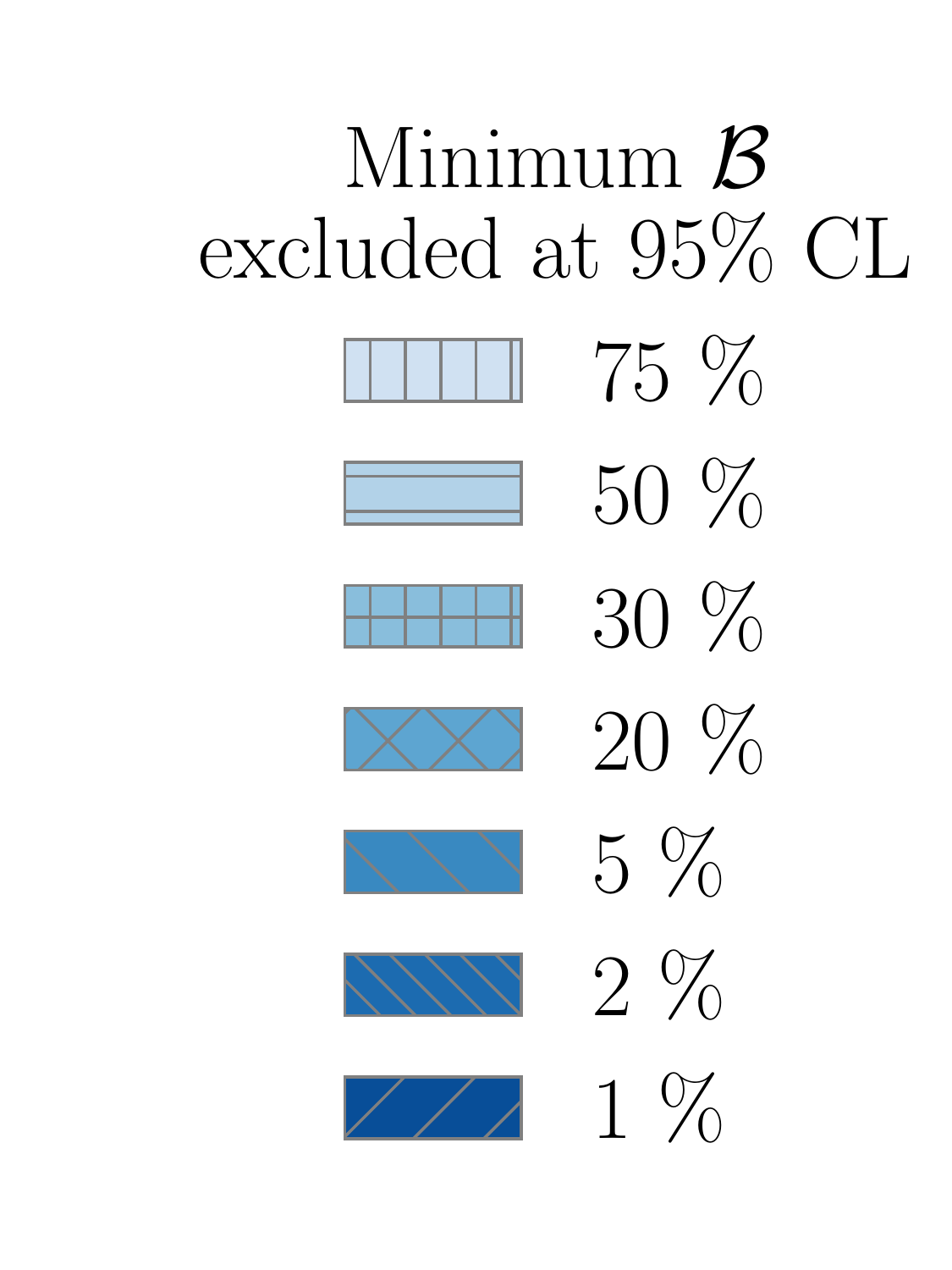} \end{minipage}
    \caption{High luminosity LHC projections for LHCb sensitivity to hidden valley pions, $\pi_V$ decaying into two jets and a single displaced vertex as a function of the $\pi_V$ mass and lifetime.~\cite{LHCb-CONF-2018-006}}
    \label{fig:lhcb_hv}
\end{figure}

The hidden valley pion decays into a light quark/antiquark pair which then hadronises into jets of SM hadrons. The LHCb analyses search for two displaced jets coming from a common displaced vertex within the LHCb acceptance, and have sensitivity for hidden valley pions with lifetimes between $1$ and $100\,\text{ps}$ and masses within the range of $20$ to $50\gevcc$. A projection of these searches to the expected coverage by LHCb with a full $300\invfb$ of data from the high luminosity LHC~\cite{LHCb-CONF-2018-006} is shown in figure~\ref{fig:lhcb_hv} where the regions covered by ATLAS and CMS are indicated in grey. As expected, LHCb has world-leading sensitivity in the low mass and lifetime parameter space of the hidden valley pion. This is due to the flexible trigger and excellent displaced vertex resolution of the detector.

A similar signature of two jets from a single displaced vertex, but now with an associated muon, has been used to search for an MSSM neutralino benchmark scenario, $\tilde{\chi}^0 \to qq\mu$, where the neutralino decays into two quarks a muon~\cite{Aaij:2016xmb}. Unlike the hidden valley analysis which required two separate jets, this analysis only requires a high multiplicity displaced vertex with an associated muon. The lifetime and mass parameter space coverage for this search is similar to the di-jet hidden valley search.

LHCb has also performed a search for two displaced jets originating from separate displaced vertices~\cite{Aaij:2016isa}, in contrast to a single displaced vertex. The chosen benchmark signal is neutralino pair production from a Higgs-like particle decay, $h \to \tilde{\chi}^0[\to qqq] \tilde{\chi}^0[\to qqq]$ where each neutralino decays through a baryon number violating process into a system of three quarks. These two decays then are detectable as jets associated with two independent displaced vertices. Similar to the single displaced vertex search, this search has sensitivity to neutralino lifetimes between $5$ and $100\,\textrm{ps}$ and masses within the range of $20$ to $60\gevcc$.

Another common signature of hidden valley models is the introduction of a confined hidden sector, similar to QCD, where jets of hidden sector hadrons are produced, which then decay into SM final states. This results in a unique experimental signature with a high multiplicity of displaced vertices or particles, rather than just the one or two displaced vertices of the previous searches. Reconstructing such a final state can be experimentally challenging, as most reconstruction algorithms make strong assumptions about particle production originating from the central collision point of an event, and the visible particles produced from these hidden valley hadron decays can have very low transverse momentum.

LHCb, however, has performed a search for such models using low mass di-muons, and the benchmark scenario $Z_V \to q_V \bar{q}_V$ where a heavy hidden valley boson $Z_V$ kinetically mixes with the photon~\cite{Aaij:2020ikh}. The neutral hidden dark vector meson, $\rho_V^0$, produced from the hidden quark/antiquark, then decays into a muon pair. Setting an average hidden valley hadron multiplicity benchmark of $10$, LHCb is able to constrain the kinetic mixing of the $Z_V$ with the photon to be less than unity in the $\rho_V^0$ mass window of $20$ to $70~\text{GeV}$ over a wide range of $\rho_V^0$ lifetimes.

\subsection{Light Dark Sectors}

The hidden valley signals of the previous section typically require production through some heavy resonance such as the Higgs boson. This is oftentimes experimentally favourable as the final visible state may have large transverse momentum, allowing for more efficient triggering. However, since LHCb is specifically designed for $b$-physics, it can also search for signals with very low transverse momentum. One such signal is light new resonances produced in the decays of $B$-hadrons. LHCb has set world-leading limits on heavy neutral leptons, $\nu_H$, from the lepton number violation decay $B^- \to \mu^- \nu_H[\to \mu^- \pi^+]$~\cite{Aaij:2014aba}. Upper limits are set on the branching fraction of this $B^-$ decay for $\nu_H$ masses between the muon-pion threshold and $5~\text{GeV}$, and lifetimes of $1$ to $1000~\text{ps}$. Heavier neutral leptons have also been searched in prompt $W^+\to\mu^+\nu_H(\mu^\pm{\rm jet})$ decays~\cite{Aaij:2020ovh} and displaced $\nu_H\to e^\pm \mu^\mp\nu$ decays~\cite{Aaij:2020iew}.

Similarly, di-muon resonance searches have been performed by LHCb in the decays $B^0 \to K^{*0} \phi[\to \mu^+ \mu^-]$ and $B^- \to K^- \phi[\to \mu^+ \mu^-]$. Here, $\phi$ is a generic new scalar particle which mixes with an off-shell Higgs boson~\cite{Aaij:2016qsm}. Limits have been set for $\phi$ lifetimes between $10^{-1}$ and $10^3~\text{ps}$ and masses between the di-muon threshold and $5~\text{GeV}$. Model specific limits are set for an inflaton benchmark scenario~\cite{Bezrukov:2009yw}, closing off almost all parameter space below masses of $5~\text{GeV}$.

\begin{figure}
    \centering
    \includegraphics[width=0.9\columnwidth]{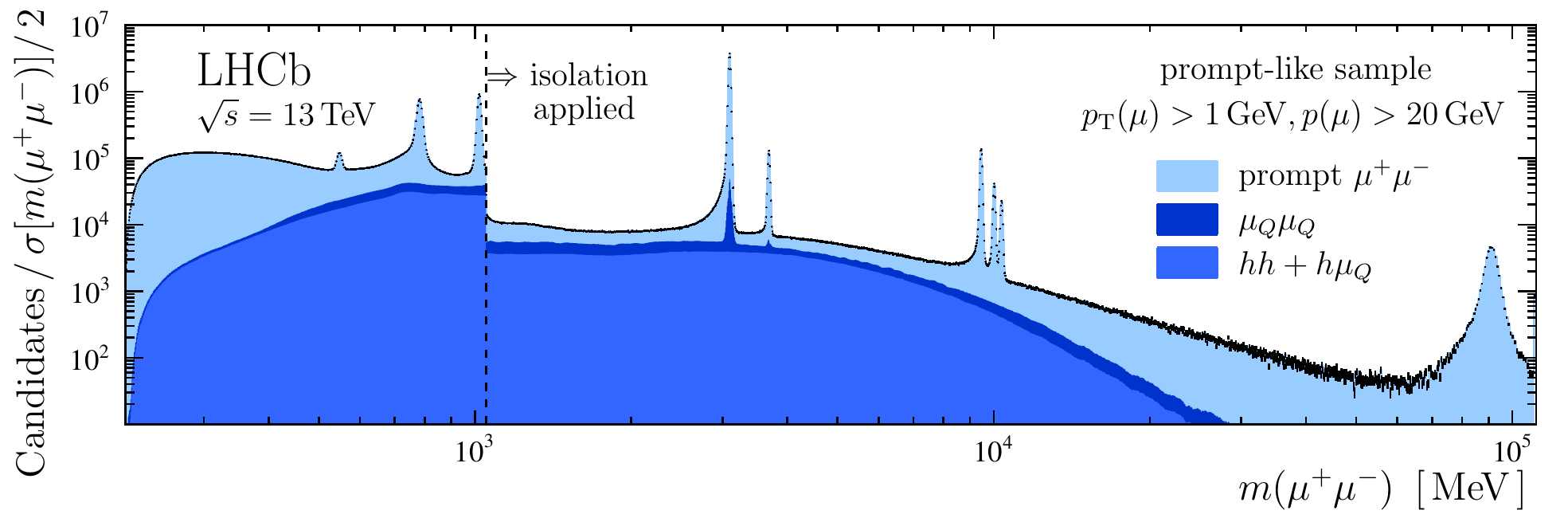} 
     \caption{Inclusive di-muon spectrum collected during 2016 by the LHCb detector and fully reconstructed with the high level trigger.~\cite{Aaij:2017rft}}
    \label{fig:lhcb_dimuon}
\end{figure}

Inclusive di-muon searches for light resonances have also been performed by LHCb, relying upon its flexible trigger. In figure~\ref{fig:lhcb_dimuon} the di-muon spectrum collected by LHCb during 2016 using a prototype trigger where the raw detector data is discarded, and only reconstructed physics objects from the trigger are kept~\cite{Aaij:2017rft}. This trigger has allowed unprecedented data samples of di-muon candidates to be collected and resulted in a number of world-leading new physics searches. A similar trigger was in place, with even lower muon momentum requirements, during the remainder of Run~2 in 2017 and 2018.

In the dark photon model, a massive dark photon kinetically mixes with the photon. 
LHCb searched for both prompt and displaced dark photons decaying to di-muons covering new parameter space for dark photon masses between the di-muon threshold and $1~\text{GeV}$ and between $10$ and $60~\text{GeV}$~\cite{Aaij:2017rft, Aaij:2019bvg}.
These results can be recast to arbitrary models with a new vector boson~\cite{Ilten:2018crw}. 
The LHCb di-muon dataset has also been used to search for new scalars produced promptly in $pp$ collisions~\cite{Aaij:2020ikh}, including the difficult region close to the $\Upsilon(nS)$ resonances~\cite{Aaij:2018xpt}.
Furthermore, LHCb searched for dimuon resonances with minimal model assumptions, including resonances that decay promptly or displaced from the proton-proton collision as well as produced in association with a $b$-jet~\cite{Aaij:2020ikh}. Displaced di-muon resonances were either required to point back to the $pp$ collision vertex or not\cite{Aaij:2020ikh}.


\section{Theoretical perspective}
\label{sec:theory}

While traditionally the search for Physics Beyond the SM is focused on probing phenomena at higher and higher energies, it is plausible that New Physics lies within the energy regime of the LHC ($M \lesssim 1 $TeV), but it manifests itself in a ``Stealth'' manner, as we described in section~\ref{sec:Introduction}.

In order to probe such scenarios, LHCb offers three crucial advantages, when compared to its main competitors, the General Purpose Detectors (GPD) ATLAS and CMS. LHCb is uniquely designed to probe the mass regime $m \lesssim 10\gevcc$, and also the displaced regime (eventually both), while also being able to trigger with lower thresholds and pick topologies that are outside of the reach of the current capabilities of the GPD. 

In what follows we present a sample of phenomenological studies where the LHCb experiment can provide the leading signatures. This list is by no means complete, and we are providing it as an appetizing sample of the exciting prospects for Beyond SM searches at the LHCb. We note that these models are supported by strong theoretical motivations, and not fabricated \emph{ex-profeso} to exploit the LHCb capabilities.
For the readers convenience we summarize the main points of each contribution in Table~\ref{tab:theory_overview}.
\begin{table}[]
\resizebox{\textwidth}{!}{
\begin{tabular}{|l|l|l|l|}
\hline
Sec.  & Main signatures   & LHCb helps with... & Example process   \\ \hline
 \ref{sec:zportal} & DV$_{\mu \mu}$                                                                                         & \begin{tabular}[c]{@{}l@{}}DV reco  \\ (vertexing, invariant mass) \end{tabular}                                                                                     & \begin{tabular}[c]{@{}l@{}}$p p \to Z \to q_V q_V $,\\$ q_V \to V^{\rm (LLP)} + X$,\\ $V \to \mu^{+} \mu^{-}$\end{tabular}                                                                  \\ \hline
 \ref{sec:hv_twin} & \begin{tabular}[c]{@{}l@{}}DV$_{\mu^+ \mu^-}$, \\ DV$_{DD}$\end{tabular}                        & \begin{tabular}[c]{@{}l@{}} DV reco \\ (soft hadrons and muons) \end{tabular}                                                                                                                 & \begin{tabular}[c]{@{}l@{}}$p p \to Z/h \to q_V q_V$,\\ $q_V^{\rm (LLP)} \to DD,  \mu^{+} \mu^{-}$\end{tabular}                                                                           \\ \hline
 \ref{sec:maria} & \begin{tabular}[c]{@{}l@{}}Rare B-decays \\ (4-6 $\mu$ in final state)\end{tabular}                    & \begin{tabular}[c]{@{}l@{}}Low $p_T$ threshold for muons\\ Software trigger,\\$\mu$ mass reconstruction\end{tabular}    & \begin{tabular}[c]{@{}l@{}}$B \to K^{+} +n(\mu^+ \mu^-)$,\\ $n=2,3$\end{tabular}  \\ \hline
 \ref{sec:compoALPS} & \begin{tabular}[c]{@{}l@{}}Pseudoscalar ($a$)\\ resonances ($\mu^+ \mu^-, c \bar{c}$)\end{tabular}    & Low mass region                                                                                                         & $p p \to a, a \to \mu^+ \mu^- / c \bar{c}$                                                                                                                                             \\ \hline
 \ref{sec:ruth} & Rare B-decays                                                                                          & \begin{tabular}[c]{@{}l@{}}Belle-II better reach than LHCb \\ due to a larger decay length allowed\end{tabular}         & $B \to K + (1|2) S, S \to f \bar{f}$                  \\ \hline
\ref{sec:coscattering} & \begin{tabular}[c]{@{}l@{}}displaced $\pi^{\pm}, j, l^{\pm}$\\ + invisibles\end{tabular}               & \begin{tabular}[c]{@{}l@{}}Soft particles\\ Veto on SL meson decays\end{tabular}                                        & \begin{tabular}[c]{@{}l@{}}$ p p \to X^{+} X_2^0 $,\\ $X^+ \to l^+ \nu, $ \\ $X_2^0 \to X_1^0 h^*, ( h^* \to b \bar{b}).$\end{tabular}                                                              \\ \hline
 \ref{sec:mesogenesis}   & \begin{tabular}[c]{@{}l@{}}Rare B-decays\\ CPV in B (D) systems\end{tabular}                          & Hadron / Meson PID                                                                                                      & \begin{tabular}[c]{@{}l@{}}$B \to \psi + bar + n M$\\ (M=meson),\\ asymmetries $A_{sl}^d, A_{sl}^s$\end{tabular}                                                                                         \\ \hline
 \ref{sec:hnl1} & \begin{tabular}[c]{@{}l@{}}DV$_{l \nu}$\\ Neutral meson oscillations\end{tabular}                      & \begin{tabular}[c]{@{}l@{}}Vertexing,\\ Single $l(=\mu, \tau)$,\\ low $p_T$ trigger \end{tabular}                     & \begin{tabular}[c]{@{}l@{}}$p p \to W^{\pm } \to l^{\pm} N$,\\ $N \to l^{\mp} l^{\pm} \nu$\end{tabular}                                                                                     \\ \hline
 \ref{sec:hnl2} & Rare B-decays                                                                                          & \begin{tabular}[c]{@{}l@{}}Reconstruction of $B$-meson decays \end{tabular}      & $B \to \mu N, N \to \mu \pi$                   \\ \hline
   \ref{sec:patrick} & \begin{tabular}[c]{@{}l@{}} 4-leptons \\ (one DV$_{ll}$,\\ one prompt pair). \end{tabular}               & \begin{tabular}[c]{@{}l@{}} $l$ invariant masses\\ Vertexing\\ $\mu$ efficiency\end{tabular}                             & \begin{tabular}[c]{@{}l@{}} 
   $p p \to \gamma / Z \to l^+ l^-$,
   \\ $l^+ \to A^{\prime} l^{\prime +} l^{\prime -}$ 
   \\ $ l,l^{\prime} = \mu, e$ (all combinations)
   \end{tabular}          \\ \hline
  \ref{sec:YotamDP} & \begin{tabular}[c]{@{}l@{}}DV$_{\mu \mu}$ \\ or prompt\\ D- decays\end{tabular}                         & \begin{tabular}[c]{@{}l@{}}Particle ID\\ Displacements\end{tabular}                                                      & \begin{tabular}[c]{@{}l@{}}$p p \to X A', A' \to \mu^+ \mu^-$\\ $D^{*} \to D^{0} A^{\prime} $\end{tabular}                                  \\ \hline
  \ref{sec:zuowei} & \begin{tabular}[c]{@{}l@{}} DV$_{\mu \mu}$ \\ Delayed Dark Photon\end{tabular}                           & \begin{tabular}[c]{@{}l@{}}Displaced $\mu$ \\ Time delay measurement \\ (with TORCH) \end{tabular}                        & \begin{tabular}[c]{@{}l@{}}$pp \to Z \to \chi \chi$ \\ $\chi \to A' \chi, A' \to \mu^+ \mu^-$\\ (A' is LLP).\end{tabular}               \\ \hline
  \ref{sec:displacedKaons} & DV$_{KK}$                                                                                              & \begin{tabular}[c]{@{}l@{}}Exclusive hadron \\ reconstruction\\ (vertexing, inv mass, \\ low pile-up, etc)\end{tabular} & \begin{tabular}[c]{@{}l@{}}$ p p \to h \to S S, S\to {\rm had}\,\,{\rm had}$\\ (S is LLP, had=$K$).\end{tabular}                                                                                             \\ \hline
  \ref{sec:lightA} & \begin{tabular}[c]{@{}l@{}}Pseudoscalar \\ ($a$) resonances \\ ($\mu^+ \mu^-, b \bar{b}$)\end{tabular} & Light masses                                                                                                            & $p p \to a, a \to \mu^+ \mu^- / \tau^+ \tau^- / b \bar{b}$                                                                                                                                   \\ \hline
  \ref{sec:flavor:cons:ALP} & \begin{tabular}[c]{@{}l@{}}Rare B-decays \\ (flavor conserving)\end{tabular}                           & PID                                                                                                                     & $B\to Ka$, $a\to 2\gamma,\ell^+\ell^-$,~MET                                                                                                                                                               \\ \hline
  \ref{sec:flavor:NOcons:ALP} & \begin{tabular}[c]{@{}l@{}}Rare B-decays\\ (with FCNCs)\end{tabular}                                    & PID                                                                                                                     & \begin{tabular}[c]{@{}l@{}}$q \to q' a$ decays\\ $b \to s a$,\,\,e.g: $B_s \to \mu+ \mu^- a$\\ $c \to u a$,\,\,e.g: $ D\to \pi^+ \pi^- \nu  \nu$\end{tabular}          \\ \hline
  \ref{sec:TM}   & \begin{tabular}[c]{@{}l@{}}$e^+ e^-$ resonance\\ with $m_{ee} = 2 m_{\mu}$.\end{tabular}               & $\ell$ invariant masses and vertexing                                                                                   & \begin{tabular}[c]{@{}l@{}}$\eta \to \gamma$ TM, \\ TM $\to e^+ e^- / \mu^+ \mu^-$  \end{tabular}                                                    \\ \hline
  \ref{sec:softbomb}  & SUEPs (multiple soft tracks)                                                                            & Low $p_T$ track threshold at VELO                                                                                       & $p p \to Z/h$, $Z \to$ hadrons.                                                                                               \\ \hline
 \ref{sec:quirks}  & Non-standard VELO tracks                                                                            & Forward coverage &  $ p p \to Q \bar{Q}$, Q=quirk.          \\ \hline
\end{tabular}
}
\caption{A summary of the main collider signatures in each theory discussion and what makes LHCb a unique experiment to probe the signals.}
\label{tab:theory_overview}
\end{table}

\subsection{Neutral Naturalness}
\label{sec:neutralnatura}

\subsubsection{Z Portal to a Confining Hidden Sector \footnote{Contributed by Ennio Salvioni.}}
\label{sec:zportal}

The existence of a hidden sector, interacting weakly with the SM, is an important possible manifestation of New Physics. In the broad range of possibilities for such theories, motivations from outstanding problems of the SM provide key guidance. In this respect, the class of models known as Neutral Naturalness (NN)~\cite{Chacko:2005pe,Burdman:2006tz} stand out as they generically predict a {\it confining} hidden sector, or ``hidden valley''\cite{Strassler:2006im}, at a scale comparable to the one of SM QCD, $0.1 \lesssim \Lambda / \mathrm{GeV} \lesssim 10$. The light hidden hadrons often give crucial signatures of NN, which may lead to discovery of the dynamics behind electroweak symmetry breaking.

Therefore, a clear path to progress is to identify representative hidden sectors with distinct phenomenology, in order to set targets for experimental searches. Here we consider a new type of confining hidden sector, where some hidden quarks are light compared to $\Lambda$ and the coupling to the SM proceeds through the (irrelevant) $Z$ and Higgs portals~\cite{Cheng:2019yai}. Originally motivated by supersymmetric NN~\cite{Cheng:2018gvu}, such setup contains in fact rather generic ingredients, and could arise in a variety of BSM contexts.

We assume a hidden sector containing one Dirac fermion $\psi$ in the fundamental representation of hidden color, with mass $m_\psi \ll \Lambda$. $\psi$ is a full singlet under the SM, but couples to the visible sector via the effective Lagrangian~\cite{Cheng:2019yai}
\begin{equation}\label{eq:EFT}
\mathcal{L}_6 = \frac{y_t^2}{2M^2} \Big( i (D_\mu H)^\dagger H\, \overline{\psi}_R \gamma^\mu \psi_R + \mathrm{h.c.}  +  \frac{\hat{\alpha}_s}{12\pi} |H|^2  \hat{G}_{\mu\nu}^a \hat{G}^{a\,\mu\nu} \Big),
\end{equation}
where $M$ is the mass of new, heavy fermion mediators charged under both the SM electroweak interactions and hidden color symmetries, which are expected in several UV scenarios. The first and second operator yield a coupling of $\psi$ to the $Z$ boson and a coupling of $\hat{g}$ to the Higgs boson, respectively, leading to
\begin{equation}  \label{eq:BRs}
\resizebox{.9\hsize}{!}{$
\mathrm{BR}(Z\to \overline{\psi}\psi) \approx 2.2\times 10^{-5} \left( \frac{2\;\mathrm{TeV}}{M} \right)^4 , \;\; \mathrm{BR}(h\to \hat{g} \hat{g}) \approx 1.3 \times 10^{-5} \left( \frac{\hat{\alpha}_s}{0.18}\right)^2 \hspace{-1mm} \left( \frac{2\;\mathrm{TeV}}{M} \right)^4.$}
\end{equation}
At the ($13$~TeV) LHC we have $\sigma_Z \approx 55$~nb and $\sigma_h \approx 49$~pb, hence $Z$ decays to the hidden sector dominate by far the phenomenology. Here we show that LHCb has the best potential to probe the associated signature of $Z$ decay to hidden jets.

\begin{figure}[!tp]
    \centering
    \raisebox{0.065\height}{\includegraphics[width=0.43\textwidth]{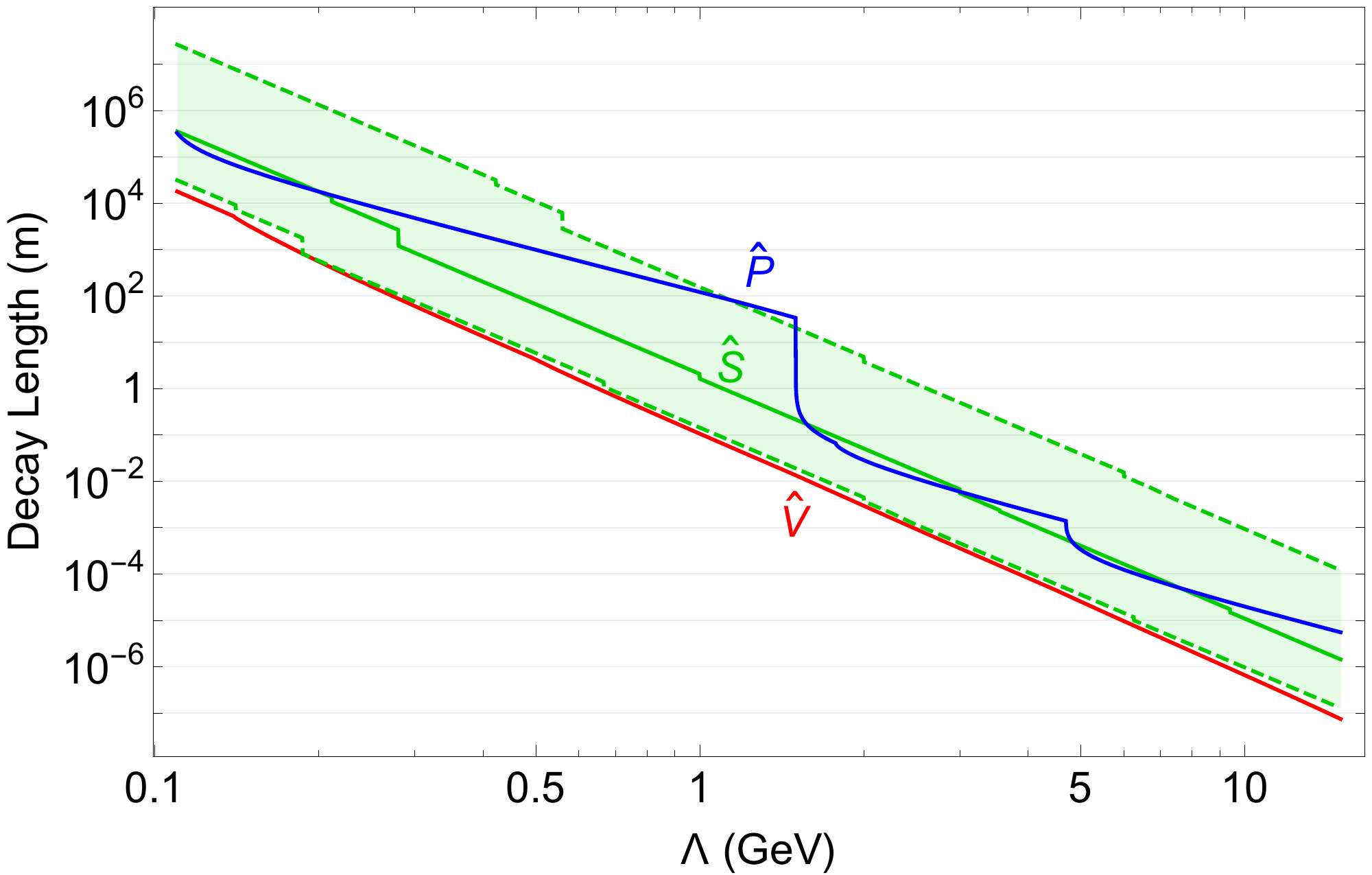}} 
\hspace{2mm}
\raisebox{0\height}{\includegraphics[width=0.44\textwidth]{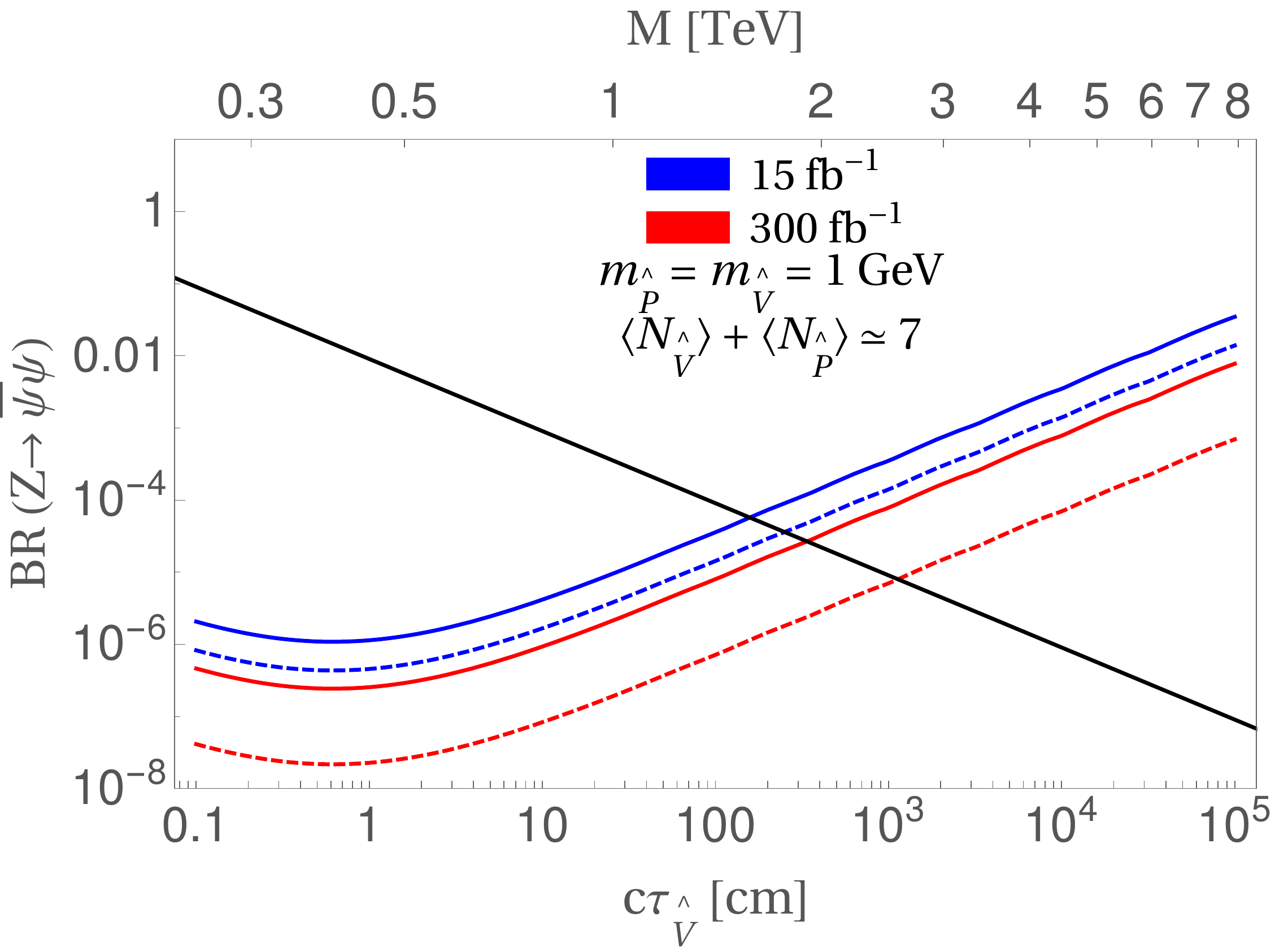}}
    \caption{{\it Left:} hidden meson lifetimes for $M = 2\;\mathrm{TeV}$, assuming $m_{\hat{P}, \hat{V}} = 2 \Lambda$ and $5\Lambda/2 < m_{\hat{S}} < 7\Lambda/2$. {\it Right:} projected limits on $\mathrm{BR}(Z\to \psi \overline{\psi})$ from a search for single $\hat{V} \to \mu\mu$ displaced vertex inside a hidden jet at LHCb. Solid~(dashed) curves correspond to standard background count~(no background). The black line corresponds to the theoretical prediction $\mathrm{BR} \approx 10^{-2} / (c\tau_{\hat{V}}/\mathrm{cm})$.}
    \label{fig:plots_Ennio}
\end{figure}

Even though one-flavor QCD does not predict any light pseudo Nambu-Goldstone bosons, we know from the lattice~\cite{Farchioni:2007dw} that the mesons are lighter than the baryons. The lightest mesons are the pseudoscalar $\hat{P}$ and (presumably) the vector $\hat{V}$, while the scalar meson $\hat{S}$ is heavier by a factor $m_{\hat{S}} / m_{\hat{P}} \approx 1.5$. The lifetimes of these three states, which all decay dominantly to SM particles through the $Z$ portal, are shown in the left panel of Fig.~\ref{fig:plots_Ennio}. If $\Lambda \sim O( \mathrm{GeV})$, on which we focus, all hidden hadrons have macroscopic lifetimes for $M\sim O(\tev)$. The $\hat{P}$ is stable on detector scales; the $\hat{V}$ decays to SM fermions by mixing with the transverse $Z$; while the $\hat{S}$, being heavier than the other two mesons, can be neglected in a first description of the hidden jets. Our hidden jets share some similarities with emerging\cite{Schwaller:2015gea} and semi-visible\cite{Cohen:2015toa} jets, but are characterized by the softer production mode and by the ``democratic'' decay of the vector meson to SM fermions. We find that LHCb can achieve the best LHC sensitivity, through the process
\begin{equation}
pp\to Z \to \psi \overline{\psi}\to j_{\rm h} j_{\rm h}\,, \qquad j_{\rm h} \ni \hat{V} \to \mu^+\mu^- \,\text{in the VELO}
\end{equation}
where $j_{\rm h}$ denotes a hidden jet (see also~\cite{Pierce:2017taw} for previous related work).

In comparison, existing searches for emerging jets at CMS~\cite{Sirunyan:2018njd} and ATLAS necessarily rely on hard cuts that suppress our signals to a negligible level.\footnote{Reference~\cite{Aad:2019tua} is a potential exception deserving further study.} Therefore we are led to consider the production of an associated object, e.g. $pp \to ZV$ with $V$ a leptonically decaying EW gauge boson, or $pp\to Zj$ with $j$ a hard jet, to ensure triggering. The price to pay is a strong reduction of the signal rate, whereas the backgrounds remain appreciable. We find that these searches cannot compete with the sensitivity of LHCb, which is singled out for its unique sensitivity to these relatively soft hidden jet events.

Detection of a $\hat{V}\to \mu^+\mu^-$ inside the VELO requires the vector meson to have transverse decay length between $6$ and $22$~mm~\cite{Ilten:2016tkc} and pseudorapidity $\eta \in [2,5]$. We simulated $Z$ decay events in Pythia8, choosing $m_{\hat{P},\hat{V}} = 1$~GeV as benchmark. We imposed that the average numbers of mesons in each jet satisfy $\left\langle N_{\hat{P}} \right\rangle / \left\langle N_{\hat{V}} \right\rangle = 1/3\,$, as expected from a counting of the spin degrees of freedom, and neglected all heavier hadrons in the shower. The simulation produces an average of $\sim7$ hidden mesons per jet. Each $\hat{V}\to \mu\mu$ displaced vertex with transverse decay length and $\eta$ as above, as well as $p_T^{\hat{V}}>1\;\mathrm{GeV}$, was assumed to be reconstructed with constant efficiency $\epsilon_{\mu\mu} = 0.5\,$. The efficiency degradation caused by overlap with other nearby tracks is not explicitly discussed in previous studies~\cite{Ilten:2016tkc,Pierce:2017taw}, but it is expected to reduce $\epsilon_{\mu\mu}$ significantly. Therefore, we rejected non-isolated events where the $\mu\mu$ vertex is accompanied by additional $\hat{V}$ visible decays with transverse length shorter than $22$~mm and $\Delta R < 0.4$. The $\hat{P}$ was assumed to be stable on detector scales. 

The SM background to such displaced $\mu\mu$ vertices is expected to be about 25 events for 15~fb$^{-1}$~\cite{Ilten:2016tkc}. To estimate the ultimate sensitivity achievable at LHCb, we also computed the constraints by assuming negligible SM background. The results are shown in the right panel of Fig.~\ref{fig:plots_Ennio}, where the solid (dashed) curves correspond to standard (negligible) background. Due to the mild typical boost factor, the sensitivity is optimal for \mbox{$c\tau_{\hat{V}} \simeq 1$~cm}, where it is expected to reach $Z$ branching ratios of $O(10^{-7})$ in the high-luminosity phase. In terms of the mediation scale, the bounds translate to $M\gtrsim 1.6\,(2.0)$~TeV for $L=15\,(300)$~fb$^{-1}$ with standard background count, and $M\gtrsim 1.8 \,(2.7)$~TeV in the background-free case.

\subsubsection{Confining Hidden Valleys and the Twin Higgs model\footnote{Contributed by Yuhsin Tsai.}}
\label{sec:hv_twin}
Hidden Valley (HV) scenarios~\cite{Strassler:2006im,Han:2007ae} consist of a sector with light (e.g.$\lsim$GeV scale) particles connected to the SM sector only via massive particles, effectively forming a barrier between the sectors. The hidden sector may confine via a non-abelian gauge symmetry, such as a \emph{dark QCD} force with a model-dependent mass spectrum. If the hidden quarks are heavier than the hidden confinement scale $\Lambda_v$, hidden hadrons have masses mainly from the hidden quarks. If the hidden quarks $q_v$ are lighter than $\Lambda_v$, the mass of hidden mesons like the vector ($\omega_v$) or pseudo-scalar ($\eta_v$) bound states can be of order $\Lambda_v$. An approximate chiral symmetry that is spontaneously broken by hidden sector confinement also leads to light hidden pions, $\pi_v$, with $m_{\pi_v}\ll\Lambda_v$. 

Depending on the portal coupling between the hidden sector and the SM sector, some of the hidden mesons can decay into SM particles with decay rates that are suppressed by the small portal coupling and can travel through a macroscopic distance before the decay. The decay length of hidden mesons can have a wide range and is model-dependent. When considering the thermal history of HV particles, cosmological constraints can sometimes provide upper bounds on the lifetime of HV mesons. For example, in HV scenarios where $\pi_v$ is the lightest hidden particle and cannot decay before the Big-Bang Nucleosynthesis, there is a upper bound on some of the hidden meson decay length to be comparable to the size of particle detectors~\cite{Li:2019ulz}. For GeV scale $\Lambda_v$, the lifetime bound can ensure some of the hidden mesons to decay inside the VELO after being produced at the LHCb.

\begin{figure}[!htp]
    \centering
    \includegraphics[width=0.49\textwidth]{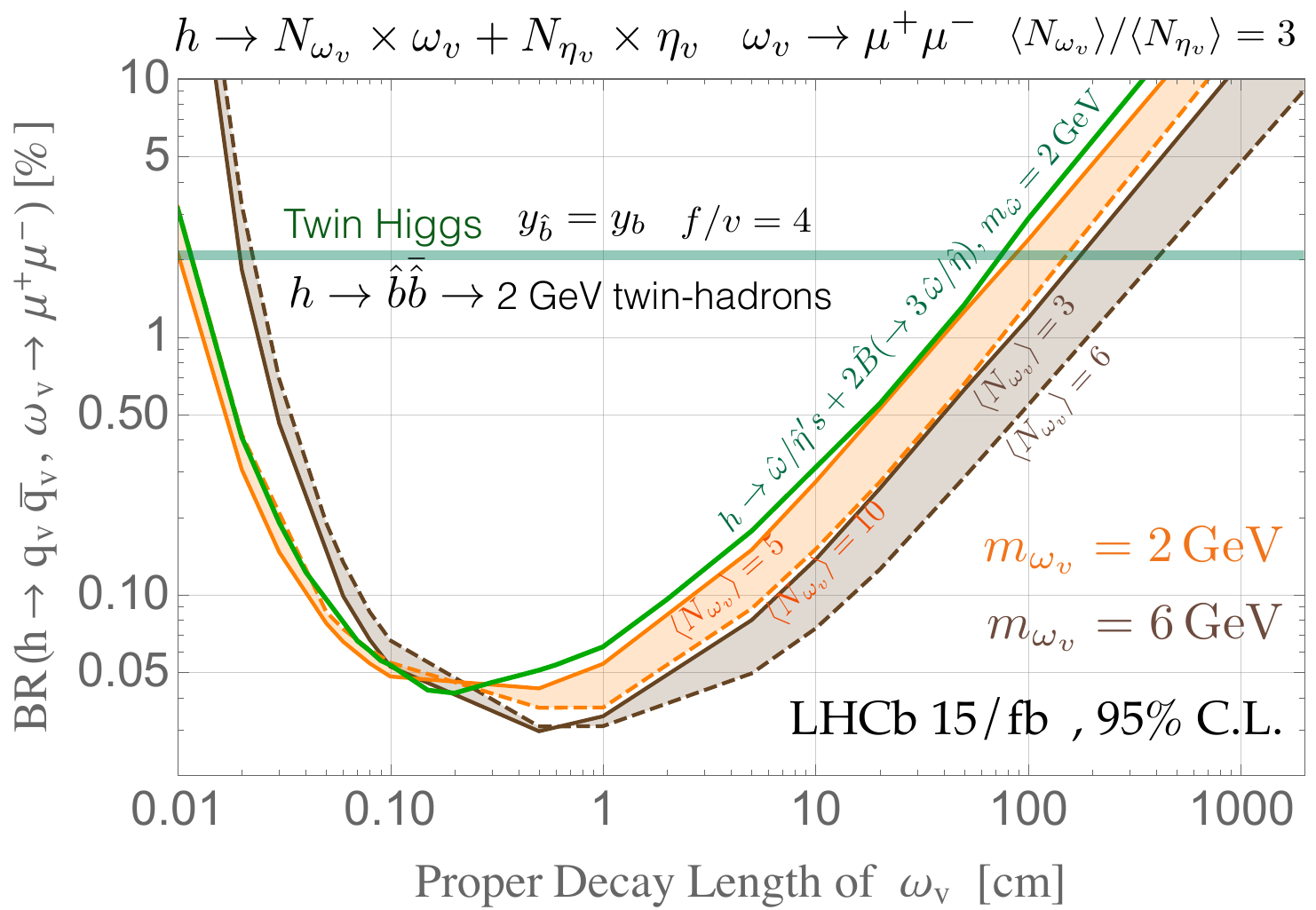}\,\,
        \includegraphics[width=0.49\textwidth]{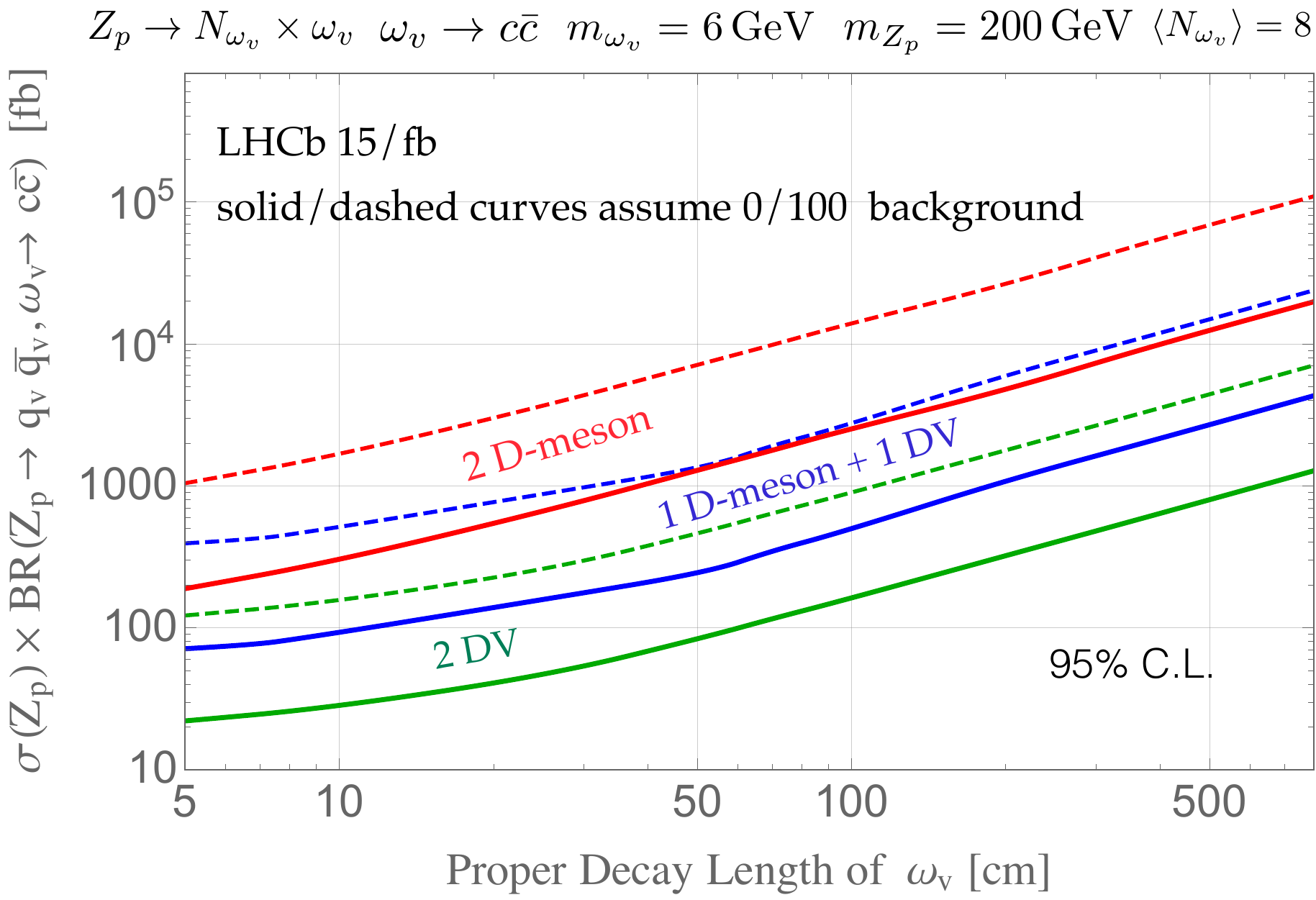}
    \caption{{\it Left}: Projected upper bounds on BR($h\to$ twin bottom quarks)
using the 1\,DV search. This process produces twin mesons $\hat{\omega}/\hat{\eta}$ followed by $\hat{\omega}\to\mu^{+}\mu^-$. Horizontal green line: prediction in a variation of the Fraternal Twin Higgs model. The band widths
correspond to the average number of hidden meson in the final state $10\leq\langle N_v\rangle\leq
30$. {\it Right}: Projected bounds from various displaced $c\bar{c}$ search
strategies. See~\cite{Pierce:2017taw} for more details.}
    \label{fig:darkshower}
\end{figure}

Hidden mesons can be produced in different ways at a collider depending on the portal coupling between the hidden and the SM sectors. It is typically
\begin{equation}
pp\to {\rm mediator}^{(*)}\to q_v\,\bar{q}_v + X\,,\quad q_v\,\bar{q}_v\to N_{\pi_v}\times\pi_v+N_{\eta_v}\times\eta_v+N_{\omega_v}\times\omega_v\,...\,,
\end{equation}
where the mediator can either be a heavy $Z'$, SM Higgs, or heavy fermions that charged under both the hidden and SM gauge symmetries, and in some cases the product can also come with additional hard SM particles ($X$, e.g.,~\cite{Cheng:2015buv}). The showering and hadronization in hidden sector
produces many hadrons, each much softer than the total energy.
Hadron multiplicity can range from a few in the absence of light
quarks to ${\mathcal O(1000)}$ \cite{Knapen:2016hky} for a long
showering window with large 't Hooft coupling. In the examples shown in Fig.~\ref{fig:darkshower}, we focus on theories averaging ${\mathcal O(10)}$ hadrons per event as in the SM QCD and mimic the showering/hadronization process using Pythia simulations. The energy spectrum and the composition of hidden hadrons in the final state can vary drastically with on the size of hidden force coupling and quark masses. Therefore, when designing searches for HV scenarios, it will be useful to focus on the property of individual hidden meson decays, such as studying the energy cuts for some examples of average $p_T$ and the multiplicity of hidden mesons, other than relying on the detailed energy distribution of the decay final states. See also \cite{Knapen:2021eip} for the study of more benchmark HV models.

At LHCb, planned upgrades will remove hardware-level triggers
entirely, dramatically improving search flexibility. Furthermore,
excellent vertex reconstruction, invariant mass resolution, particle
identification, and a lower pile-up environment make LHCb well
suited for exotic soft long-lived particle searches. Previous HV
searches at LHCb~\cite{Aaij:2016isa,Aaij:2017mic} have provided useful constraints by focusing on final states with only two hard $\pi_v$. It will be helpful to extend the search for softer final states and lighter hidden mesons that better describe the nature of confining HV models.

Fig.~\ref{fig:darkshower} shows two examples of the projection of HV constraints from the LHCb search. In the left panel, we show results for
exotic Higgs decay as brown (vector meson $m_{\omega_v}$=6 GeV) and orange
($m_{\omega_v}$=2 GeV) bands. We adapt the energy and angular cuts proposed for the dark photon search~\cite{Ilten:2016tkc} and focus on signals of displaced muons. We also explore an incarnation of the
Twin Higgs model~\cite{Chacko:2005pe,Craig:2015pha} that solves the Higgs little hierarchy problem. We focus on $g g\to h\to \hat{b}\bar{\hat{b}}$, with
subsequent decays $\hat{b}\to \hat{c} \bar{\hat{c}} \hat{s}$.  This
produces $\hat{\omega}/\hat{\eta}$ comprised of the
light flavor, which can decay as $\hat{\omega}\to \mu^+ \mu^-$
through a kinetic mixing $\epsilon$ between the SM and twin
photons. Although the
light twin mesons are primarily produced by heavy twin $B$ meson
decays (rather than showering/hadronization), the multiplicity is low and the kinematic
distributions are similar to the showering case, giving a comparable reach (green curve). The prediction in the Twin Higgs scenario (green
line) demonstrates LHCb's ability to probe such models for
$c\tau_{\hat{\omega}}\lsim 1\,$m. From the experiment side, \cite{Aaij:2020ikh} has studied $\mathcal{O}({\rm GeV})$ scale HV mesons decaying into $\mu^+\mu^-$ and shows the plausibility of probing light HV mesons at LHCb.

In the right plot, we consider a heavy $U(1)'$ gauge boson $Z_{p}$ coupled to both SM quarks $q$ and HV quarks $q_{v}$. The resulting hidden meson $\omega_v$ can dominantly decay into $c$-quarks. In this case, subsequent SM hadronization often produces
two $D$ mesons, $D_{(s)}^0$, $\bar D_{(s)}^0$ or $D^\pm$. The
non-negligible lifetimes of charmed hadrons create an additional
separation between the DV from the two $D$ meson decays, producing
two vertices with large separation from the primary vertex and a small but
significant separation from each other. Resolving the secondary
vertices should be straightforward as the position resolution in the
VELO is $\mathcal{O}(10)\mu$m while
$c\tau_D\sim\mathcal{O}(100)\mu$m. Several
strategies to identify $D$ mesons at LHCb exist, some without full
reconstruction \cite{Aaij:2015yqa}. Given the sizable probability
for one $D$ meson to decay to $\geq\,3$ charged tracks that can be
well-reconstructed in the VELO, we consider the following
increasingly inclusive search strategies: (1) Two nearby reconstructed displaced $D$ mesons (2D). (2) One displaced reconstructed $D$ meson and one DV with $\geq\,3$ charged tracks nearby (1D1V).
(3) Two DV, each with $\geq 3$ charged tracks, near each other (2V). Using detailed information of the $D$-meson decays, there is a chance to identify the decay of hidden meson $\omega_v$ into SM $D$-mesons. The LHCb is the only experiment that can perform such a search.

\subsection{Composite Higgs}
\subsubsection{Novel B-decay signatures of light scalars at high energy facilities\footnote{{Contributed by Maria Ramos.}}}
\label{sec:maria}

In non-minimal composite Higgs models~\cite{Gripaios:2009pe,Vecchi:2013bja,Sanz:2015sua,Chala:2016ykx,Balkin:2017aep,DaRold:2019ccj,Ramos:2019qqa}, the Higgs boson and new singlet scalars $a_1$ and $a_2$ can emerge as pseudo Nambu-Goldstone bosons of the same strongly interacting sector that confines near the TeV. Hence, their masses are naturally at or below the EW scale. Likewise, they couple to heavy resonances of the composite sector such as flavour-violating vector bosons $V$, which are promising candidates to explain the lepton flavour universality (LFU) anomalies ~\cite{Niehoff:2015bfa,Niehoff:2015iaa,Carmona:2015ena,Megias:2016bde,GarciaGarcia:2016nvr,Sannino:2017utc,Carmona:2017fsn,Chala:2018igk}. At low energies, such vector triggers rare $B$ meson decays into the light scalars. 
Before electroweak symmetry breaking (EWSB), the relevant interactions to study this process read: 
\begin{equation}\label{eq:lag}\nonumber
 L \supset\, \frac{1}{2}m_V^2 V_\mu V^\mu + \frac{1}{2}m_1^2 a_1^2 + \frac{1}{2}m_2^2 
a_2^2 + m_{12} a_2 a_1^2 + V^\mu \left[g_{12} a_1\overleftrightarrow{\partial_\mu} a_2 + 
g_{qq}(\overline{q_L}\gamma_\mu q_L + \text{h.c.})\right]~,
\end{equation}
where $m_V \gg m_{1,2}\sim \mathcal{O}(1)$ GeV and $m_2 > m_1$. Assuming that $V$ interacts mostly with the third generation quarks, after EWSB 
it couples to 
$\overline{b_L}s_L+\text{h.c.}$ with strength
\begin{equation}
g_{sb}\equiv g_{qq} V_{ts}^{\rm CKM} V_{tb}^{\rm CKM}\sim 0.04\, g_{qq}~.
\end{equation}
We consider the regime in which the light scalars are muonphilic or, similarly to the Higgs boson, couple to leptons according to their masses. %
In this case, processes such as $B^0_s \to a_1 a_2$ 
\textit{can} lead to a final state with four muons. The corresponding signal has been studied experimentally at the LHCb with a collected luminosity of $3~{\rm fb}^{-1}$ and $\sqrt{s} = 8$ TeV; the most stringent limit being $\mathcal{B} \left(B_s^0 \to 2\mu^+2\mu^-\right) < 2.5 \times 10^{-9}$~\cite{Aaij:2016kfs}.

There are, however, compelling reasons to consider alternative decays~\cite{Blance:2019ixw}. First, the partial decay width of $a_2\to a_1 a_1$ can easily dominate over the leptonic one (due to the Yukawa suppression). This gives rise to \textit{six} muon final states instead. Second, since the scalars couple to the mediator as a vector current, the two body decay width scales as $\Gamma \sim (m_2^2 - m_1^2)/m_B$. It therefore vanishes in the limit of \textit{quasi}-degenerate scalars. In order to probe the model in this limit, one should rather study three body decays of $B$ with the emission of an extra meson, such as  $B^+ \to K^+ a_1 a_2 \to K^+ 3\mu^+ 3\mu^-$.
None of these signals have been studied experimentally. We have therefore proposed the first dedicated analyses to search for these channels at the LHCb. (Remarkably, not even the $B^+ \to K^+ 2\mu^+ 2\mu^-$ channel has been probed yet; such process was studied in Ref.~\cite{Chala:2019vzu}.)

Due to the large multiplicity of muons in the final state, the SM backgrounds are negligible. They mostly arise from the productions of $J/\Psi$ and $\phi$ resonances which subsequently decay to muons; we remove them by requiring that no zero charge muon pair has an invariant mass in the range $[0.95,1.09]\cup[3.0,3.2]$ GeV. Being the analysis essentially background-free, we can estimate the upper limits on the branching ratio of the new processes at $\sqrt{s} = 14$ GeV from the upper limit on $\mathcal{B}\left(B_s^0\to 2\mu^+2\mu^-\right)$, obtained in Ref.~\cite{Aaij:2016kfs}. With this aim, we have compared the efficiencies for selecting events in the $B_s^0\to 3\mu^+3\mu^-$ channel with that for $B_s^0\to 2\mu^+2\mu^-$. Typically, the former is nearly one order of magnitude smaller than the latter.

\begin{figure}
\centering
\includegraphics[scale=0.6]{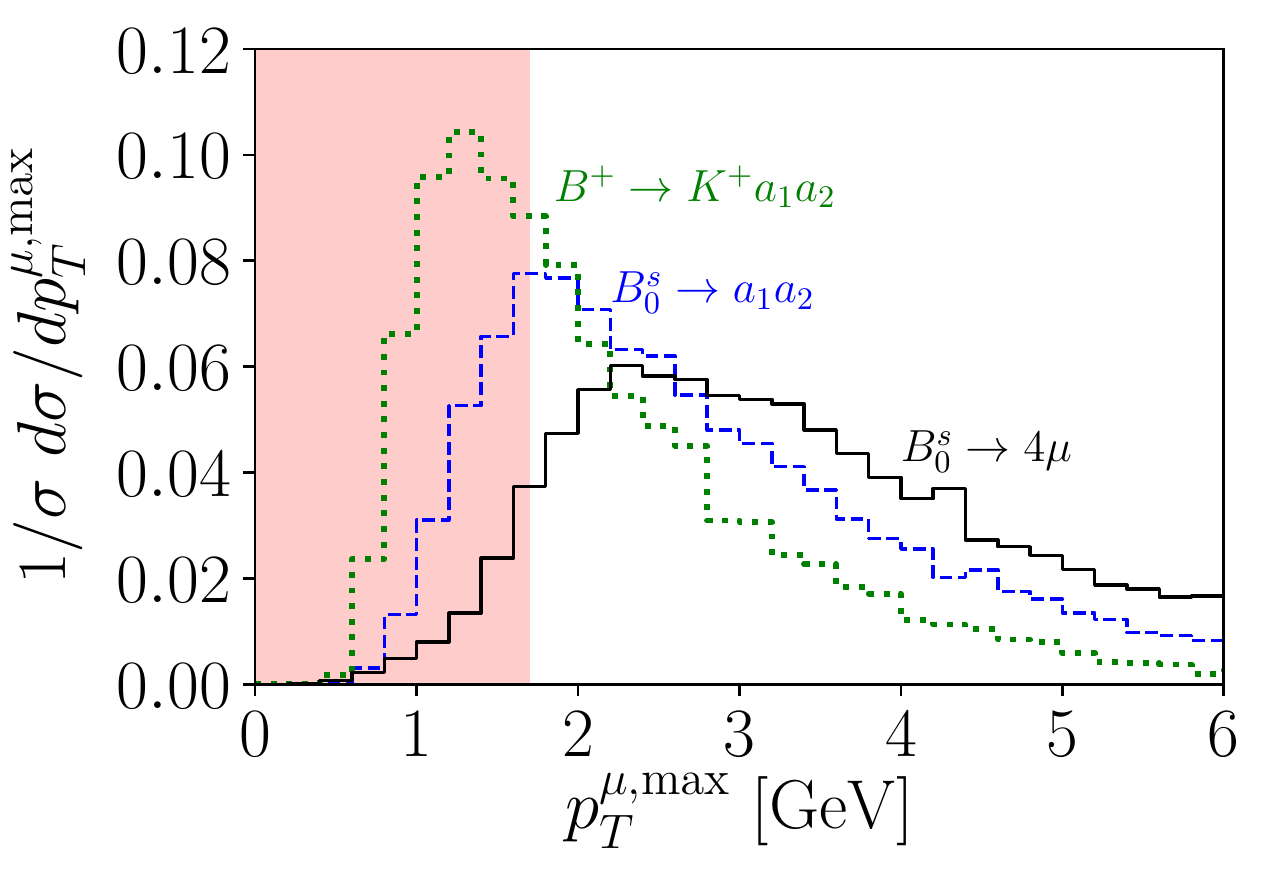}
\includegraphics[scale=0.6]{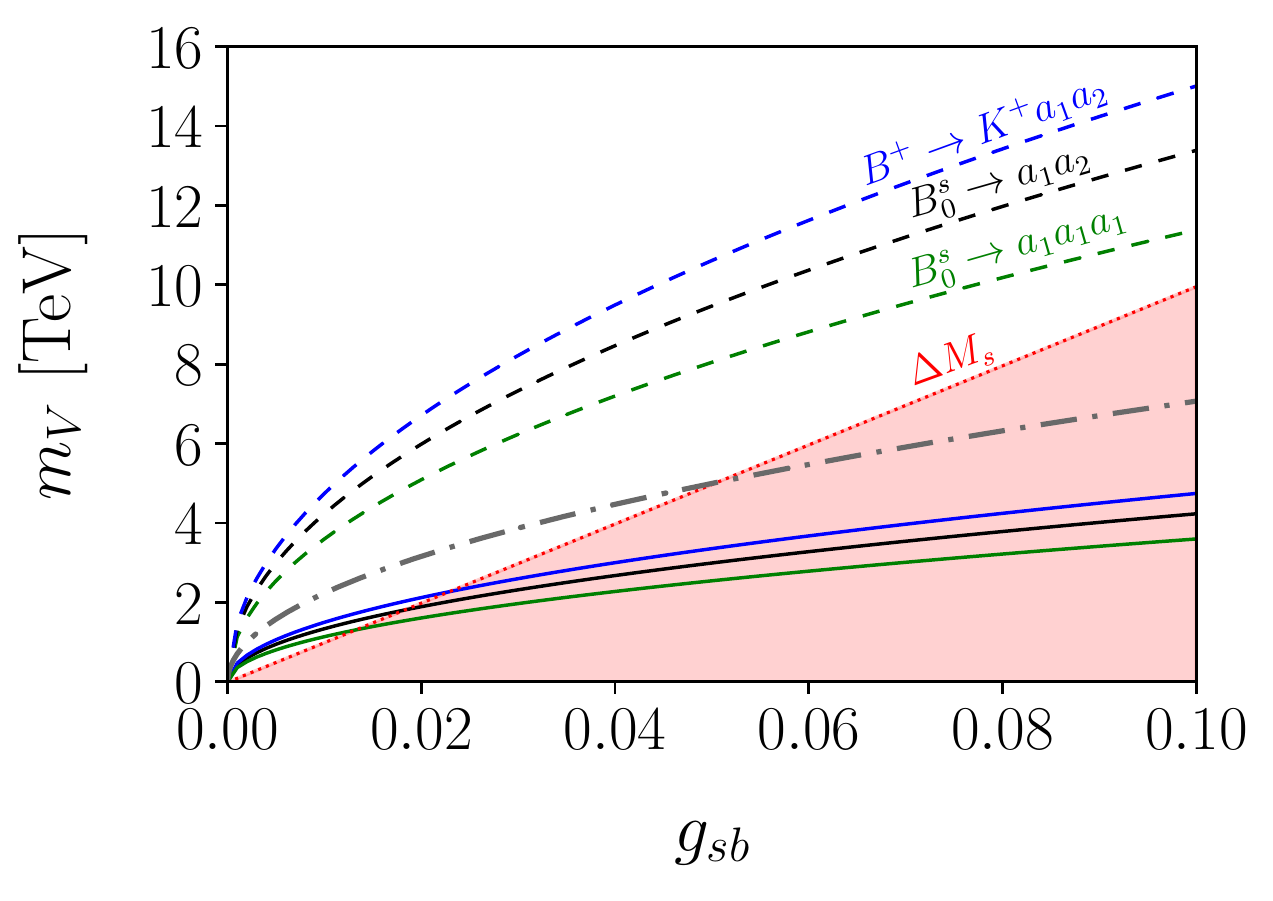}
\caption{\textit{(Left)} Normalized distribution of the transverse momentum of the hardest muon for $B_0^s \to a_1 a_2$ and $B^+\to K^+ a_1 a_2$, with $m_{1} = 1$ GeV and $m_{2} = 2.5$ GeV. These distributions are compared with that of $B_0^s \to 2\mu^+2\mu^-$. \textit{(Right)} Maximum value of $m_V$ that can be tested in the 
searches for $B_s^0\to3\mu^+ 3\mu^-$ and $B^+\to K^+ 3\mu^+ 3\mu^-$ 
at the current run of the
LHCb (solid lines) and for Upgrade II (dashed lines). The red dotted line 
delimits the area excluded by measurements of $\Delta M_s$. In the  
dash-dotted line the anomalies in $R_K$ and $R_{K^*}$ can be explained at the $1\sigma$ level~\cite{DiLuzio:2017fdq}. We have fixed $g_{12} = 0.5$ and $m_1 = 1.2$ GeV. We have set $m_2 = 2.0$ GeV for $B_0^s 
\rightarrow a_1 a_2$ and $B^+ \rightarrow K^+ a_1 a_2$. For $B_0^s 
\rightarrow a_1 a_1 a_1$, we have fixed instead $m_2 = m_{12} = 5$ GeV.}
\label{fig:Bto6mu}
\end{figure}

The explanation for the smaller efficiencies in the six muons process lies in the selection cuts that we impose in the analysis, namely: \textit{(1)} at least one muon is required to have $p_T > 1.7$ GeV, to ensure that the events pass the same hardware trigger used at $\sqrt{s} = 8$ TeV; \textit{(2)} all muon tracks are required to have $p_T > 0.5$ GeV and lie in the pseudorapidity volume $2.5<\eta<5.0$; and \textit{(3)} all muon tracks are required to have total momentum larger than $2.5$ GeV. 
Since the initial energy is shared among a larger number of tracks, the six muons process produces more events not satisfying \textit{(1)}; see the left panel of Fig.~\ref{fig:Bto6mu}. In addition, there are more events with at least one track with $p_T < 0.5$ GeV which is therefore not detected.

For the $B^+\to K^+ 3\mu^+3\mu^-$ channel, on top of the selection criteria above, we require the presence of a charged kaon satisfying the requirements in \textit{(2)}. We have found that, in spite of the smaller efficiency from the occurrence of an extra track, the limits on this channel can be actually stronger than on the decay mode without the extra meson. 
This is due to the fact that the $B^+$ production cross section is $\sim 4$ times larger than that of $B_s^0$. 
This strongly motivates searches for $B^+\to K^+ 3\mu^+3\mu^-$ which is \textit{the} key signature of the model when $m_1 \sim m_2$.

In the right panel of Fig.~\ref{fig:Bto6mu}, we represent the expected upper limits in the plane $(g_{sb},m_V)$. Prospects for the Upgrade II of the detector are also shown. Interestingly, our proposed analyses outperform bounds from meson mixing $\Delta M_s$, and can probe entirely the region in which the anomalies in LFU can be explained.

\sloppy Overall, our study motivates new searches for $B_s^0\to 3\mu^+3\mu^-$,~$B^0\to 3\mu^+3\mu^-$,~$B^+\to K^+ 3\mu^+3\mu^-$ and $B_s^0\to K^{0*}3\mu^+3\mu^-$ at the LHCb. We demonstrate that this facility could probe branching ratios as small as $6.0,~1.6,~5.9$ and $18\times 10^{-9}$ for the aforementioned channels, respectively. These sensitivity estimates are conservative, as we assume no changes to the trigger or tracking performance in the upgrades of the LHCb; the removal of the hardware trigger and a lower $p_T$ threshold for the muons identification would certainly improve the signal efficiencies. 
Finally, if a signal is observed in these channels, the masses of the scalar particles involved could be fairly well reconstructed ; different algorithms for specific kinematic regions are presented in Ref.~\cite{Blance:2019ixw}.

\subsubsection{ALPs from composite Higgs models\footnote{Contributed by Diogo Buarque Franzosi, Gabriele Ferretti, Giacomo Cacciapaglia, Thomas Flacke, Xabier Cid Vidal, Carlos Vázquez Sierra.}}
\label{sec:compoALPS}
\newcommand{\be}{\begin{equation}}
\newcommand{\ee}{\end{equation}}
\newcommand{\MeV}{\mbox{ ${\mathrm{MeV}}$}}
\newcommand{\GeV}{\mbox{ ${\mathrm{GeV}}$}}
\newcommand{\TeV}{\mbox{ ${\mathrm{TeV}}$}}
\def\fig#1{{fig.~(\ref{#1})}}
\newcommand{\mg}{\textsc{MG5\_aMC@NLO}}

Fundamental Composite Higgs (FCH) based on a new fermionic confining gauge theory $\mathcal{G}$ in 4-dimensions can address the hierarchy problem~\cite{Cacciapaglia:2020kgq}, the strong-CP problem~\cite{Gherghetta:2020ofz}, introduce dark matter candidates, explain the fermion flavour structure and the observed LHCb anomalies~\cite{Chala:2018igk}. 
In FCH a condensate of hyperfermions $\psi$ charged under $\mathcal{G}$ as well as under $SU(2)_L$ spontaneously breaks the EW symmetry dynamically generating the EW scale $v$ with a little hierarchy with respect to the composite scale $f$ via the vacuum misalignment mechanism~\cite{Dugan:1984hq}, $v=f\sin\theta$, with $\theta$ being the so-called misalignment angle. 

In this class of models it is typically necessary that the top quark mass is generated by the mechanism of partial compositeness (PC), where a composite fermionic state mixes with the elementary top quark. One 
option is to incorporate new hyperfermions $\chi$ charged under QCD, hypercharge and hypercolor $\mathcal{G}$ (in a different representation than that of $\psi$). The top partners are then states formed as $\psi\psi\chi$ or $\chi\chi\psi$ states.
A detailed study\cite{Ferretti:2013kya} found 12 different \emph{models} M1, ..., M12 fulfilling minimum requirements. Each model represents a specific hypercolor gauge group and hyperfermion representations of $\psi$ and $\chi$. 
An unavoidable consequence of such construction is the appearance of several types of resonances similarly to QCD. Among those are a number of light pseudo-Nambu-Goldstone-bosons (pNGB). 
One of them, $a$, is associated to an anomaly free U(1) symmetry, is light and couples to the gluons via the Wess-Zumino-Witten term (see e.g. ~\cite{Cacciapaglia:2019bqz,Cacciapaglia:2017iws,Belyaev:2016ftv}).

Interactions of $a$ can be parametrized by a typical axion-like-particle (ALP) lagrangian
\begin{equation}
 \mathcal{L}_{\mbox{eff}}\ \supset  \frac{a}{16\pi^2 f}\left(g_i^2 K_{i} 
F_{i\, , \mu\nu} \widetilde{F_i}^{\mu\nu} \right) -i\sum_f C_f\frac{ m_\psi}{f}a\bar{\psi}_f\gamma^5\psi_f
\end{equation}
where $\widetilde{F_i}^{\mu\nu}=\frac{1}{2}\epsilon^{\mu\nu\rho\sigma} F_{i,\rho\sigma}$, and the $i$ index parametrizes the SM gauge group with coupling constant $g_i$. 
For simplicity we assume that the state $\eta'$ associated to the anomalous U(1) combination decouples.
In this case, for models M1-M12 the $a$ interactions with bosons parametrized by $K_i$ are fixed by the anomaly coefficients. 
The coupling to top quark $C_t$ depends on the representation of the dominant top partner responsible for the top mass, here we consider the model with largest value for illustration. 
The other fermions masses might be generated by mixing with other heavier states or via interactions with a $\langle \psi\psi\rangle$ condensate à là extended-Technicolor. 
Here we assume the latter, which fixes the $a$ couplings to other (not top) fermions only by the non-anomalous U(1) charges. 
A table with the values of the coupling coefficients is provided in~\cite{Cacciapaglia:2019bqz}.

Interestingly, if $a$ is light enough ($m_a\lesssim 40\GeV$), LHCb might have an advantage over the other LHC experiments, as shown in the di-muon search~\cite{Haisch:2016hzu, Aaij:2020ikh}.  
Here we report preliminary results on  the $a\to \mu^+\mu^-, c \bar{c}$ and leave  $a\to \gamma\gamma, \tau^+ \tau^-$ for a future study. 

\begin{figure}
\centering\includegraphics[width=0.38\textwidth]{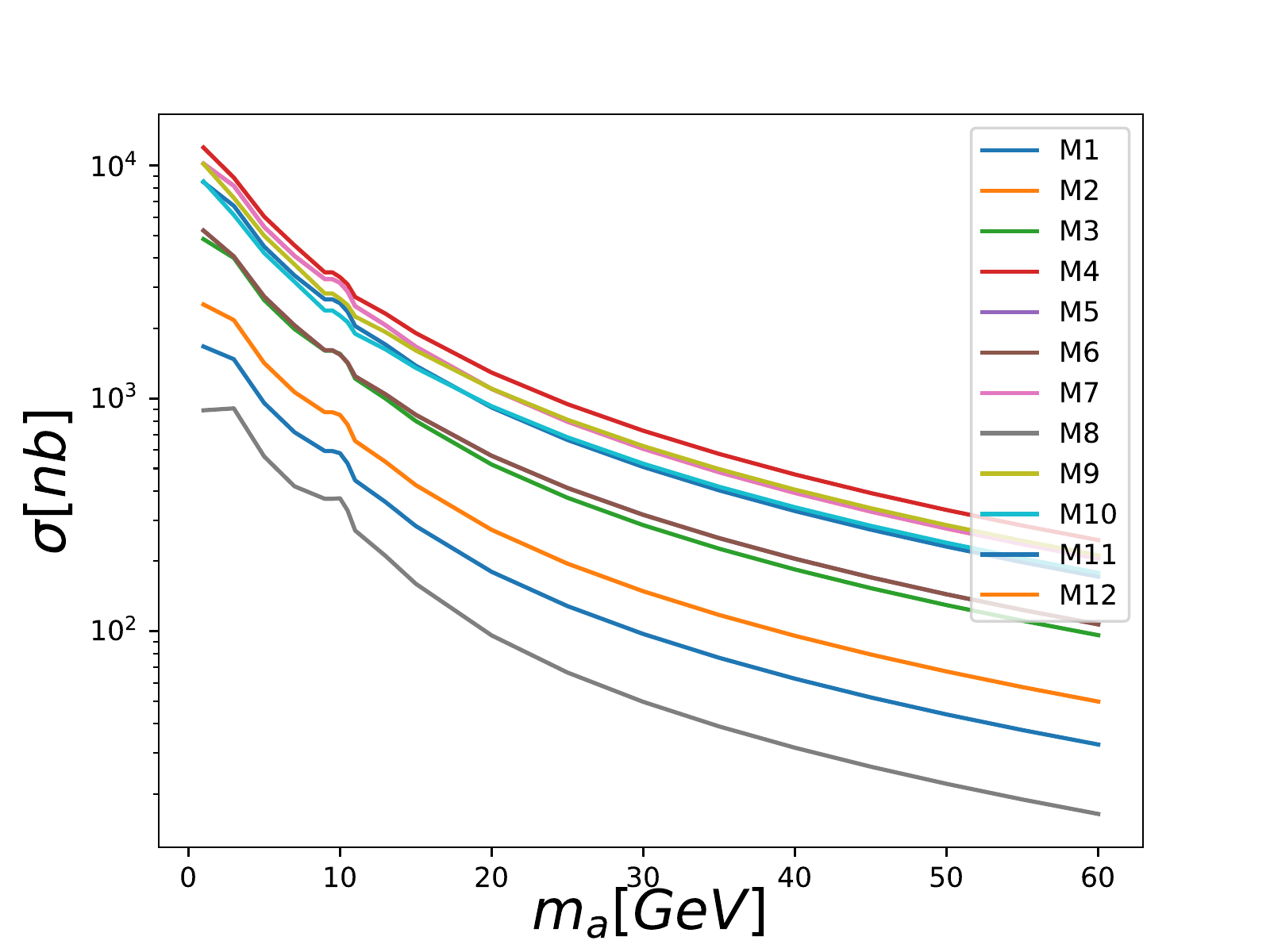}
\includegraphics[width=0.56\textwidth]{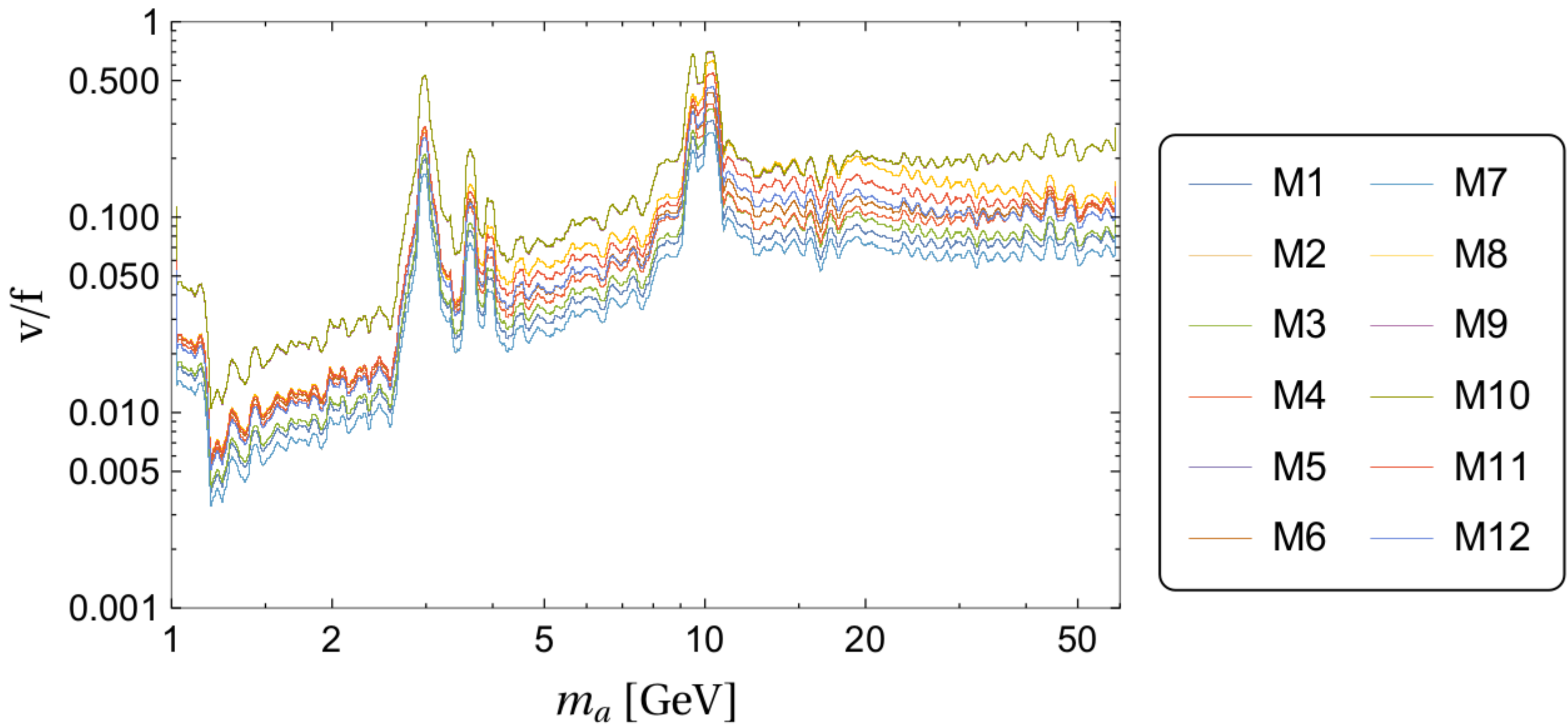}
\caption{\emph{Left:} Cross section for the different models. \emph{Right:} Exclusion from di-muon search.}
\label{fig:xs_mumu}
\end{figure}

The production cross-sections for a pseudoscalar at NNLO in QCD at the 14 TeV LHC in each of the 12 models are shown in  \fig{fig:xs_mumu} (left), and were obtained using HIGLU~\cite{Spira:1996if} and the NNPFD 3.1 set containing LHCb data~\cite{Bertone:2018dse}, setting the renormalization and factorization scales to $\mu=m_a$. We note that for such low masses a very large scale dependence is present. Partial widths are computed at NLO using~\cite{Spira:1996if}.  

For the $\mu^+ \mu^-$ exclusion we reinterpret the limits on $\sin\theta_H$ of the 2HDM model~\cite{Aaij:2020ikh} in terms of $v/f$. 
The resulting bounds, shown in \fig{fig:xs_mumu}(right), are quite strong and surpass the indirect bounds from EWPO, $v/f \lesssim 0.2$. 
The $c\bar{c}$ channel is only relevant for very low masses $3.8\GeV \lesssim m_a \lesssim 6\GeV$. It might be motivated in scenarios where the muon gets its mass from a different mechanism, and $a$ couples only to quarks of up families, similar to the scenarios considered in~\cite{Carmona:2021seb}.

To estimate the LHCb sensitivity in this channel, we simulate signal events using \mg~\cite{Alwall:2011uj} with the Higgs Characterization model~\cite{Artoisenet:2013puc} and pass them through \pythia ~\cite{Sjostrand:2014zea} for showering and hadronization. 
Similarly for the SM background. We use these MC events to estimate efficiencies of event selections described below.

LHCb can identify $D^{\pm} \to K^{\mp} 2\pi^{\pm}$  with high efficiency. 
We select events with at least one $D^+$ and one $D^-$ each one decaying in this mode, and with the 6 decay products  $K^\pm$ and 2 $\pi^\pm$ within LHCb coverage, $2<\eta<5$ and $p_T>0.25\GeV$. 
We can then assume  ${m_{D^+ D^-} = m_a \pm 20}$ MeV~\cite{Aaij:2016yip}. 
The signal efficiency drops rapidly for $m_a\gtrsim 550\MeV$ due to the opening of other decay channels with extra pions for instance. 
\begin{figure}
\centering\includegraphics[width=0.375\textwidth]{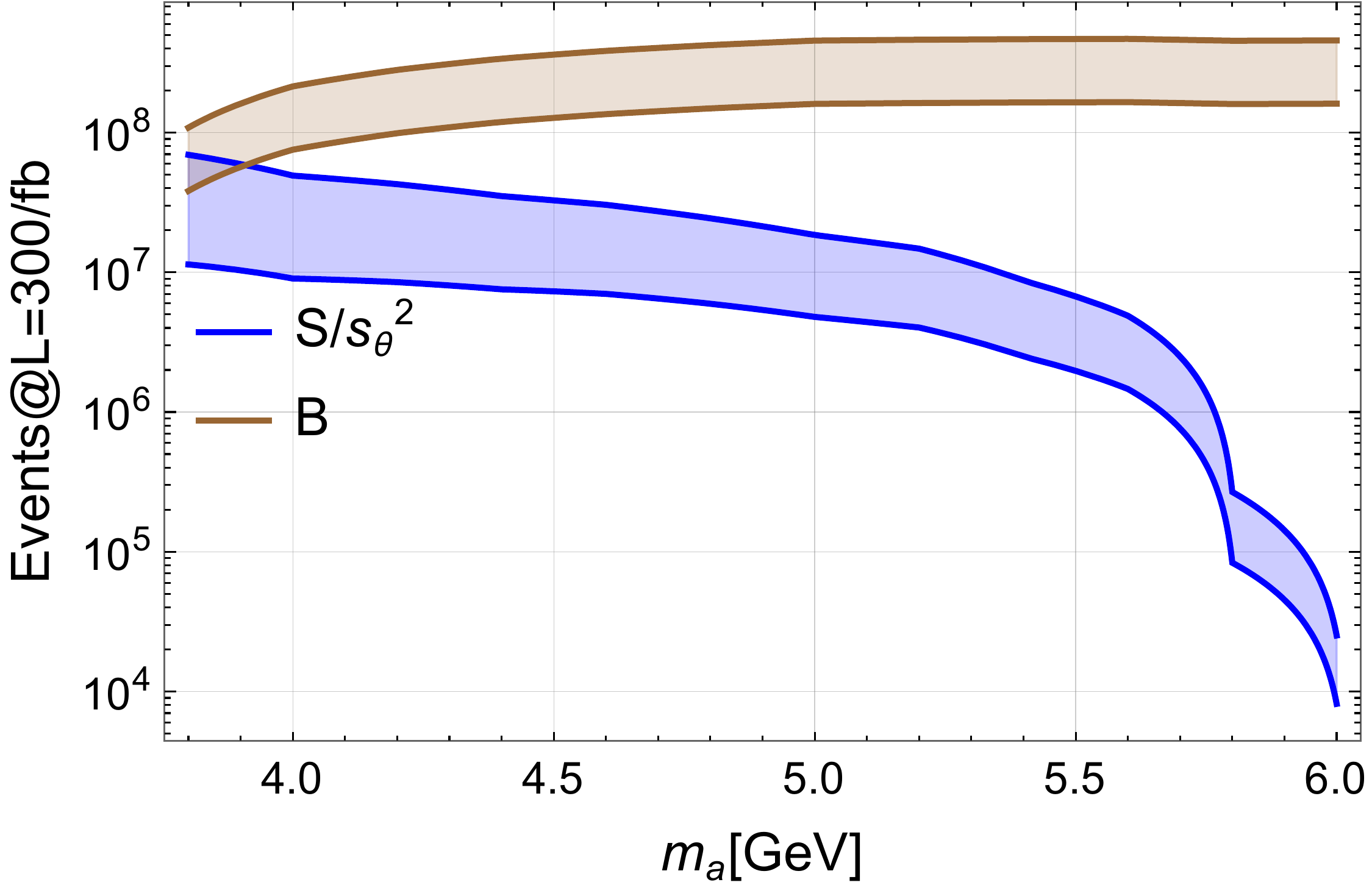}\quad\qquad
\includegraphics[width=0.385\textwidth]{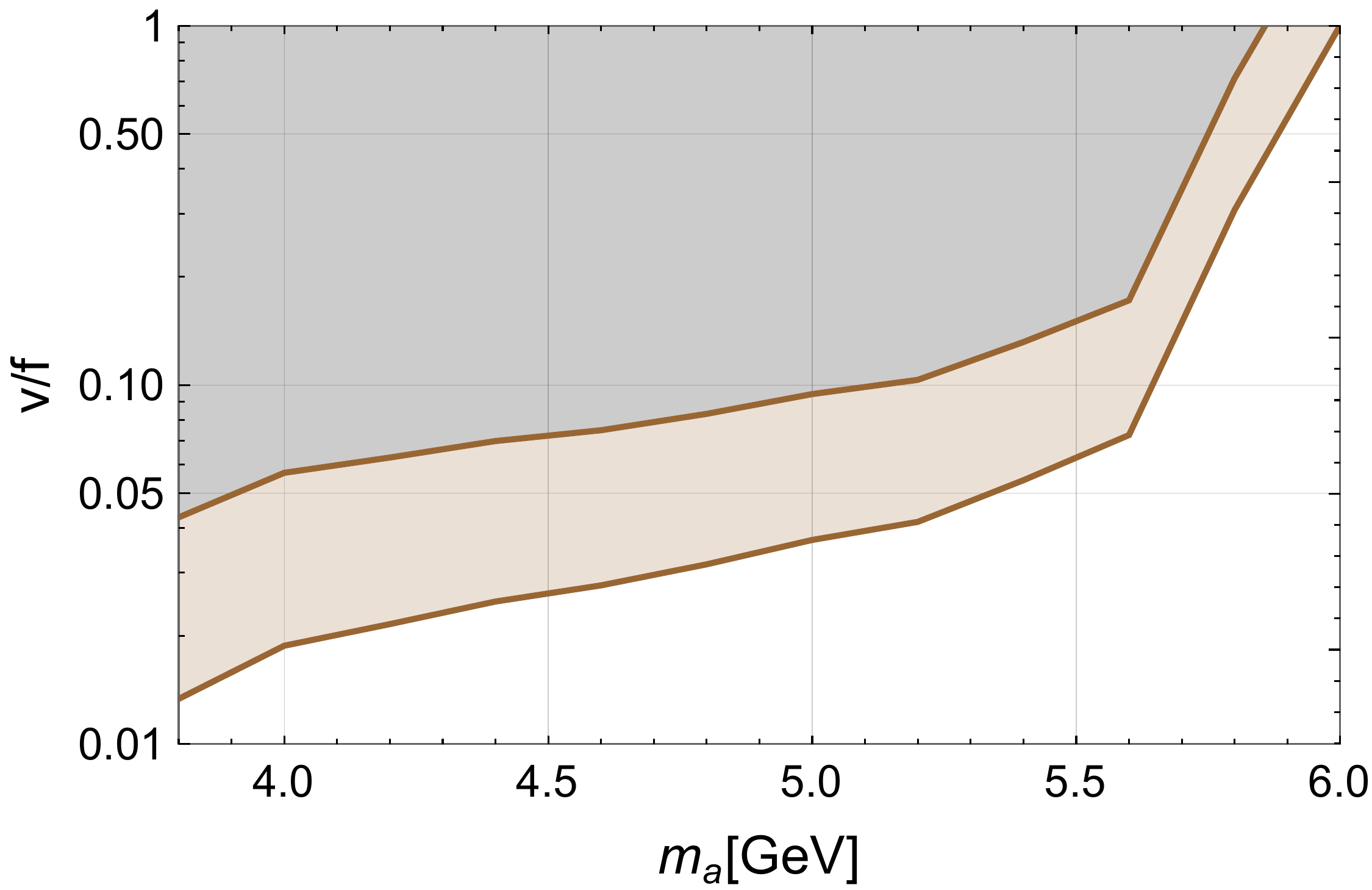}
\caption{Signal (M1) and background expected number of events after selections for 300 fb$^{-1}$. 
The error bands for the background are taken from ~\cite{dEnterria:2016ids} and for the signal it is estimated by a 7-point renormalization and factorization scale dependence.
\emph{Right:} Estimated exclusion region on $v/f$.
}
\label{fig:cc}
\end{figure}
The combination of efficiencies, cross sections (\fig{fig:xs_mumu} for the signal and $\sigma_{SM}=7.1\pm 3.4$mb for the SM~\cite{dEnterria:2016ids}), branching ratios of $a\to c\bar{c}$ and of $D\to K\pi\pi$, and the luminosity, gives us the number of expected events for signal $S$ and background $B$, shown for $L= 300/fb$ on the left panel of \fig{fig:cc}. 
The obtained bounds using $S/\sqrt{B}=2$  on $v/f$ for M1 are shown in \fig{fig:cc}(right). The brown band represents the uncertainty on the exclusion. 

It can be noticed that LHCb can play an important role in looking for pseudo-scalar particles at low masses, and that not well explored decay channels like $a\to c\bar{c}$ can be relevant. In our future publication we will show that also the $a\to \tau^+\tau^-$ can give relevant constraints. 

\subsection{Dark Sectors}
\subsubsection{Probing dark sectors with long-lived particles\footnote{Contributed by Ruth Schaefer.}}
\label{sec:ruth}
    We explore the capability of Belle II to detect long lived scalars produced in $B$ meson decays and compare it with LHCb~\cite{Filimonova:2019tuy}. Long-lived scalars could serve as mediators to a dark sector and as such might mediate a new force between the Standard Model and a dark matter candidate. The interaction is described by a Higgs portal Lagrangian
    \begin{align}
    	\mathcal{L}&= \mathcal{L}_{SM} - \frac{1}{2} m_{\phi}^2\phi^2 - \lambda_3\,|H|^2\phi.
    \end{align}
    Here, we introduce a new scalar particle $\phi$ that is a singlet under the Standard Model gauge group and couples to the Standard Model only via the Higgs boson. After electroweak symmetry breaking, the pre-EWSB Higgs boson $H^0$ and the scalar $\phi$ mix with a mixing angle $\theta$ into the physical scalar $S$ and the observed 125\,GeV Higgs boson $h$. The couplings of these scalars to fermions are then described by
    \begin{align}
        \mathcal{L}_Y &= -\sum_f \frac{m_f}{v}\left(\cos\theta \bar{f}f h + \sin\theta \bar{f}f h\right).
    \end{align}
    At Belle II, we consider decays of the form $B\to K S, S\to f$ where the $B$ meson is either a $B^\pm$, a $B^0$ or $\overline{B^0}$, the $K$ stands for any relevant charged or neutral kaon resonance, and the final state $f$ is a pair of muons, charged pions or kaons, or tau leptons. Decays to hadronic final states can be described by non-perturbative methods~\cite{Winkler:2018qyg}, valid up to a scalar mass of about 2\,GeV. The scalar $S$ inherits flavour hierarchical couplings to fermions from the Higgs boson. In $b\to s$ transitions it couples mostly to the top quark in the loop, and its decay is suppressed by the Yukawa couplings of the light decay products.
    
    To estimate the probeable parameter space of this model at Belle II, we calculate the number of decays we can expect to happen within Belle II's Central Drift Chamber (CDC) to be
    \begin{align}
        N_{f\bar{f}} &= N_{B\bar{B}} \times 1.93\,\mathcal{B}(B \to K S) \mathcal{B}(S \to f\bar{f}) \times \frac{1}{2}\int d\vartheta_0\, \frac{\sin\vartheta_0}{d_S}\int dr\,e^{- \tfrac{r}{d_S}},
    \end{align}
    where $N_{B\bar{B}}$ is the number of $B$ mesons expected to be produced at Belle II for a given luminosity, $r$ and $\vartheta_0$ are the radial distance of $S$ from the $e^+e^-$ collision point and its polar momentum direction in the rest frame of the $B$ meson and are used to integrate over the detector volume, and $d_S$ is the decay length of $S$ in the lab frame. The factor 1.93 takes into account the effect of the different lifetimes of the charged and neutral $B$ mesons at Belle~II. 
    
\begin{figure}[t]
        \centering
        \includegraphics[width=0.57\textwidth]{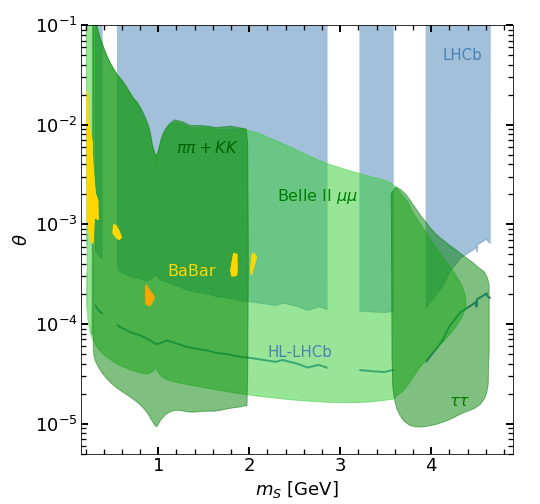}
        \caption{The projected region in mass $m_S$ and mixing angle $\theta$ that Belle II will be able to probe (green), compared with the regions excluded by BaBar \cite{Lees:2015rxq} (yellow, orange) and LHCb \cite{Aaij:2016qsm} (blue) as well as projections for LHCb at the High Luminosity LHC \cite{Bondarenko:2019vrb} (blue line). Figure take from Ref.~\cite{Filimonova:2019tuy}.}
        \label{ruth:fig:result}
    \end{figure}

    In figure \ref{ruth:fig:result}, we show the sensitivity of different flavour experiments to dark scalars. For Belle II, the experimental reach with 50\,ab$^{-1}$ of data corresponds to $N_{f\bar f}>3$, i.e., at least 3 such events occur for the three final states mentioned before. In comparison to the Belle II projections, we show the current \cite{Aaij:2016qsm} and projected \cite{Bondarenko:2019vrb} reach of scalar decays with prompt and displaced muons at LHCb, as well as BaBar results \cite{Lees:2015rxq} for displaced muons (yellow) and pions (orange).
    
    The decay of $S$ to mesons is enhanced around 1\,GeV because of the scalar resonance $f_0(980)$ \cite{Winkler:2018qyg}, making pion or kaon pairs a much stronger probe than muons in that region. At large masses of $S$, the decay to tau leptons is greatly enhanced due to their larger Yukawa coupling. $D$ meson pairs would work in a similar fashion, but the decay rates are much harder to predict due to the presence of charmonium resonances, resulting in a higher uncertainty on all decay modes for scalar masses above the $D\bar{D}$ production threshold.
    
    Compared with LHCb, the greater projected reach of Belle II comes mostly from the fact that the $B$ mesons (and thus the scalars that are produced in their decays) are much less boosted at Belle II. This gives the scalars more time to decay within the detector, thus probing longer lifetimes and smaller couplings. An additional advantage of Belle II is the cleaner environment of an $e^+e^-$ collider, where it is possible to detect pairs of pions, kaons and tau leptons in addition to muons. It is also important to note that the reach is greatly enhanced by considering all relevant kaon resonances in $B\to K$ decays. Compared with the inclusive search at BaBar, the full reconstruction of these kaons from their decay products at Belle II significantly increases the sensitivity.
    
    Similar to Belle II, the sensitivity to dark scalars at LHCb will benefit from detecting final states other than muons. In particular, final states with displaced vertices of pion or tau lepton pairs promise greater sensitivities due to their enhanced couplings to the dark scalar. While the reconstruction of tau leptons poses a challenge for LHCb, existing searches for $B^0_{s/d}\to\tau^+\tau^-$ \cite{Aaij:2017xqt} and $B^+\to K^+\mu^-\tau^+$ \cite{Aaij:2020mqb} suggest that searches for $S\to\tau^+\tau^-$ are possible. Extending the reach to dark scalars with a longer lifetime is another very interesting new research direction. In the current LHCb search for $B \to K \mu^+\mu^-$~\cite{Aaij:2016qsm}, displaced vertices are only detected within the vertex locator (VELO). By searching for downstream tracks in the outer tracking chambers, LHCb could detect long-lived particles decaying at a larger distance from the primary vertex. This would increase LHCb’s sensitivity to scalars with longer lifetimes, thereby probing smaller couplings.

\subsubsection{LHC probes of co-scattering dark matter\footnote{Contributed by Susanne Westhoff.}}
\label{sec:coscattering}

\begin{figure}[t!]
    \centering
    \raisebox{0.6cm}{\includegraphics[width=0.45\textwidth]{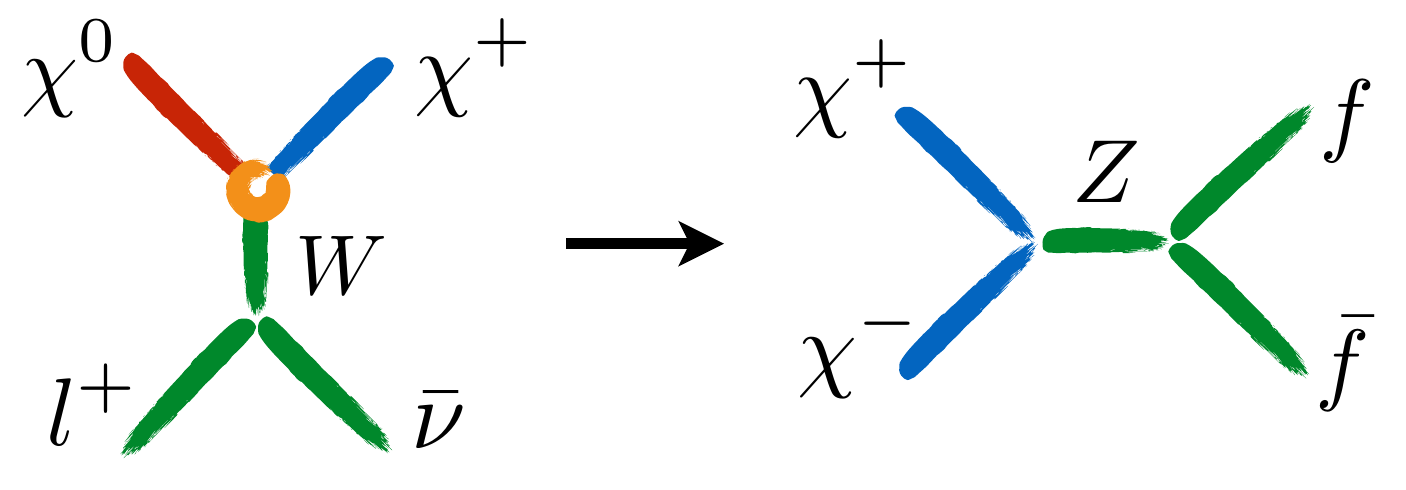}}\hspace*{1.4cm} \includegraphics[width=0.3\textwidth]{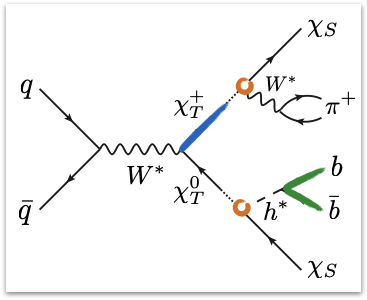}
    \caption{Left: Dark matter freeze-out through co-scattering in the early universe. Right: Dark matter produced in association with soft displaced b-jets at the LHC.}
    \label{fig1}
\end{figure}

\noindent Feebly interacting dark matter has very distinct features from a thermal WIMP, affecting its thermal history just as much as laboratory searches. Here we focus on dark matter candidates around the weak scale with couplings to visible matter in the range
\begin{align}\label{eq:coupling}
10^{-3} > g_\chi > 10^{-8}\,.
\end{align}
In such scenarios, dark matter freeze-out is often driven by co-annihilation or co-scattering off a dark partner particle~\cite{DAgnolo:2017dbv,Garny:2017rxs}. An example is sketched in Fig.~\ref{fig1}, left: Fermion dark matter, $\chi^0$, co-scatters to produce a heavier dark partner, $\chi^+$, which pair-annihilates and reduces the dark matter abundance.
 The relative number density of $\chi^\pm$ and $\chi^0$ around the freeze-out temperature $T_f$ scales exponentially with the mass difference
\begin{align}\label{eq:number-densities}
\frac{n_c}{n_0} \sim \exp(- \Delta m/T_f)\,,\qquad \Delta m = m_c - m_0\,.
\end{align}
Efficient co-scattering at freeze-out typically requires~\cite{Bharucha:2018pfu,Filimonova:2018qdc}
\begin{align}\label{eq:deltam}
   10\gev < \Delta m < 40\gev \ll m_c\,,
\end{align}
predicting a compressed spectrum of dark particles. The decay of the dark partners, $\chi^\pm \to \chi^0 X$, is doubly suppressed: first, by the small mass difference $\Delta m$ and second, by the small coupling $g_\chi$. The proper decay length scales as
\begin{align}
c\tau_c \sim \frac{1}{g_\chi^2}\left(\frac{m_c}{\Delta m}\right)^n,
\end{align}
where $g_\chi$ and $\Delta m$ are set by the observed relic dark matter abundance and $n$ is a model-dependent number. Remarkably, co-scattering predicts macroscopic decay lengths that are accessible at current collider experiments. In general, we expect to see \emph{displaced soft leptons or jets} from suppressed decays of dark partners.

To quantify the search potential of the LHC, we consider a concrete model for co-scattering dark matter with SM-singlet fermion $\chi_S^0$ and a weak triplet fermion field $\chi_T$~\cite{Filimonova:2018qdc}, similar to the wino-bino scenario in the MSSM~\cite{ArkaniHamed:2006mb}. Typical signatures are a displaced vertex of b-jets (see Fig.~\ref{fig1}, right) or displaced leptons or jets,
\begin{align}\label{eq:signal-process}
    pp & \to W^\ast \to \chi_T^+\chi_T^0 \to (\chi_S^0\pi^+)(\chi_S^0 b\bar b)\,,\\\nonumber
    pp & \to Z^\ast / \gamma^\ast \to \chi_T^+\chi_T^- \to (\chi_S^0\ell^+\nu)(\chi_S^0\ell^-\bar{\nu})\,,\ (\chi_S^0\,\rm{jet(s)})(\chi_S^0\,\rm{jet(s)})\,,
\end{align}
both in association with missing energy~\cite{Nagata:2015pra,Bharucha:2018pfu,Filimonova:2018qdc}. A third interesting signature is to observe the dark partner directly as a disappearing track or as a track that decays (semi-)hadronically and leaves a displaced vertex.

At ATLAS and CMS, the main challenge to detect these signatures is the softness of the leptons or jets. Current displaced lepton searches rely on triggers with a transverse momentum cut of $p_T(\ell) > 40\gev$~\cite{Khachatryan:2014mea}. For smaller transverse momenta, the soft-lepton background from meson decays steeply increases, ultimately overwhelming the trigger capacities. New triggers are required to allow to detect soft displaced leptons, for instance by implementing an additional selection criterion based on missing energy. A dedicated simulation with a lower cut of $p_T(\ell) > 20\gev$ shows that CMS has the potential to detect soft leptons from dark partners with decay lengths $2\,\text{mm} < c\tau_c < 1\,\text{m}$ in Run-2 data~\cite{Blekman:2020hwr}.
 For short lifetimes and/or high masses of the dark partner, the sensitivity is limited by the large background from (semi-)leptonic meson decays. For long lifetimes, i.e., large displacements, the sensitivity is limited by the geometric acceptance of the CMS vertex detector. Dedicated searches for soft displaced b-jet pairs have not been performed at ATLAS or CMS so far.
LHCb has the potential to overcome these limitations in several ways. Here we list the main features that can increase the sensitivity to signals with soft displaced leptons or displaced vertices of $b$-jets.
\begin{itemize}
\item Thanks to the efficient trigger for soft tracks, LHCb is sensitive to particles with very low momenta.
\item In searches for soft displaced leptons, background from (semi-)leptonic meson decays can be reduced by vetoeing hadronic activity around the leptons.
\item Due to the different detector geometry, LHCb could be sensitive to different decay lengths of the dark partners than ATLAS and CMS. 
\end{itemize}
To assess LHCb's sensitivity to soft displaced particles, we suggest to study the impact of each of these features for the dark matter signatures outlined above.

Due to the comparably small acceptance at LHCb, total event rates are lower than at ATLAS and CMS. Whether the sensitivity is statistically or systematically limited can only be addressed in concrete models. In our benchmark model, the dark partners decay through weak interactions. For b-jet pairs, the branching ratio can be near 100\%; for leptons, the branching ratio ranges around $1/9$ per flavor. In many scenarios with hidden sectors that directly couple to leptons, the dark particles typically decay to $100\%$ into leptons. Examples are feebly coupled leptophilic dark matter~\cite{Fox:2008kb,Freitas:2014jla,Evans:2016zau,Barducci:2018esg,Junius:2019dci} or models with heavy neutral leptons~\cite{Cai:2017mow,Drewes:2019fou}. Soft displaced leptons or jets can also arise in decay chains involving tau decays, for instance in models with pair-produced supersymmetric staus~\cite{Evans:2016zau}. While we have focused here on compressed spectra, the main features apply to such scenarios as well. By searching for soft displaced particles, LHCb can explore an entire class of hidden sectors and gain important information on the nature of weak-scale dark matter.

\subsection{Dark Matter and Baryogenesis}
\subsubsection{Mesogenesis: Baryogenesis and Dark Matter from Mesons\footnote{Contributed by Gonzalo Alonso-Álvarez, Gilly Elor and Miguel Escudero.
}}
\label{sec:mesogenesis}
\newcommand{\Br}{\text{Br} \left( B \to \psi + \mathcal{B}+\mathcal{M} \right)}
\newcommand{\bea}{\begin{eqnarray}}
\newcommand{\eea}{\end{eqnarray}}

\newcommand{\GE}[1]{{\color{red}{[GE: #1]}}}
\newcommand{\GEc}[1]{{\color{red}{ #1}}}
\newcommand{\ME}[1]{{\bf \color{blue}{[ME: #1]}}}
\newcommand{\GA}[1]{{\bf \color{magenta}{[GA: #1]}}}

Although many baryogenesis mechanisms have been proposed, the vast majority of them remain experimentally challenging to verify due to the inaccessibly high scales and very massive particles involved.
This has led to the common lore that there is little hope to test baryogenesis experimentally.
That said, it has recently been proposed that low scale testable baryogenesis and dark matter production may be achieved by leveraging the CP violation in Standard Model meson systems~\cite{Elor:2018twp,Elor:2020tkc} -- a paradigm known as \textit{Mesogenesis}. 

In particular, the scenario of~\cite{Elor:2018twp}: $B$-Mesogenesis, predicts distinct signals at $B$ factories and hadron collider experiments such as LHCb.
The two most characteristic features of $B$-Mesogenesis are: \textit{i)} a large positive enhancement of the semileptonic-leptonic asymmetries in neutral $B$ meson decays, and \textit{ii)} the presence of a partially invisible and seemingly baryon number-violating decay mode of $b$-flavoured hadrons with a branching fraction $> 10^{-4}$. 
As is thoroughly investigated in~\cite{Alonso-Alvarez:2020}, existing collider experiments will be able to conclusively assess the viability of this paradigm.
 
\subsubsection*{Cosmological Dynamics and Particle Physics Model}
In the early Universe, $B$-Mesogenesis~\cite{Elor:2018twp} relies on the existence of an out-of-equilibrium population of $b$ and $\bar{b}$ quarks.
These are assumed to originate from the decay of a inflaton-like or modulus scalar field $\Phi$ with a mass $M_\Phi\gtrsim 11$~GeV.
If the $b$ quarks are produced at temperatures $T_R \sim 5-100 \, \text{MeV}$, i.e. after the QCD phase transition but before BBN, they quickly hadronize producing charged and neutral $B$ mesons.
The produced $B^0_{s,d}$ mesons  undergo particle anti-particle oscillations before subsequently decaying. 
The relevant effect for our purposes is
CP violation in mixing, which can be measured via the \emph{semileptonic CP asymmetries} (also known as CP asymmetries in flavor-specific decays).
The asymmetries are defined as
\begin{equation}\label{eq:semileptonic}
A^q_{\mathrm{sl}} = \frac{\Gamma(\bar{B}_q^0\to \ell^+ \, X) - \Gamma(B_q^0\to \ell^- X)}{\Gamma(\bar{B}_q^0\to \ell^+ \, X)+\Gamma(B_q^0\to \ell^- X)} =  -\left| \frac{\Gamma_{12}^q}{M_{12}^q} \right| \sin \phi^q_{12}\,,
\end{equation}
The matrix elements $\Gamma_{12}^q$ and $M_{12}^q$ can be related to the physical width and mass differences as $\Delta \Gamma_q = 2|\Gamma_{12}^q| \cos \phi_{12}^q$ and $\Delta M_q = 2|M_{12}^q|$.
A sufficiently large and \textit{positive} semileptonic asymmetry provides the necessary CP violation required to generate the baryon asymmetry of the Universe.

\begin{figure}[t]
\centering
\includegraphics[width=0.72\linewidth]{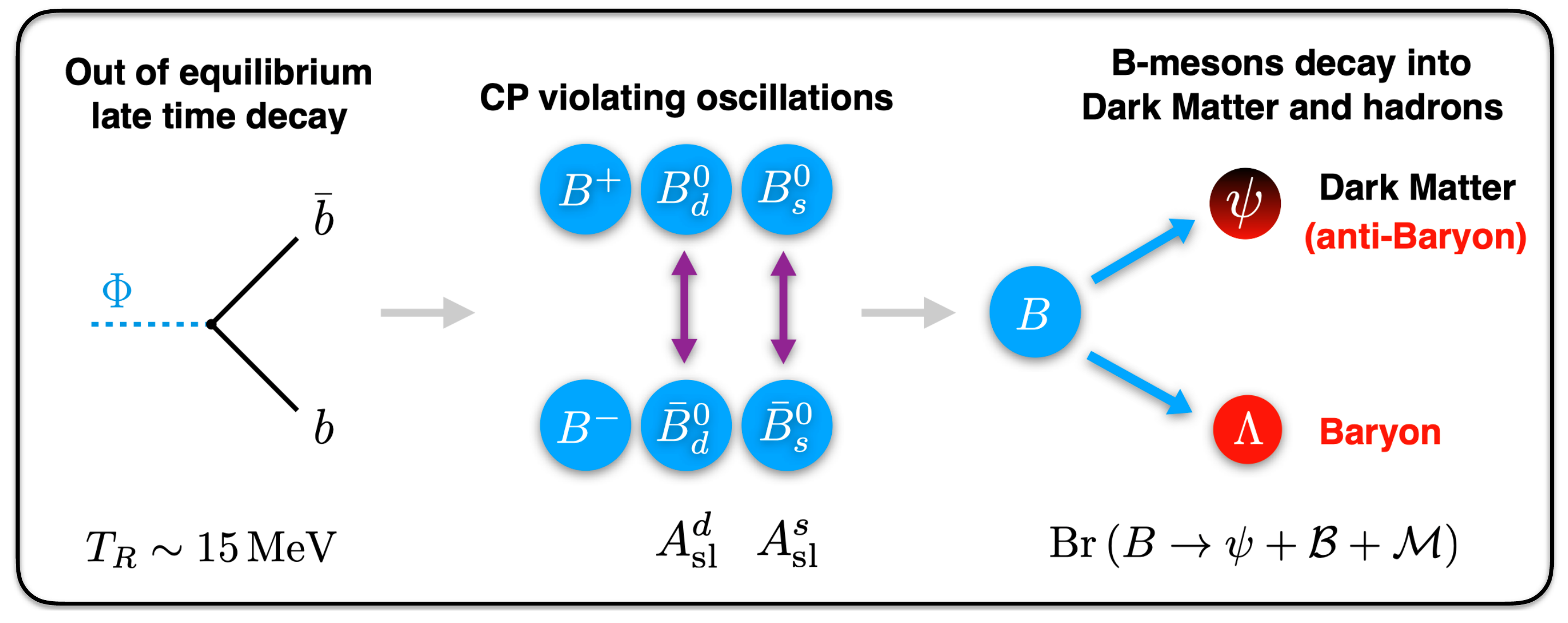} 
\vspace{-0.1cm}
\caption{
Depiction of the $B$-Mesogenesis mechanism (adapted from~\cite{Elor:2018twp}, see also~\cite{Alonso-Alvarez:2020}). }
\label{fig:cartoon}
\end{figure}

Rather than invoking baryon number violation, in this model the \emph{visible} baryon number asymmetry is exactly compensated by an \emph{invisible} one in the dark sector.
This way, global baryon number is exactly conserved and, as a consequence, the DM abundance becomes linked to the baryon asymmetry of the Universe.
In order to transmit a baryon asymmetry to the dark sector, a fermion $\psi$ which is neutral under the SM gauge group but carries one unit of baryon number is introduced.
Interactions between this dark fermion and the SM are mediated by a  heavy colored triplet scalar baryon $Y$ with SM quantum numbers $(3,1,-1/3)$ or $(3,1,2/3)$. This particle can arise in Supersymmetric realizations of the mechanism~\cite{Alonso-Alvarez:2019fym} and leads to an interaction of the type
\bea \label{eq:Y_couplings}
 \mathcal{O}_{ud} =  u \, d\, b \, \psi \,, 
\eea
where $u$ and $d$ are right handed quarks in the mass basis which may come in any flavor combination.
Collider constraints~\cite{Alonso-Alvarez:2020,Sirunyan:2019ctn,Aad:2020aze} on colored scalars force the mass of $Y$ to be at least greater than $M_Y  > 1.2$~TeV, and so it can be integrated out generating the four-fermion effective operator in Eq.~\eqref{eq:Y_couplings}. 
There, we have required one of the quarks to be of $b$-type, while the flavor of the other two is unspecified.
Such an operator can trigger the decay of the $\bar{b}$ quark within the $B$ meson.
The parton-level process $\bar{b} \rightarrow \psi + u + d $ translates into a hadronic decay $B \rightarrow \psi \, + \mathcal{B} + \mathcal{M}$, where $\mathcal{B}$ is a single SM baryon and $\mathcal{M}$ stands for any number of accompanying mesons.
Kinematic considerations restrict the mass of $\psi$, and consequently the missing energy, to lie in the $m_\psi \in 0.94-4.34$ GeV range.
This decay, occurring after a CP asymmetry has been generated via $B^0-\bar{B}^0$ oscillations, produces an equal and opposite baryon asymmetry between the SM and the dark sectors.

The early Universe dynamics and particle content described in the previous paragraphs are summarized in Fig.~\ref{fig:cartoon}.
The resulting baryon asymmetry is obtained by numerically solving the set of coupled Boltzmann equations \cite{Elor:2018twp,Aitken:2017wie,Nelson:2019fln} governing the evolution of the number densities of the relevant particles $\Phi$, $B_q^0$, $\bar{B}_q^0$ and the DM.
The final yield factorizes into the form
\begin{equation}
\label{eq:Y3p2}
\!\!\!\! Y_B = Y_B^{\rm obs} \left( \frac{\mathrm{Br}(B^0\rightarrow\mathcal{B}+\mathrm{DM}+\mathcal{M})}{10^{-3}} \right) \left( \frac{11\,\mathrm{GeV}}{M_\Phi} \right) \left( \alpha_d(T_{R})\,\frac{A^d_{\rm SL}}{10^{-3}} + \alpha_s(T_{R})\,\frac{A^s_{\rm SL}}{10^{-3}} \right) \,,
\end{equation}
where $0 \leq \alpha_{d,s} \leq 1.4$ are temperature-dependent functions that peak at a temperature $T_R\sim 15$~MeV, see~\cite{Alonso-Alvarez:2020}.
As discussed above, the asymmetry is directly related to the semileptonic asymmetries in the neutral $B$ meson systems and the exotic decay of $B$ mesons into a final state with a single SM baryon and missing energy.
Crucially, both these quantities are experimental observables that can be searched for at $e^+ e^-$ and hadron colliders.

\subsubsection{Collider Implications of Baryogenesis and DM from B Mesons\footnote{Contributed by Gonzalo Alonso-Álvarez, Gilly Elor and Miguel Escudero.
}}
As highlighted in Eq.~\eqref{eq:Y3p2}, there are two clear predictions of the $B$-Mesogenesis mechanism that can be thoroughly tested in existing and upcoming experiments:
\begin{enumerate}
    \item A large ($A_{\rm sl}^q > 10^{-5}$) and positive semileptonic asymmetry in the $B^0_{s,d}$ systems.
    \item A large ($> 10^{-4}$) inclusive branching fraction for the exotic process $B \rightarrow \psi + \mathcal{B} + \mathcal{M}$.
\end{enumerate}
The quantitative prediction is that the product of these two quantities should be at the $ 10^{-6}$ level.
For instance, for the value $A_{\rm sl}^s \simeq 2 \times 10^{-5}$ expected in the SM~\cite{Lenz:2019lvd}, the mechanism works provided an exotic branching fraction $> 2.5\,\%$.
These large branching fractions are still experimentally allowed due to the lack of dedicated searches for baryon number violating $b$-hadron decays.
The branching fraction could, however, be as low as $10^{-4}$ if new physics were to enhance the value of the semileptonic asymmetries to saturate experimental constraints, which currently sit at the $10^{-4}$ level.
LHCb can greatly contribute to experimentally clarify this situation, as we detail in what follows.

\subsubsection*{CP Violation in the $B_q^0$ meson system}

\noindent\textit{Semileptonic Asymmetries.}
In the left panel of Fig.~\ref{fig:CP_violation}, the region of the $A_{\rm sl}^d$-$A_{\rm sl}^s$ plane where baryogenesis can be achieved is highlighted in red.
Current direct measurements of the semileptonic asymmetries imply that $\Br \gtrsim 10^{-4}$, while global fits including all relevant CPV data suggests $\Br \gtrsim 10^{-3}$.
Here, we expect LHCb to play a fundamental role in the future, as after the Upgrade II ($300\,\text{fb}^{-1}$) sensitivities are forecasted to reach $\delta A_{\rm sl}^d \sim 3\times 10^{-4} $ and $\delta A_{\rm sl}^s \sim 2\times 10^{-4}$~\cite{Cerri:2018ypt}.
Measurements consistent with $A_{\rm sl}^q<0$ would further constrain the range of $\Br$, while a preference for positive $A_{sl}$ could be seen as a clear indication of $B$-Mesogenesis.

\medskip
\noindent\textit{CP violation in $b\to c\bar{c} s$ decays.}
As showcased in Eq.~\eqref{eq:semileptonic}, the sign of the semileptonic asymmetries is dictated by the phase $\phi_{12}^q$, which thus plays a fundamental role in determining the sign of the baryon asymmetry.
Importantly, this phase can be accessed with measurements in processes with interference between $B_q^0$ oscillations and decay, such as $B_s^0 \to J/\psi \, \phi$.
The case of $\phi_{12}^s$ is interesting as penguin diagrams are subleading and $\phi_{12}^s\simeq \phi_s^{c\bar{c} s} \pm 0.01$~\cite{Barel:2020jvf}.
Thus, the sign and magnitude of $\phi_{s}^{c\bar{c}s}$ can be important for probing the mechanism.
This is explicitly showcased in the right panel of Fig.~\ref{fig:CP_violation}, where the relevant measurements are shown in the $\Delta \Gamma_s$-$\phi_s^{c\bar{c} s}$ plane.
The red-shaded region corresponds to $\phi_s^{c \bar{c} s} < 0$, as is required to generate a positive $A_{\rm sl}^s$.
The expected reach of LHCb is in this regard promising: statistical sensitivities of $\sim 0.003$ in $\phi_s^{c\bar{c}s}$ are projected with $300\,\text{fb}^{-1}$~\cite{Cerri:2018ypt}.

\subsubsection*{Exotic $b$-flavored Hadron Decays}
\begin{figure*}[t]
\centering
\begin{tabular}{cc}
		\label{fig:asl}
		\hspace{-0.4cm}\includegraphics[width=0.45\textwidth]{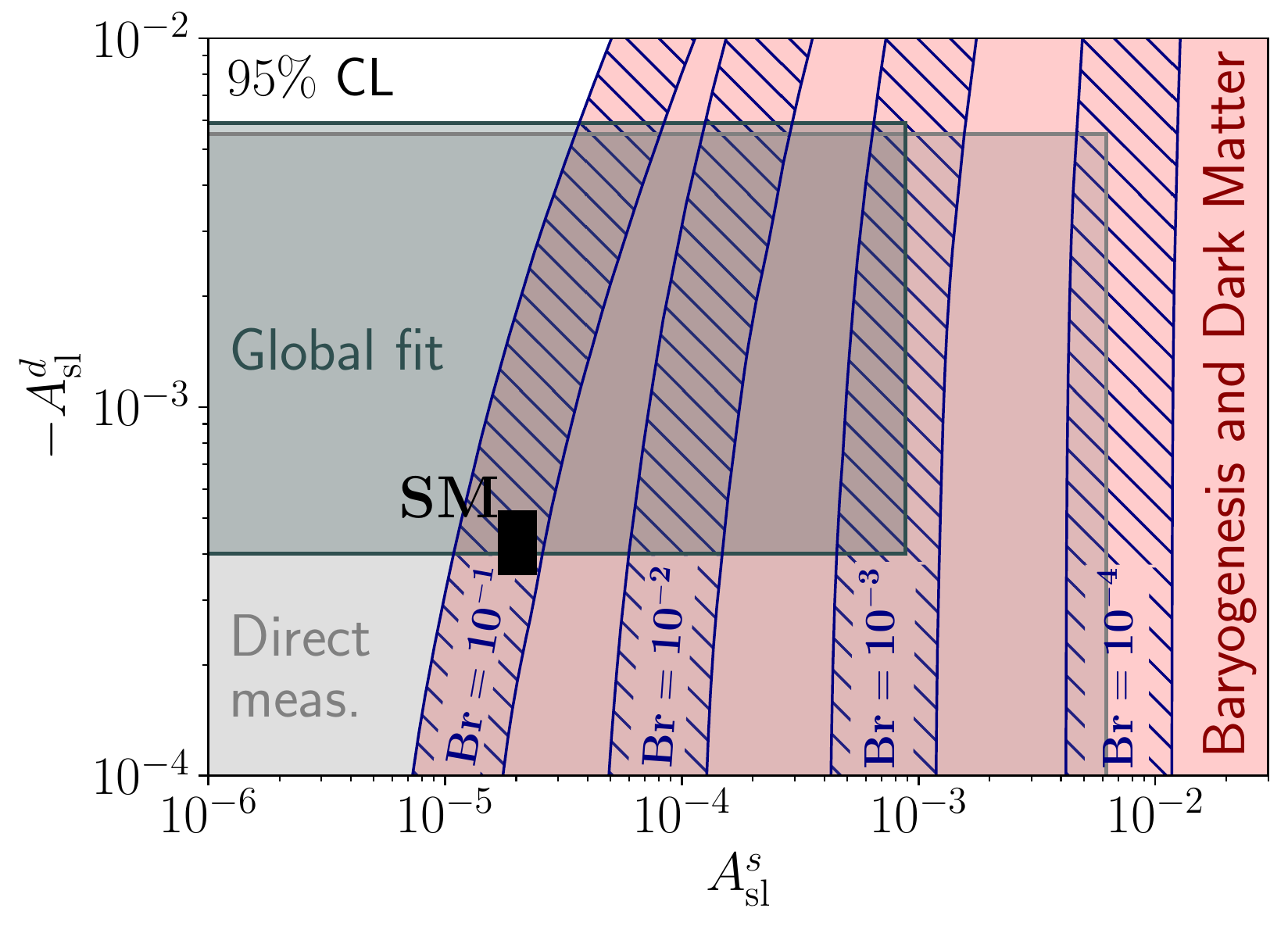}
&
		\label{fig:DeltaGamma}
		\includegraphics[width=0.45\textwidth]{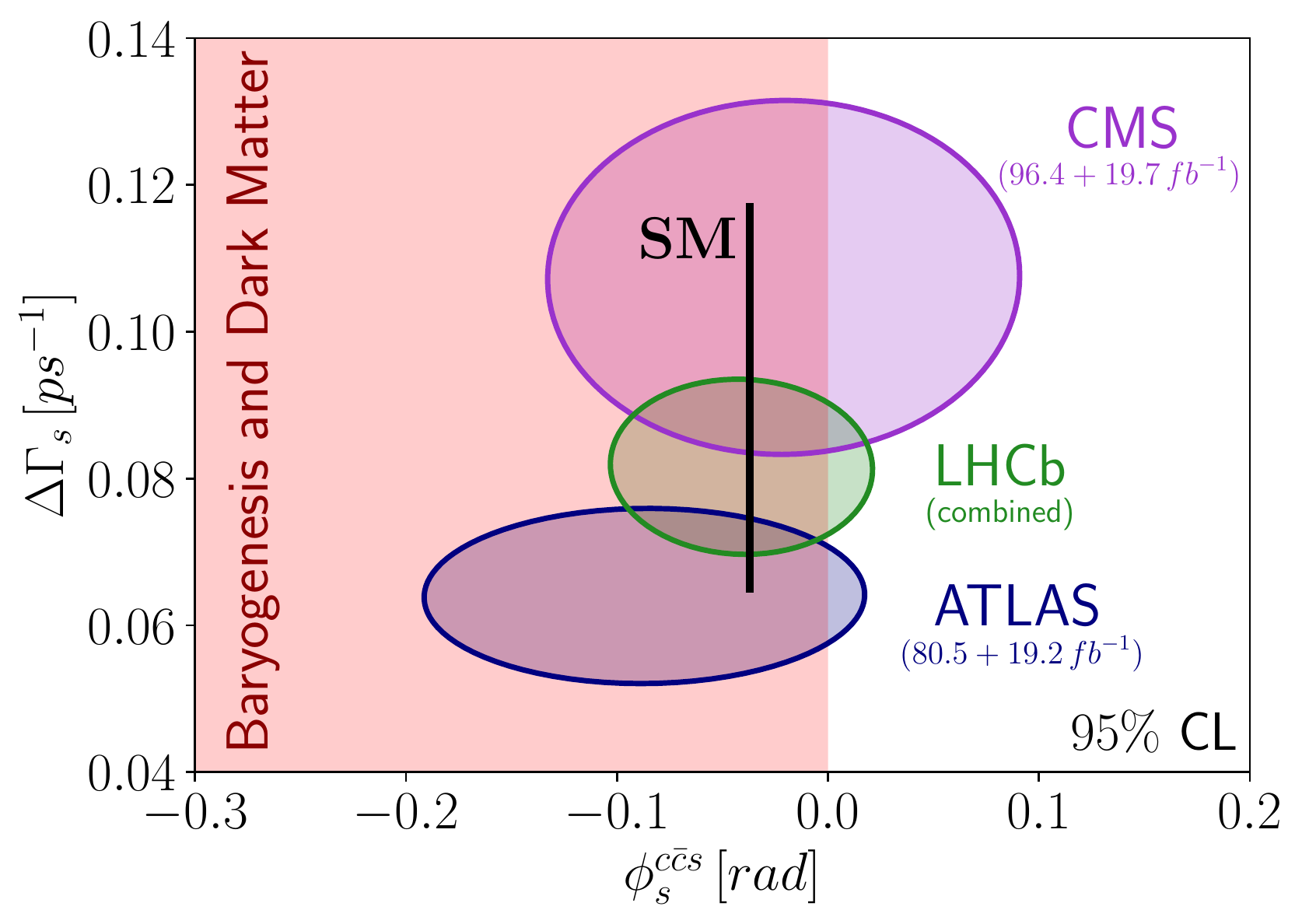}
\end{tabular}
\vspace{-0.4cm}
\caption{\textit{Left panel:} Parameter space in the $A_{\rm sl}^d$-$A_{\rm sl}^s$ plane. We show SM predictions from~\cite{Lenz:2019lvd}, current world averages~\cite{pdg}, and results from global fits (\href{http://www.utfit.org/UTfit/ResultsSummer2018NP}{UTfit})~\cite{Bona:2007vi}. \textit{Right panel:} Experimental landscape in the $\Delta \Gamma_s$-$\phi_s^{c\bar{c}s}$ plane.
Current measurements from LHCb~\cite{Aaij:2019vot}, ATLAS~\cite{Aad:2020jfw}, and CMS~\cite{Sirunyan:2020vke} are shown.
The regions highlighted in red allow for successful baryogenesis and dark matter production as described in the text.}
\label{fig:CP_violation}
\end{figure*}

The relevant branching fraction for $B$-Mesogenesis in Eq.~\ref{eq:Y3p2} and Fig.~\ref{fig:CP_violation} is an inclusive rate containing contributions from all possible channels.
At present, the current bound on such inclusive branching fraction is $\lesssim 10\,\%$ and arises from inclusive measurements of $B$ meson decays~\cite{Alonso-Alvarez:2020,pdg}.
However, to date no searches on exclusive channels such as $B\to \psi + \mathcal{B}$ have been performed.
Considering the sensitivities in similar processes involving missing energy like $B \to K \nu \bar{\nu}$~\cite{Lees:2013kla,Lutz:2013ftz,Lees:2019lme}, we expect $B$-factories to be able to probe branching ratios down to $\sim 10^{-5}$.
Given that recent phase space estimations~\cite{Alonso-Alvarez:2020} suggest a relationship $\text{Br} \left( B \to \psi + \mathcal{B} \right) \gtrsim (0.01-0.1)\, \Br$ between exclusive and inclusive channels, $B$ factories may be able to fully test this mechanism. 

Even though final states containing missing energy are challenging for hadron colliders, there are several avenues that can be followed to constrain $B$-Mesogenesis at LHCb thanks to the exquisite $b$-reconstruction efficiency and the relatively clean signal environment:
\begin{itemize}
    \item One can search for processes such as $B^+ \to \psi + \mathcal{B}^\star$, where $\mathcal{B}^\star$ is an excited baryon that decays promptly~\cite{Brea:2020}. $\mathcal{B}^\star$ would be easier to tag as the decay track will not be displaced. Importantly, the exact mass of $\psi$ would not be fully reconstructed and so one would require a handle on irreducible SM backgrounds such as $B^+ \to \bar{\mathcal{B}} + n$.
    \item A possibility that was recently put forward in~\cite{Poluektov:2019trg} relies on the fact that $B_s^0$ oscillate to use the reconstructed $\mathcal{B}$ baryon in order to have a handle on the $\psi$ mass and new non-standard decay rate. Although Ref.~\cite{Poluektov:2019trg} did not fully study the sensitivity to $\Br$, it is an interesting future possibility.
    \item  $b$-flavored Baryons ($\mathcal{B}_b$) are only produced at hadron colliders and they will also posses a non-standard decay mode, $\mathcal{B}_b \to \psi + \mathcal{M}$, which one may try to search for. This is challenging as it would require knowledge of the initial energy of $\mathcal{B}_b$. Nonetheless, there are studies showing that in principle this could be achieved~\cite{Stone:2014mza}.
\end{itemize}

\subsubsection*{Other indirect signals}

A TeV-scale color-triplet scalar mediator has potential implications for SUSY-like LHC searches involving jets and missing energy~\cite{Alonso-Alvarez:2020,Aitken:2017wie}.
Depending on the flavor structure of the couplings, modifications to neutral meson mixing can be induced, including an enhancement of CP-violating quantities like the $A_{\rm sl}^q$~\cite{Alonso-Alvarez:2019fym,Alonso-Alvarez:2020}.
Interestingly and due to the existence of new light states, modifications of the decay width mixing $\Gamma_{12}$ in $B$ mesons are possible, contrary to what is commonly assumed.
Finally, the exact characterization of the dark sector can lead to DM annihilation or decay signals, or even long-lived decays at collider experiments~\cite{Alonso-Alvarez:2019fym} which may be interesting for CODEX-b~\cite{Gligorov:2017nwh}.

\subsubsection*{CP Violation in $D^\pm$ Decays}
As discussed in the introduction, Mesogenesis is also possible by leveraging the CP violation in other SM meson systems; in $D$-Mesogenesis~\cite{Elor:2020tkc} the CPV of charged $D^\pm$ decays to an odd number of final state charged pions, improved measurement of both the branching fraction and the charge asymmetry in such channels will further constrain or discover this mechanism.

\subsection{Neutrino Masses }
\subsubsection{Heavy neutral leptons from Drell-Yan production\footnote{Contributed by Jan Hajer.}}
\label{sec:hnl1}
{
\makeatletter\DeclareRobustCommand*{\unit}[2][]{\begingroup\def\0{#1}\expandafter\endgroup\ifx\0\@empty\else#1\,\fi\ifthenelse{\boolean{mmode}}{\mathrm{#2}}{#2}}\makeatother

The neutrinos of the SM are the only purely left-handed fermions.
The resulting absence of a mass term contradicts experimental observations.
The most straight forward solution is the inclusion of three right handed neutrinos $\nu_{Ri}$ \cite{Minkowski:1977sc, Mohapatra:1979ia}
\begin{equation}
\mathcal L_{\nu^{}_R} = - {\lambda_{ai}} \overline \ell_a \varepsilon H \nu^{}_{Ri} - \frac{1}{2} \overline{\nu_R^c}_i {M_{ij}} \nu^{}_{Rj} + \text{h.c.} \ ,
\end{equation}
where $\lambda_{ai}$ is a Yukawa coupling and $M_{ij}$ is the Majorana mass.
After electroweak symmetry breaking the Higgs $H$ expectation value $v$ generates the Dirac mass ${m_{ai}} = v \lambda_{ai}$.
These two mass terms generate the light neutrino masses via the seesaw mechanism
\begin{align}
m_\nu &= - m_{ai} M_{ij}^{-1} m_{bj}^T = - \theta_{ai} M_{ij} \theta_{bj}^T \ , &
\theta_{ai} &= m_{aj} M_{ij}^{-1} \ , &
U_{ai}^2 &= |\theta_{ai}|^2 \ , &
U_a^2 &= \sum_i U_{ai}^2 \ .
\end{align}
The heavy neutrino mass eigenstates $N_i$ couple to the heavy SM bosons via
\begin{equation}
\mathcal L \supset - \frac{m_W}{v} \overline N \theta^*_a \gamma^\mu e_{La} W^+_\mu
- \frac{m_Z}{\sqrt 2 v} \overline N \theta^*_a \gamma^\mu \nu^{}_{La} Z_\mu
- \frac{M}{v} \theta_a h \overline{\nu^{}_L}_\alpha N + \text{h.c.} \ .
\end{equation}

A minimal phenomenologically relevant model of this type is the Neutrino Minimal Standard Model ($\nu$MSM) \cite{Asaka:2005pn, Shaposhnikov:2006nn}.
In which, the $B-L$ symmetry of the SM, broken by the Majorana mass term is approximately restored when the $\nu^{}_{Ri}$ form pseudo-Dirac pairs $\nu^{}_{Ri} + \nu_{Rj}^c$ in the limit of small $M_{ij}$.
The matrices of the Majorana mass and the Yukawa coupling take the form
\begin{align}
M_{ij} &= M \operatorname{diag}\left(1-\mu, 1+\mu, \mu^\prime\right) \ , &
\lambda_{ai} &= \begin{pmatrix}
\lambda_e + \epsilon_e & i(\lambda_e-\epsilon_e) & \epsilon_e^\prime \\
\lambda_\mu + \epsilon_\mu & i(\lambda_\mu-\epsilon_\mu) & \epsilon_\mu^\prime \\
\lambda_\tau + \epsilon_\tau & i(\lambda_\tau-\epsilon_\tau) & \epsilon_\tau^\prime \\
\end{pmatrix} \ ,
\end{align}
where $\epsilon$, $\epsilon^\prime$, $\mu$, and $\mu^\prime$ are small $B-L$ violating parameter  \cite{Drewes:2018gkc}.
Hence, the heavy neutrino mass spectrum features an almost mass degenerate pseudo Dirac pair with coupling $\order(\lambda)$ and a lighter $\order(\unit{keV})$ dark matter candidate with coupling $\order(\epsilon^\prime)$.
Furthermore, the $\nu$MSM can explain neutrino oscillation data and baryogenesis via leptogenesis.
In order to ensure comparability between the various experiments it is usually assumed that a single heavy neutrino $N$, called heavy neutral lepton (HNL), coupling to one of the lepton flavours is detectable.
\begin{figure}
\includegraphics[width=0.98\linewidth]{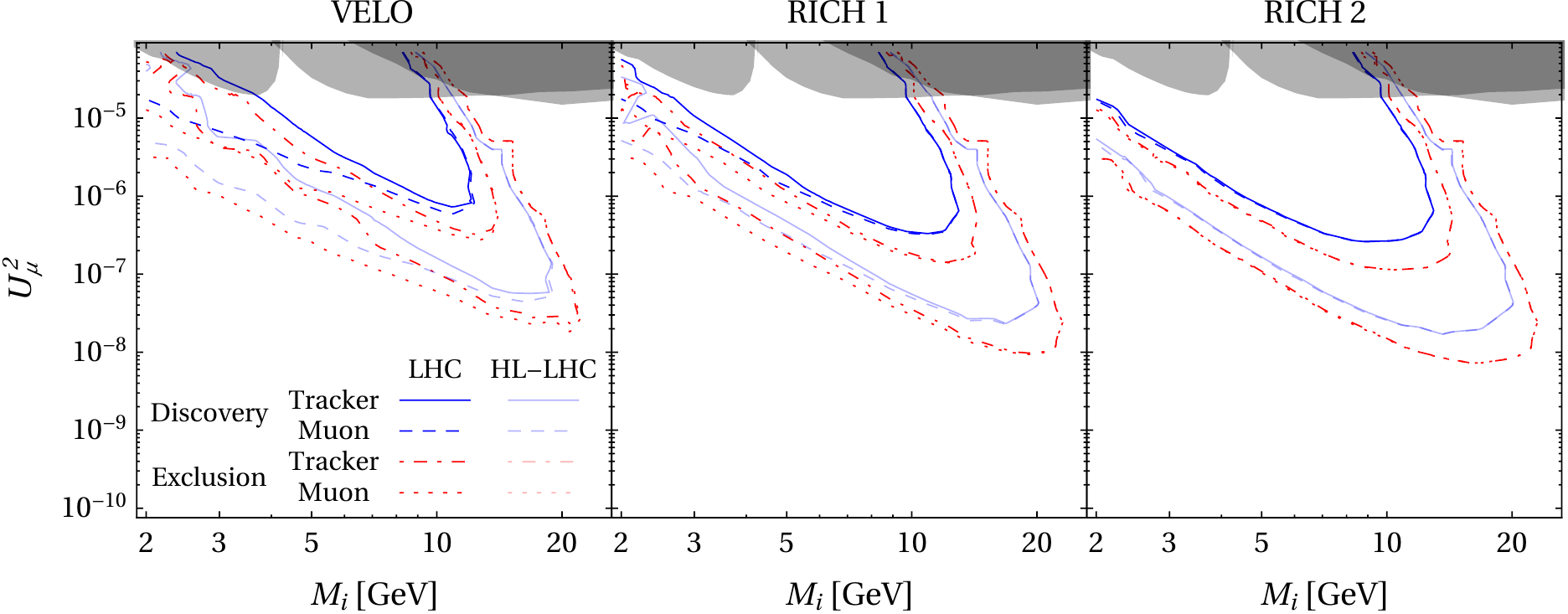}
\caption{
Comparison of the discovery (blue) and exclusions (red) reaches for HNLs with pure muon mixing $U_\mu^2$ at the LHCb detector with \unit[13 and 14]{TeV} after collecting \unit[30 and 380]{fb$^{-1}$} of data.
Here we assume that secondary vertices originating from HNL decays happening within the first half of the VELO or before the RICH 1 or 2 can be reliably identified.
The dashed and dotted curves include muons identified only identified in the muon system.
} \label{fig:HNL comparison of reconstruction efficiencies}
\end{figure}
\begin{figure}
\subfigure[$U_e^2$]{\label{Ue}\includegraphics[width=.3\linewidth]{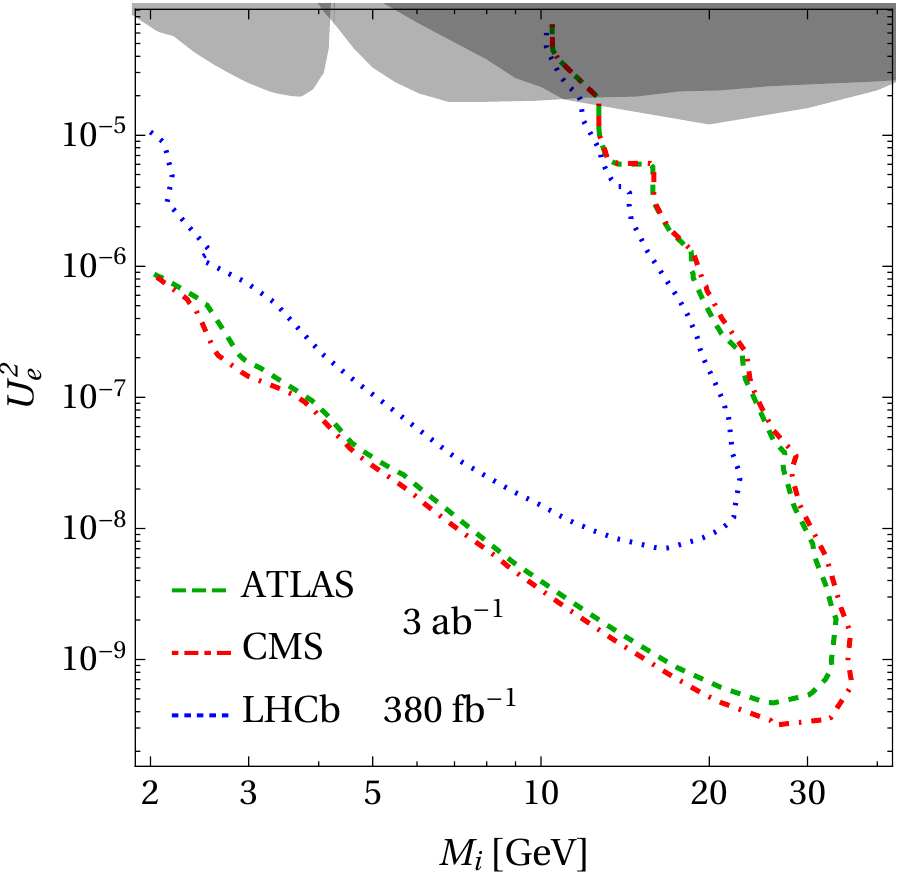}}\hfill
\subfigure[$U_\mu^2$]{\label{Umu}\includegraphics[width=.3\linewidth]{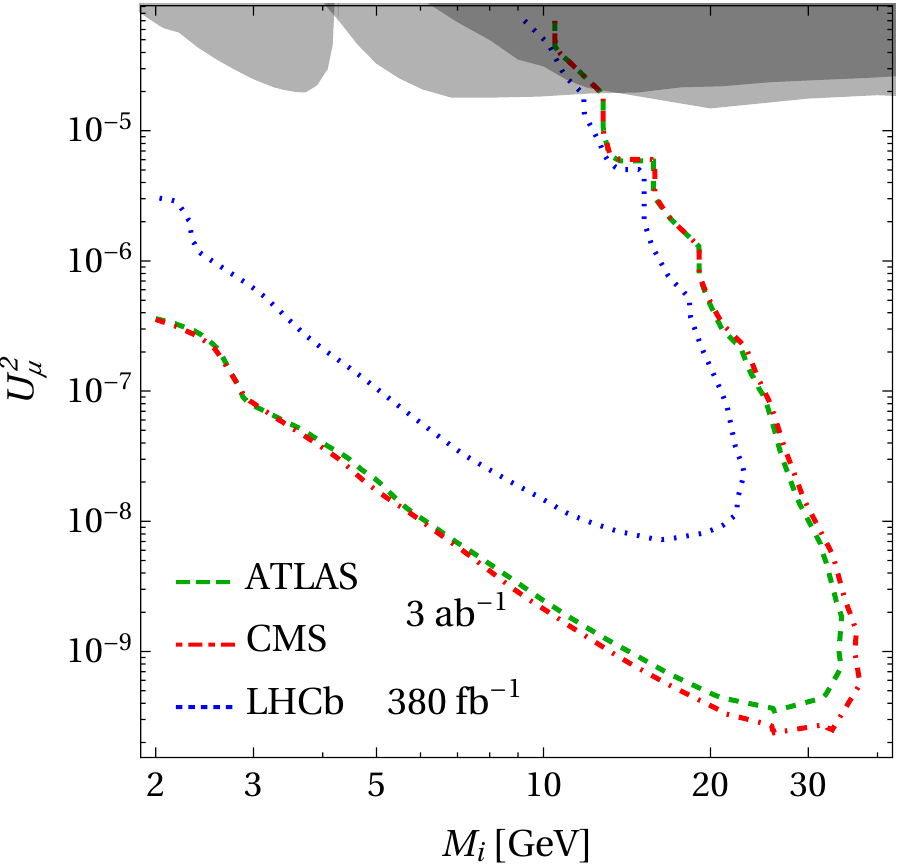}}\hfill
\subfigure[$U_\tau^2$]{\label{Utau}\includegraphics[width=.3\linewidth]{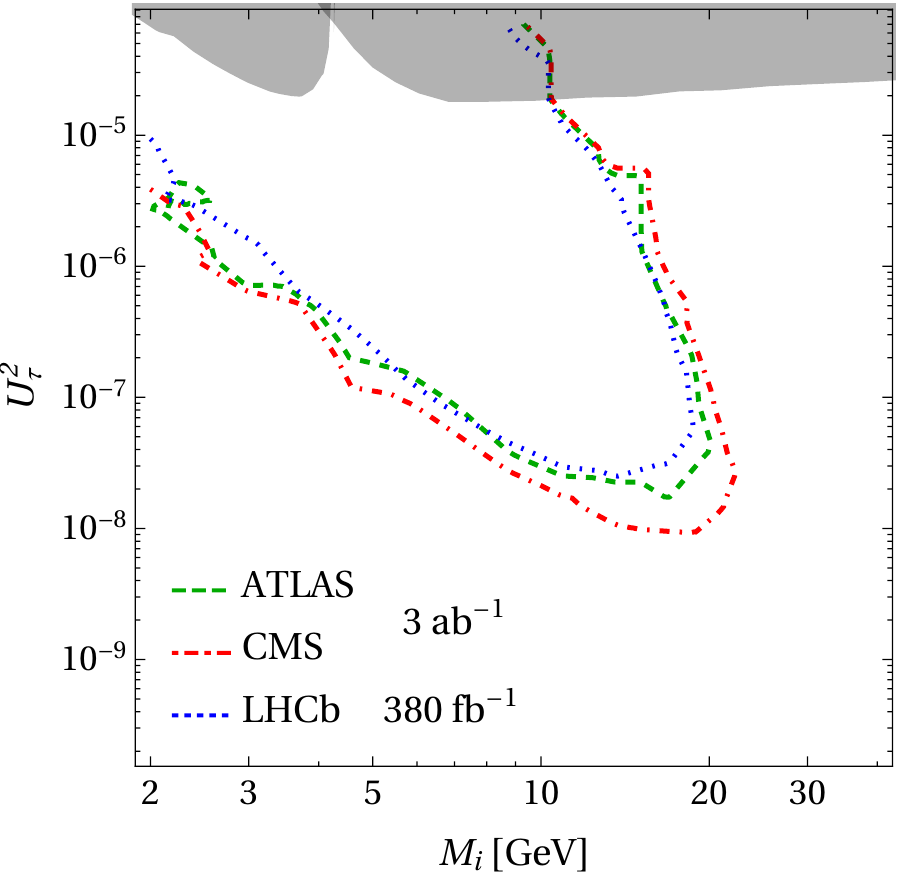}}
\caption{
Comparison of the maximal exclusion reach for HNLs at ATLAS and CMS with \unit[3]{ab$^{-1}$} and LHCb with \unit[380]{fb$^{-1}$} for pure $U_e^2$, $U_\mu^2$, and $U_\tau^2$ mixing in \subref{Ue}, \subref{Umu}, and \subref{Utau}, respectively.
While the larger luminosity collected by ATLAS and CMS results in a better sensitivity compared to LHCb the softer trigger requirements for $\tau$ leptons allow the LHCb detector to gain a sensitivity competitive with ATLAS and CMS for HNLs with pure $U_\tau^2$ mixing.
Here we have assumed \unit[50, 140, and 170]{GeV} as minimal $p_T$ requirement for the single $\tau$ trigger at LHCb, ATLAS and CMS, respectively \cite{ATL-DAQ-PUB-2017-001, Aaij:2017rft}.
} \label{fig:HNL comparison maximal exclusion reach}
\end{figure}

At colliders the HNL can be produced in association with a charged lepton via a $W$-boson.
The decay into one neutral and two oppositely charged leptons is mediated via a $W$- or $Z$-boson.
This signature $pp \to W^\pm \to l^\mp_\text{hard} (N \xrightarrow{\text{displaced}} l^\mp (W^\pm \to l^\pm \nu))$ allows to trigger on the first lepton and search for a secondary vertex \cite{Drewes:2019fou}.
Due to the potentially long lifetime it is crucial to identify displaced vertices as far away from the primary vertex as possible e.g.\ consider muon chamber only muons \cite{Bobrovskyi:2011vx, Bobrovskyi:2012dc}.
Based on the LHCb detector design it is interesting to explore the potential sensitivity reach under the assumption that vertex reconstruction without relying on the VELO or even the first RICH is reliably possible.
A comparison of the potential sensitivities is shown in figure \ref{fig:HNL comparison of reconstruction efficiencies}.
The necessity to trigger on a single charged lepton opens an interesting avenue for searches at LHCb.
The comparably smaller luminosity allows for fairly soft transverse momentum $p_T$ requirements for the single lepton trigger.
In figure \ref{fig:HNL comparison maximal exclusion reach} we compare the maximal exclusion reach at the LHC main detectors.
The low $p_T$ requirement of the single $\tau$ trigger allows the LHCb experiment to be competitive with ATLAS and CMS despite the reduced luminosity.
}

\subsubsection{Heavy neutral leptons from Meson decays\footnote{Contributed by Inar Timiryasov and Alexey Boyarsky.}}
\label{sec:hnl2}
\newcommand{\BrI}{\mathrm{Br}}
\newcommand{\decayI}{\mathrm{dec}}
\newcommand{\events}{\mathrm{events}}
\renewcommand{\min}{\mathrm{min}}
\renewcommand{\max}{\mathrm{max}}

The LHCb search for HNLs in~\cite{Aaij:2014aba} targeted the rare $B \to \mu N$ production channel followed by $ N \to \mu \pi$. Estimates of the future LHCb sensitivity for HNLs produced in $W$ bosons decays were discussed in~\cite{Antusch:2017hhu} and in Section~\ref{sec:hnl1}.  In~\cite{Boiarska:2019jcw}, we have updated the sensitivity estimates by including the HNLs produced in inclusive $B$ meson decays and decaying into different final states.

\paragraph{Sensitivity estimates.}
\label{sec:sensitivity-estimates}
The number of detected events is given by
\begin{equation}
  \label{eq:n-events-w}
  N_{\events} = N_{\rm parent} \cdot \BrI\cdot P_{\decayI}\cdot \epsilon,
\end{equation}
where $N_{\rm parent}$ is the number of parent particles ($W$ bosons or
$B$ mesons)\footnote{Other production channels  (via $Z$ and Higgs bosons, Drell-Yan processes, gluon fusion) are
  subdominant for the masses under consideration~\cite{Gago:2015vma,Degrande:2016aje,Das:2017rsu,Ruiz:2017yyf,Abada:2018sfh,Das:2017zjc,Das:2016hof}.}, see Table~\ref{tab:parent}; $\BrI$ is the
branching fraction of (semi-)leptonic $B$ mesons decays;
$P_{\decayI}$ is the HNL decay probability
 \begin{equation}
   P_{\decayI} = e^{-l_{\min}/c\tau \langle \gamma_{N}\rangle}-e^{-l_{\max}/c\tau \langle \gamma_{N}\rangle}
   \label{eq:Pdecay}
\end{equation}
with $l_\min$ and $l_\max$ determined by the geometry of a tracker, and
$\langle \gamma_N\rangle$ being the average $\gamma$ factor of the
HNL.\footnote{For the HNLs produced from $B$ mesons we define $ \langle \gamma_{N}\rangle = \sqrt{\langle \gamma_{X}\rangle^{2} (\gamma^{\text{rest}}_{N})^{2}-1},$ where $\gamma_X$ is the $\gamma$ factor of a parent meson that moves in the direction of the decay volume, and $\gamma^{\text{rest}}_{N}$ is the $\gamma$ factor of the HNL in the rest frame of the meson.}
Finally, the parameter $\epsilon$ is the \emph{efficiency}---the fraction of all HNL decays that occurred inside the decay volume (between $l_\min$ and $l_\max$) that have passed the selection criteria and were successfully reconstructed.

  \bigskip
We assume that the searches are background free, which is guaranteed by the selection criteria for a DV event within the search schemes we consider. The sensitivity curve is determined by the condition $N_\events \approx 3$ (95\% confidence limit), corresponding to the exclusion limit. The lower boundary of such a curve is determined by the regime $l_{\min} \ll c\tau_{N}\langle \gamma_{N}\rangle$ where the number of events scales approximately as $N_\events \propto U^4$ which allows to find the $U^2$ value for the lower boundary from Eqs.~\eqref{eq:n-events-w}--\eqref{eq:Pdecay}. The lower boundary assuming $N_{\text{events}}>3$ can be obtained from the curve for $N_{\text{events}} = 3$ by a simple rescaling.
The upper boundary is determined by $l_{\text{min}}>c\tau_{N}\langle \gamma_{N}\rangle$, so that $P_{\decayI} \approx \exp\left[-l_{\min}/(c\tau \gamma_{N})\right]$, and hence $N_{\text{events}}$ is exponentially sensitive to $\gamma_{N}$. Therefore the HNL momentum distribution becomes important with the most energetic HNLs determining the exact shape of the boundary. By assuming that all of the HNLs are produced with the average $\gamma$ factor, we \emph{underestimate} the position of the upper boundary and, as a result, the maximal mass probed, which is defined as the intersection of the lower and upper bounds of the sensitivity.

\begin{table*}[t]
  \begin{tabular}{l|c|c|c}
  \small
    Parent/Experiment & $l_\min, \ l_\max$ & Cross-section & Number   \\
    \hline {\small$W$ @  ATLAS/CMS, Short DV }& 0.4~cm, 30~cm~\cite{Cottin:2018nms} & $\sigma_W \simeq 193$~nb~\cite{Aad:2016naf} & $5\times 10^{11}$  \\
    {\small$W$ @ CMS, Long DV }& 2~cm, 300~cm & $\sigma_W \simeq 193$~nb~\cite{Aad:2016naf} &$5\times 10^{11}$  \\
    {\small $B$  @ LHCb} & 2~cm, 60~cm~\cite{Antusch:2017hhu} &$\sigma_{b\bar b} \simeq 1.3\cdot 10^8$\,pb~\cite{LHCb-bbbar} & $4.9\cdot 10^{13}$ \\
  \end{tabular}
  \caption[Number of parent particles]{Number of parent particles ($W$ bosons or $B$  mesons), their average energy $\langle E\rangle$ and corresponding parameters of the DV search schemes used for the estimates. All numbers are given for the high luminosity phase of the LHC ($\mathcal{L} = 3000~\text{fb}^{-1}$ for CMS/ATLAS and $\mathcal{L} = 380\text{ fb}^{-1}$ for the LHCb) and can be proportionally rescaled for other luminosities. The average energy of $B$ mesons is $84$~GeV~\cite{Cacciari:1998it} and we used their spectrum  at $\sqrt{s} = 13\text{ TeV}$ in the pseudorapidity range of LHCb as provided by FONLL simulations~\cite{Cacciari:1998it}.}
  \label{tab:parent}
\end{table*}

\paragraph{DV at LHCb.}
\label{sec:lhcb}
We identify an HNL decay event at LHCb as a DV event if it passes the following selection criteria adapted from~\cite{Aaij:2014aba}:
\begin{itemize}
    \item Decay products must be produced in pseudorapidity range $2<\eta<5$.
    \item The single muon must have $p_{T} > 1.64\text{ GeV}$ to pass the trigger.
    \item Hadrons must have $p > 2\text{ GeV}$ and $p_{T} > 1.1\text{ GeV}$ to be tracked.
    \item Lepton products of HNL decay should have $p > 3\text{ GeV}$ and $p_T > 0.75\text{ GeV}$.
\end{itemize}
Following~\cite{Aaij:2014aba}, we estimate the corresponding efficiency as
$\epsilon \sim 10^{-2}$ for all visible decay channels.

The main parameters for the LHCb experiment are given in Table~\ref{tab:parent}. We notice that at the energies of the LHC for large masses of the HNL ($m_{N} \simeq 3\text{ GeV}$ in the case of the mixing with $\nu_{e/\mu}$ and $m_{N} \simeq 2\text{ GeV}$ in the case of the mixing with $\nu_{\tau}$) the main production channel is the 2-body leptonic decays of the $B_{c}$ mesons (see, e.g.,~\cite{Bondarenko:2018ptm,SHiP:2018xqw}). This makes possible to probe HNL masses up to $m_{B_{c}}\approx 6.3\text{ GeV}$.

\begin{figure}[h!]
  \begin{minipage}{0.45\textwidth}
    \centering
    \includegraphics[width=\textwidth,draft=false]{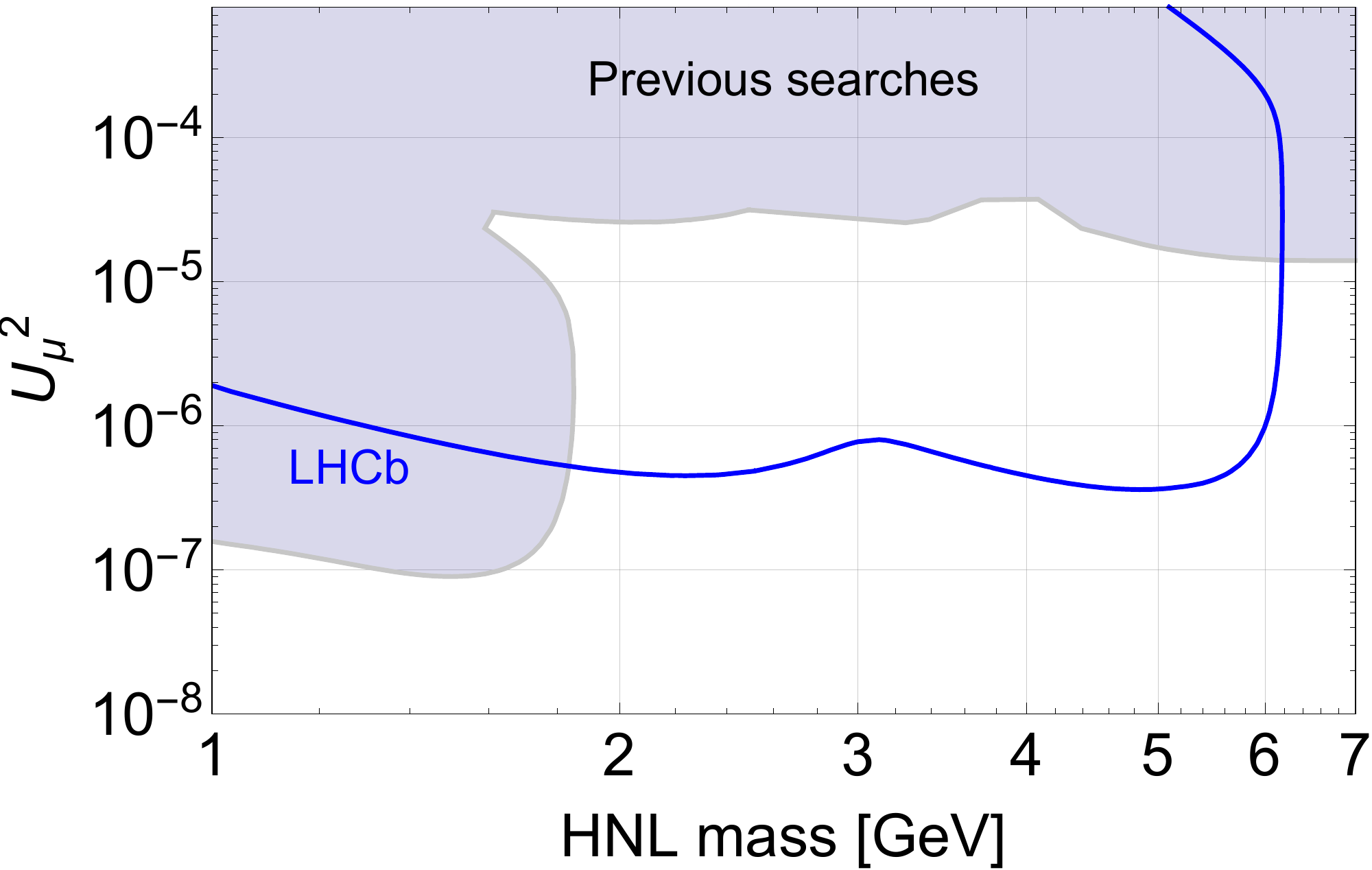}
    \end{minipage}~~\begin{minipage}{0.45\textwidth}
    \centering
    \includegraphics[width=\textwidth,draft=false]{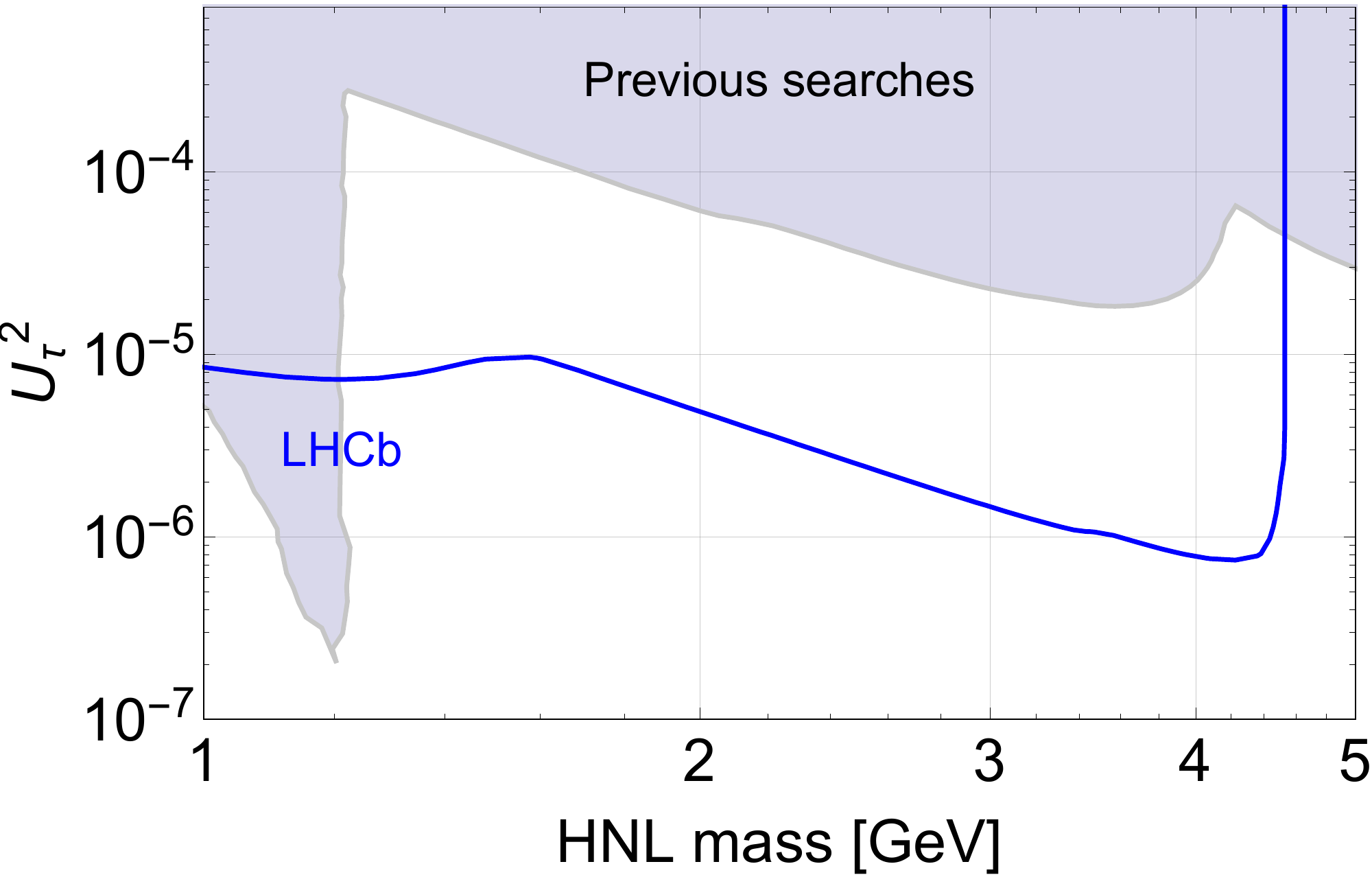}
    \end{minipage}
    \caption{The sensitivity of DV searches at LHCb in the high luminosity phase with $\mathcal L = \unit[300]{fb^{-1}}$. 
    Bounds from the previous experiments are based on~\cite{Alekhin:2015byh}.}
    \label{fig:LHCb}
\end{figure}

The mixing angle $U^{2}$ at the lower bound of the sensitivity is given by
\begin{equation}
U^{2}_{\text{lb}} = 2.6\cdot 10^{-6} \sqrt{\frac{300}{\mathcal{L} [\text{fb}^{-1}]}\frac{10^{-2}}{\epsilon}  \frac{c\tau_{N}}{1\text{ m}}   \frac{\langle\gamma_{N}\rangle}{\sum_{B}f_{b\to B}\text{Br}_{B\to N+X}}}.
\label{eq:lower-bound-lhcb}
\end{equation}
The corresponding sensitivity plot is shown in Fig.~\ref{fig:LHCb}.

\paragraph{Production of HNLs from B mesons.}
In order to estimate the number of detected events one needs the production branching ratio $\BrI$ and the branching ratios of the HNL decays.
entering Eq.~(1) of the main text. These branching ratios as functions of the mass of the HNL are shown in Fig.~\ref{fig:LHCb}.

\subsection{Dark Photons }
\label{sec:dp}
\subsubsection{Dark photons in multi-lepton searches\footnote{Contributed by Patrick Foldenauer.}}
\label{sec:patrick}
\newcommand{\Umt}{$U(1)_{L_\mu-L_\tau}$\xspace}
\newcommand{\Ubl}{$U(1)_{B-L}$\xspace}
\newcommand{\Ubli}[1]{$U(1)_{B-3L_\text{#1}}$\xspace}

Dark Photons with GeV-scale masses have become very popular in the literature since it was realised that they can help to explain the observed excess in the cosmic ray position spectrum~\cite{ArkaniHamed:2008qn,Cheung:2009qd} as well as the excess in the muon anomalous magnetic moment $(g-2)_\mu$~\cite{Pospelov:2008zw}.
The most general Lagrangian for a new $U(1)_X$ gauge boson, allowed by gauge symmetry, reads
\begin{equation}
    \mathcal{L} = -\frac{1}{4} X_{\mu\nu} X^{\mu\nu} -\frac{\epsilon}{2} B_{\mu\nu}  X^{\mu\nu} - \frac{M_X^2}{2} X_\mu X^\mu- g_X\, j_X^\mu\, X_\mu\,.
\end{equation}
In the minimal (secluded) dark photon model~\cite{Okun:1982xi,Holdom:1985ag,Pospelov:2008zw} there are no fermions charged under the new symmetry, i.e.~$j_X^\mu=0$. In this scenario interactions of the dark photon mass eigenstate $A'$ with SM fields are entirely determined by the interactions of the SM photon with an effective mass of $M_{A'}$. Due to its excellent muon reconstruction efficiency LHCb sets the leading constraints on such minimal dark photons over a large range of dark photon masses from resonance searches in both prompt and displaced dimuon signatures~\cite{Aaij:2019bvg}. In the future, LHCb will be able to set stringent limits based on dielectron resonance searches in the process  $D^{*0}\to D^0 (A'\to e^+e^-)$~\cite{Ilten:2015hya}.

\paragraph{Anomaly-free dark photon models}

Beyond the minimal secluded model  ($j_x^\mu=0$) there are a number of new gauge symmetries with charged SM fields that only require the addition of right-handed neutrinos in order to cancel gauge anomalies. This class of models is made up of the groups \Ubl, $U(1)_{L_\mu-L_e}, U(1)_{L_e-L_\mu}$, $U(1)_{L_\mu-L_\tau}$ and combinations thereof, with the gauge currents,
\begin{align}
j_{B-L}^\mu&=  \frac{1}{3}\bar Q \gamma^\mu Q 
          + \frac{1}{3}\bar u_R\gamma^\mu u_R 
          + \frac{1}{3}\bar d_R\gamma^\mu d_R
          - \bar L \gamma^\mu L 
          - \bar \ell_R\gamma^\mu \ell_R
          - \bar \nu_R\gamma^\mu \nu_R \,, \\ 
j_{\, i-j}^\mu&= \bar L_i \gamma^\mu L_i 
          + \bar \ell_{R,i}\gamma^\mu \ell_{R,i} +  \bar \nu_{R,i}\gamma^\mu \nu_{R,i} 
          - \bar L_j \gamma^\mu L_j -\bar\ell_{R,j}\gamma^\mu \ell_{R,j} -\bar \nu_{R,j}\gamma^\mu \nu_{R,j} \,.
\label{eq:fermion_currents}
\end{align}
While resonance searches in the dimuon channel and in the dielectron channel in charm meson decays are optimal for minimal secluded dark photons, searches for the associated gauge bosons of this broader class of minimal anomaly-free $U(1)$ models often require complementary strategies due to their special coupling structure~\cite{Ilten:2018crw,Bauer:2018onh,Bauer:2020itv}.

\begin{figure}[ht]
\begin{center}
\includegraphics[width=.4\textwidth]{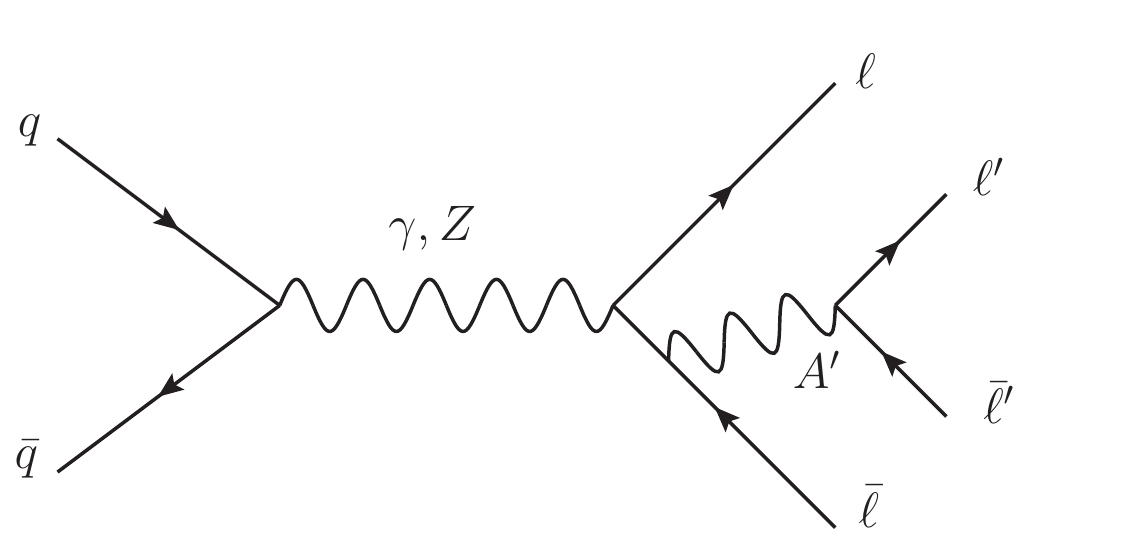}%
\end{center}
\caption{\label{fig:4leps} Four lepton final state due to dark photon production in final state radiation.}
\end{figure}

\begin{figure}
\centering
\includegraphics[width=.42\textwidth]{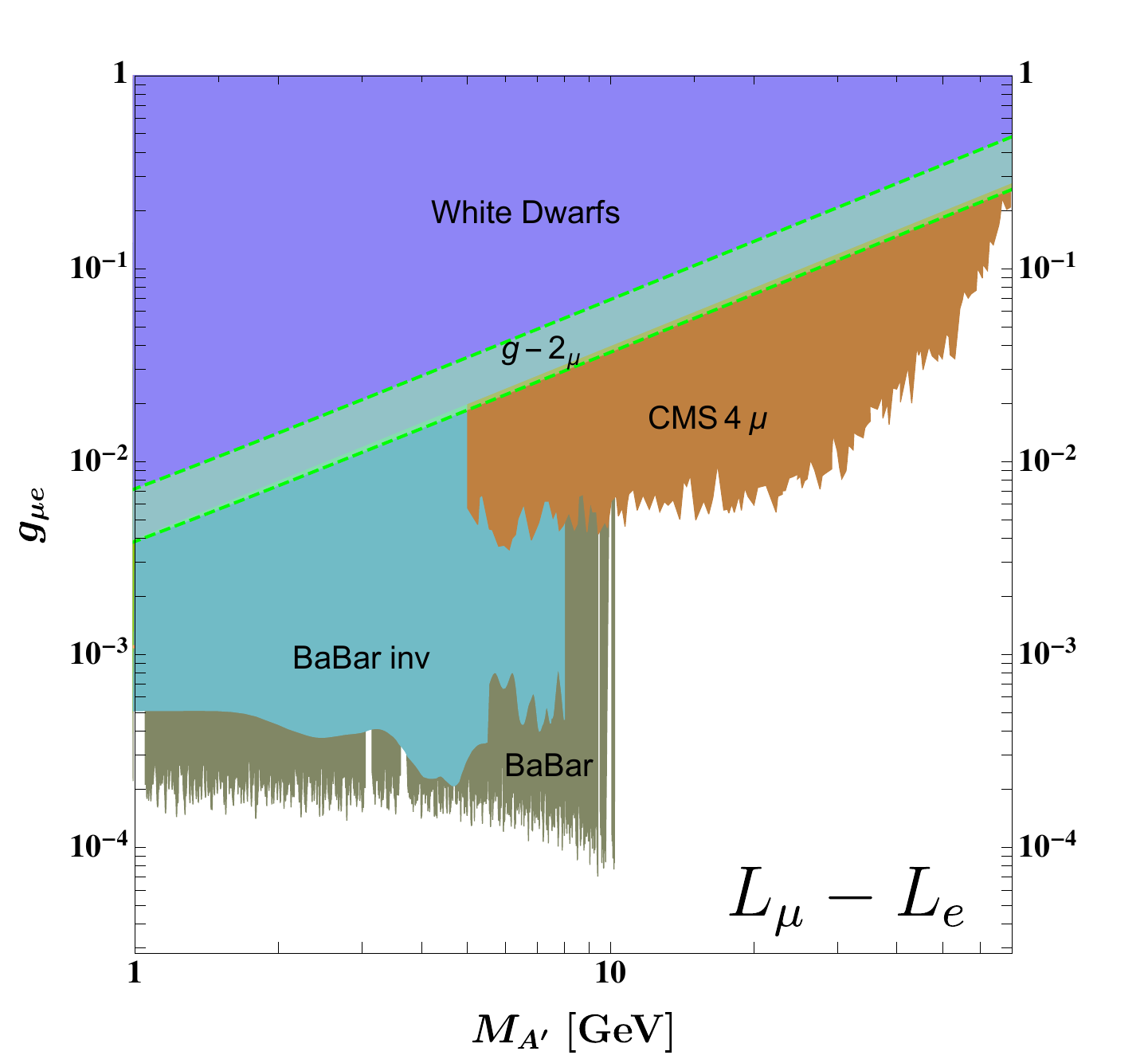}\quad
\includegraphics[width=.42\textwidth]{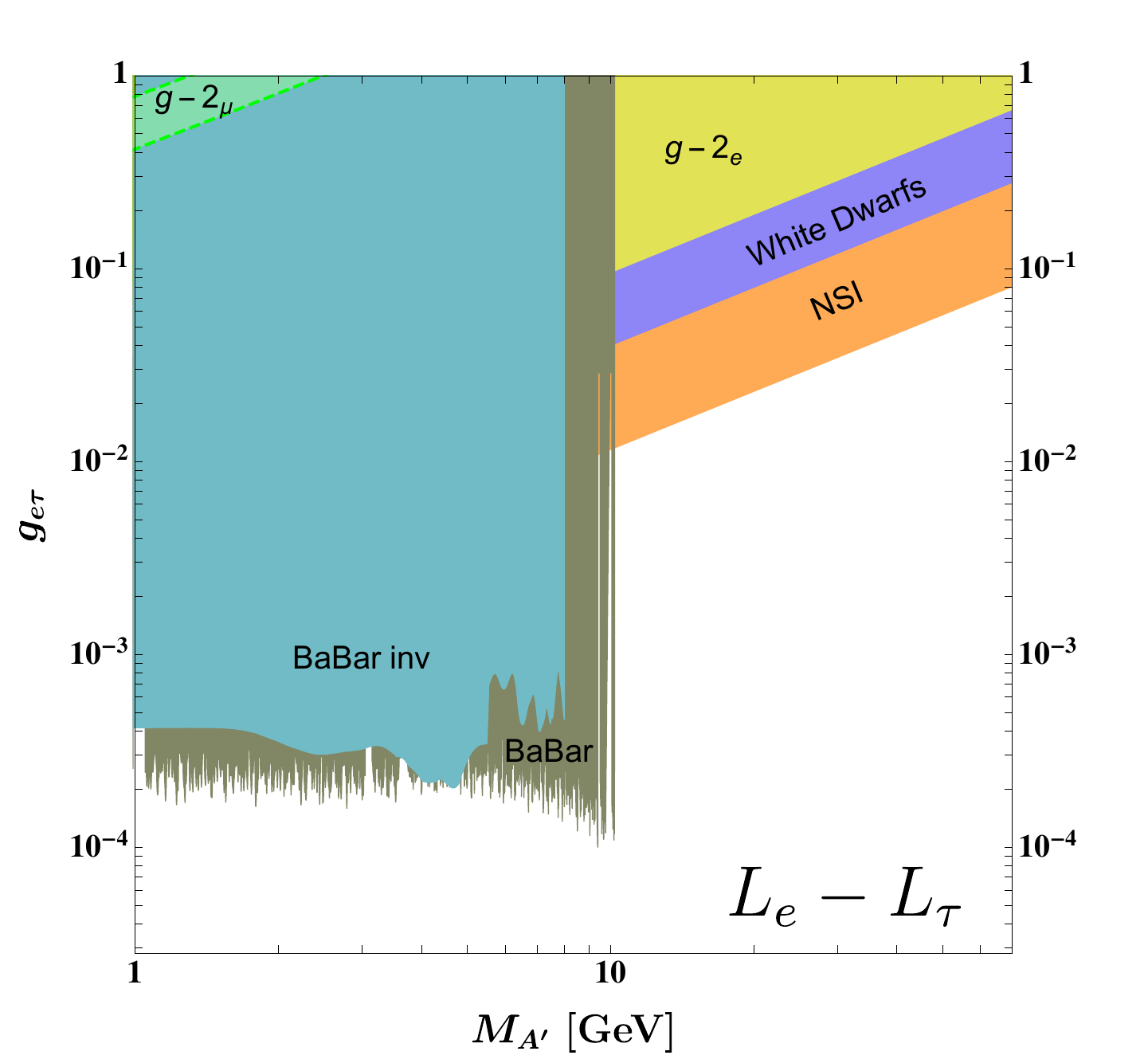}
\caption{\label{fig:lim_lilj} Constraints (from~\cite{Foldenauer:2019dai}) on GeV-mass dark photons of gauged $U(1)_{L_\mu-L_e}$  ({\it left}) and $U(1)_{L_e-L_\tau}$ ({\it right}) from BaBar~\cite{Lees:2014xha,Lees:2017lec}, CMS~\cite{Sirunyan:2018nnz}, White Dwarf cooling~\cite{Bauer:2018onh} and NSI~\cite{Coloma:2020gfv}.}
\end{figure}

In particular, the associated gauge bosons $A'$ of the anomaly-free lepton groups $U(1)_{L_i-L_j}$ cannot be produced in $pp$ collisions or meson decays directly due to the lack of couplings to baryons. Instead searches for these at the LHC would rather have to be focused on four-lepton signatures, where the $A'$ is produced from final state radiation off a lepton (cf.~Fig.~\ref{fig:4leps}). Such a search strategy has been previously employed by CMS~\cite{Sirunyan:2018nnz} in the four-muon channel. The CMS search yields the leading constraint for muon-philic gauge bosons with masses $M_{A'}\gtrsim 10$ GeV (see e.g. left panel of Fig.~\ref{fig:lim_lilj}). Due to the excellent vertexing capabilities of LHCb as well as its high muon reconstruction efficiency, a similar search at LHCb could in particular target long-lived $A'$ and yield sensitivities to much smaller gauge couplings. 

Similarly, a resonance search in a four-electron final state at LHCb could be sensitive to unconstrained parameter space of electro-philic gauge bosons, if couplings to muons are absent. For example, in the right panel of Fig.~\ref{fig:lim_lilj} it can be seen that the parameter space of $U(1)_{L_e-L_\tau}$ is rather unconstrained for boson masses of $M_{A'}\gtrsim 10$ GeV where gauge couplings of up to $g\sim0.1$ are still allowed. 
This region of parameter space provides a blind spot for traditional dark photon searches at $e^+e^+$ collider experiments like Belle and Babar, since these operate at a center of mass energy of $\sqrt{s}=10.58$ GeV corresponding to the $\Upsilon(4S)$ meson resonance. 
Although reconstruction of electrons at LHCb is traditionally more challenging than of muons (see Sec.~\ref{sec:PID}), due to the relatively unconstrained parameter space in such electro-philic models LHCb could fill the gap and provide leading constraints for  $M_{A'}\gtrsim 10$ GeV. Especially, due to its excellent displaced vertex reconstruction~\cite{Alexander:2018png} four-electron searches could also be a prime search channel for long-lived electro-philic dark photons.\footnote{Similar considerations hold for the related groups \Ubli{$e$}, $U(1)_{B-2L_e-L_\tau}$ or $U(1)_{B-2L_\tau-L_e}$. There, due to the dark photon coupling to quarks, even searches in the dielectron channel at LHCb could provide competitive constraints for  $M_{A'}\gtrsim 10$ GeV as well as for long-lived $A'$.}

In summary, resonance searches in four-lepton final states at LHCb can probe non-minimal dark photons complementary to existing searches.
These prospects call for dedicated sensitivity studies of such searches at LHCb in order to realise its full potential for dark photons in LHC run 3.
It will be particularly important to systematically study all possible combinations of such four-lepton channels (including $\ell^+\ell^-\ell^{\prime+}\ell^{\prime-}$ with $\ell\neq\ell'$) to fully determine the flavour structure of the associated gauge group.

\subsubsection{Minimal Dark photons\footnote{Contributed by Yotam Soreq, Michael Williams.}}
\label{sec:YotamDP}
\newcommand{\ys}[1]{\textbf{\textcolor{blue}{[YS: #1]}}}

Two methods for probing dark photon at LHCb are:
\begin{itemize}
    \item An inclusive search in the $\mu^+\mu^-$ final state~\cite{Ilten:2016tkc,Aaij:2017rft,Aaij:2019bvg,Aaij:2020ikh}, {\em i.e.}
    \begin{align}
         p p \to X A' \to X  \mu^+\mu^- \, .
    \end{align}
    \item An exclusive search using $D^*\to D^0 A'$ decays in the $e^+e^-$  final state~\cite{Ilten:2015hya}
    \begin{align}
         D^* \to D^0 (A' \to e^+e^-)  \, .
    \end{align}
\end{itemize}

\begin{figure}[!htp]
    \centering
    \includegraphics[width=0.48\textwidth]{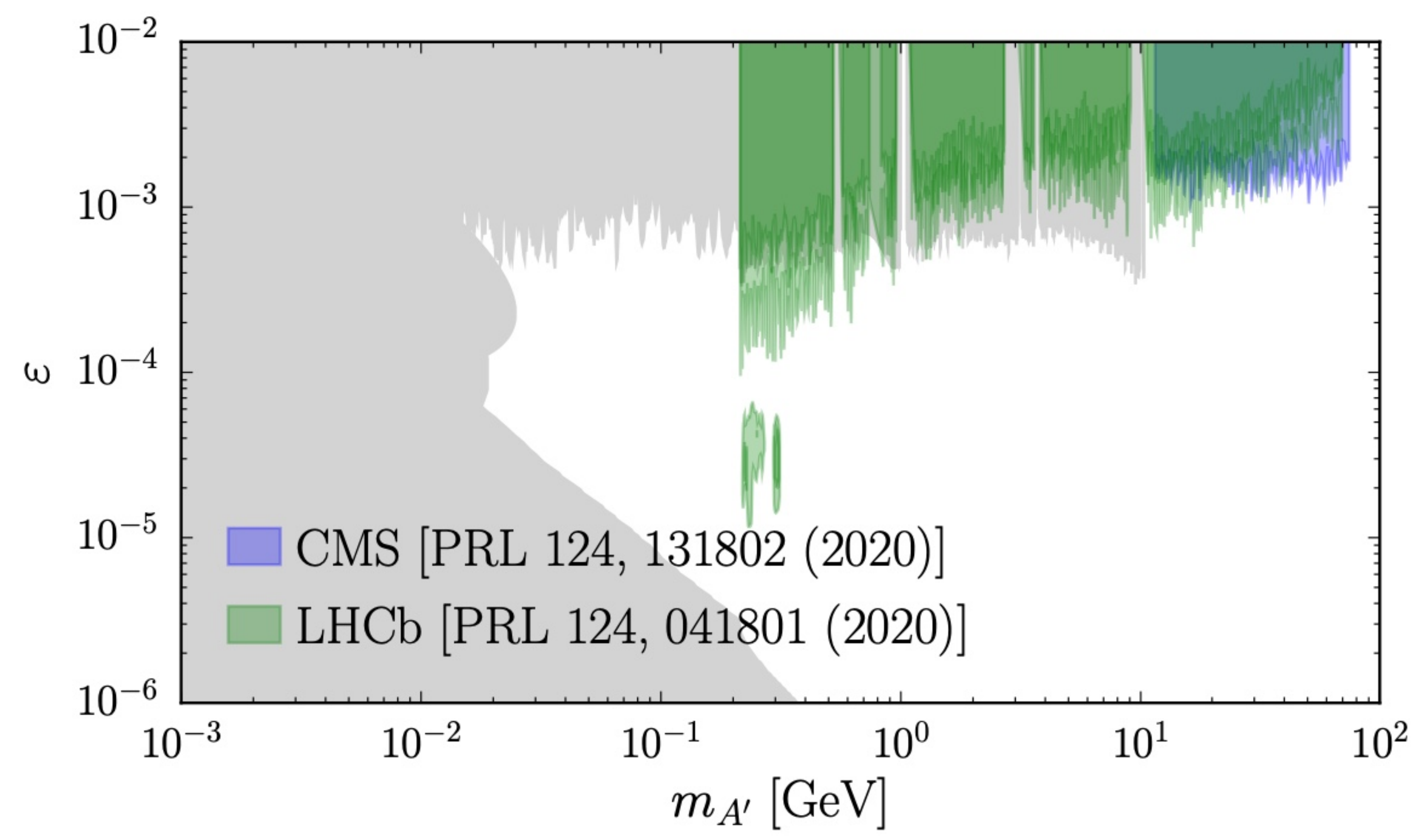}
    \includegraphics[width=0.48\textwidth]{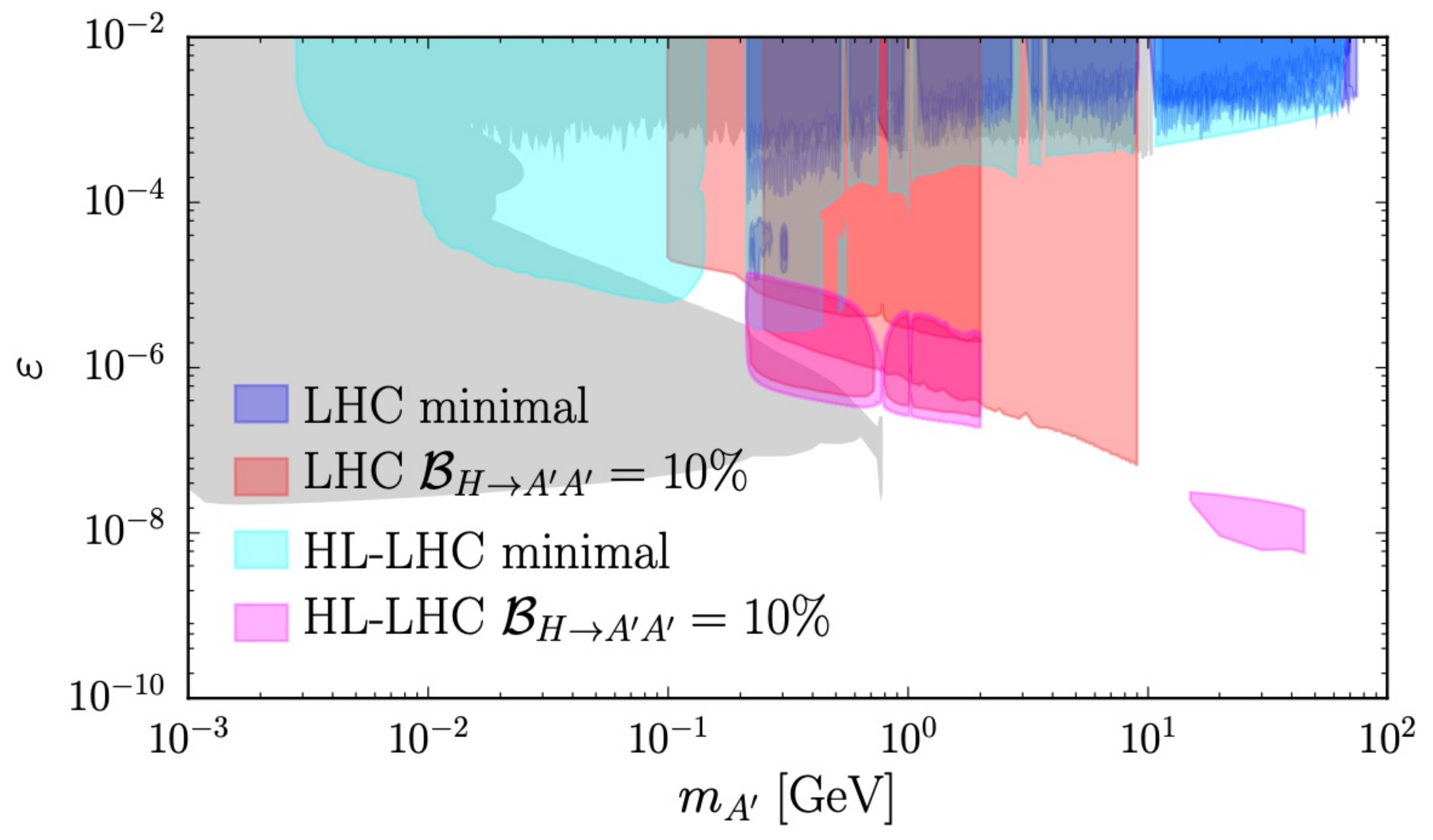}
    \caption{Current (left) and future expected (right) exclusion regions from searches for dark photons, including LHCb and other experiments. Plots produced using \textsc{DarkCast}~\cite{Ilten:2018crw}. Note that both the first LHCb search~\cite{Aaij:2017rft} and its subsequent update~\cite{Aaij:2019bvg} are shown as overlapping green regions.}
    \label{fig:LHCbAp}
\end{figure}

The dark photon primarily couples to the electromagnetic current  when it is much lighter than the $Z$ boson. The rate of dark photon production and its decay $A'\to\mu^+\mu^-$ can be fully determined using a data-driven process based on measuring the rate of  
off-shell photon production, $\gamma^*\to\mu^+\mu^-$. 
This approach greatly simplifies the search.  
The inclusive analysis of $A'\to\mu^+\mu^-$ has  two signal regions depending on the decay location. If the muons produced in the dark photon decay are experimentally consistent with originating from the proton-proton collision, we consider this a prompt $A'$ decay. Otherwise, we consider it as a displaced dark photon decay. See Sec.~\ref{sec:LHCbmuon} for more details of the muon search.  Figure~\ref{fig:LHCbAp} shows the current and future prospect of bounds on the $A'$ parameter space (mass versus mixing strength). 
As we can see, LHCb can probe the parameter space that is not accessible by any other experiment. This data-driven search can be also implemented using the $A'\to e^+e^-$ channel that is similar to the True-Muonium search (see Sec.~\ref{sec:TM}). 

The $D^*\to D^0 A'$ search~\cite{Ilten:2015hya} leverages the large charm-meson production rate at LHCb.
There are two search strategies, a dispalced search, which is almost background free, and a prompt (resonant decay) search. 
Moreover, since the $D^*$ is a narrow resonance, its mass can serve as a kinematical anchor resulting in improved $e^+e^-$ invariant mass resolution, thus, better sensitivity to dark photons. 
As we can see from Fig.~\ref{fig:LHCbAp}, this search can explore dark photons with masses below the pion mass and mixing strengths in the range  $\epsilon^2 \gtrsim 10^{-10}$, which is in the gap between the $B$-factory and the beam-dump searches. 

\begin{figure}[!htp]
    \centering
    \includegraphics[width=0.49\textwidth]{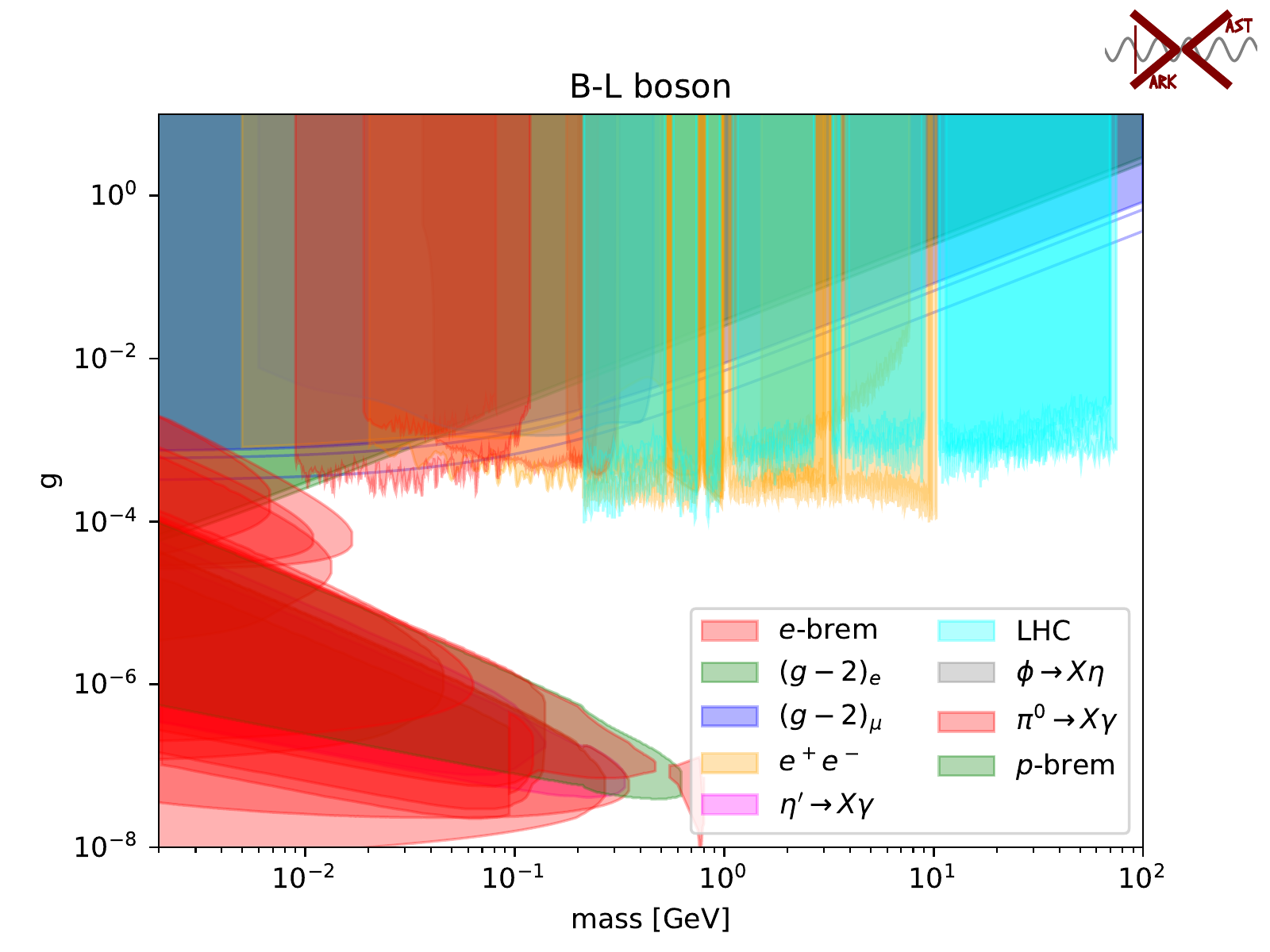} 
    \caption{Recast of searches for dark photons to $B-L$ models, containing a leptophobic $B$ boson (see text for details). The plot was produced using \textsc{DarkCast}~\cite{Ilten:2018crw}.}
    \label{fig:DarkCast}
\end{figure}

Finally, the searches for dark photons can probe other vector particles with different couplings to the SM, such as $B-L$, a leptophobic $B$ boson, or  a vector that mediates a protophobic force.
A simple recasting framework for laboratory dark photon searches was developed in Ref.~\cite{Ilten:2018crw}, \textsc{DarkCast}.
This framework includes a data-driven method for determining hadronic decay rates at the QCD scale, which cannot be calculated from first princple without a dedicated lattice calculation. 
This approach can easily be generalized to any massive gauge boson with vector couplings to the Standard Model fermions, and software to perform any such recasting is provided at \url{https://gitlab.com/philten/darkcast}.
In Fig.~\ref{fig:DarkCast}, we show the current bounds on the $B-L$ and $B$ bosons based on \textsc{DarkCast}.

\subsubsection{Long Lived Dark photons\footnote{Contributed by Zuowei Liu, Mingxuan Du, Van Que Tran.}}
\label{sec:zuowei}

Dark photons can naturally arise in models in which 
a hidden sector abelian gauge boson has 
either a kinetic mixing (see previous section) 
or a Stueckelberg mass mixing
\cite{Kors:2004dx, 
Feldman:2006ce, 
Feldman:2007wj} 
 with the SM hypercharge gauge boson.  
 Long-lived dark photons (LLDPs) with a mass 
 in the MeV-GeV scale are interesting
 new physics experimental targets.

However, LHC searches for such LLDPs suffer 
from the fact that the same coupling that makes 
the dark photon long-lived suppresses the production rate. 
Here we present a model where the LHC production rate 
can be decoupled from the decay process, 
with interesting prospects at LHCb \cite{Du:2019mlc}.

The standard model is extended by a hidden sector 
that contains two abelian gauge bosons $X_\mu$ and $C_\mu$, 
with gauge groups $U(1)_F$ and $U(1)_W$ respectively, 
and one Dirac fermion $\psi$ charged under both groups 
with gauge couplings $g_F$ and $g_W$. 
The interactions between the hidden sector 
and the standard model are mediated by 
the Stueckelberg mass terms \cite{Du:2019mlc}; 
the relevant Lagrangian is given by
\begin{equation}
{\cal L} = -\frac{1}{4}  X_{\mu\nu}^2 - \frac{1}{4}  C_{\mu\nu}^2  
- \frac{1}{2} ( \partial_\mu \sigma_1 +  m_{1}\epsilon_1 B_{\mu} +  m_{1} X_{\mu} )^2 
- \frac{1}{2} ( \partial_\mu \sigma_2 +  m_{2}\epsilon_2 B_{\mu} +  m_{2} C_{\mu} )^2,
\end{equation}
where $\sigma_1$ and $\sigma_2$ are the axion fields,
and $m_1$, $m_2$, $m_{1}\epsilon_1$, 
and $m_{2}\epsilon_2$ are mass terms in 
the Stueckelberg mechanism.

Due to the extension of the gauge group, 
the neutral gauge boson mass matrix is enlarged to be 
a 4 by 4 matrix which can be diagonalized by 
an orthogonal transformation
${\cal O}$.  
The mass {basis}
$E= ( Z', A', Z, A)$ {is}
related to the gauge {basis} 
$V= ( C,X, B, A^3)$
via 
$E_i={\cal O}_{ji} V_{j}$.
In the mass basis, 
$A$ is the SM photon, 
$Z$ is the SM $Z$ boson, 
$A'$ is the dark photon 
and $Z'$ is the 
{heavy boson.}
Diagonalization of the mass matrix leads to 
interactions between $Z/A$ %
to the hidden fermion $\psi$, %
and also interactions between $Z'/A'$ %
and SM fermions $f$; 
both $\psi-Z/A$ and $f-Z'/A'$ couplings are 
suppressed by the small $\epsilon_1$ and 
$\epsilon_2$ parameters, 
{and vanish in the $\epsilon_1=0=\epsilon_2$ limit.}

At the LHC,
for a sufficient large value of $\epsilon_2$ and gauge coupling $g_F$, 
a large number of dark photons 
can be radiated off from the hidden sector particle $\psi$, 
where  $\psi$ is 
pair-produced via $pp \to Z/Z' \to \bar \psi \psi$.
Assuming $\epsilon_1 \ll 1$ and $m_{A'} < 2 m_{\psi}$, 
the dark photon can travel a 
macroscopic distance away from its production point 
and then decay into a pair of SM particles. 
The muonic decay mode of $A'$ can be searched for at LHCb 
via displaced dimuon signal \cite{Ilten:2016tkc, Aaij:2017rft, Aaij:2019bvg}.

To analyze the displaced dimuon signal at LHCb, we choose a benchmark point in which 
$m_2 = 700$ GeV, $m_{\psi} = 5$ GeV, $\epsilon_2 = 0.01$,  $g_F = 1.5$ and $g_W = 1$. 
In this benchmark point, the $\bar \psi \psi$ production cross section is about $4.3$ pb 
and dominated by the $Z$-boson exchange channel. 
We use \verb|MG5_aMC@NLO| \cite{Alwall:2014hca} to generate the events for each model point on the 
$\epsilon_1$-$m_{A'}$ plane, which are then 
passed to  \verb|Pythia 8| \cite{Sjostrand:2014zea, Carloni:2010tw, Carloni:2011kk} 
for showering (including showering in hidden sector) 
and hadronization.
We follow the LLDP search criteria in 
Ref.\ \cite{Aaij:2017rft, Aaij:2019bvg} to analyze the signal.
The background event is estimated to be $B = 25$ at ${\cal L} = 15\, \rm{fb}^{-1}$ 
\cite{Ilten:2016tkc}.
We compute the exclusion region by demanding that 
$S/\sqrt{B} > 2$ 
where $S$ is the signal event number. 

\begin{figure}[t!]
\centering
        \includegraphics[width=0.49\textwidth]{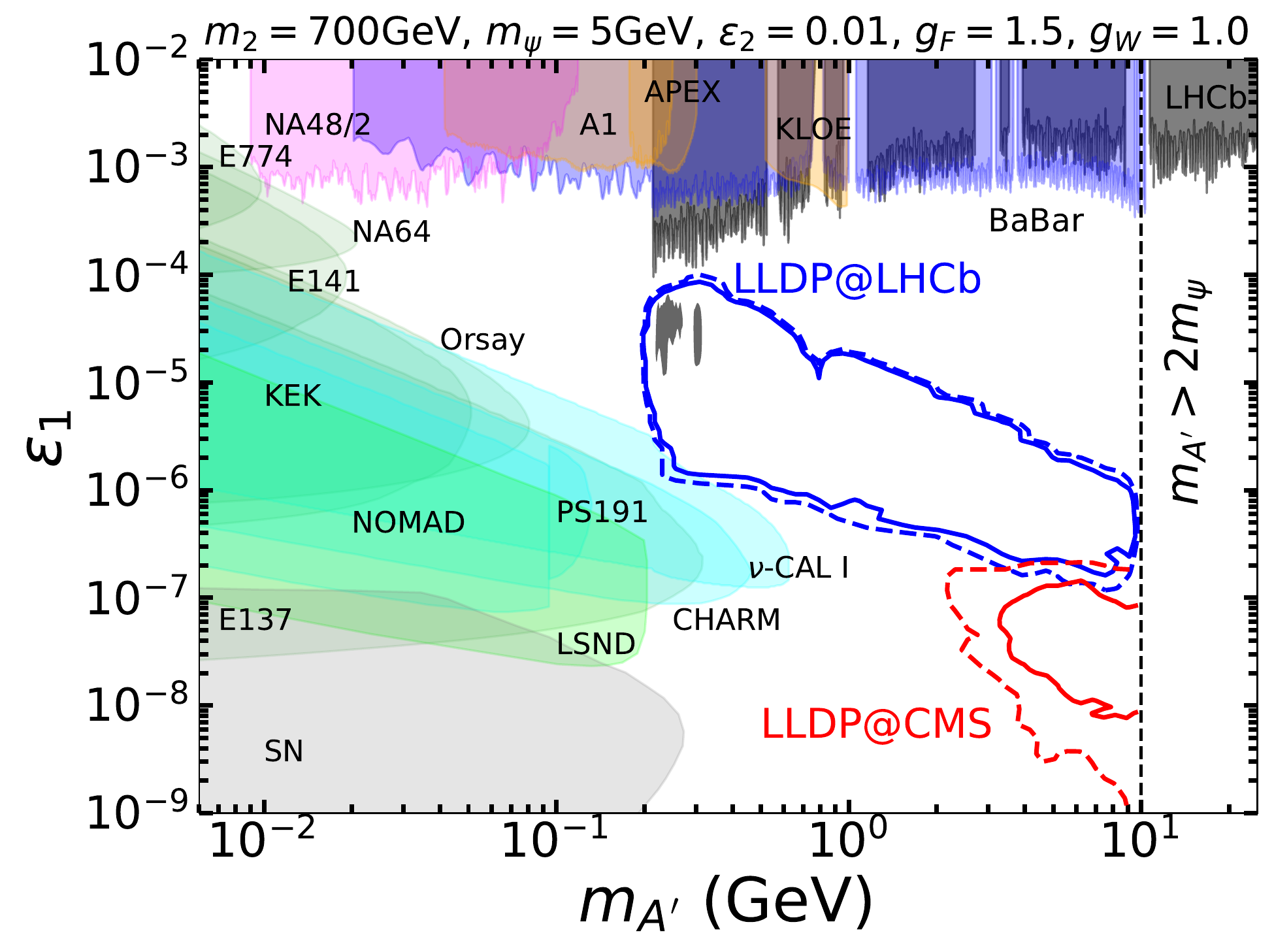}
\caption{LHC current and future sensitivity contours
to the LLDP parameter space spanned by 
$\epsilon_1$ and $m_{A'}$. 
The blue solid and dashed contours indicate the regions probed by the 
LHCb with $5.5$ $\rm{fb}^{-1}$ and $15$ $\rm{fb}^{-1}$ data respectively. 
The regions probed by the future precision timing detector 
MTD CMS \cite{CMStiming} 
with $250$ ${\rm fb}^{-1}$ and $3000$ ${\rm fb}^{-1}$ data 
are shown as red solid and dashed contours respectively.
The gray islands at $\epsilon_1 \sim (10^{-4}-10^{-5})$ are 
the LHCb exclusion regions for the {conventional} dark photon scenario \cite{Aaij:2019bvg}. 
Various experimental constraints on the {conventional} dark photon scenario 
are shown as color shaded regions. 
Reproduced from Ref.\ \cite{Du:2019mlc}. 
}
\label{fig:LHCb_zl}
\end{figure}

Fig.\ ({\ref{fig:LHCb_zl}}) shows the exclusion region 
and future sensitivity 
in the parameter space spanned by $\epsilon_1$ and $m_{A'}$ 
from the LHCb displaced dimuon search. 
With the current luminosity $5.5$ ${\rm fb}^{-1}$, 
LHCb can probe the parameter space of our model: 
$200\, {\rm MeV} < m_{A'} < 9 \, {\rm GeV}$ 
and $ 2 \times 10^{-7} <  \epsilon_1 < 6 \times 10^{-5}$, 
shown as the blue solid contour. 
The exclusion region in the {conventional} dark photon scenario is, 
however, much smaller, 
which is shown as two small gray islands at 
$\epsilon_1 \sim (10^{-4}-10^{-5})$. 
Thus, in our model, 
a significantly larger region of parameter space 
than the {conventional} dark photon model can be probed  
by the current LLDP search at LHCb. 
If $15$ $\rm{fb}^{-1}$ integrated luminosity can be 
accumulated in the LHC Run 3 data, 
LHCb can probe the parameter space: 
$200 \, {\rm MeV} < m_{A'} < 10 \, {\rm GeV}$ and $ 10^{-7} <  \epsilon_1 < 10^{-4}$, 
shown as the blue dashed contour.

Furthermore, if the produced LLDP is moving non-relativistically 
and has a long lifetime, 
a significant time delay $\Delta t$ \cite{Liu:2018wte}
can be measured by the precision timing detectors 
proposed to be installed 
at LHCb \cite{LHCb-PII-EoI}, CMS \cite{CMStiming} and ATLAS \cite{Allaire:2018bof, Allaire:2019ioj}. 
A time delay signal from the LLDPs model \cite{Du:2019mlc}  has been investigated
using 
the future timing detector at CMS (MTD CMS) 
which has a $\sim$30 ps timing resolution \cite{CMStiming}. 
The major background due to pile-up can be suppressed 
to be negligible if a significant time delay is required, 
$\Delta t > 1.2$ ns. 
The sensitivity region spanned by 
$\epsilon_1$ and $m_{A'}$ from the MTD CMS
with $250$ ${\rm fb}^{-1}$ and $3000$ ${\rm fb}^{-1}$ data 
are shown in Fig.\ ({\ref{fig:LHCb_zl}}) as red solid and dashed contours respectively. 
The MTD CMS probes the parameter space with 
a somewhat smaller $\epsilon_1$  
than the LHCb di-muon search, as shown in 
Fig.\ ({\ref{fig:LHCb_zl}}).

{A time delay signal from the LLDPs can also be searched for 
at the future TORCH detector, 
a precision timing detector 
proposed to be installed at the next upgrade of LHCb (see Sec.~\ref{lhcb-future}). 
Because the TORCH detector 
is located $\sim 10$ m  away from the interaction point, 
it can probe a slightly different LLDP parameter space 
than the MTD CMS detector. 
Also, 
because the TORCH detector has a better timing resolution, $15$ ps per track, 
the time delay searches of LLDPs 
at LHCb is expected to be more promising.}

\let\bea\undefined
\let\eea\undefined

\subsection{Light new scalars}
\subsubsection{Exotic Higgs Decay\footnote{Contributed by José Zurita.}}
\label{sec:displacedKaons}
Coupling light new degrees of freedom to the Standard Model while being consistent with the existing experimental data can be done through the so called ``portals" (see Ref.~\cite{Beacham:2019nyx} for a review). In this section we focus on the so-called Higgs portal, which adds a new scalar singlet $S$ to the Standard Model\footnote{Portal couplings have been studied in a variety of physics cases: dark matter~\cite{Pospelov:2007mp,Krnjaic:2015mbs,Dev:2017dui}, SUSY models~\cite{Fayet:1974pd}, cosmic inflation~\cite{Shaposhnikov:2006xi,Bezrukov:2009yw}, and cosmological solutions to the Higgs hierarchy problem~\cite{Graham:2015cka}.}.

In this setup the $h \to SS$ decays yield \emph{Exotic Higgs decays}, namely final states of the Higgs boson not present in the Standard Model.\footnote{Finding such decay modes would be an indisputable sign of new physics, see Ref.~\cite{Curtin:2013fra} for a review. }. The relevant Lagrangian can be written (after electroweak symmetry breaking) as
\be
{\cal L} \supset - \lambda_{SSh} h S^2 - \sin \theta \frac{m_f}{v} S f \bar{f}
\ee
where $h$ is the SM-like Higgs with a mass of approximately 125.09 GeV and S is the new scalar, and $\theta$ is the mixing angle between both scalars, constrained to be small due to Higgs data. Since $S$ has the total width of a SM-like Higgs boson scaled by $\sin \theta^2$ it is likely that $S$ is long-lived. There are direct constraints from displaced jets (from ATLAS \cite{Aaboud:2018aqj,Aaboud:2019opc,Aad:2019tua}, CMS \cite{Sirunyan:2018vlw} and LHCb\footnote{see Sec.~\ref{sec:review} and Ref.~\cite{Aaij:2017mic}.}) but they are only valid if  $m_S \gtrsim 10$ GeV. In addition, there is an upper limit on the maximum ``invisible" (i.e: non-SM like) branching fraction of $h$, being of 19 (26)\% from CMS\cite{Sirunyan:2018owy} (ATLAS\cite{Aaboud:2019rtt}).\footnote{A recent ATLAS study\cite{ATLAS:2020cjb} updates this value to 13 \%.}. An important advantage of searching for exotic Higgs decays is that the production cross section is fixed, being approximately 55 pb~\cite{Anastasiou:2016cez}\footnote{Alternatively one could look for off-shell Higgs effects in exotic decays of B-mesons, which we do not pursue here. We refer the reader to Sections \ref{sec:maria} and \ref{sec:ruth}.}. 

LHCb provides unique advantages to tackle the displaced decay of $S$ into exclusive hadronic final states, due to their exquisite vertexing precision  and also for being the only LHC experiment with hadronic particle identification capabilities\footnote{If $S$ lives longer than about a meter and most of the decays happen outside the detector, it falls in the domain of new external detectors such as MATHUSLA\cite{Curtin:2018mvb}, CODEX-b\cite{Gligorov:2017nwh} and FASER\cite{Feng:2017uoz}.}. More details about these advantages are given in Sec.~\ref{sec:experimental}. The specific branching fractions into hadronic final states\footnote{We refer the reader to Ref.~\cite{Winkler:2018qyg} for details of the calculation.} would depend on the flavour structure. 

To assess the LHCb sensitivity to a light $S$ decaying into SM hadrons, we perform a study of displaced $S\to K^+K^-$. Events are generated using Pythia 8.1~\cite{Sjostrand:2007gs} using the SM Higgs production module and forcing the $H \to SS \to K^+ K^- K^+ K^-$ branching ratio to be unity. Our search strategy is based on selecting $K^{\pm}$ candidates ($p_T > 0.5$ GeV, $2 < \eta < 5$ and reconstructing S candidates from those kaons, satisfying $d (K,K) < 0.1$ mm and $p_T(S) > 10$ GeV. The S vertex must point to the primary vertex with impact parameter larger than 0.1 mm, and needs to occur in $2 < \rho  < 25$ and $z < 400$ mm, where $\rho$ is the radial distance (perpendicular to the beam axis $z$) in centimeters. Our analysis applies an isolation criteria for muons, and also vetoes specific $m_{KK}$ region to account for $k$ misidentification into other hadrons (for details see  Ref.~\cite{CidVidal:2019urm}.). We then classify our event into 8 signal regions, depending on whether we a) reconstruct 1 or 2 S candidates b) apply kaon isolation and c) if $6 <\rho <10 $ or $14 < \rho < 25$ for each candidate, and report the strongest constrain among the 8 regions. We note that while here we focus on charged kaons, our strategy is general and applicable to any other charged hadron (e.g: $D, \pi,$ etc).

\begin{figure}[!htp]
    \centering
        \includegraphics[width=0.79\textwidth]{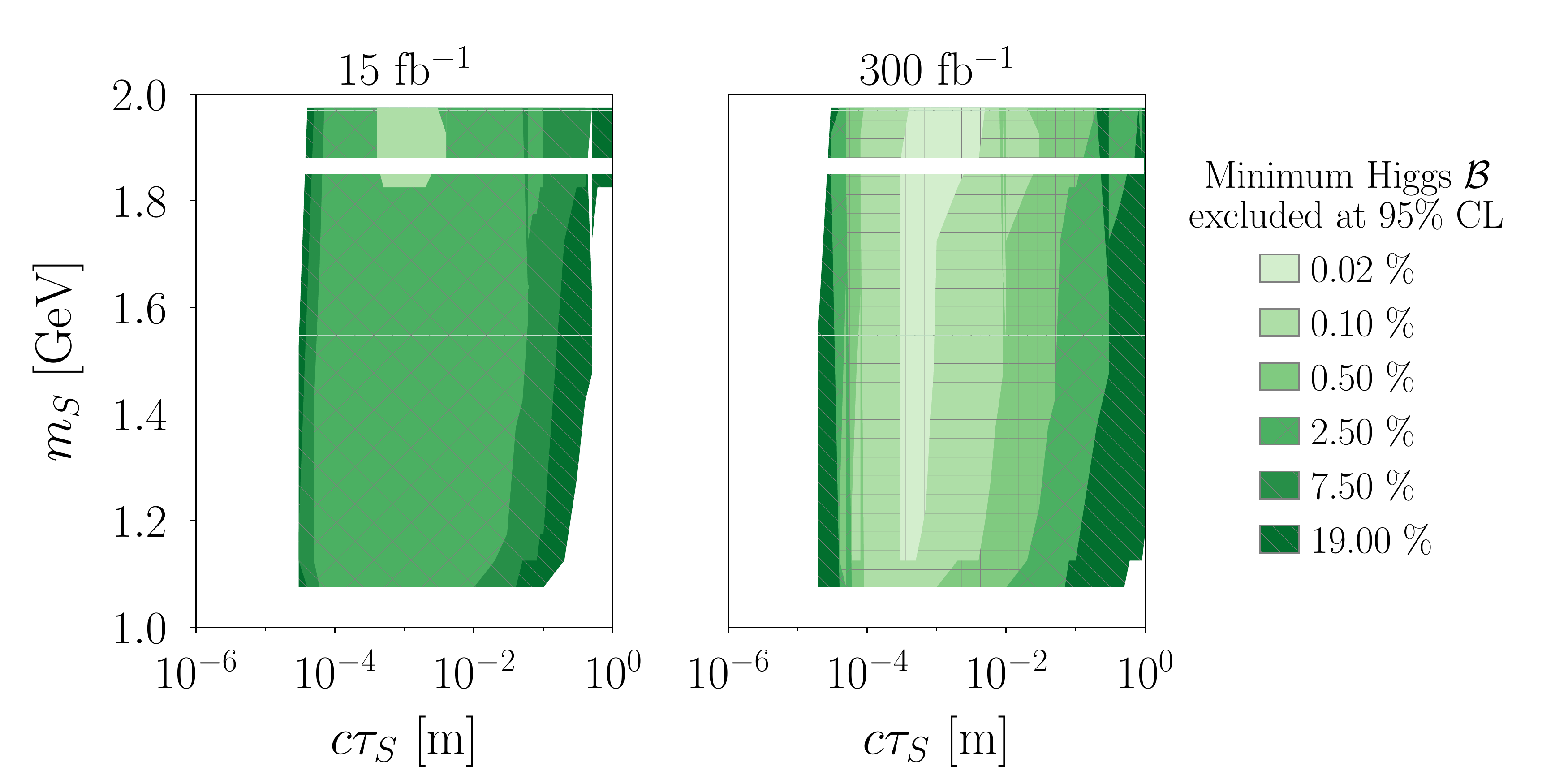}
    \caption{Exotic Higgs branching ratio excluded at the 95 \% C. L at LHCb for a total integrated luminosity of 15 fb$^{-1}$ (left) and 300  fb$^{-1}$ (right) Results are presented in a model-independent manner in terms of the $S$ scalar mass and lifetime (from Ref.~\cite{CidVidal:2019urm}).}
    \label{fig:EHD_mod_ind}
\end{figure}

\begin{figure}[!htp]
    \centering
        \includegraphics[width=0.79\textwidth]{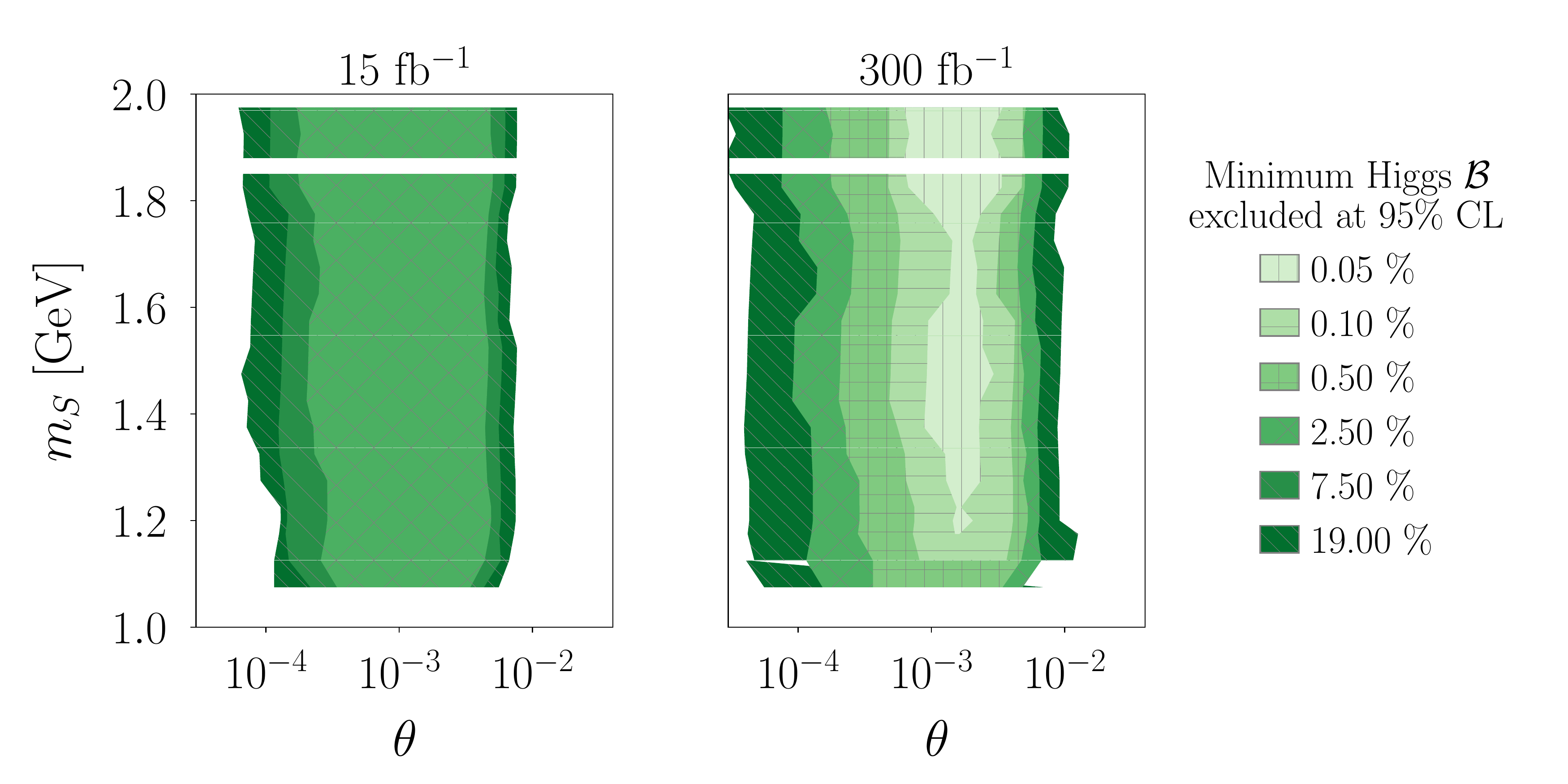}
    \caption{Exotic Higgs branching ratio excluded at the 95 \% C. L at LHCb for a total integrated luminosity of 15 fb$^{-1}$ (left) and 300  fb$^{-1}$ (right) Results are presented for a model with a hadrophilic (leptophobic) $S$, following \cite{Batell:2018fqo} (plot from Ref.~\cite{CidVidal:2019urm}).}
    \label{fig:EHD_hadro}
\end{figure}

We present model-independent results in the $m_S-c\tau$ plane (figure~\ref{fig:EHD_mod_ind}), and for the hadrophilic Higgs portal (figure~\ref{fig:EHD_hadro}) in the $\theta-m_S$ plane.\footnote{The MFV Higgs portal (results can be found in Ref.~\cite{CidVidal:2019urm}) is strongly constrained from the $B \to K \mu \mu$ LHCb search \cite{Aaij:2016qsm,Aaij:2015tna}, however for the leptophobic (hadrophilic case) those constraints do not apply.} We show the minimum exotic Higgs branching fraction that can be excluded at the 95 \% C.L. 
Confronting our study with the expected ATLAS and CMS ``invisible" Higgs search we see that a) LHCb can go below the 2.5 \% expected reach in several regions of the parameter space and b) in case of an excess in that search, LHCb has the potential to \emph{characterize} the signal, which would be quite hard (if not impossible) at ATLAS and CMS.


To summarize, the LHCb can help to identify long-lived scalar decays into exclusive hadrons, allowing for thorough (and unique) exploration of the mass regime $m_S \lesssim 10$ GeV which can not be directly tested by other LHC experiments, thanks to its hadron ID and vertexing.

\subsubsection{Single (pseudo-)scalar production\footnote{Contributed by Jernej F. Kamenik.}}
\label{sec:lightA}

Several well-motivated extensions of the SM include new (pseudo)scalar particles with mass below the electroweak scale. A well-known example in the context of supersymmetry is the next-to-minimal supersymmetric SM, where a light pseudo-scalar state can arise as a result of an approximate global $U(1)_R$ symmetry~\cite{Dobrescu:2000yn}. Non-supersymmetric extensions featuring a light (pseudo)scalars include Little Higgs models, hidden valley scenarios (see Ref.~\cite{Curtin:2013fra}, and references therein for details), and simplified models where a (complex) singlet scalar is coupled to the Higgs potential of the SM or the 2HDM. 

Light (pseudo-)scalars can be searched via various collider signatures such as exotic decays of the 125\,GeV Higgs $h$, radiative decays of bottomonium, direct production from $pp$ collisions in association with $b$-jets and also inclusively, where the main production mode is usually gluon-gluon fusion.

Despite the significantly lower luminosity collected with respect to ATLAS and CMS, LHCb has already proven to be capable of placing world-best limits for low-mass (pseudo)scalars produced in gluon-gluon fusion~\cite{Haisch:2016hzu, Haisch:2018kqx}, by searching simply for resonant pairs of opposite-sign muons~\cite{Aaij:2017rft, Aaij:2018xpt} (see Sec.~\ref{sec:review}). Indeed, a large fraction of these light particles are produced with large boosts at the LHC and end up in the LHCb acceptance. On top of that, the LHCb detector is capable of triggering on muons with low transverse momenta, greatly enhancing its acceptance to the low invariant mass di-muon signature with respect to ATLAS and CMS (see Sec.~\ref{sec:trigger}). Below we illustrate this using two specific model examples.

Working within an effective theory description below the EW breaking scale ($v \simeq 246 \, {\rm GeV}$), the relevant Lagrangian is given by
\be \label{eq:simplifiedmodel}
\begin{split}
\mathcal L \supset \frac{1}{2} \left[ (\partial P)^2 - m^2_P P^2 + (\partial S)^2 - m_S^2 S^2 \right] 
-  \sum_f \frac{m_f}{v} \left( i \kappa^f_P P \bar f \gamma_5 f +\kappa^f_S S \bar f f \right) \,,
\end{split}
\ee
One can easily match the above interactions to more complete EW  descriptions above the weak scale, such as multi-Higgs models. 
We have assumed that the new spin-0 particles $P,S$ couple to all SM fermions~$f$ in a flavour-conserving way and that their interactions conserve CP, which renders the coefficients $\kappa^f_{P,S}$ real (see Ref.~\cite{Haisch:2016hzu} for more details). 

The simplified model  (\ref{eq:simplifiedmodel}) is valid as long as the new  scalar $S$ does not mix strongly with the SM Higgs boson and there are no additional light degrees of freedom  below the EW scale. In such a case the model dependence associated to the full Higgs sector is encoded in the portal couplings $\kappa^f_{P,S}$.  The simplest choice of couplings is universal~$\kappa^f_{P,S} = \kappa_{P,S}$ and realised in singlet scalar extensions of the SM Higgs sector. Within the decoupling (and approximate $U(1)_{\rm PQ}$) limit of the 2HDM type II (2HDMII)~(see~e.g.~Refs.~\cite{Gunion:1989we,Branco:2011iw} for 2HDM reviews), a light 2HDM pseudoscalar $P=A$ emerges and one has instead $\kappa^{e,\mu,\tau,d,s,b}_{P} = t_\beta $, $\kappa_{P}^{u,c,t} = 1/t_\beta$ with $t_\beta \equiv \tan \beta$  denoting the ratio of vacuum expectation values of the two Higgs doublets. Another interesting possibility is the 2HDMII+S scenario. In this model a complex scalar singlet is added to the 2HDMII Higgs potential. The singlet field couples only to the two Higgs doublets but has no direct Yukawa couplings, acquiring all of its couplings to SM fermions through its mixing with the Higgs doublets. A light pseudoscalar $P=a$ can again arise in such a setup from the admixture of the 2HDM pseudoscalar and the imaginary part of the complex singlet. The corresponding mixing angle will be denoted by $\theta$, and defined such  that for $\theta \to 0$ the mass eigenstate $a$ becomes exactly singlet-like. In this setup, the effective couplings become $\kappa^{e,\mu,\tau,d,s,b}_{P} = s_\theta t_\beta $, $\kappa_{P}^{u,c,t} = s_\theta / t_\beta$\,, where $s_\theta \equiv \sin \theta$. For more model examples see Refs.~\cite{Haisch:2016hzu, Haisch:2018kqx}.

In the  mass range  of interest and under the assumption that the couplings $\kappa^f_{P,S}$ are approximately universal, the mediators~$P,S$ decay dominantly to $b\bar b$ (for $m_{P,S} > 2 m_b$), $c\bar c$ and $\tau^+ \tau^-$. Somewhat suppressed are instead the  $\mu^+\mu^-$ and $\gamma \gamma$ branching ratios. These features are illustrated in Fig.~1 of Ref.~\cite{Haisch:2016hzu}. From these plots it is also evident that for $\big | \kappa_{P,S}^{f} \big | \lesssim \mathcal O(1)$ the new resonances are narrow with the total decay widths~$\lsim 1 \, {\rm MeV}$.
The relative suppression of the clean $\mu^+\mu^-$ and $\gamma \gamma$ final states turns out to be of no big concern in practice given the sizeable production rates of light spin-0 states at the LHC. From Fig.~2 of Ref.~\cite{Haisch:2016hzu}, one sees that the inclusive cross sections at $8 \, {\rm TeV}$  for a  scalar or pseudoscalar of ${\cal O} ( 10 \, {\rm GeV})$ mass range from a few~${\rm nb}$ to tens of ${\rm nb}$.  
At LHCb, narrow resonances in dimuon decays are the most promising signals to look for these particles.
\begin{figure}
\centering
\includegraphics[width=0.3\textwidth]{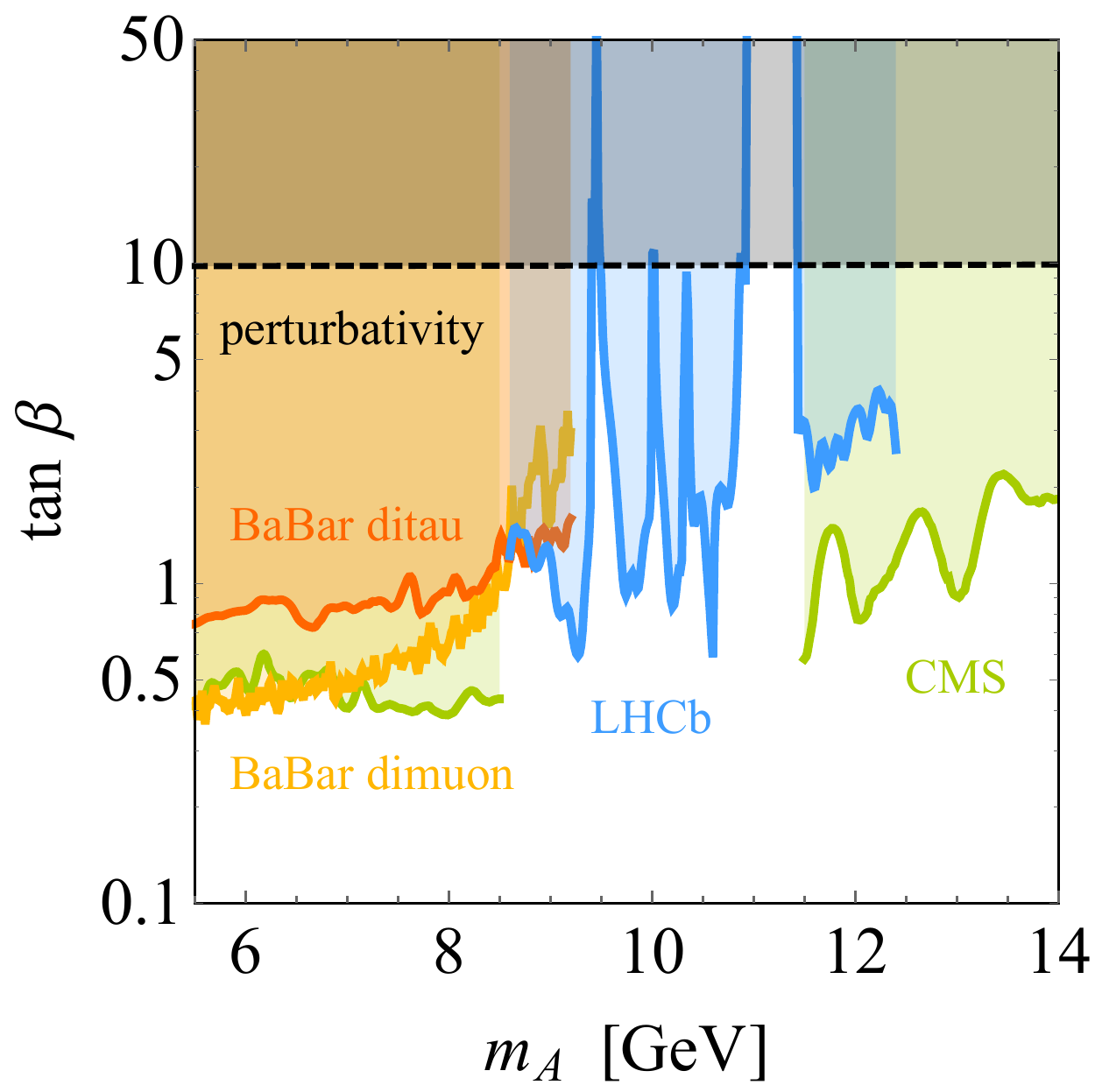}
\includegraphics[width=0.45\textwidth]{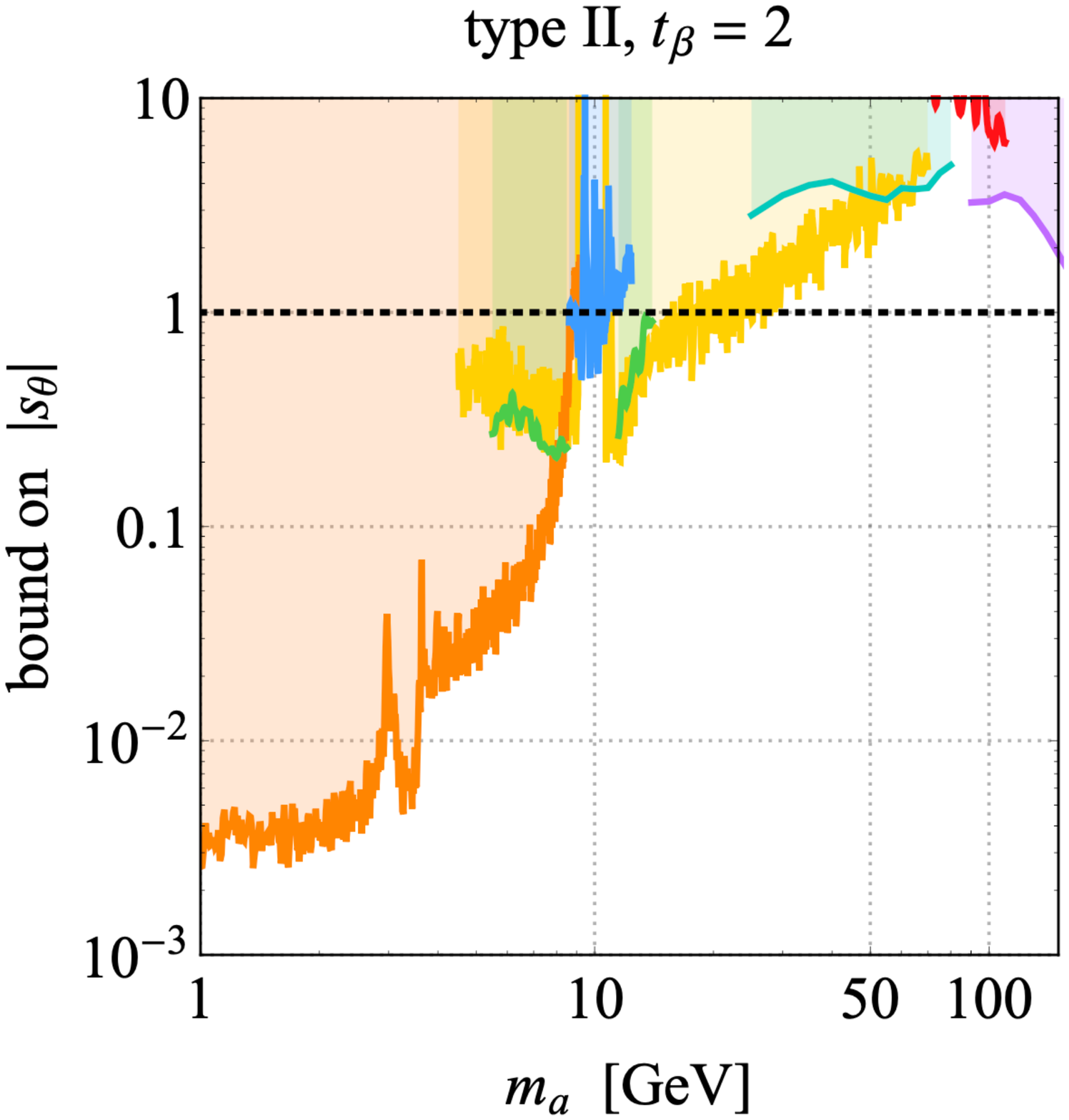}
\vspace{2mm}
\caption{\label{fig:3} {\bf Left:} 95\%~CL bounds on $\tan \beta$ in the 2HDMII scenario. The blue, green, and orange curves follow from $\Upsilon (n)$ production by LHCb~\cite{Aaij:2015awa, Haisch:2016hzu}, a CMS dimuon resonance search~\cite{Chatrchyan:2012am}, and the BaBar 90\%~CL limit on radiative $\Upsilon (1)$ decays in the dimuon~\cite{Lees:2012iw} and ditau~\cite{Lees:2012te} channels. The perturbativity bound on $\tan \beta$ is also shown~(black dashed). All shaded regions are excluded. For further details see Ref.~\cite{Haisch:2016hzu}. {\bf Right:}  Limits on $|s_\theta|$ in the 2HDMII+S with $t_\beta = 2$. The green, blue and yellow regions are the same as in the left panel, while the additional turquoise, red, purple and brown exclusions correspond to the CMS searches for $pp \to a b \bar b \to\tau^+ \tau^-  b \bar b$~\cite{Khachatryan:2015baw}, $pp \to a \to \gamma \gamma$~\cite{CMS:2017yta}, $pp \to a \to \tau^+ \tau^-$~\cite{CMS:2016rjp} and the LHCb inclusive dimuon cross section~\cite{Aaij:2017rft}. The dashed black lines indicate $|s_\theta| = 1$ and coloured regions are excluded at 95\%~CL apart from the orange and yellow contours which only hold at 90\%~CL. For further details see Ref.~\cite{Haisch:2018kqx}.
}
\end{figure}

In order to interpret experimental bounds on ${\cal O} ( 10 \, {\rm GeV})$ di-muon resonances in terms of new spin-0 states described by Eq.~\eqref{eq:simplifiedmodel} one needs to consider non-perturbative effects due to the presence of bottomonium resonances and the $b\bar b$ threshold. In particular, close to the $b\bar b$ threshold a perturbative description of the production and the decay of the new resonances breaks down. In this region we can however approximate the $b\bar b$ contributions to the $P,S$ widths by a sum over exclusive states interpolated to the continuum sufficiently above threshold~\cite{Drees:1989du,Baumgart:2012pj}. We can assume that the dominant contributions to production and the total width arise from the mixing of the new spin-0 mediators with bottomonium states. In particular, $P$ will mix with the  six $\eta_b (n)$ states, while $S$ will mix with the three $\chi_b(n)$ resonances, see Ref.~\cite{Haisch:2016hzu} for a detailed discussion.

Taking all of these considerations into account, as two examples of the important complementarity of LHCb di-muon searches with other experimental strategies, in Fig.~\ref{fig:3} we present combined experimental constraints on the parameter space of the explicit 2HDMII and 2HDMII+S model realizations of Eq.~\eqref{eq:simplifiedmodel}. 
In the 2HDMII case the constraints can be shown as a function of $\tan\beta$ on the pseudoscalar mass $m_A$, while in the 2HDMII+S setup for a fixed value of $t_\beta$, the measurements set limits on $|s_\theta|$ as a function of $m_a$. 

From Fig.~\ref{fig:3} and also other similar plots in Refs.~\cite{Haisch:2016hzu, Haisch:2018kqx}, we see that the existing collider constraints on models with light (pseudo)scalars are in general not very strong. The inclusive  dimuon~\cite{Ilten:2016tkc,Aaij:2017rft} cross section measurements that specifically focus on the $10-70\,{\rm GeV}$ mass region are competitive and being a worthwhile scientific goal of LHCb. 

\subsection{Axion-Like Particles}
If the UV model of new physics contains a spontaneously broken global U(1) symmetry, this gives rise to an Axion-like-particle~(ALP), a light pseudoscalar $a$ that is the (pseudo-)Nambu-Goldstone boson of the broken symmetry.  The ALP couples to the SM fields through dimension 5 operators, so that at $\mu\sim m_b$ the most general ALP interaction Lagrangian  is given by (up to higher dimension operators)
\begin{align}
\label{eq:La:QCDaxion}
{\cal L}_{a,{\rm int}} & = c_{GG}  \frac{a}{f_a} \frac{\alpha_s}{8 \pi} G_{\mu \nu} \tilde{G}^{\mu \nu} +  c_{\gamma\gamma}\frac{a}{f_a} \frac{\alpha_{\rm em}}{8 \pi} F_{\mu \nu} \tilde{F}^{\mu \nu}  +\frac{\partial_\mu a}{2 f_a} \overline{f}_i \gamma^\mu \left[ C^V_{ij} + C^A_{ij} \gamma_5 \right] f_j \, .
\end{align}
Conventionally, $c_{GG}$ is absorbed into the definition of the ALP decay constant $f_a$, i.e., in \eqref{eq:La:QCDaxion} $c_{GG}=1$ without loss of generality. The coupling to photons is $c_{\gamma\gamma}\sim {\mathcal O}(1)$ in the most common ALP models. Finally, in \eqref{eq:La:QCDaxion} we kept the fully general flavor structure of ALP couplings to the SM fermions, $f_i=u,d,s,c,b,e,\mu, \tau$ (the top quark as well as $W,Z, h$ were integrated out).  Because ALP is a pseudo Goldstone boson, its mass, $m_a$, is much smaller than the scale at which the global symmetry is broken,  i.e., $m_a \ll f_a$.

Since spontaneously broken symmetries are easy to come by, ALPs are obtained in many new physics models. Perhaps the most celebrated example of an ALP is the QCD axion which solves the strong-CP problem~\cite{Peccei:1977hh,Peccei:1977ur,Weinberg:1977ma,Wilczek:1977pj,Zhitnitsky:1980tq,Dine:1981rt}. The mass of the axion is entirely due to the anomalous coupling to gluons, and is given by~\cite{Gorghetto:2018ocs}, 
\begin{align}
m_a = 5.691(51) \mu eV \left( \frac{10^{12} \GeV}{f_a} \right) \, .
\end{align}
For the ``invisible'' axion the decay constant is $f_a \gg 10^6 \GeV$~\cite{Georgi:1986df}, in which case the axion is much lighter than an eV and essentially massless for collider purposes. Alternatively, $m_a$ can be introduced as an explicit breaking of the shift symmetry, with $m_a$ treated as a completely free parameter. If $m_a$ is in the 100s MeV to few GeV range the ALP mass becomes an important parameter also in collider processes. For small enough masses, below ${\mathcal O}({\rm keV})$ \cite{Calibbi:2020jvd}, ALPs are stable on cosmological time scales and can be valid dark matter candidates~(DM)~\cite{Abbott:1982af,Preskill:1982cy,Dine:1982ah}. Furthermore, ALPs can be a portal to DM and/or a dark sector~\cite{Nomura:2008ru,Freytsis:2010ne,Dolan:2014ska,Hochberg:2018rjs}, or be related to the cosmological solutions to the electroweak hierarchy problem \cite{Graham:2015cka}.

\subsubsection{Probing the flavor conserving ALP couplings \footnote{Contributed by Yotam Soreq.}}
\label{sec:flavor:cons:ALP}

We consider an ALP with mass $m_a$ in the MeV-to-GeV scale and a dominant coupling to gluons.
Its interaction term is given by
\begin{align}
    \mathcal{L} 
    \supset
    -\frac{4\pi\alpha_s}{\Lambda} a G_{\mu\nu} \tilde{G}^{\mu\nu} \, ,
\end{align}
where  $\Lambda$ is the UV cutoff (the ALP decay constant is $f_a=-\Lambda/32\pi^2$) and $G_{\mu\nu}$ is the gluon field strength and $\tilde{G}_{\mu\nu}$ its dual.
In this mass range the ALP phenomenology is challenging because of non perturbative QCD effects.
In particular, for  an ALP with a mass of $1\,{\rm GeV} \lesssim m_a \lesssim 3\,{\rm GeV}$ the ALP hadronic decay rates cannot be estimated neither by perturbative QCD (which holds for higher masses) nor by chiral perturbation theory (which hold for lighter ALPs). 
The hadronic decay rates of this scenario were estimated in Ref.~\cite{Aloni:2018vki} by using a data driven approach. 

\begin{figure}[!htp]
    \centering
    \includegraphics[width=0.49\textwidth]{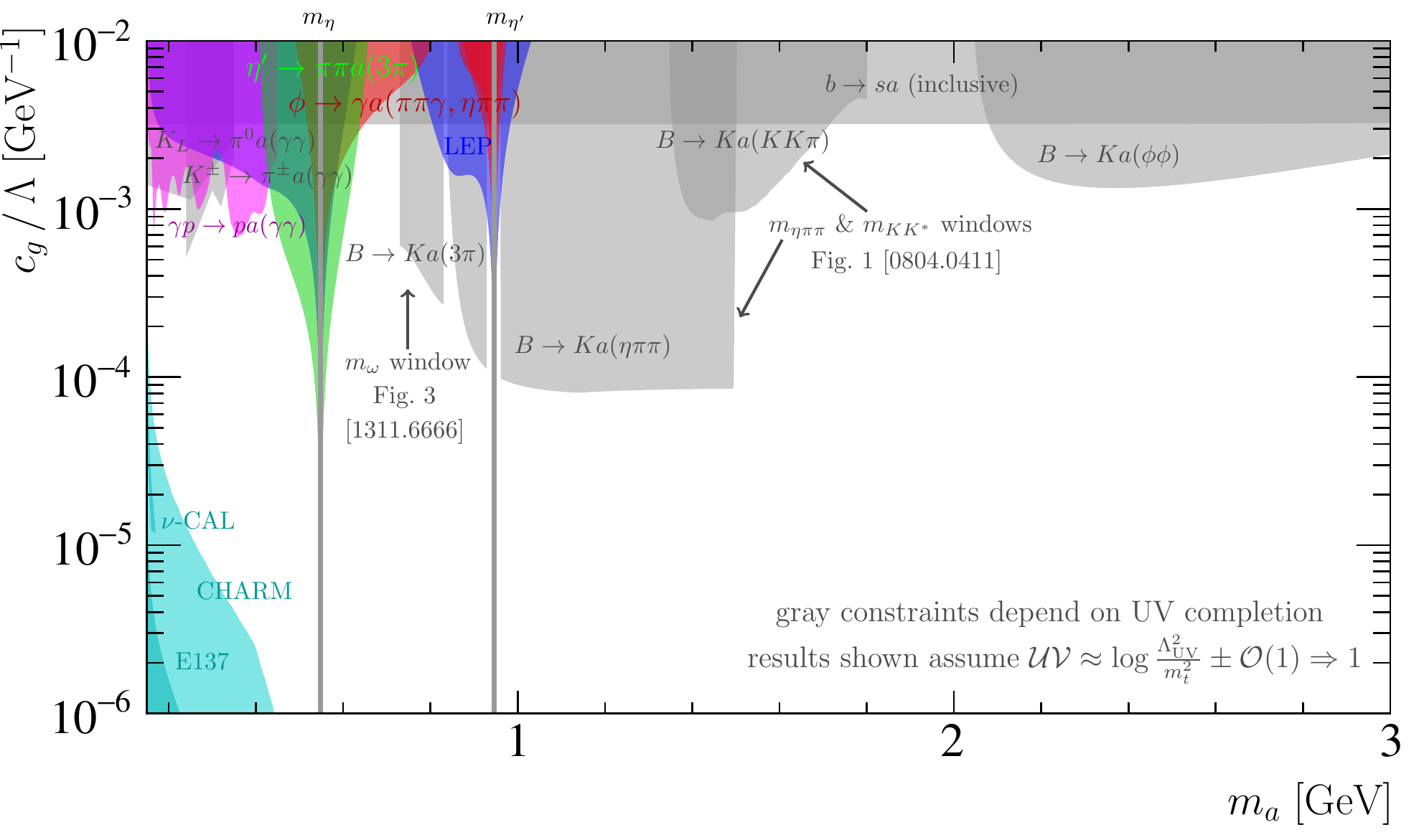}
    \caption{The bounds of $a G_{\mu\nu} \tilde{G}^{\mu\nu}$ as function of $m_a$ from different processes, adopted from Ref.~\cite{Aloni:2018vki}.
    }
    \label{fig:ALPGGtil}
\end{figure}

The phenomenology of this scenario and the current bound from different experimental probes were derived in Refs.~\cite{Marciano:2016yhf,Jaeckel:2015jla,Dobrich:2015jyk,Izaguirre:2016dfi,Knapen:2016moh,Mariotti:2017vtv,Bauer:2017ris,CidVidal:2018blh,Bauer:2018uxu,Harland-Lang:2019zur,Ebadi:2019gij,Mimasu:2014nea,Brivio:2017ije,Aloni:2018vki,Aloni:2019ruo}. 
For $m_a > 700\,{\rm MeV}$ one of the most sensitive ALP production mechanisms is $b\to s a$ decays, see \textit{e.g.}~\cite{Batell:2009jf}.
These transitions were studied at the two-loop level in Ref.~\cite{Chakraborty:2021wda}, where it was found that some of the existing limits are enhanced.
In Fig.~\ref{fig:ALPGGtil} we plot the current bounds on $a G_{\mu\nu} \tilde{G}^{\mu\nu}$ from the different probes. 

\subsubsection{Probing the flavor violating ALP couplings \footnote{Contributed by Jure Zupan.}}
\label{sec:flavor:NOcons:ALP}
If the ALP has flavor violating couplings, $C^{V,A}_{ij}\ne 0$, the FCNC decays such as $B^+\to K^+ a$, $K^+\to \pi^+ a$ or $\mu^+\to e^+ a$ can lead to a discovery of an ALP. To shorten the notation we follow \cite{MartinCamalich:2020dfe,Calibbi:2020jvd} and introduce
\begin{align}
\label{eq:FVA}
F^{V,A}_{ i j} & = \frac{2 f_a}{C^{V,A}_{i j}} \, , & F_{i j} & = \frac{2 f_a}{\sqrt{|C^{V}_{i j}|^2 + |C^{A}_{i j}|^2}} \, .  
\end{align}
We can distinguish two different regimes for sizes of $F_{ij}$ relative to $f_a$. These lead to very different phenomenology and possible search strategies. 

{\bf Maximal flavor violation.} If all flavor violating couplings to the SM fermions are large, $C^{V,A}_{i j}\sim {\mathcal O}(1)$ so that $F_{ij}\sim f_a$, then FCNC transitions lead to stringent bounds on the axion decay constant. The maximally flavor violating ALP is stable on collider timescales and escapes the detector, resulting in a missing energy signature. The case of effectively massless ALPs was considered in detail in Ref.~\cite{MartinCamalich:2020dfe}, with the most stringent constraints arising from $K^+\to \pi^+ a$ decay giving a bound $f_a\gtrsim 10^{12}$ GeV.  The bound is somewhat relaxed for heavier ALPs. For $m_a >m_K$ the most stringent constraint is  due to $B\to K^{(*)} a $ decays, which would contribute to the $B\to K^{(*)} \nu\bar \nu$ searches. The absence of a signal at Belle and BaBar translates to $f_a\gtrsim 10^{8}$ GeV. (A proper recast of experimental results for $m_a\ne 0$ has not been performed yet so the quoted bound is only approximate. 
Furthermore, for $m_a\lesssim m_B$ the ALP may already start decaying inside the detector, leading to displaced vertices signatures.) 

The maximal flavor violating ALPs can be searched for in $P_1\to P_2 a$ or $P_1\to  V_2 a$, with $P_{1,2} (V_2)$ the pseudoscalar (vector) mesons, with $a$ escaping the detector. The possible decays that have been considered in Ref.~\cite{MartinCamalich:2020dfe} (see also Refs.~\cite{Calibbi:2016hwq,Bauer:2019gfk,Cornella:2019uxs,Albrecht:2019zul,Calibbi:2020jvd,Davoudiasl:2021haa,Carmona:2021seb}) are, for $s\to d a$ transitions: $K^+\to \pi^+a$, $K^+\to \pi^+\pi^0 a$, $\Lambda\to n a$, $\Sigma^+\to p a$, $\Xi\to \Sigma a$, $\Xi^0\to \Lambda a$; for $c\to u a$ transitions: $D^+\to \pi^+a$, $\Lambda_c\to p a$; for $b\to s a$ transitions: $B^\to K^{(*)} a$, $\Lambda_b \to \Lambda a$, $B_s\to \mu^+\mu^- a$; for $b\to d a$ transitions: $B^+\to \pi^+ a$, $B\to \rho a$, $\Lambda_b\to n a$, $B\to \mu^+\mu^-a$. 

Several potentially interesting channels are lacking any experimental analyses so far. For example, there is no  experimental analysis of  $c\to u a$ transitions that are sensitive to the axial-vector coupling, i.e.,  there are no $D \to \pi \pi X_{\rm inv}$ or $D \to \rho X_{\rm inv}$,  $X_{\rm inv}=\nu \overline{\nu},a$, searches. One could also search for a $c\to u a$ signal in $D_s\to K a$, $D_s\to K^* a$ decays, all of which could be performed at Belle II and BESIII. 
 Potentially, LHCb could also probe these couplings using decay chains, such as $B^-\to D^0\pi^-$ followed by $D^0\to \rho^0a$, which results in three charged pions + MET and two displaced vertices.  The lack of such analyses means that there is at present no bound from meson decays on axial $cu$ couplings to the axion. Similarly, there is at present no publicly available experimental analysis that bounds the $B \to \rho a$ decays. 
 The recast bounds  on $B \to K^{(*)}a, B \to \pi a$ from \cite{MartinCamalich:2020dfe} could be easily improved by dedicated experimental searches using already collected data. At LHCb one could measure the $B\to K^* a$ and $B\to \rho a$ branching ratios using the decay chains such as $\bar B_s^{0**}\to K^+ B^-$ or $\bar B^{0**}\to \pi^+ B^-$ followed by  $B^-\to K^{*-}(\to K_S \pi^-)a $, or $\bar B_s^{0**}\to K_S \bar B^0$ followed by $\bar B^0 \to \bar K^{*0} a, \rho^0 a$ \cite{Stone:2014mza}.
 One could also attempt
 more challenging decay chain measurements such as $B_s^*\to B_s\gamma$, followed by $B_s\to \phi a$ or $B_s\to K^* a$.  

{\bf Suppressed flavor violation.} In this scenario ALPs couple predominantly to the SM through flavor universal couplings, i.e., $C^{V,A}_{i j} \ll 1 $ and thus $F_{ij}\gg f_a$ for $i\ne j$. The ALP can still be searched for using the $P_1\to P_2 a$ or $P_1\to  V_2 a$ decays, but for an ALP that decays inside the detector either promptly or via displaced vertices. The dominant decay channel depends on the flavor structure, as well as on which of the decay channels are kinematically allowed. A comprehensive search analysis would thus have to include all of the possible decays: the decay to two photons, $a\to \gamma\gamma$; the decays to leptons, $a\to e^+e^-, \mu^+\mu^-,\tau^+\tau^-$; as well as hadronic decays, $a\to gg, q\bar q$. At low ALP masses the hadronic decays will be dominated by two body decays, $a\to \pi\pi, \rho \pi, K\bar K$, while at larger ALP masses the $a$ decays to multihadron final states would need to be considered as well, since these are not expected to be suppressed. The relative branching ratios for the case where ALP couplings to gluons dominate, were discussed in Section \ref{sec:flavor:cons:ALP}.

\subsection{True Muonium\footnote{Contributed by Yotam Soreq.}}
\label{sec:TM}
\newcommand{\TM}{\ensuremath{\mathcal{T\!M}}\xspace}
\newcommand{\sig}{\ensuremath{\sigma_\mathrm{stat}}\xspace}

The True-Muonium~(\TM) is a $\mu^+\mu^-$ bound state,  hypothesized in 1971~\cite{Hughes:1971}, and yet to be experimentally observed.
In Ref.~\cite{CidVidal:2019qub}, it was shown that the LHCb experiment has the potential to discover the \TM in its lowest spin-1 state  with a statistical significance exceeding $5$ standard deviations using the expected 15\,fb$^{-1}$. 

The most promising \TM state in terms of discovery potential is the $1^2S_1$. 
This spin-1 state is similar to a dark photon with a mass of $m_{\TM}\approx 2m_\mu$ and a kinetic mixing of $\epsilon = \alpha^2/2\approx 2.7 \times 10^{-5}$. 
Thus, its phenomenology is similar to that of the dark photon, which has been subject to large number of recent studies, \textit{e.g.}~Refs.~\cite{Essig:2013lka,Alexander:2016aln,Beacham:2019nyx}, see also Sec.~\ref{sec:dp} in this paper, allowing us to use these developments in the search for the \TM at LHCb. 

The \TM $1^2S_1$ state has a lifetime of $\tau_{\TM}\approx 1.8 \times 10^{-12}\,$s and it dominantly decays to $e^+e^-$ with a branching ratio of ${\rm BR}(\TM\to e^+e^-)\approx 98\%\,$. 
Since LHCb is a forward detector, a large boost is expected for \TM falling within its acceptance, resulting in a relatively long proper propagation distance of
$0.53\,$mm, which typically produces a displaced $e^+e^-$ vertex at LHCb. 

As in the case of dark photon (with the same mass), the dominant production mechanism for \TM is $\eta\to \gamma \TM\to\gamma e^+e^-$ decays. 
Thus, two search strategies are considered. 
First, an inclusive search for a $e^+e^-$ displaced resonance with the \TM lifetime. 
Second, a search for a $e^+e^-$ resonance where the electron pair and a photon provide access to the $\eta$ mass. 
The advantage of the second option is having lower backgrounds. 
In both cases, the signal yield can be normalized by a data-driven method from the $\gamma^*\to e^-e^+$ rates, similarly to the case of $A'\to\mu^+\mu^-$~\cite{Ilten:2016tkc}. 
The main difference between the \TM and the dark photon cases is that the \TM can suffer dissociation as a result of its interaction with the material, with a typical length of $7.7\,$mm.
This results in a loss of $\sim50\,\%$ of the \TM signal. 

\begin{figure}[!htp]
    \centering
    \includegraphics[width=0.49\textwidth]{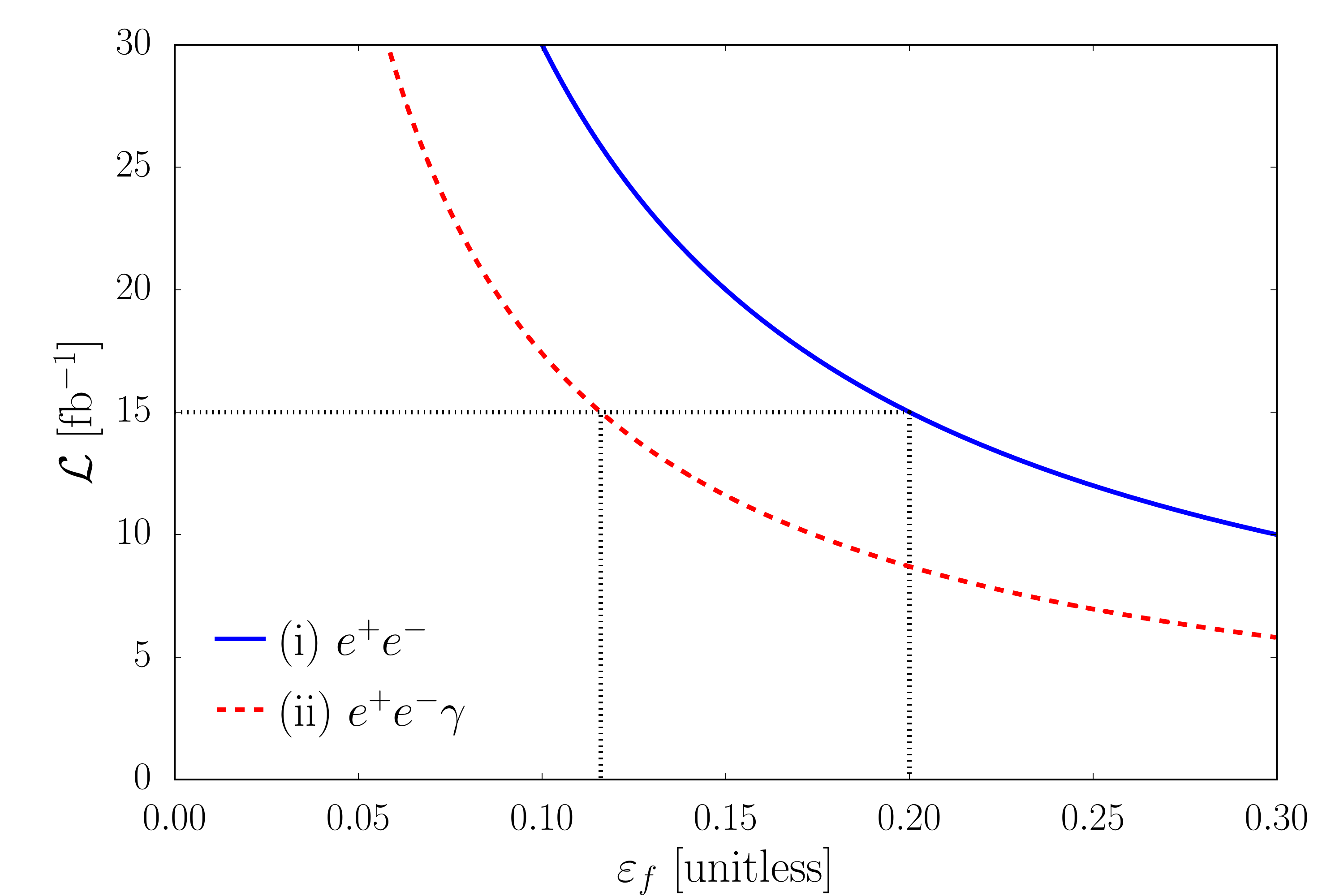}
    \caption{The required integrated luminosity for a $5\,\sig$ discovery of \TM as function of the final reconstruction efficiency, $\varepsilon_f$ for the proposed (blue) $e^+ e^-$ and (red) $e^+ e^-\gamma$ searches. The plot is adopted from Ref.~\cite{CidVidal:2019qub}}
    \label{fig:TM}
\end{figure}

The expected discovery potential in the two search modes as function of the final reconstruction efficiency, $\varepsilon_f$, is plot in Fig.~\ref{fig:TM}. 
As we can see, for 10\,\% to 20\,\% reconstruction efficiencies we expect 5 statistical standard deviation significance with an integrated luminosity of $15\,$fb$^{-1}$.

\subsection{Soft Bombs/SUEPs/Dark Showers\footnote{Contributed by Simon Knapen.}}
\label{sec:softbomb}
In this section we consider strongly coupled, quasi-conformal hidden sectors. They are examples of hidden valley models~\cite{Strassler:2006im,Strassler:2006ri} and are characterized by a high multiplicity of very soft final states. Hence they were originally labeled ``soft bombs'', though this terminology has been replaced by the experimental collaborations by the less controversial ``Soft Unclustered Energy Patterns'' (SUEP). In the context of the LHC, this possibility was first pointed out by Strassler in 2008 \cite{Strassler:2008bv} and its phenomenology was studied in more detail in Refs.~\cite{Knapen:2016hky,Cesarotti:2020ngq}.

Theoretically, models of this type are characterized by a large `t Hooft coupling and a quasi-conformal behavior, such that the gauge coupling remains strong over a large energy range. Fairly little is known about theories of this kind, and we can only make qualitative statements: Most notably, the fragmentation process in the dark sector is extremely efficient, and the soft and collinear enhancements that produce a jet-like structure in standard model QCD are expected to be absent. This results in a spherical configuration for which the final states have momenta of the order of the hadronization scale of the model \cite{Strassler:2008bv,Polchinski:2002jw,Hofman:2008ar,Lin:2007fa}. The energy distributions are approximately thermal, and can be parametrized by an effective Hagedorn temperature~\cite{Hagedorn:1965st}, $T$, a feature which is confirmed by AdS/CFT calculations \cite{Hatta:2008qx}. The shower is therefore characterized by 3 parameters: its total energy ($M$), the mass of the lightest meson, which ought of the order of the hadronization scale ($m$) and the temperature parameter $T$. The ratio of $T/m$ ought to be $\mathcal{O}(1)$ and varying it gives some systematic way of sampling the unknowns associated the strong dynamics. (Calling this a measure of the ``theory uncertainty'' however is probably exaggerating the level of theoretical control provided by this approach.) It is important to note that these are still idealized calculations, and one expects uncalculable, $\mathcal{O}(1)$ deviations from both sphericity and the Boltzman behavior of the final state momenta. See \cite{Cesarotti:2020uod} and \cite{Alimena:2019zri} for studies which explore what may happen, though still utilizing toy models. Within this framework however, one expects all hidden sector mesons to be semi-relativistic, and the overall parton multiplicity is roughly $\sim m/M$.

\begin{figure}
    \centering
    \includegraphics{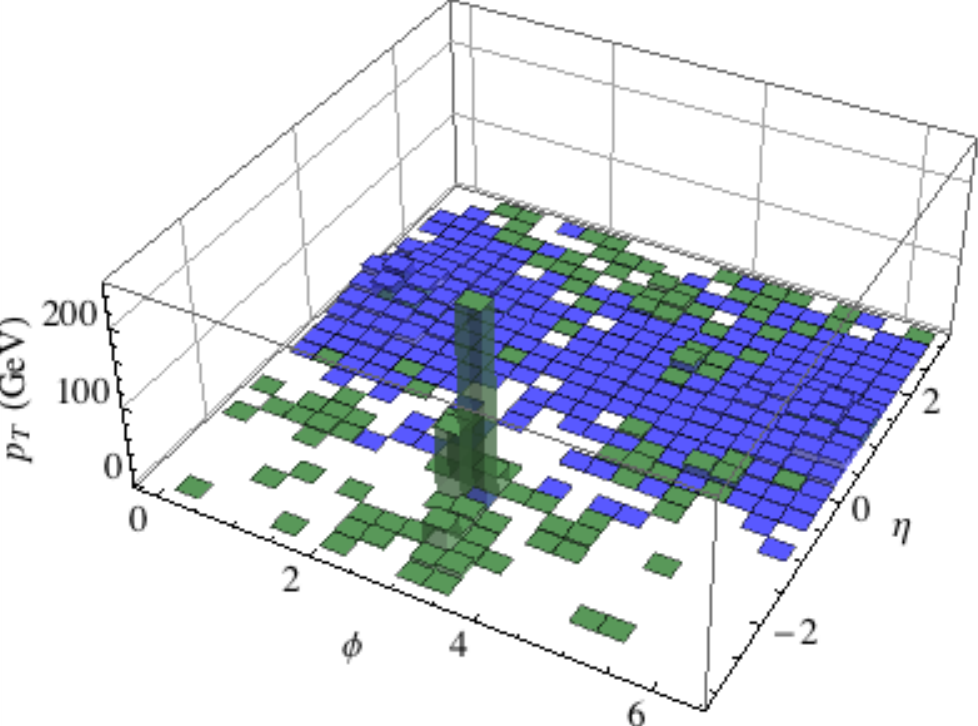}
    \caption{Lego plot of SUEP (blue) recoiling against an ISR jet (green).\label{suepeventdisplay} }
    \label{fig:my_label}
\end{figure}

 The hadron spectra in such a theory is unknown at the moment, and we are forced to work with a toy model of a single low laying hadron ($\phi$), which decays back to the SM. We identify one such ``hadron'' $\phi$ to each parton, in the sense explained above. Its lifetime and branching ratios are also poorly constrained. A fairly safe option would be to have $\phi$ decay to a light dark photon, which subsequently decays back to the SM, though other options may be feasible as well. 
 A soft bomb or SUEP may be initiated most easily through the decay of a heavy, neutral state. This may either be a $Z'$ boson or an exotic scalar, or it could be the SM Higgs itself. The event generation can be done with a dedicate pythia 8 plugin, based on Ref.~\cite{Knapen:2016hky}. It can be found at \url{https://gitlab.com/simonknapen/suep_generator}. An example event display is shown in Fig.~\ref{suepeventdisplay}. SUEP events are characterized by a very large multiplicity of soft tracks. The low $p_T$ threshold of the VELO may therefore provide unique sensitivity as compared to ATLAS and CMS.

\subsection{Quirks\footnote{Contributed by Matthew Low.}}
\label{sec:quirks}
Quirks are particles that are charged under a new confining force but where the confinement scale $\Lambda$ is far below the quirk mass $m_Q$ and there are no light particles charged under the new confining force~\cite{Kang:2008ea}.
As a result, when quirk and anti-quirk pairs are produced they remain bound by a flux tube of the confining force that acts like a string with length $\ell_{\rm str} \sim m_Q / \Lambda^2$.
This is unlike QCD where light quarks from the vacuum allow the QCD flux tube to break apart and form color-neutral hadrons.

There is no apriori expected value for $\Lambda$ but we can expect that LHC detectors may have sensitivity on scales of $\ell_{\rm str} \sim \mu{\rm m} - {\rm m}$ which corresponds to $\Lambda \sim 100~{\rm eV} - 1~{\rm MeV}$.\footnote{It turns out that the LHC detectors are actually sensitive to length scales of $\ell_{\rm str} \sim 100~{\rm km}$, corresponding to $\Lambda \sim 1~{\rm eV}$, due to their precise tracking capabilities \cite{Farina:2017cts}.}
Due to the string-like force between quirk and anti-quirk pairs, their dynamics are very different from standard particles over much of this parameter range.  Thus while the signal is very striking, it is also very difficult to search for.

Strong bounds for quirks exist for quirks for $\Lambda \lesssim 50~{\rm eV}$~\cite{Farina:2017cts} and a sensitive search has been proposed for $\Lambda \sim 1 - 30~{\rm keV}$~\cite{Knapen:2017kly}.  The range in-between of $\Lambda \sim 50~{\rm eV} - 1~{\rm keV}$ is comparatively weakly probed with the ATLAS and CMS detectors.  This parameter range corresponds to $\ell_{\rm str} \sim 1~{\rm cm} - 10~{\rm m}$ which is a region where LHCb may have sensitivity.
\\\\
{\bf Currrent Bounds.} The current bounds for quirks are cast in the $m_Q$ vs. $\Lambda$ plane and typically shown for a colored quirk with the same Standard Model gauge charges as an up-type quirk.  It is assumed that this quirk hadronizes into an electrically-charged quirk-hadron.  The production cross section is proportional to the size of the representation of the new confining force.  Typically we take the new confining group to be SU$(N_c)$ and compute limits for $N_c = 2$.

\begin{figure}
\begin{center}
\includegraphics[width=0.4\textwidth]{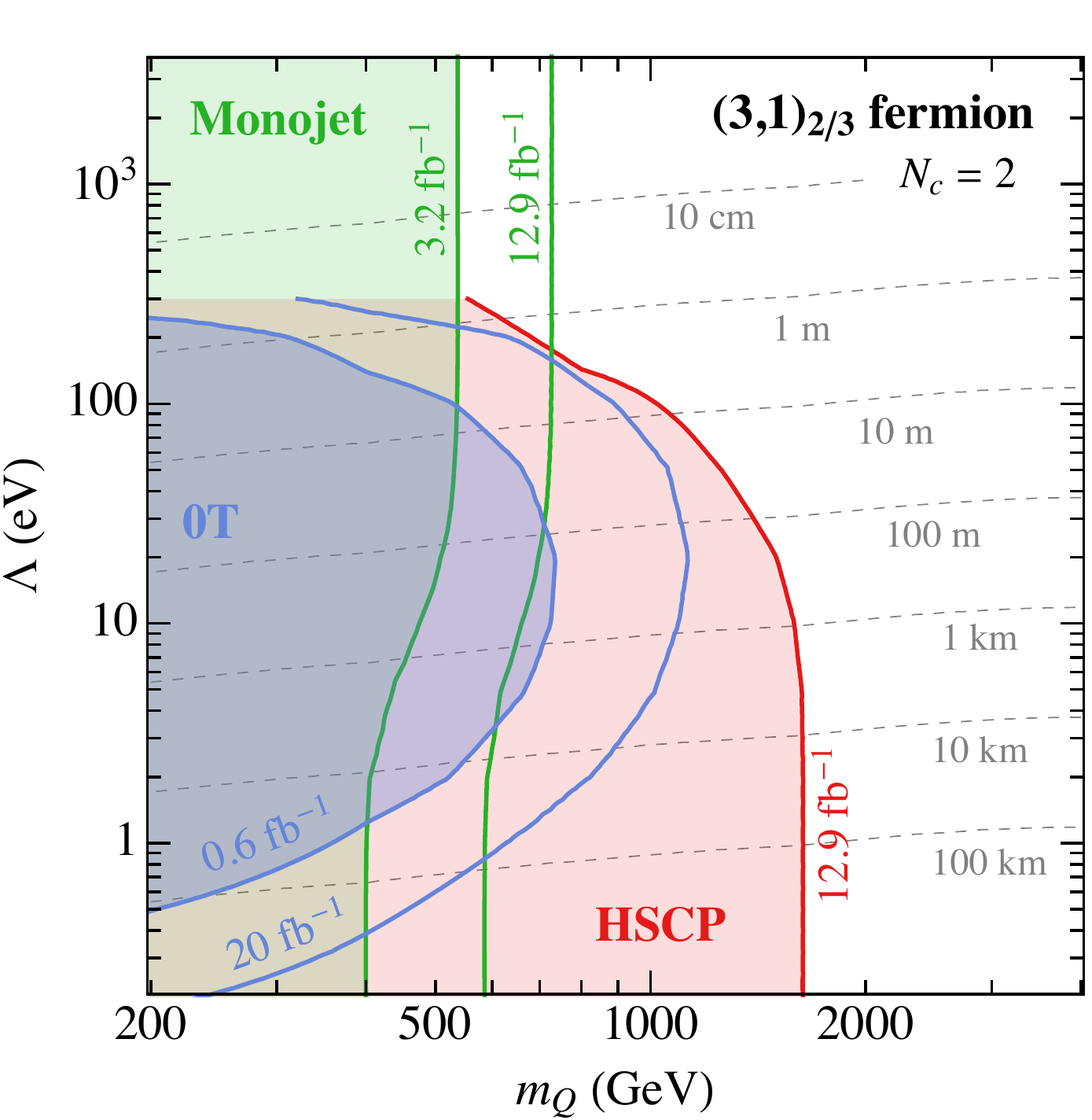} \quad\quad\quad
\includegraphics[width=0.45\textwidth]{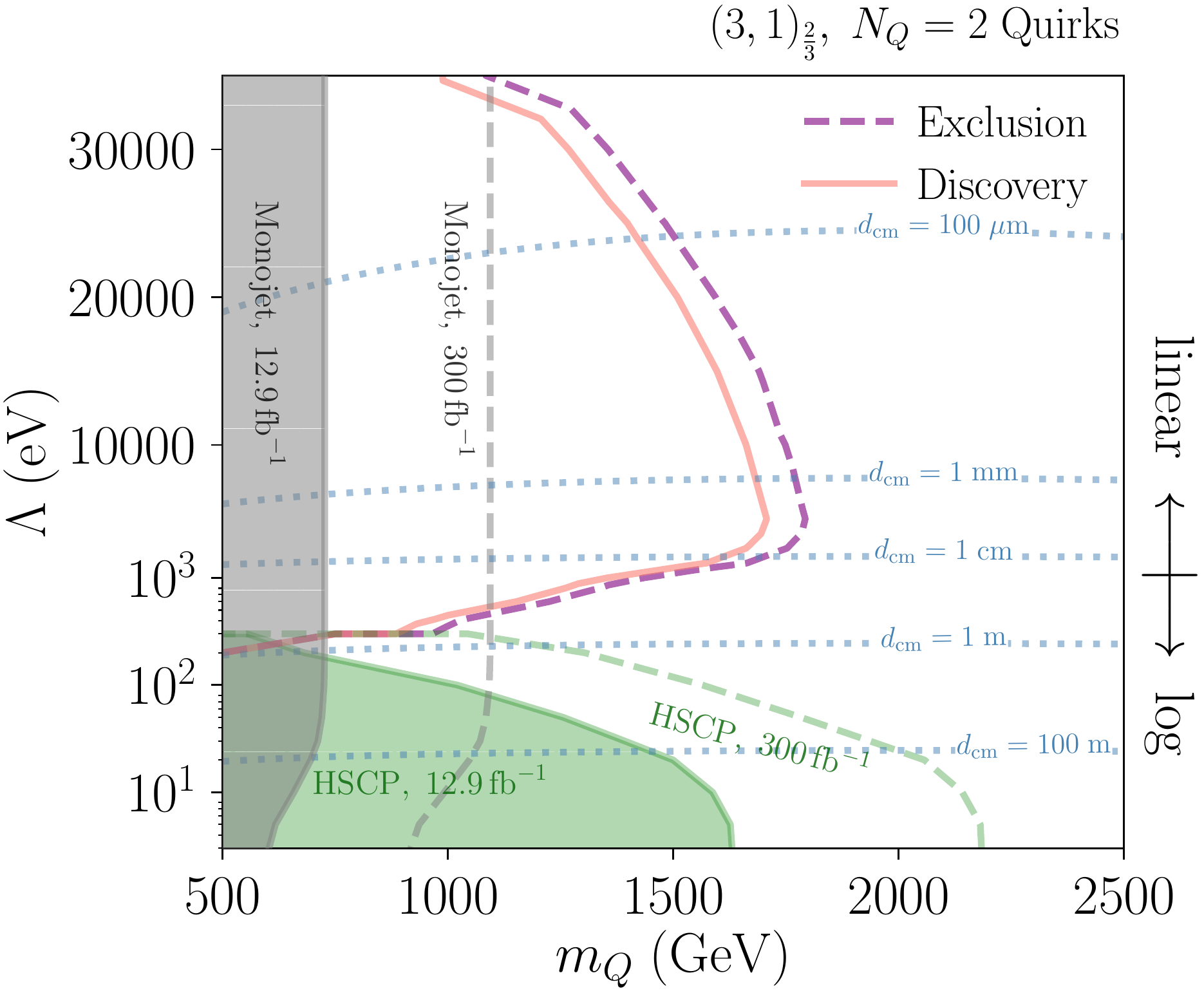}
\caption{Recast limits on quirks from~\cite{Farina:2017cts} (left) and projected limits on quirks from~\cite{Knapen:2017kly} (right).}
\label{fig:existing}
\end{center}
\end{figure}

In Ref.~\cite{Farina:2017cts} the authors recasted existing LHC searches into the quirk parameter space.  For relatively low $\Lambda$ values or long string lengths, the string-like force plays a minor role and the quirks appear similar to heavy stable charged particles (HSCP). Searches for HSCPs are very sensitive and the bounds on quirks are correspondingly strong.  Around $\Lambda \sim 50~{\rm eV}$ the string-like force has a sizable effect and the limits weaken.   For $\Lambda$ values above $50~{\rm eV}$ monojet searches can be recast.  Monojet searches simply look for missing energy and hard initial radiation so they suffer from a higher background and lower signal, resulting in weaker limits.  The bounds are shown in Fig.~\ref{fig:existing} (left).

In Ref.~\cite{Knapen:2017kly} the authors proposed a more sensitive search for the range $\Lambda \sim 1-30~{\rm keV}$.  In that range the effective string length between quirks and anti-quirks is $\ell_{\rm str} \sim 100~\mu{\rm m} - 1~{\rm cm}$ so that the two particles cross each other multiple times as the traverse the tracker.  The proposal is to look for pairs of tracker hits that lie in a plane.  In order to trigger on the events and for the quirks to be displaced away from the beam pipe this search requires an initially radiated jet with $p_T > 200~{\rm GeV}$.  Above $\sim 30~{\rm keV}$ the quirk and anti-quirk are separated by $\lesssim 10~\mu{\rm m}$ and the hits are not reliably resolved resulting in diminishing sensitivity.
As part of the event selection the authors make a cut on the effective quirk and anti-quirk separation of $1~{\rm cm}$ because above these separation values the backgrounds rapidly increase.  This is the cause for the fall in sensitivity below $\sim 1~{\rm keV}$.
\\\\\\
{\bf Search at LHCb.} From Fig.~\ref{fig:existing} we see that even including existing limits~\cite{Farina:2017cts} and proposed searches~\cite{Knapen:2017kly} there are not strong bounds for $\Lambda \sim 50~{\rm eV} - 1~{\rm keV}$ despite the fact that this length scale corresponds to very exotic signatures at the LHC.  In this range, the LHCb experiment may be able to set competitive bounds thanks to several unique features of the detector.

\begin{figure}
\begin{center}
\includegraphics[width=0.6\textwidth]{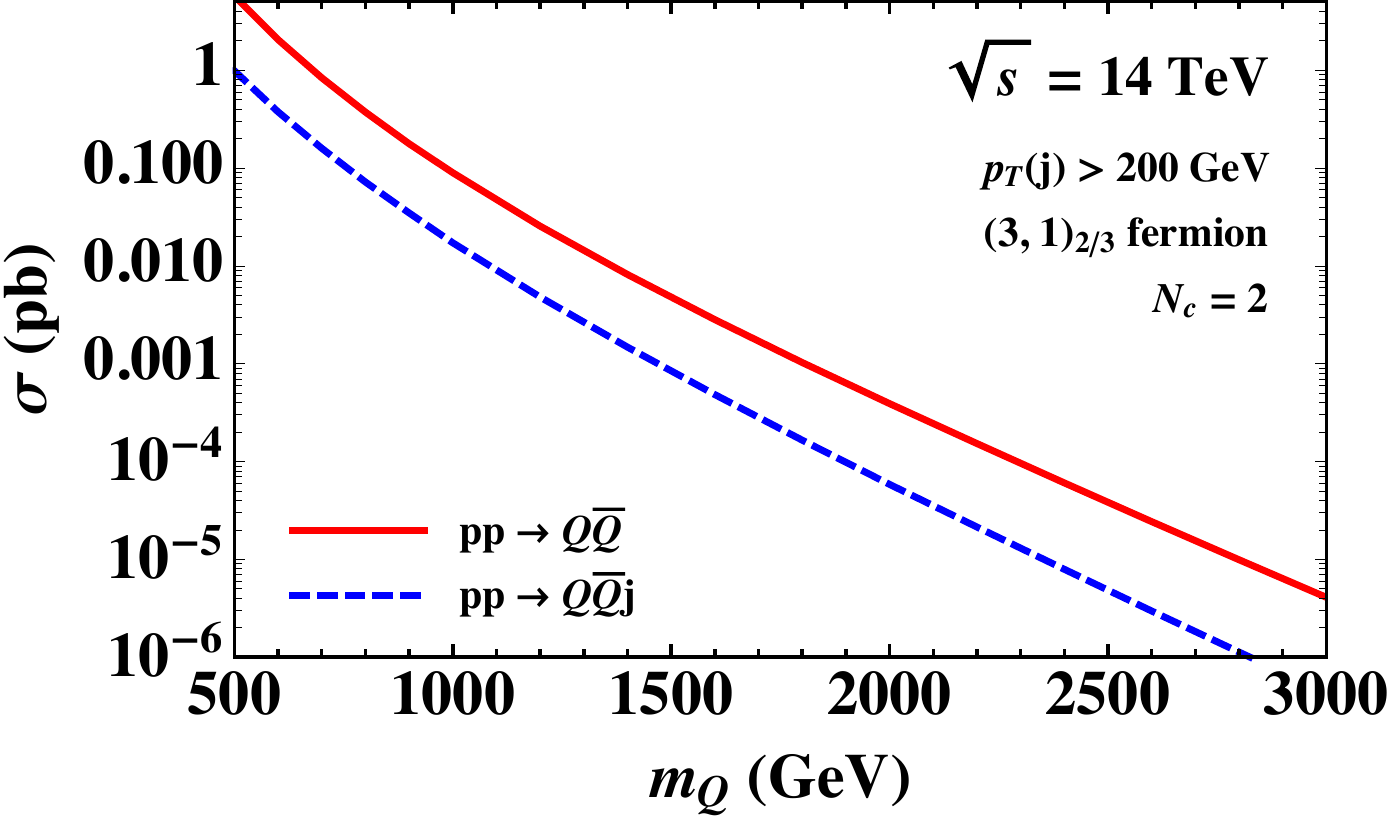}
\caption{Cross sections for a quirk $Q$ with up-type Standard Model gauge charges and $N_c=2$ for $pp \to Q\bar{Q}$ (red) and for $pp \to Q\bar{Q}j$ (blue).}
\label{fig:rates}
\end{center}
\end{figure}

One advantage that LHCb has is that it covers a more forward region than the ATLAS and CMS detectors do.  For quirk searches above $\sim 50~{\rm eV}$ all the methods rely on the quirk and anti-quirk system recoiling against some initial state radiation.  This requirement reduces the signal rate.  Without initial state radiation we expect the quirk and anti-quirk to oscillate around the beam line either in the forward or backward direction since the system will still have a longitudinal boost.  Fig.~\ref{fig:rates} compares the rates with and without initial state radiation (with $p_T > 200~{\rm GeV}$ as a reference).

We propose that LHCb looks for quirk and anti-quirk pairs traveling in the forward direction with minimal transverse recoil.  The signal would then consist of pairs of hits in each layer of the VELO.  These hits will not necessarily be at larger radii for more forward VELO layers which is unlike standard particle tracks.  The quirk and anti-quirk, however, should leave hits at opposite $\phi$ coordinates which will vary very little between the VELO layers.  This $\phi$ information can be utilized to reduce backgrounds. Similar to HSCP searches at LHCb there should be no corresponding signal in the RICH detectors.

We expect the sensitivity for such a search to peak when the quirk and anti-quirk hit each layer of the VELO.  This should be the case for effective string lengths inside the radius of the VELO, $\ell_{\rm str} \lesssim 1~{\rm m}$.  The VELO also excludes an inner radius of $1~{\rm cm}$ so the search sensitivity should sharply fall off for $\ell_{\rm str} \lesssim 1~{\rm cm}$.  The reach of this search would depend on a detailed simulation of the signal acceptance and backgrounds, however, we expect reasonable bounds for $\Lambda \sim 1 - 10~{\rm eV}$.

\section{Conclusions}
\label{sec:conclusions}
The LHCb experiment has great potential to investigate BSM signatures stemming from models of Stealth physics. In the first part of this paper, we have provided a compendium of the unique capabilities of the LHCb detector and its upgrades, including a detailed discussion of a set of final states of interest for Stealth signatures.  

This guide to the experimental capabilities of LHCb in Stealth physics is complemented by a non-exhaustive theoretical review of the SM extensions that LHCb has the potential to explore. These include ideas such as neutral naturalness, composite Higgs models, dark Sectors, heavy neutrinos and axion-like particles. The collider phenomenology of these models is briefly laid out and some sensitivity projections are provided, showcasing the LHCb potential in comparison to other experiments.

At present, few of these discovery opportunities have been explored at LHCb and with this paper we aim to reinforce the discussion, especially across the theoretical and experimental communities. It is imperative to do this now as most of these searches rely on special online selections that have to be implemented before data-taking. We hope this document will foster the growth of the LHCb Stealth program.

\section{Acknowledgements}
\label{sec:acknowledgements}
We thank the participants in the ``Stealth Physics at LHCb'' workshop, held in Santiago de Compostela in February 2020\cite{stealth_workshop}, for many fruitful discussions and for triggering many ideas that were developed here. We also thank the LHC Long-Lived Particle community, and the LPCC LHC Long-lived Particle and Dark Matter working groups for fostering and nurturing the interest in the ideas here discussed. We kindly acknowledge Vladimir Gligorov, Stephen Farry and Niels Tuning for kindly reading and providing comments to this document. This work has been supported by MINECO (Spain) through the Ram\'{o}n y Cajal program RYC-2016-
20073 and by XuntaGAL under the ED431F 2018/01 project. It has also received
financial support from XuntaGAL (Centro singular de investigaci\'{o}n de Galicia accreditation 2019-2022), by European Union ERDF, by the “Mar\'{i}a de Maeztu” Units of Excellence
program MDM-2016-0692 and the Spanish Research State Agency, and by the {\it Generalitat Valenciana} (Spain) through the {\it plan GenT} program (CIDEGENT/2019/068).

\addcontentsline{toc}{section}{References}
\bibliographystyle{bibtexFile}

\bibliography{main}

\ifx\mcitethebibliography\mciteundefinedmacro
\PackageError{LHCb.bst}{mciteplus.sty has not been loaded}
{This bibstyle requires the use of the mciteplus package.}\fi
\providecommand{\href}[2]{#2}
\begin{mcitethebibliography}{100}
\mciteSetBstSublistMode{n}
\mciteSetBstMaxWidthForm{subitem}{\alph{mcitesubitemcount})}
\mciteSetBstSublistLabelBeginEnd{\mcitemaxwidthsubitemform\space}
{\relax}{\relax}

\bibitem{Allanach:2016yth}
B.~C. Allanach, \ifthenelse{\boolean{articletitles}}{\emph{{Beyond the Standard
  Model Lectures for the 2016 European School of High-Energy Physics}}, }{} in
  {\em {2016 European School of High-Energy Physics}},
  \href{https://doi.org/10.23730/CYRSP-2017-005.123}{ 2016},
  \href{http://arxiv.org/abs/1609.02015}{{\normalfont\ttfamily
  arXiv:1609.02015}}\relax
\mciteBstWouldAddEndPuncttrue
\mciteSetBstMidEndSepPunct{\mcitedefaultmidpunct}
{\mcitedefaultendpunct}{\mcitedefaultseppunct}\relax
\EndOfBibitem
\bibitem{Alimena:2019zri}
J.~Alimena {\em et~al.}, \ifthenelse{\boolean{articletitles}}{\emph{{Searching
  for long-lived particles beyond the Standard Model at the Large Hadron
  Collider}}, }{}\href{https://doi.org/10.1088/1361-6471/ab4574}{J.\ Phys.\ G
  \textbf{47} (2020) 090501},
  \href{http://arxiv.org/abs/1903.04497}{{\normalfont\ttfamily
  arXiv:1903.04497}}\relax
\mciteBstWouldAddEndPuncttrue
\mciteSetBstMidEndSepPunct{\mcitedefaultmidpunct}
{\mcitedefaultendpunct}{\mcitedefaultseppunct}\relax
\EndOfBibitem
\bibitem{Alves:2008zz}
LHCb collaboration, A.~A. Alves, Jr.\ {\em et~al.},
  \ifthenelse{\boolean{articletitles}}{\emph{{The LHCb Detector at the LHC}},
  }{}\href{https://doi.org/10.1088/1748-0221/3/08/S08005}{JINST \textbf{3}
  (2008) S08005}\relax
\mciteBstWouldAddEndPuncttrue
\mciteSetBstMidEndSepPunct{\mcitedefaultmidpunct}
{\mcitedefaultendpunct}{\mcitedefaultseppunct}\relax
\EndOfBibitem
\bibitem{LHCb-TDR-013}
LHCb collaboration, \ifthenelse{\boolean{articletitles}}{\emph{{LHCb VELO
  Upgrade Technical Design Report}}}{}  {CERN-LHCC-2013-021}, CERN, Geneva,
  2013\relax
\mciteBstWouldAddEndPuncttrue
\mciteSetBstMidEndSepPunct{\mcitedefaultmidpunct}
{\mcitedefaultendpunct}{\mcitedefaultseppunct}\relax
\EndOfBibitem
\bibitem{LHCb-TDR-015}
LHCb collaboration, \ifthenelse{\boolean{articletitles}}{\emph{{LHCb Tracker
  Upgrade Technical Design Report}}, }{}
  \href{http://cdsweb.cern.ch/search?p=CERN-LHCC-2014-001&f=reportnumber&action_search=Search&c=LHCb}
  {CERN-LHCC-2014-001}, 2014\relax
\mciteBstWouldAddEndPuncttrue
\mciteSetBstMidEndSepPunct{\mcitedefaultmidpunct}
{\mcitedefaultendpunct}{\mcitedefaultseppunct}\relax
\EndOfBibitem
\bibitem{LHCb-TDR-016}
LHCb collaboration, \ifthenelse{\boolean{articletitles}}{\emph{{LHCb Trigger
  and Online Technical Design Report}}, }{}
  \href{http://cdsweb.cern.ch/search?p=CERN-LHCC-2014-016&f=reportnumber&action_search=Search&c=LHCb}
  {CERN-LHCC-2014-016}, 2014\relax
\mciteBstWouldAddEndPuncttrue
\mciteSetBstMidEndSepPunct{\mcitedefaultmidpunct}
{\mcitedefaultendpunct}{\mcitedefaultseppunct}\relax
\EndOfBibitem
\bibitem{Follin:2014nva}
F.~Follin and D.~Jacquet,
  \ifthenelse{\boolean{articletitles}}{\emph{{Implementation and experience
  with luminosity levelling with offset beam}}, }{} in {\em {ICFA Mini-Workshop
  on Beam-Beam Effects in Hadron Colliders}},
  \href{https://doi.org/10.5170/CERN-2014-004.183}{ 2014},
  \href{http://arxiv.org/abs/1410.3667}{{\normalfont\ttfamily
  arXiv:1410.3667}}\relax
\mciteBstWouldAddEndPuncttrue
\mciteSetBstMidEndSepPunct{\mcitedefaultmidpunct}
{\mcitedefaultendpunct}{\mcitedefaultseppunct}\relax
\EndOfBibitem
\bibitem{Chatrchyan:1704291}
CMS collaboration, S.~Chatrchyan {\em et~al.},
  \ifthenelse{\boolean{articletitles}}{\emph{{Description and performance of
  track and primary-vertex reconstruction with the CMS tracker}},
  }{}\href{https://doi.org/10.1088/1748-0221/9/10/P10009}{JINST \textbf{9}
  (2014) P10009. 80 p},
  \href{http://arxiv.org/abs/1405.6569}{{\normalfont\ttfamily
  arXiv:1405.6569}}\relax
\mciteBstWouldAddEndPuncttrue
\mciteSetBstMidEndSepPunct{\mcitedefaultmidpunct}
{\mcitedefaultendpunct}{\mcitedefaultseppunct}\relax
\EndOfBibitem
\bibitem{LHCb:2001aa}
LHCb collaboration, \ifthenelse{\boolean{articletitles}}{\emph{{LHCb VELO TDR:
  Vertex locator. Technical design report}}}{} , 2001\relax
\mciteBstWouldAddEndPuncttrue
\mciteSetBstMidEndSepPunct{\mcitedefaultmidpunct}
{\mcitedefaultendpunct}{\mcitedefaultseppunct}\relax
\EndOfBibitem
\bibitem{LHCbCollaboration:2013bkh}
LHCb collaboration, I.~Bediaga,
  \ifthenelse{\boolean{articletitles}}{\emph{{LHCb VELO Upgrade Technical
  Design Report}}}{} , 2013\relax
\mciteBstWouldAddEndPuncttrue
\mciteSetBstMidEndSepPunct{\mcitedefaultmidpunct}
{\mcitedefaultendpunct}{\mcitedefaultseppunct}\relax
\EndOfBibitem
\bibitem{LHCb-DP-2014-002}
LHCb collaboration, R.~Aaij {\em et~al.},
  \ifthenelse{\boolean{articletitles}}{\emph{{LHCb detector performance}},
  }{}\href{https://doi.org/10.1142/S0217751X15300227}{Int.\ J.\ Mod.\ Phys.\
  \textbf{A30} (2015) 1530022},
  \href{http://arxiv.org/abs/1412.6352}{{\normalfont\ttfamily
  arXiv:1412.6352}}\relax
\mciteBstWouldAddEndPuncttrue
\mciteSetBstMidEndSepPunct{\mcitedefaultmidpunct}
{\mcitedefaultendpunct}{\mcitedefaultseppunct}\relax
\EndOfBibitem
\bibitem{Aaij:2017rft}
LHCb collaboration, R.~Aaij {\em et~al.},
  \ifthenelse{\boolean{articletitles}}{\emph{{Search for Dark Photons Produced
  in 13 TeV $pp$ Collisions}},
  }{}\href{https://doi.org/10.1103/PhysRevLett.120.061801}{Phys.\ Rev.\ Lett.\
  \textbf{120} (2018) 061801},
  \href{http://arxiv.org/abs/1710.02867}{{\normalfont\ttfamily
  arXiv:1710.02867}}\relax
\mciteBstWouldAddEndPuncttrue
\mciteSetBstMidEndSepPunct{\mcitedefaultmidpunct}
{\mcitedefaultendpunct}{\mcitedefaultseppunct}\relax
\EndOfBibitem
\bibitem{schedule}
\ifthenelse{\boolean{articletitles}}{\emph{Longer term lhc schedule}, }{}
  \url{https://lhc-commissioning.web.cern.ch/schedule/LHC-long-term.htm}.
\newblock Accessed: 26/05/2021\relax
\mciteBstWouldAddEndPuncttrue
\mciteSetBstMidEndSepPunct{\mcitedefaultmidpunct}
{\mcitedefaultendpunct}{\mcitedefaultseppunct}\relax
\EndOfBibitem
\bibitem{LHCb-TDR-012}
LHCb collaboration, \ifthenelse{\boolean{articletitles}}{\emph{{Framework TDR
  for the LHCb Upgrade: Technical Design Report}}, }{}
  \href{http://cdsweb.cern.ch/search?p=CERN-LHCC-2012-007&f=reportnumber&action_search=Search&c=LHCb}
  {CERN-LHCC-2012-007}, 2012\relax
\mciteBstWouldAddEndPuncttrue
\mciteSetBstMidEndSepPunct{\mcitedefaultmidpunct}
{\mcitedefaultendpunct}{\mcitedefaultseppunct}\relax
\EndOfBibitem
\bibitem{LHCb-PII-Physics}
LHCb collaboration, \ifthenelse{\boolean{articletitles}}{\emph{{Physics case
  for an LHCb Upgrade II --- Opportunities in flavour physics, and beyond, in
  the HL-LHC era}},
  }{}\href{http://arxiv.org/abs/1808.08865}{{\normalfont\ttfamily
  arXiv:1808.08865}}\relax
\mciteBstWouldAddEndPuncttrue
\mciteSetBstMidEndSepPunct{\mcitedefaultmidpunct}
{\mcitedefaultendpunct}{\mcitedefaultseppunct}\relax
\EndOfBibitem
\bibitem{Nakada:1100545}
LHCb collaboration, T.~Nakada, O.~Ullaland, and W.~Witzelling,
  \ifthenelse{\boolean{articletitles}}{\emph{{Expression of Interest for an
  LHCb Upgrade}}}{},
  \href{https://cds.cern.ch/record/1100545}{CERN-LHCC-2008-007. LHCC-G-139},
  CERN, Geneva, 2008\relax
\mciteBstWouldAddEndPuncttrue
\mciteSetBstMidEndSepPunct{\mcitedefaultmidpunct}
{\mcitedefaultendpunct}{\mcitedefaultseppunct}\relax
\EndOfBibitem
\bibitem{CERN-LHCC-2011-001}
 \ifthenelse{\boolean{articletitles}}{\emph{{Letter of Intent for the LHCb
  Upgrade}}}{}, \href{https://cds.cern.ch/record/1333091}{CERN-LHCC-2011-001.
  LHCC-I-018}, CERN, Geneva, 2011\relax
\mciteBstWouldAddEndPuncttrue
\mciteSetBstMidEndSepPunct{\mcitedefaultmidpunct}
{\mcitedefaultendpunct}{\mcitedefaultseppunct}\relax
\EndOfBibitem
\bibitem{LHCb-TDR-014}
LHCb collaboration, \ifthenelse{\boolean{articletitles}}{\emph{{LHCb PID
  Upgrade Technical Design Report}}, }{}
  \href{http://cdsweb.cern.ch/search?p=CERN-LHCC-2013-022&f=reportnumber&action_search=Search&c=LHCb}
  {CERN-LHCC-2013-022}, 2013\relax
\mciteBstWouldAddEndPuncttrue
\mciteSetBstMidEndSepPunct{\mcitedefaultmidpunct}
{\mcitedefaultendpunct}{\mcitedefaultseppunct}\relax
\EndOfBibitem
\bibitem{LHCb-TDR-010}
LHCb collaboration, \ifthenelse{\boolean{articletitles}}{\emph{{LHCb trigger
  system: Technical Design Report}}, }{}
  \href{http://cdsweb.cern.ch/search?p=CERN-LHCC-2003-031&f=reportnumber&action_search=Search&c=LHCb}
  {CERN-LHCC-2003-031}, 2003\relax
\mciteBstWouldAddEndPuncttrue
\mciteSetBstMidEndSepPunct{\mcitedefaultmidpunct}
{\mcitedefaultendpunct}{\mcitedefaultseppunct}\relax
\EndOfBibitem
\bibitem{Aaij:2019zbu}
R.~Aaij {\em et~al.}, \ifthenelse{\boolean{articletitles}}{\emph{{Allen: A high
  level trigger on GPUs for LHCb}},
  }{}\href{https://doi.org/10.1007/s41781-020-00039-7}{Comput.\ Softw.\ Big
  Sci.\  \textbf{4} (2020) 7},
  \href{http://arxiv.org/abs/1912.09161}{{\normalfont\ttfamily
  arXiv:1912.09161}}\relax
\mciteBstWouldAddEndPuncttrue
\mciteSetBstMidEndSepPunct{\mcitedefaultmidpunct}
{\mcitedefaultendpunct}{\mcitedefaultseppunct}\relax
\EndOfBibitem
\bibitem{LHCb-PII-EoI}
LHCb collaboration, \ifthenelse{\boolean{articletitles}}{\emph{{Expression of
  Interest for a Phase-II LHCb Upgrade: Opportunities in flavour physics, and
  beyond, in the HL-LHC era}}, }{}
  \href{http://cdsweb.cern.ch/search?p=CERN-LHCC-2017-003&f=reportnumber&action_search=Search&c=LHCb}
  {CERN-LHCC-2017-003}, 2017\relax
\mciteBstWouldAddEndPuncttrue
\mciteSetBstMidEndSepPunct{\mcitedefaultmidpunct}
{\mcitedefaultendpunct}{\mcitedefaultseppunct}\relax
\EndOfBibitem
\bibitem{Koppenburg:2059944}
P.~Koppenburg and V.~Vagnoni,
  \ifthenelse{\boolean{articletitles}}{\emph{{Precision physics with
  heavy-flavoured hadrons. Precision physics with heavy-flavoured hadrons}},
  }{}\href{https://doi.org/10.1142/9789814644150_0002}{Adv.\ Ser.\ Dir.\ High
  Energy Phys.\  \textbf{23} (2015) 31},
  \href{http://arxiv.org/abs/1510.04466}{{\normalfont\ttfamily
  arXiv:1510.04466}}, Comments: 31 pages, 10 figures in 60 Years of CERN
  Experiments and Discoveries, Advanced Series on Directions in High Energy
  Physics 23 (2015), World Scientific Publishing\relax
\mciteBstWouldAddEndPuncttrue
\mciteSetBstMidEndSepPunct{\mcitedefaultmidpunct}
{\mcitedefaultendpunct}{\mcitedefaultseppunct}\relax
\EndOfBibitem
\bibitem{Bhasin:2729016}
S.~Bhasin {\em et~al.},  \ifthenelse{\boolean{articletitles}}{\emph{{TORCH
  physics performance : improving low-momentum PID performance during Upgrade
  IB and beyond}}}{},
  \href{https://cds.cern.ch/record/2729016}{LHCb-PUB-2020-006.
  CERN-LHCb-PUB-2020-006}, CERN, Geneva, 2020\relax
\mciteBstWouldAddEndPuncttrue
\mciteSetBstMidEndSepPunct{\mcitedefaultmidpunct}
{\mcitedefaultendpunct}{\mcitedefaultseppunct}\relax
\EndOfBibitem
\bibitem{LHCb-TDR-10}
LHCb collaboration, \ifthenelse{\boolean{articletitles}}{\emph{{LHCb trigger
  system technical design report}}, }{}
  \href{http://cdsweb.cern.ch/search?p=CERN-LHCC-2003-031&f=reportnumber&action_search=Search&c=LHCb}
  {CERN-LHCC-2003-031}, 2003\relax
\mciteBstWouldAddEndPuncttrue
\mciteSetBstMidEndSepPunct{\mcitedefaultmidpunct}
{\mcitedefaultendpunct}{\mcitedefaultseppunct}\relax
\EndOfBibitem
\bibitem{Aaij:2012me}
R.~Aaij {\em et~al.}, \ifthenelse{\boolean{articletitles}}{\emph{{The LHCb
  Trigger and its Performance in 2011}},
  }{}\href{https://doi.org/10.1088/1748-0221/8/04/P04022}{JINST \textbf{8}
  (2013) P04022}, \href{http://arxiv.org/abs/1211.3055}{{\normalfont\ttfamily
  arXiv:1211.3055}}\relax
\mciteBstWouldAddEndPuncttrue
\mciteSetBstMidEndSepPunct{\mcitedefaultmidpunct}
{\mcitedefaultendpunct}{\mcitedefaultseppunct}\relax
\EndOfBibitem
\bibitem{Albrecht:2013fba}
LHCb HLT project, J.~Albrecht, V.~V. Gligorov, G.~Raven, and S.~Tolk,
  \ifthenelse{\boolean{articletitles}}{\emph{{Performance of the LHCb High
  Level Trigger in 2012}},
  }{}\href{https://doi.org/10.1088/1742-6596/513/1/012001}{J.\ Phys.\ Conf.\
  Ser.\  \textbf{513} (2014) 012001},
  \href{http://arxiv.org/abs/1310.8544}{{\normalfont\ttfamily
  arXiv:1310.8544}}\relax
\mciteBstWouldAddEndPuncttrue
\mciteSetBstMidEndSepPunct{\mcitedefaultmidpunct}
{\mcitedefaultendpunct}{\mcitedefaultseppunct}\relax
\EndOfBibitem
\bibitem{Aaij:2018jht}
LHCb collaboration, R.~Aaij {\em et~al.},
  \ifthenelse{\boolean{articletitles}}{\emph{{Design and performance of the
  LHCb trigger and full real-time reconstruction in Run 2 of the LHC}},
  }{}\href{https://doi.org/10.1088/1748-0221/14/04/P04013}{JINST \textbf{14}
  (2019) P04013}, \href{http://arxiv.org/abs/1812.10790}{{\normalfont\ttfamily
  arXiv:1812.10790}}\relax
\mciteBstWouldAddEndPuncttrue
\mciteSetBstMidEndSepPunct{\mcitedefaultmidpunct}
{\mcitedefaultendpunct}{\mcitedefaultseppunct}\relax
\EndOfBibitem
\bibitem{Dujany:2015lxd}
G.~Dujany and B.~Storaci, \ifthenelse{\boolean{articletitles}}{\emph{{Real-time
  alignment and calibration of the LHCb Detector in Run II}},
  }{}\href{https://doi.org/10.1088/1742-6596/664/8/082010}{J.\ Phys.\ Conf.\
  Ser.\  \textbf{664} (2015) 082010}\relax
\mciteBstWouldAddEndPuncttrue
\mciteSetBstMidEndSepPunct{\mcitedefaultmidpunct}
{\mcitedefaultendpunct}{\mcitedefaultseppunct}\relax
\EndOfBibitem
\bibitem{LHCb-DP-2016-001}
R.~Aaij {\em et~al.}, \ifthenelse{\boolean{articletitles}}{\emph{{Tesla: an
  application for real-time data analysis in High Energy Physics}},
  }{}\href{https://doi.org/10.1016/j.cpc.2016.07.022}{Comput.\ Phys.\ Commun.\
  \textbf{208} (2016) 35},
  \href{http://arxiv.org/abs/1604.05596}{{\normalfont\ttfamily
  arXiv:1604.05596}}\relax
\mciteBstWouldAddEndPuncttrue
\mciteSetBstMidEndSepPunct{\mcitedefaultmidpunct}
{\mcitedefaultendpunct}{\mcitedefaultseppunct}\relax
\EndOfBibitem
\bibitem{Benson:2015yzo}
S.~Benson, V.~V. Gligorov, M.~A. Vesterinen, and M.~Williams,
  \ifthenelse{\boolean{articletitles}}{\emph{{The LHCb Turbo Stream}},
  }{}\href{https://doi.org/10.1088/1742-6596/664/8/082004}{J.\ Phys.\ Conf.\
  Ser.\  \textbf{664} (2015) 082004}\relax
\mciteBstWouldAddEndPuncttrue
\mciteSetBstMidEndSepPunct{\mcitedefaultmidpunct}
{\mcitedefaultendpunct}{\mcitedefaultseppunct}\relax
\EndOfBibitem
\bibitem{LHCb-PAPER-2019-006}
LHCb collaboration, R.~Aaij {\em et~al.},
  \ifthenelse{\boolean{articletitles}}{\emph{{Observation of \CP violation in
  charm decays}},
  }{}\href{https://doi.org/10.1103/PhysRevLett.122.211803}{Phys.\ Rev.\ Lett.\
  \textbf{122} (2019) 211803},
  \href{http://arxiv.org/abs/1903.08726}{{\normalfont\ttfamily
  arXiv:1903.08726}}\relax
\mciteBstWouldAddEndPuncttrue
\mciteSetBstMidEndSepPunct{\mcitedefaultmidpunct}
{\mcitedefaultendpunct}{\mcitedefaultseppunct}\relax
\EndOfBibitem
\bibitem{LHCb-TDR-017}
LHCb collaboration, \ifthenelse{\boolean{articletitles}}{\emph{{LHCb Upgrade
  Software and Computing}}, }{}
  \href{http://cdsweb.cern.ch/search?p=CERN-LHCC-2018-007&f=reportnumber&action_search=Search&c=LHCb}
  {CERN-LHCC-2018-007}, 2018\relax
\mciteBstWouldAddEndPuncttrue
\mciteSetBstMidEndSepPunct{\mcitedefaultmidpunct}
{\mcitedefaultendpunct}{\mcitedefaultseppunct}\relax
\EndOfBibitem
\bibitem{LHCb-TDR-018}
LHCb collaboration, \ifthenelse{\boolean{articletitles}}{\emph{{Computing Model
  of the Upgrade LHCb experiment}}, }{}
  \href{http://cdsweb.cern.ch/search?p=CERN-LHCC-2018-014&f=reportnumber&action_search=Search&c=LHCb}
  {CERN-LHCC-2018-014}, 2018\relax
\mciteBstWouldAddEndPuncttrue
\mciteSetBstMidEndSepPunct{\mcitedefaultmidpunct}
{\mcitedefaultendpunct}{\mcitedefaultseppunct}\relax
\EndOfBibitem
\bibitem{LHCb-TDR-021}
LHCb collaboration,  \ifthenelse{\boolean{articletitles}}{\emph{{LHCb Upgrade
  GPU High Level Trigger Technical Design Report}}}{},
  \href{https://cds.cern.ch/record/2717938}{CERN-LHCC-2020-006. LHCB-TDR-021},
  CERN, Geneva, 2020\relax
\mciteBstWouldAddEndPuncttrue
\mciteSetBstMidEndSepPunct{\mcitedefaultmidpunct}
{\mcitedefaultendpunct}{\mcitedefaultseppunct}\relax
\EndOfBibitem
\bibitem{Aaij:2021mzf}
LHCb, R.~Aaij {\em et~al.}, \ifthenelse{\boolean{articletitles}}{\emph{{A
  Comparison of CPU and GPU implementations for the LHCb Experiment Run 3
  Trigger}}, }{}\href{http://arxiv.org/abs/2105.04031}{{\normalfont\ttfamily
  arXiv:2105.04031}}\relax
\mciteBstWouldAddEndPuncttrue
\mciteSetBstMidEndSepPunct{\mcitedefaultmidpunct}
{\mcitedefaultendpunct}{\mcitedefaultseppunct}\relax
\EndOfBibitem
\bibitem{Bos:1070314}
E.~Bos and E.~Rodrigues, \ifthenelse{\boolean{articletitles}}{\emph{{The LHCb
  Track Extrapolator Tools}}}{} , CERN, Geneva, 2007\relax
\mciteBstWouldAddEndPuncttrue
\mciteSetBstMidEndSepPunct{\mcitedefaultmidpunct}
{\mcitedefaultendpunct}{\mcitedefaultseppunct}\relax
\EndOfBibitem
\bibitem{Cacciari:2008gp}
M.~Cacciari, G.~P. Salam, and G.~Soyez,
  \ifthenelse{\boolean{articletitles}}{\emph{{The anti-$k_t$ jet clustering
  algorithm}}, }{}\href{https://doi.org/10.1088/1126-6708/2008/04/063}{JHEP
  \textbf{04} (2008) 063},
  \href{http://arxiv.org/abs/0802.1189}{{\normalfont\ttfamily
  arXiv:0802.1189}}\relax
\mciteBstWouldAddEndPuncttrue
\mciteSetBstMidEndSepPunct{\mcitedefaultmidpunct}
{\mcitedefaultendpunct}{\mcitedefaultseppunct}\relax
\EndOfBibitem
\bibitem{Cacciari:2011ma}
M.~Cacciari, G.~P. Salam, and G.~Soyez,
  \ifthenelse{\boolean{articletitles}}{\emph{{FastJet User Manual}},
  }{}\href{https://doi.org/10.1140/epjc/s10052-012-1896-2}{Eur.\ Phys.\ J.\ C
  \textbf{72} (2012) 1896},
  \href{http://arxiv.org/abs/1111.6097}{{\normalfont\ttfamily
  arXiv:1111.6097}}\relax
\mciteBstWouldAddEndPuncttrue
\mciteSetBstMidEndSepPunct{\mcitedefaultmidpunct}
{\mcitedefaultendpunct}{\mcitedefaultseppunct}\relax
\EndOfBibitem
\bibitem{Aaij:2013lla}
LHCb collaboration, R.~Aaij {\em et~al.},
  \ifthenelse{\boolean{articletitles}}{\emph{{Search for rare
  $B^0_{(s)}\rightarrow \mu^+ \mu^- \mu^+ \mu^-$ decays}},
  }{}\href{https://doi.org/10.1103/PhysRevLett.110.211801}{Phys.\ Rev.\ Lett.\
  \textbf{110} (2013) 211801},
  \href{http://arxiv.org/abs/1303.1092}{{\normalfont\ttfamily
  arXiv:1303.1092}}\relax
\mciteBstWouldAddEndPuncttrue
\mciteSetBstMidEndSepPunct{\mcitedefaultmidpunct}
{\mcitedefaultendpunct}{\mcitedefaultseppunct}\relax
\EndOfBibitem
\bibitem{Aaij:2015yqa}
LHCb collaboration, R.~Aaij {\em et~al.},
  \ifthenelse{\boolean{articletitles}}{\emph{{Identification of beauty and
  charm quark jets at LHCb}},
  }{}\href{https://doi.org/10.1088/1748-0221/10/06/P06013}{JINST \textbf{10}
  (2015) P06013}, \href{http://arxiv.org/abs/1504.07670}{{\normalfont\ttfamily
  arXiv:1504.07670}}\relax
\mciteBstWouldAddEndPuncttrue
\mciteSetBstMidEndSepPunct{\mcitedefaultmidpunct}
{\mcitedefaultendpunct}{\mcitedefaultseppunct}\relax
\EndOfBibitem
\bibitem{Aaij:2014nma}
LHCb collaboration, R.~Aaij {\em et~al.},
  \ifthenelse{\boolean{articletitles}}{\emph{{Search for long-lived particles
  decaying to jet pairs}},
  }{}\href{https://doi.org/10.1140/epjc/s10052-015-3344-6}{Eur.\ Phys.\ J.\
  \textbf{C75} (2015) 152},
  \href{http://arxiv.org/abs/1412.3021}{{\normalfont\ttfamily
  arXiv:1412.3021}}\relax
\mciteBstWouldAddEndPuncttrue
\mciteSetBstMidEndSepPunct{\mcitedefaultmidpunct}
{\mcitedefaultendpunct}{\mcitedefaultseppunct}\relax
\EndOfBibitem
\bibitem{Aaij:2017fak}
LHCb collaboration, R.~Aaij {\em et~al.},
  \ifthenelse{\boolean{articletitles}}{\emph{{Study of J/\ensuremath{\psi}
  Production in Jets}},
  }{}\href{https://doi.org/10.1103/PhysRevLett.118.192001}{Phys.\ Rev.\ Lett.\
  \textbf{118} (2017) 192001},
  \href{http://arxiv.org/abs/1701.05116}{{\normalfont\ttfamily
  arXiv:1701.05116}}\relax
\mciteBstWouldAddEndPuncttrue
\mciteSetBstMidEndSepPunct{\mcitedefaultmidpunct}
{\mcitedefaultendpunct}{\mcitedefaultseppunct}\relax
\EndOfBibitem
\bibitem{Komiske:2017ubm}
P.~T. Komiske, E.~M. Metodiev, B.~Nachman, and M.~D. Schwartz,
  \ifthenelse{\boolean{articletitles}}{\emph{{Pileup Mitigation with Machine
  Learning (PUMML)}}, }{}\href{https://doi.org/10.1007/JHEP12(2017)051}{JHEP
  \textbf{12} (2017) 051},
  \href{http://arxiv.org/abs/1707.08600}{{\normalfont\ttfamily
  arXiv:1707.08600}}\relax
\mciteBstWouldAddEndPuncttrue
\mciteSetBstMidEndSepPunct{\mcitedefaultmidpunct}
{\mcitedefaultendpunct}{\mcitedefaultseppunct}\relax
\EndOfBibitem
\bibitem{GNN}
J.~Pata {\em et~al.}, \ifthenelse{\boolean{articletitles}}{\emph{{MLPF:
  Efficient machine-learned particle-flow reconstruction using graph neural
  networks}}, }{}\href{http://arxiv.org/abs/2101.08578}{{\normalfont\ttfamily
  arXiv:2101.08578}}\relax
\mciteBstWouldAddEndPuncttrue
\mciteSetBstMidEndSepPunct{\mcitedefaultmidpunct}
{\mcitedefaultendpunct}{\mcitedefaultseppunct}\relax
\EndOfBibitem
\bibitem{GNN2}
X.~Ju and B.~Nachman, \ifthenelse{\boolean{articletitles}}{\emph{{Supervised
  Jet Clustering with Graph Neural Networks for Lorentz Boosted Bosons}},
  }{}\href{https://doi.org/10.1103/PhysRevD.102.075014}{Phys.\ Rev.\ D
  \textbf{102} (2020) 075014},
  \href{http://arxiv.org/abs/2008.06064}{{\normalfont\ttfamily
  arXiv:2008.06064}}\relax
\mciteBstWouldAddEndPuncttrue
\mciteSetBstMidEndSepPunct{\mcitedefaultmidpunct}
{\mcitedefaultendpunct}{\mcitedefaultseppunct}\relax
\EndOfBibitem
\bibitem{CMSTagging}
CMS collaboration, \ifthenelse{\boolean{articletitles}}{\emph{Identification of
  heavy-flavour jets with the {CMS} detector in pp collisions at 13 {TeV}},
  }{}\href{https://doi.org/10.1088/1748-0221/13/05/p05011}{Journal of
  Instrumentation \textbf{13} (2018) P05011}\relax
\mciteBstWouldAddEndPuncttrue
\mciteSetBstMidEndSepPunct{\mcitedefaultmidpunct}
{\mcitedefaultendpunct}{\mcitedefaultseppunct}\relax
\EndOfBibitem
\bibitem{Aaij:2020mqb}
LHCb collaboration, R.~Aaij {\em et~al.},
  \ifthenelse{\boolean{articletitles}}{\emph{{Search for the lepton flavour
  violating decay $B^+ \rightarrow K^+ \mu^- \tau^+$ using $B_{s2}^{*0}$
  decays}}, }{}\href{https://doi.org/10.1007/JHEP06(2020)129}{JHEP \textbf{06}
  (2020) 129}, \href{http://arxiv.org/abs/2003.04352}{{\normalfont\ttfamily
  arXiv:2003.04352}}\relax
\mciteBstWouldAddEndPuncttrue
\mciteSetBstMidEndSepPunct{\mcitedefaultmidpunct}
{\mcitedefaultendpunct}{\mcitedefaultseppunct}\relax
\EndOfBibitem
\bibitem{Likhomanenko:2015aba}
T.~Likhomanenko {\em et~al.}, \ifthenelse{\boolean{articletitles}}{\emph{{LHCb
  Topological Trigger Reoptimization}},
  }{}\href{https://doi.org/10.1088/1742-6596/664/8/082025}{J.\ Phys.\ Conf.\
  Ser.\  \textbf{664} (2015) 082025},
  \href{http://arxiv.org/abs/1510.00572}{{\normalfont\ttfamily
  arXiv:1510.00572}}\relax
\mciteBstWouldAddEndPuncttrue
\mciteSetBstMidEndSepPunct{\mcitedefaultmidpunct}
{\mcitedefaultendpunct}{\mcitedefaultseppunct}\relax
\EndOfBibitem
\bibitem{Aaij:2017deq}
LHCb collaboration, R.~Aaij {\em et~al.},
  \ifthenelse{\boolean{articletitles}}{\emph{{Test of Lepton Flavor
  Universality by the measurement of the $B^0 \to D^{*-} \tau^+ \nu_{\tau}$
  branching fraction using three-prong $\tau$ decays}},
  }{}\href{https://doi.org/10.1103/PhysRevD.97.072013}{Phys.\ Rev.\
  \textbf{D97} (2018) 072013},
  \href{http://arxiv.org/abs/1711.02505}{{\normalfont\ttfamily
  arXiv:1711.02505}}\relax
\mciteBstWouldAddEndPuncttrue
\mciteSetBstMidEndSepPunct{\mcitedefaultmidpunct}
{\mcitedefaultendpunct}{\mcitedefaultseppunct}\relax
\EndOfBibitem
\bibitem{LHCb-PUB-2017-006}
C.~Fitzpatrick {\em et~al.},
  \ifthenelse{\boolean{articletitles}}{\emph{{Upgrade trigger: Bandwidth
  strategy proposal}}}{},
  \href{https://cds.cern.ch/record/2244313}{LHCb-PUB-2017-006.
  CERN-LHCb-PUB-2017-006}, CERN, Geneva, 2017\relax
\mciteBstWouldAddEndPuncttrue
\mciteSetBstMidEndSepPunct{\mcitedefaultmidpunct}
{\mcitedefaultendpunct}{\mcitedefaultseppunct}\relax
\EndOfBibitem
\bibitem{LHCb-PUB-2019-013}
A.~Alfonso~Albero {\em et~al.},
  \ifthenelse{\boolean{articletitles}}{\emph{{Upgrade trigger selection
  studies}}}{}, \href{https://cds.cern.ch/record/2688423}{LHCb-PUB-2019-013.
  CERN-LHCb-PUB-2019-013}, CERN, Geneva, 2019\relax
\mciteBstWouldAddEndPuncttrue
\mciteSetBstMidEndSepPunct{\mcitedefaultmidpunct}
{\mcitedefaultendpunct}{\mcitedefaultseppunct}\relax
\EndOfBibitem
\bibitem{Aaij:2017mic}
LHCb collaboration, R.~Aaij {\em et~al.},
  \ifthenelse{\boolean{articletitles}}{\emph{{Updated search for long-lived
  particles decaying to jet pairs}},
  }{}\href{https://doi.org/10.1140/epjc/s10052-017-5178-x}{Eur.\ Phys.\ J.\
  \textbf{C77} (2017) 812},
  \href{http://arxiv.org/abs/1705.07332}{{\normalfont\ttfamily
  arXiv:1705.07332}}\relax
\mciteBstWouldAddEndPuncttrue
\mciteSetBstMidEndSepPunct{\mcitedefaultmidpunct}
{\mcitedefaultendpunct}{\mcitedefaultseppunct}\relax
\EndOfBibitem
\bibitem{Marzani:2017mva}
S.~Marzani, L.~Schunk, and G.~Soyez,
  \ifthenelse{\boolean{articletitles}}{\emph{{A study of jet mass distributions
  with grooming}}, }{}\href{https://doi.org/10.1007/JHEP07(2017)132}{JHEP
  \textbf{07} (2017) 132},
  \href{http://arxiv.org/abs/1704.02210}{{\normalfont\ttfamily
  arXiv:1704.02210}}\relax
\mciteBstWouldAddEndPuncttrue
\mciteSetBstMidEndSepPunct{\mcitedefaultmidpunct}
{\mcitedefaultendpunct}{\mcitedefaultseppunct}\relax
\EndOfBibitem
\bibitem{Larkoski:2014wba}
A.~J. Larkoski, S.~Marzani, G.~Soyez, and J.~Thaler,
  \ifthenelse{\boolean{articletitles}}{\emph{{Soft Drop}},
  }{}\href{https://doi.org/10.1007/JHEP05(2014)146}{JHEP \textbf{05} (2014)
  146}, \href{http://arxiv.org/abs/1402.2657}{{\normalfont\ttfamily
  arXiv:1402.2657}}\relax
\mciteBstWouldAddEndPuncttrue
\mciteSetBstMidEndSepPunct{\mcitedefaultmidpunct}
{\mcitedefaultendpunct}{\mcitedefaultseppunct}\relax
\EndOfBibitem
\bibitem{Aaij:2016xmb}
LHCb collaboration, R.~Aaij {\em et~al.},
  \ifthenelse{\boolean{articletitles}}{\emph{{Search for massive long-lived
  particles decaying semileptonically in the LHCb detector}},
  }{}\href{https://doi.org/10.1140/epjc/s10052-017-4744-6}{Eur.\ Phys.\ J.\
  \textbf{C77} (2017) 224},
  \href{http://arxiv.org/abs/1612.00945}{{\normalfont\ttfamily
  arXiv:1612.00945}}\relax
\mciteBstWouldAddEndPuncttrue
\mciteSetBstMidEndSepPunct{\mcitedefaultmidpunct}
{\mcitedefaultendpunct}{\mcitedefaultseppunct}\relax
\EndOfBibitem
\bibitem{LHCb-CONF-2018-006}
{LHCb collaboration}, \ifthenelse{\boolean{articletitles}}{\emph{{Prospects for
  searches for long-lived particles after the LHCb detector upgrades}}, }{}
  \href{http://cdsweb.cern.ch/search?p=LHCb-CONF-2018-006&f=reportnumber&action_search=Search&c=LHCb+Conference+Contributions}
  {LHCb-CONF-2018-006}, {2018}\relax
\mciteBstWouldAddEndPuncttrue
\mciteSetBstMidEndSepPunct{\mcitedefaultmidpunct}
{\mcitedefaultendpunct}{\mcitedefaultseppunct}\relax
\EndOfBibitem
\bibitem{CidVidal:2019urm}
X.~Cid~Vidal, Y.~Tsai, and J.~Zurita,
  \ifthenelse{\boolean{articletitles}}{\emph{{Exclusive displaced hadronic
  signatures in the LHC forward region}},
  }{}\href{https://doi.org/10.1007/JHEP01(2020)115}{JHEP \textbf{01} (2020)
  115}, \href{http://arxiv.org/abs/1910.05225}{{\normalfont\ttfamily
  arXiv:1910.05225}}\relax
\mciteBstWouldAddEndPuncttrue
\mciteSetBstMidEndSepPunct{\mcitedefaultmidpunct}
{\mcitedefaultendpunct}{\mcitedefaultseppunct}\relax
\EndOfBibitem
\bibitem{Williams:2011aza}
M.~Williams {\em et~al.},  \ifthenelse{\boolean{articletitles}}{\emph{{The HLT2
  Topological Lines}}}{},
  \href{https://cds.cern.ch/record/1323557}{LHCb-PUB-2011-002.
  CERN-LHCb-PUB-2011-002}, CERN, Geneva, 2011\relax
\mciteBstWouldAddEndPuncttrue
\mciteSetBstMidEndSepPunct{\mcitedefaultmidpunct}
{\mcitedefaultendpunct}{\mcitedefaultseppunct}\relax
\EndOfBibitem
\bibitem{Pierce:2017taw}
A.~Pierce, B.~Shakya, Y.~Tsai, and Y.~Zhao,
  \ifthenelse{\boolean{articletitles}}{\emph{{Searching for confining hidden
  valleys at LHCb, ATLAS, and CMS}},
  }{}\href{https://doi.org/10.1103/PhysRevD.97.095033}{Phys.\ Rev.\ D
  \textbf{97} (2018) 095033},
  \href{http://arxiv.org/abs/1708.05389}{{\normalfont\ttfamily
  arXiv:1708.05389}}\relax
\mciteBstWouldAddEndPuncttrue
\mciteSetBstMidEndSepPunct{\mcitedefaultmidpunct}
{\mcitedefaultendpunct}{\mcitedefaultseppunct}\relax
\EndOfBibitem
\bibitem{Aaij:2016qsm}
LHCb collaboration, R.~Aaij {\em et~al.},
  \ifthenelse{\boolean{articletitles}}{\emph{{Search for long-lived scalar
  particles in $B^+ \to K^+ \chi (\mu^+\mu^-)$ decays}},
  }{}\href{https://doi.org/10.1103/PhysRevD.95.071101}{Phys.\ Rev.\ D
  \textbf{95} (2017) 071101},
  \href{http://arxiv.org/abs/1612.07818}{{\normalfont\ttfamily
  arXiv:1612.07818}}\relax
\mciteBstWouldAddEndPuncttrue
\mciteSetBstMidEndSepPunct{\mcitedefaultmidpunct}
{\mcitedefaultendpunct}{\mcitedefaultseppunct}\relax
\EndOfBibitem
\bibitem{Aaij:2015tna}
LHCb collaboration, R.~Aaij {\em et~al.},
  \ifthenelse{\boolean{articletitles}}{\emph{{Search for hidden-sector bosons
  in $B^0 \!\to K^{*0}\mu^+\mu^-$ decays}},
  }{}\href{https://doi.org/10.1103/PhysRevLett.115.161802}{Phys.\ Rev.\ Lett.\
  \textbf{115} (2015) 161802},
  \href{http://arxiv.org/abs/1508.04094}{{\normalfont\ttfamily
  arXiv:1508.04094}}\relax
\mciteBstWouldAddEndPuncttrue
\mciteSetBstMidEndSepPunct{\mcitedefaultmidpunct}
{\mcitedefaultendpunct}{\mcitedefaultseppunct}\relax
\EndOfBibitem
\bibitem{Aaij:2018xpt}
LHCb collaboration, R.~Aaij {\em et~al.},
  \ifthenelse{\boolean{articletitles}}{\emph{{Search for a dimuon resonance in
  the $\Upsilon$ mass region}},
  }{}\href{https://doi.org/10.1007/JHEP09(2018)147}{JHEP \textbf{09} (2018)
  147}, \href{http://arxiv.org/abs/1805.09820}{{\normalfont\ttfamily
  arXiv:1805.09820}}\relax
\mciteBstWouldAddEndPuncttrue
\mciteSetBstMidEndSepPunct{\mcitedefaultmidpunct}
{\mcitedefaultendpunct}{\mcitedefaultseppunct}\relax
\EndOfBibitem
\bibitem{Aaij:2019bvg}
LHCb collaboration, R.~Aaij {\em et~al.},
  \ifthenelse{\boolean{articletitles}}{\emph{{Search for $A'\to\mu^+\mu^-$
  Decays}}, }{}\href{https://doi.org/10.1103/PhysRevLett.124.041801}{Phys.\
  Rev.\ Lett.\  \textbf{124} (2020) 041801},
  \href{http://arxiv.org/abs/1910.06926}{{\normalfont\ttfamily
  arXiv:1910.06926}}\relax
\mciteBstWouldAddEndPuncttrue
\mciteSetBstMidEndSepPunct{\mcitedefaultmidpunct}
{\mcitedefaultendpunct}{\mcitedefaultseppunct}\relax
\EndOfBibitem
\bibitem{LHCb-DP-2014-001}
R.~Aaij {\em et~al.}, \ifthenelse{\boolean{articletitles}}{\emph{{Performance
  of the LHCb Vertex Locator}},
  }{}\href{https://doi.org/10.1088/1748-0221/9/09/P09007}{JINST \textbf{9}
  (2014) P09007}, \href{http://arxiv.org/abs/1405.7808}{{\normalfont\ttfamily
  arXiv:1405.7808}}\relax
\mciteBstWouldAddEndPuncttrue
\mciteSetBstMidEndSepPunct{\mcitedefaultmidpunct}
{\mcitedefaultendpunct}{\mcitedefaultseppunct}\relax
\EndOfBibitem
\bibitem{Aaij:2017xqt}
LHCb collaboration, R.~Aaij {\em et~al.},
  \ifthenelse{\boolean{articletitles}}{\emph{{Search for the decays
  $B_s^0\to\tau^+\tau^-$ and $B^0\to\tau^+\tau^-$}},
  }{}\href{https://doi.org/10.1103/PhysRevLett.118.251802}{Phys.\ Rev.\ Lett.\
  \textbf{118} (2017) 251802},
  \href{http://arxiv.org/abs/1703.02508}{{\normalfont\ttfamily
  arXiv:1703.02508}}\relax
\mciteBstWouldAddEndPuncttrue
\mciteSetBstMidEndSepPunct{\mcitedefaultmidpunct}
{\mcitedefaultendpunct}{\mcitedefaultseppunct}\relax
\EndOfBibitem
\bibitem{Aaij:2624023}
LHCb collaboration, R.~Aaij {\em et~al.},
  \ifthenelse{\boolean{articletitles}}{\emph{{Measurement of
  $Z\rightarrow\tau^+\tau^-$ production in proton-proton collisions at
  $\sqrt{s} = 8$ TeV}}, }{}\href{https://doi.org/10.1007/JHEP09(2018)159}{JHEP
  \textbf{09} (2018) 159},
  \href{http://arxiv.org/abs/1806.05008}{{\normalfont\ttfamily
  arXiv:1806.05008}}\relax
\mciteBstWouldAddEndPuncttrue
\mciteSetBstMidEndSepPunct{\mcitedefaultmidpunct}
{\mcitedefaultendpunct}{\mcitedefaultseppunct}\relax
\EndOfBibitem
\bibitem{Benson:2314368}
S.~Benson and A.~Puig~Navarro,
  \ifthenelse{\boolean{articletitles}}{\emph{{Triggering $B_s^0 \to
  \gamma\gamma$ at LHCb}}}{},
  \href{http://cds.cern.ch/record/2314368}{LHCb-PUB-2018-006.
  CERN-LHCb-PUB-2018-006}, CERN, Geneva, 2018\relax
\mciteBstWouldAddEndPuncttrue
\mciteSetBstMidEndSepPunct{\mcitedefaultmidpunct}
{\mcitedefaultendpunct}{\mcitedefaultseppunct}\relax
\EndOfBibitem
\bibitem{CidCasaisPhotonTrigger}
S.~Benson, A.~Casais~Vidal, X.~Cid~Vidal, and A.~Puig~Navarro,
  \ifthenelse{\boolean{articletitles}}{\emph{{Real-time discrimination of
  photon pairs using machine learning at the LHC}},
  }{}\href{https://doi.org/10.21468/SciPostPhys.7.5.062}{SciPost Phys.\
  \textbf{7} (2019) 62}\relax
\mciteBstWouldAddEndPuncttrue
\mciteSetBstMidEndSepPunct{\mcitedefaultmidpunct}
{\mcitedefaultendpunct}{\mcitedefaultseppunct}\relax
\EndOfBibitem
\bibitem{LHCb-PAPER-2017-024}
LHCb collaboration, R.~Aaij {\em et~al.},
  \ifthenelse{\boolean{articletitles}}{\emph{{First observation of forward
  \mbox{\decay{\Z}{b\bquarkbar}} production in \proton\proton collisions at
  \mbox{$\sqs=$8\tev}}},
  }{}\href{https://doi.org/10.1016/j.physletb.2017.11.066}{Phys.\ Lett.\
  \textbf{B776} (2018) 430},
  \href{http://arxiv.org/abs/1709.03458}{{\normalfont\ttfamily
  arXiv:1709.03458}}\relax
\mciteBstWouldAddEndPuncttrue
\mciteSetBstMidEndSepPunct{\mcitedefaultmidpunct}
{\mcitedefaultendpunct}{\mcitedefaultseppunct}\relax
\EndOfBibitem
\bibitem{Sirunyan:2016iap}
CMS collaboration, A.~M. Sirunyan {\em et~al.},
  \ifthenelse{\boolean{articletitles}}{\emph{{Search for dijet resonances in
  proton\textendash{}proton collisions at $\sqrt{s}$ = 13 TeV and constraints
  on dark matter and other models}},
  }{}\href{https://doi.org/10.1016/j.physletb.2017.02.012}{Phys.\ Lett.\ B
  \textbf{769} (2017) 520},
  \href{http://arxiv.org/abs/1611.03568}{{\normalfont\ttfamily
  arXiv:1611.03568}}, [Erratum: Phys.Lett.B 772, 882--883 (2017)]\relax
\mciteBstWouldAddEndPuncttrue
\mciteSetBstMidEndSepPunct{\mcitedefaultmidpunct}
{\mcitedefaultendpunct}{\mcitedefaultseppunct}\relax
\EndOfBibitem
\bibitem{Khachatryan:2016ecr}
CMS collaboration, V.~Khachatryan {\em et~al.},
  \ifthenelse{\boolean{articletitles}}{\emph{{Search for narrow resonances in
  dijet final states at $\sqrt(s)=$ 8 TeV with the novel CMS technique of data
  scouting}}, }{}\href{https://doi.org/10.1103/PhysRevLett.117.031802}{Phys.\
  Rev.\ Lett.\  \textbf{117} (2016) 031802},
  \href{http://arxiv.org/abs/1604.08907}{{\normalfont\ttfamily
  arXiv:1604.08907}}\relax
\mciteBstWouldAddEndPuncttrue
\mciteSetBstMidEndSepPunct{\mcitedefaultmidpunct}
{\mcitedefaultendpunct}{\mcitedefaultseppunct}\relax
\EndOfBibitem
\bibitem{Aaboud:2017yvp}
ATLAS collaboration, M.~Aaboud {\em et~al.},
  \ifthenelse{\boolean{articletitles}}{\emph{{Search for new phenomena in dijet
  events using 37 fb$^{-1}$ of $pp$ collision data collected at $\sqrt{s}=$13
  TeV with the ATLAS detector}},
  }{}\href{https://doi.org/10.1103/PhysRevD.96.052004}{Phys.\ Rev.\ D
  \textbf{96} (2017) 052004},
  \href{http://arxiv.org/abs/1703.09127}{{\normalfont\ttfamily
  arXiv:1703.09127}}\relax
\mciteBstWouldAddEndPuncttrue
\mciteSetBstMidEndSepPunct{\mcitedefaultmidpunct}
{\mcitedefaultendpunct}{\mcitedefaultseppunct}\relax
\EndOfBibitem
\bibitem{Aaboud:2018fzt}
ATLAS collaboration, M.~Aaboud {\em et~al.},
  \ifthenelse{\boolean{articletitles}}{\emph{{Search for low-mass dijet
  resonances using trigger-level jets with the ATLAS detector in $pp$
  collisions at $\sqrt{s}=13$ TeV}},
  }{}\href{https://doi.org/10.1103/PhysRevLett.121.081801}{Phys.\ Rev.\ Lett.\
  \textbf{121} (2018) 081801},
  \href{http://arxiv.org/abs/1804.03496}{{\normalfont\ttfamily
  arXiv:1804.03496}}\relax
\mciteBstWouldAddEndPuncttrue
\mciteSetBstMidEndSepPunct{\mcitedefaultmidpunct}
{\mcitedefaultendpunct}{\mcitedefaultseppunct}\relax
\EndOfBibitem
\bibitem{Sirunyan:2018xlo}
CMS collaboration, A.~M. Sirunyan {\em et~al.},
  \ifthenelse{\boolean{articletitles}}{\emph{{Search for narrow and broad dijet
  resonances in proton-proton collisions at $ \sqrt{s}=13 $ TeV and constraints
  on dark matter mediators and other new particles}},
  }{}\href{https://doi.org/10.1007/JHEP08(2018)130}{JHEP \textbf{08} (2018)
  130}, \href{http://arxiv.org/abs/1806.00843}{{\normalfont\ttfamily
  arXiv:1806.00843}}\relax
\mciteBstWouldAddEndPuncttrue
\mciteSetBstMidEndSepPunct{\mcitedefaultmidpunct}
{\mcitedefaultendpunct}{\mcitedefaultseppunct}\relax
\EndOfBibitem
\bibitem{Aaboud:2018tqo}
ATLAS collaboration, M.~Aaboud {\em et~al.},
  \ifthenelse{\boolean{articletitles}}{\emph{{Search for resonances in the mass
  distribution of jet pairs with one or two jets identified as $b$-jets in
  proton-proton collisions at $\sqrt{s}=13$ TeV with the ATLAS detector}},
  }{}\href{https://doi.org/10.1103/PhysRevD.98.032016}{Phys.\ Rev.\ D
  \textbf{98} (2018) 032016},
  \href{http://arxiv.org/abs/1805.09299}{{\normalfont\ttfamily
  arXiv:1805.09299}}\relax
\mciteBstWouldAddEndPuncttrue
\mciteSetBstMidEndSepPunct{\mcitedefaultmidpunct}
{\mcitedefaultendpunct}{\mcitedefaultseppunct}\relax
\EndOfBibitem
\bibitem{Sirunyan:2018pas}
CMS collaboration, A.~M. Sirunyan {\em et~al.},
  \ifthenelse{\boolean{articletitles}}{\emph{{Search for narrow resonances in
  the b-tagged dijet mass spectrum in proton-proton collisions at $\sqrt{s} =$
  8 TeV}}, }{}\href{https://doi.org/10.1103/PhysRevLett.120.201801}{Phys.\
  Rev.\ Lett.\  \textbf{120} (2018) 201801},
  \href{http://arxiv.org/abs/1802.06149}{{\normalfont\ttfamily
  arXiv:1802.06149}}\relax
\mciteBstWouldAddEndPuncttrue
\mciteSetBstMidEndSepPunct{\mcitedefaultmidpunct}
{\mcitedefaultendpunct}{\mcitedefaultseppunct}\relax
\EndOfBibitem
\bibitem{Sirunyan:2018ikr}
CMS collaboration, A.~M. Sirunyan {\em et~al.},
  \ifthenelse{\boolean{articletitles}}{\emph{{Search for low-mass resonances
  decaying into bottom quark-antiquark pairs in proton-proton collisions at
  $\sqrt{s} =$ 13 TeV}},
  }{}\href{https://doi.org/10.1103/PhysRevD.99.012005}{Phys.\ Rev.\ D
  \textbf{99} (2019) 012005},
  \href{http://arxiv.org/abs/1810.11822}{{\normalfont\ttfamily
  arXiv:1810.11822}}\relax
\mciteBstWouldAddEndPuncttrue
\mciteSetBstMidEndSepPunct{\mcitedefaultmidpunct}
{\mcitedefaultendpunct}{\mcitedefaultseppunct}\relax
\EndOfBibitem
\bibitem{Sirunyan:2019qia}
CMS collaboration, A.~M. Sirunyan {\em et~al.},
  \ifthenelse{\boolean{articletitles}}{\emph{{A search for the standard model
  Higgs boson decaying to charm quarks}},
  }{}\href{https://doi.org/10.1007/JHEP03(2020)131}{JHEP \textbf{03} (2020)
  131}, \href{http://arxiv.org/abs/1912.01662}{{\normalfont\ttfamily
  arXiv:1912.01662}}\relax
\mciteBstWouldAddEndPuncttrue
\mciteSetBstMidEndSepPunct{\mcitedefaultmidpunct}
{\mcitedefaultendpunct}{\mcitedefaultseppunct}\relax
\EndOfBibitem
\bibitem{Aaboud:2018fhh}
ATLAS collaboration, M.~Aaboud {\em et~al.},
  \ifthenelse{\boolean{articletitles}}{\emph{{Search for the Decay of the Higgs
  Boson to Charm Quarks with the ATLAS Experiment}},
  }{}\href{https://doi.org/10.1103/PhysRevLett.120.211802}{Phys.\ Rev.\ Lett.\
  \textbf{120} (2018) 211802},
  \href{http://arxiv.org/abs/1802.04329}{{\normalfont\ttfamily
  arXiv:1802.04329}}\relax
\mciteBstWouldAddEndPuncttrue
\mciteSetBstMidEndSepPunct{\mcitedefaultmidpunct}
{\mcitedefaultendpunct}{\mcitedefaultseppunct}\relax
\EndOfBibitem
\bibitem{LHCb-CONF-2016-005}
{LHCb collaboration}, \ifthenelse{\boolean{articletitles}}{\emph{{LHCb dimuon
  and charm mass distributions}}, }{}
  \href{http://cdsweb.cern.ch/search?p=LHCb-CONF-2016-005&f=reportnumber&action_search=Search&c=LHCb+Conference+Contributions}
  {LHCb-CONF-2016-005}, {2016}\relax
\mciteBstWouldAddEndPuncttrue
\mciteSetBstMidEndSepPunct{\mcitedefaultmidpunct}
{\mcitedefaultendpunct}{\mcitedefaultseppunct}\relax
\EndOfBibitem
\bibitem{LHCb-CONF-2016-006}
{LHCb collaboration}, \ifthenelse{\boolean{articletitles}}{\emph{{Search for
  $\H\to \bbbar$ or $\ccbar$ in association with a $\W$ or $\Z$ boson in the
  forward region of $pp$ collisions}}, }{}
  \href{http://cdsweb.cern.ch/search?p=LHCb-CONF-2016-006&f=reportnumber&action_search=Search&c=LHCb+Conference+Contributions}
  {LHCb-CONF-2016-006}, {2016}\relax
\mciteBstWouldAddEndPuncttrue
\mciteSetBstMidEndSepPunct{\mcitedefaultmidpunct}
{\mcitedefaultendpunct}{\mcitedefaultseppunct}\relax
\EndOfBibitem
\bibitem{Asquith:2018igt}
R.~Kogler {\em et~al.}, \ifthenelse{\boolean{articletitles}}{\emph{{Jet
  Substructure at the Large Hadron Collider: Experimental Review}},
  }{}\href{https://doi.org/10.1103/RevModPhys.91.045003}{Rev.\ Mod.\ Phys.\
  \textbf{91} (2019) 045003},
  \href{http://arxiv.org/abs/1803.06991}{{\normalfont\ttfamily
  arXiv:1803.06991}}\relax
\mciteBstWouldAddEndPuncttrue
\mciteSetBstMidEndSepPunct{\mcitedefaultmidpunct}
{\mcitedefaultendpunct}{\mcitedefaultseppunct}\relax
\EndOfBibitem
\bibitem{Plehn:2009rk}
T.~Plehn, G.~P. Salam, and M.~Spannowsky,
  \ifthenelse{\boolean{articletitles}}{\emph{{Fat Jets for a Light Higgs}},
  }{}\href{https://doi.org/10.1103/PhysRevLett.104.111801}{Phys.\ Rev.\ Lett.\
  \textbf{104} (2010) 111801},
  \href{http://arxiv.org/abs/0910.5472}{{\normalfont\ttfamily
  arXiv:0910.5472}}\relax
\mciteBstWouldAddEndPuncttrue
\mciteSetBstMidEndSepPunct{\mcitedefaultmidpunct}
{\mcitedefaultendpunct}{\mcitedefaultseppunct}\relax
\EndOfBibitem
\bibitem{LHCb-PAPER-2015-002}
LHCb collaboration, R.~Aaij {\em et~al.},
  \ifthenelse{\boolean{articletitles}}{\emph{{Search for long-lived heavy
  charged particles using a ring-imaging Cherenkov technique at LHCb}},
  }{}\href{https://doi.org/10.1140/epjc/s10052-015-3809-7}{Eur.\ Phys.\ J.\
  \textbf{C75} (2015) 595},
  \href{http://arxiv.org/abs/1506.09173}{{\normalfont\ttfamily
  arXiv:1506.09173}}\relax
\mciteBstWouldAddEndPuncttrue
\mciteSetBstMidEndSepPunct{\mcitedefaultmidpunct}
{\mcitedefaultendpunct}{\mcitedefaultseppunct}\relax
\EndOfBibitem
\bibitem{Knapen:2016hky}
S.~Knapen, S.~Pagan~Griso, M.~Papucci, and D.~J. Robinson,
  \ifthenelse{\boolean{articletitles}}{\emph{{Triggering Soft Bombs at the
  LHC}}, }{}\href{http://arxiv.org/abs/1612.00850}{{\normalfont\ttfamily
  arXiv:1612.00850}}\relax
\mciteBstWouldAddEndPuncttrue
\mciteSetBstMidEndSepPunct{\mcitedefaultmidpunct}
{\mcitedefaultendpunct}{\mcitedefaultseppunct}\relax
\EndOfBibitem
\bibitem{Aaij:2017vbb}
LHCb collaboration, R.~Aaij {\em et~al.},
  \ifthenelse{\boolean{articletitles}}{\emph{{Test of lepton universality with
  $B^{0} \rightarrow K^{*0}\ell^{+}\ell^{-}$ decays}},
  }{}\href{https://doi.org/10.1007/JHEP08(2017)055}{JHEP \textbf{08} (2017)
  055}, \href{http://arxiv.org/abs/1705.05802}{{\normalfont\ttfamily
  arXiv:1705.05802}}\relax
\mciteBstWouldAddEndPuncttrue
\mciteSetBstMidEndSepPunct{\mcitedefaultmidpunct}
{\mcitedefaultendpunct}{\mcitedefaultseppunct}\relax
\EndOfBibitem
\bibitem{Aaij:2014ora}
LHCb collaboration, R.~Aaij {\em et~al.},
  \ifthenelse{\boolean{articletitles}}{\emph{{Test of lepton universality using
  $B^{+}\rightarrow K^{+}\ell^{+}\ell^{-}$ decays}},
  }{}\href{https://doi.org/10.1103/PhysRevLett.113.151601}{Phys.\ Rev.\ Lett.\
  \textbf{113} (2014) 151601},
  \href{http://arxiv.org/abs/1406.6482}{{\normalfont\ttfamily
  arXiv:1406.6482}}\relax
\mciteBstWouldAddEndPuncttrue
\mciteSetBstMidEndSepPunct{\mcitedefaultmidpunct}
{\mcitedefaultendpunct}{\mcitedefaultseppunct}\relax
\EndOfBibitem
\bibitem{Aaij:2014tfa}
LHCb collaboration, R.~Aaij {\em et~al.},
  \ifthenelse{\boolean{articletitles}}{\emph{{Angular analysis of charged and
  neutral $B \to K \mu^+\mu^-$ decays}},
  }{}\href{https://doi.org/10.1007/JHEP05(2014)082}{JHEP \textbf{05} (2014)
  082}, \href{http://arxiv.org/abs/1403.8045}{{\normalfont\ttfamily
  arXiv:1403.8045}}\relax
\mciteBstWouldAddEndPuncttrue
\mciteSetBstMidEndSepPunct{\mcitedefaultmidpunct}
{\mcitedefaultendpunct}{\mcitedefaultseppunct}\relax
\EndOfBibitem
\bibitem{Aaij:2020ruw}
LHCb collaboration, R.~Aaij {\em et~al.},
  \ifthenelse{\boolean{articletitles}}{\emph{{Angular analysis of the
  $B^{+}\rightarrow K^{\ast+}\mu^{+}\mu^{-}$ decay}},
  }{}\href{http://arxiv.org/abs/2012.13241}{{\normalfont\ttfamily
  arXiv:2012.13241}}\relax
\mciteBstWouldAddEndPuncttrue
\mciteSetBstMidEndSepPunct{\mcitedefaultmidpunct}
{\mcitedefaultendpunct}{\mcitedefaultseppunct}\relax
\EndOfBibitem
\bibitem{Aaij:2019wad}
LHCb collaboration, R.~Aaij {\em et~al.},
  \ifthenelse{\boolean{articletitles}}{\emph{{Search for lepton-universality
  violation in $B^+\to K^+\ell^+\ell^-$ decays}},
  }{}\href{https://doi.org/10.1103/PhysRevLett.122.191801}{Phys.\ Rev.\ Lett.\
  \textbf{122} (2019) 191801},
  \href{http://arxiv.org/abs/1903.09252}{{\normalfont\ttfamily
  arXiv:1903.09252}}\relax
\mciteBstWouldAddEndPuncttrue
\mciteSetBstMidEndSepPunct{\mcitedefaultmidpunct}
{\mcitedefaultendpunct}{\mcitedefaultseppunct}\relax
\EndOfBibitem
\bibitem{Aaij:2021vac}
LHCb collaboration, R.~Aaij {\em et~al.},
  \ifthenelse{\boolean{articletitles}}{\emph{{Test of lepton universality in
  beauty-quark decays}},
  }{}\href{http://arxiv.org/abs/2103.11769}{{\normalfont\ttfamily
  arXiv:2103.11769}}\relax
\mciteBstWouldAddEndPuncttrue
\mciteSetBstMidEndSepPunct{\mcitedefaultmidpunct}
{\mcitedefaultendpunct}{\mcitedefaultseppunct}\relax
\EndOfBibitem
\bibitem{Aaij:2013mga}
LHCb collaboration, R.~Aaij {\em et~al.},
  \ifthenelse{\boolean{articletitles}}{\emph{{Prompt charm production in pp
  collisions at sqrt(s)=7 TeV}},
  }{}\href{https://doi.org/10.1016/j.nuclphysb.2013.02.010}{Nucl.\ Phys.\
  \textbf{B871} (2013) 1},
  \href{http://arxiv.org/abs/1302.2864}{{\normalfont\ttfamily
  arXiv:1302.2864}}\relax
\mciteBstWouldAddEndPuncttrue
\mciteSetBstMidEndSepPunct{\mcitedefaultmidpunct}
{\mcitedefaultendpunct}{\mcitedefaultseppunct}\relax
\EndOfBibitem
\bibitem{Aaij:2015bpa}
LHCb collaboration, R.~Aaij {\em et~al.},
  \ifthenelse{\boolean{articletitles}}{\emph{{Measurements of prompt charm
  production cross-sections in $pp$ collisions at $ \sqrt{s}=13 $ TeV}},
  }{}\href{https://doi.org/10.1007/JHEP03(2016)159, 10.1007/JHEP09(2016)013,
  10.1007/JHEP05(2017)074}{JHEP \textbf{03} (2016) 159},
  \href{http://arxiv.org/abs/1510.01707}{{\normalfont\ttfamily
  arXiv:1510.01707}}, [Erratum: JHEP09,013(2016); Erratum:
  JHEP05,074(2017)]\relax
\mciteBstWouldAddEndPuncttrue
\mciteSetBstMidEndSepPunct{\mcitedefaultmidpunct}
{\mcitedefaultendpunct}{\mcitedefaultseppunct}\relax
\EndOfBibitem
\bibitem{Gauld:2015yia}
R.~Gauld, J.~Rojo, L.~Rottoli, and J.~Talbert,
  \ifthenelse{\boolean{articletitles}}{\emph{{Charm production in the forward
  region: constraints on the small-x gluon and backgrounds for neutrino
  astronomy}}, }{}\href{https://doi.org/10.1007/JHEP11(2015)009}{JHEP
  \textbf{11} (2015) 009},
  \href{http://arxiv.org/abs/1506.08025}{{\normalfont\ttfamily
  arXiv:1506.08025}}\relax
\mciteBstWouldAddEndPuncttrue
\mciteSetBstMidEndSepPunct{\mcitedefaultmidpunct}
{\mcitedefaultendpunct}{\mcitedefaultseppunct}\relax
\EndOfBibitem
\bibitem{Aaij:2018svt}
LHCb collaboration, R.~Aaij {\em et~al.},
  \ifthenelse{\boolean{articletitles}}{\emph{{Measurement of Antiproton
  Production in ${\rm p He}$ Collisions at $\sqrt{s_{NN}}=110$ GeV}},
  }{}\href{https://doi.org/10.1103/PhysRevLett.121.222001}{Phys.\ Rev.\ Lett.\
  \textbf{121} (2018) 222001},
  \href{http://arxiv.org/abs/1808.06127}{{\normalfont\ttfamily
  arXiv:1808.06127}}\relax
\mciteBstWouldAddEndPuncttrue
\mciteSetBstMidEndSepPunct{\mcitedefaultmidpunct}
{\mcitedefaultendpunct}{\mcitedefaultseppunct}\relax
\EndOfBibitem
\bibitem{Giesen:2015ufa}
G.~Giesen {\em et~al.}, \ifthenelse{\boolean{articletitles}}{\emph{{AMS-02
  antiprotons, at last! Secondary astrophysical component and immediate
  implications for Dark Matter}},
  }{}\href{https://doi.org/10.1088/1475-7516/2015/09/023,
  10.1088/1475-7516/2015/9/023}{JCAP \textbf{1509} (2015) 023},
  \href{http://arxiv.org/abs/1504.04276}{{\normalfont\ttfamily
  arXiv:1504.04276}}\relax
\mciteBstWouldAddEndPuncttrue
\mciteSetBstMidEndSepPunct{\mcitedefaultmidpunct}
{\mcitedefaultendpunct}{\mcitedefaultseppunct}\relax
\EndOfBibitem
\bibitem{Aaij:2016isa}
LHCb collaboration, R.~Aaij {\em et~al.},
  \ifthenelse{\boolean{articletitles}}{\emph{{Search for Higgs-like bosons
  decaying into long-lived exotic particles}},
  }{}\href{https://doi.org/10.1140/epjc/s10052-016-4489-7}{Eur.\ Phys.\ J.\ C
  \textbf{76} (2016) 664},
  \href{http://arxiv.org/abs/1609.03124}{{\normalfont\ttfamily
  arXiv:1609.03124}}\relax
\mciteBstWouldAddEndPuncttrue
\mciteSetBstMidEndSepPunct{\mcitedefaultmidpunct}
{\mcitedefaultendpunct}{\mcitedefaultseppunct}\relax
\EndOfBibitem
\bibitem{Aaij:2020ikh}
LHCb collaboration, R.~Aaij {\em et~al.},
  \ifthenelse{\boolean{articletitles}}{\emph{{Searches for low-mass dimuon
  resonances}}, }{}\href{https://doi.org/10.1007/JHEP10(2020)156}{JHEP
  \textbf{10} (2020) 156},
  \href{http://arxiv.org/abs/2007.03923}{{\normalfont\ttfamily
  arXiv:2007.03923}}\relax
\mciteBstWouldAddEndPuncttrue
\mciteSetBstMidEndSepPunct{\mcitedefaultmidpunct}
{\mcitedefaultendpunct}{\mcitedefaultseppunct}\relax
\EndOfBibitem
\bibitem{Aaij:2014aba}
LHCb collaboration, R.~Aaij {\em et~al.},
  \ifthenelse{\boolean{articletitles}}{\emph{{Search for Majorana neutrinos in
  $B^- \to \pi^+\mu^-\mu^-$ decays}},
  }{}\href{https://doi.org/10.1103/PhysRevLett.112.131802}{Phys.\ Rev.\ Lett.\
  \textbf{112} (2014) 131802},
  \href{http://arxiv.org/abs/1401.5361}{{\normalfont\ttfamily
  arXiv:1401.5361}}\relax
\mciteBstWouldAddEndPuncttrue
\mciteSetBstMidEndSepPunct{\mcitedefaultmidpunct}
{\mcitedefaultendpunct}{\mcitedefaultseppunct}\relax
\EndOfBibitem
\bibitem{Aaij:2020ovh}
LHCb collaboration, R.~Aaij {\em et~al.},
  \ifthenelse{\boolean{articletitles}}{\emph{{Search for heavy neutral leptons
  in $W^+\to\mu^{+}\mu^{\pm}\text{jet}$ decays}},
  }{}\href{https://doi.org/10.1140/epjc/s10052-021-08973-5}{Eur.\ Phys.\ J.\ C
  \textbf{81} (2021) 248},
  \href{http://arxiv.org/abs/2011.05263}{{\normalfont\ttfamily
  arXiv:2011.05263}}\relax
\mciteBstWouldAddEndPuncttrue
\mciteSetBstMidEndSepPunct{\mcitedefaultmidpunct}
{\mcitedefaultendpunct}{\mcitedefaultseppunct}\relax
\EndOfBibitem
\bibitem{Aaij:2020iew}
LHCb collaboration, R.~Aaij {\em et~al.},
  \ifthenelse{\boolean{articletitles}}{\emph{{Search for long-lived particles
  decaying to $e^\pm \mu^\mp \nu$}},
  }{}\href{https://doi.org/10.1140/epjc/s10052-021-08994-0}{Eur.\ Phys.\ J.\ C
  \textbf{81} (2021) 261},
  \href{http://arxiv.org/abs/2012.02696}{{\normalfont\ttfamily
  arXiv:2012.02696}}\relax
\mciteBstWouldAddEndPuncttrue
\mciteSetBstMidEndSepPunct{\mcitedefaultmidpunct}
{\mcitedefaultendpunct}{\mcitedefaultseppunct}\relax
\EndOfBibitem
\bibitem{Bezrukov:2009yw}
F.~Bezrukov and D.~Gorbunov, \ifthenelse{\boolean{articletitles}}{\emph{{Light
  inflaton Hunter's Guide}},
  }{}\href{https://doi.org/10.1007/JHEP05(2010)010}{JHEP \textbf{05} (2010)
  010}, \href{http://arxiv.org/abs/0912.0390}{{\normalfont\ttfamily
  arXiv:0912.0390}}\relax
\mciteBstWouldAddEndPuncttrue
\mciteSetBstMidEndSepPunct{\mcitedefaultmidpunct}
{\mcitedefaultendpunct}{\mcitedefaultseppunct}\relax
\EndOfBibitem
\bibitem{Ilten:2018crw}
P.~Ilten, Y.~Soreq, M.~Williams, and W.~Xue,
  \ifthenelse{\boolean{articletitles}}{\emph{{Serendipity in dark photon
  searches}}, }{}\href{https://doi.org/10.1007/JHEP06(2018)004}{JHEP
  \textbf{06} (2018) 004},
  \href{http://arxiv.org/abs/1801.04847}{{\normalfont\ttfamily
  arXiv:1801.04847}}\relax
\mciteBstWouldAddEndPuncttrue
\mciteSetBstMidEndSepPunct{\mcitedefaultmidpunct}
{\mcitedefaultendpunct}{\mcitedefaultseppunct}\relax
\EndOfBibitem
\bibitem{Chacko:2005pe}
Z.~Chacko, H.-S. Goh, and R.~Harnik,
  \ifthenelse{\boolean{articletitles}}{\emph{{The Twin Higgs: Natural
  electroweak breaking from mirror symmetry}},
  }{}\href{https://doi.org/10.1103/PhysRevLett.96.231802}{Phys.\ Rev.\ Lett.\
  \textbf{96} (2006) 231802},
  \href{http://arxiv.org/abs/hep-ph/0506256}{{\normalfont\ttfamily
  arXiv:hep-ph/0506256}}\relax
\mciteBstWouldAddEndPuncttrue
\mciteSetBstMidEndSepPunct{\mcitedefaultmidpunct}
{\mcitedefaultendpunct}{\mcitedefaultseppunct}\relax
\EndOfBibitem
\bibitem{Burdman:2006tz}
G.~Burdman, Z.~Chacko, H.-S. Goh, and R.~Harnik,
  \ifthenelse{\boolean{articletitles}}{\emph{{Folded supersymmetry and the LEP
  paradox}}, }{}\href{https://doi.org/10.1088/1126-6708/2007/02/009}{JHEP
  \textbf{02} (2007) 009},
  \href{http://arxiv.org/abs/hep-ph/0609152}{{\normalfont\ttfamily
  arXiv:hep-ph/0609152}}\relax
\mciteBstWouldAddEndPuncttrue
\mciteSetBstMidEndSepPunct{\mcitedefaultmidpunct}
{\mcitedefaultendpunct}{\mcitedefaultseppunct}\relax
\EndOfBibitem
\bibitem{Strassler:2006im}
M.~J. Strassler and K.~M. Zurek,
  \ifthenelse{\boolean{articletitles}}{\emph{{Echoes of a hidden valley at
  hadron colliders}},
  }{}\href{https://doi.org/10.1016/j.physletb.2007.06.055}{Phys.\ Lett.\
  \textbf{B651} (2007) 374},
  \href{http://arxiv.org/abs/hep-ph/0604261}{{\normalfont\ttfamily
  arXiv:hep-ph/0604261}}\relax
\mciteBstWouldAddEndPuncttrue
\mciteSetBstMidEndSepPunct{\mcitedefaultmidpunct}
{\mcitedefaultendpunct}{\mcitedefaultseppunct}\relax
\EndOfBibitem
\bibitem{Cheng:2019yai}
H.-C. Cheng, L.~Li, E.~Salvioni, and C.~B. Verhaaren,
  \ifthenelse{\boolean{articletitles}}{\emph{{Light Hidden Mesons through the Z
  Portal}}, }{}\href{https://doi.org/10.1007/JHEP11(2019)031}{JHEP \textbf{11}
  (2019) 031}, \href{http://arxiv.org/abs/1906.02198}{{\normalfont\ttfamily
  arXiv:1906.02198}}\relax
\mciteBstWouldAddEndPuncttrue
\mciteSetBstMidEndSepPunct{\mcitedefaultmidpunct}
{\mcitedefaultendpunct}{\mcitedefaultseppunct}\relax
\EndOfBibitem
\bibitem{Cheng:2018gvu}
H.-C. Cheng, L.~Li, E.~Salvioni, and C.~B. Verhaaren,
  \ifthenelse{\boolean{articletitles}}{\emph{{Singlet Scalar Top Partners from
  Accidental Supersymmetry}},
  }{}\href{https://doi.org/10.1007/JHEP05(2018)057}{JHEP \textbf{05} (2018)
  057}, \href{http://arxiv.org/abs/1803.03651}{{\normalfont\ttfamily
  arXiv:1803.03651}}\relax
\mciteBstWouldAddEndPuncttrue
\mciteSetBstMidEndSepPunct{\mcitedefaultmidpunct}
{\mcitedefaultendpunct}{\mcitedefaultseppunct}\relax
\EndOfBibitem
\bibitem{Farchioni:2007dw}
F.~Farchioni {\em et~al.}, \ifthenelse{\boolean{articletitles}}{\emph{{Hadron
  masses in QCD with one quark flavour}},
  }{}\href{https://doi.org/10.1140/epjc/s10052-007-0394-4}{Eur.\ Phys.\ J.\ C
  \textbf{52} (2007) 305},
  \href{http://arxiv.org/abs/0706.1131}{{\normalfont\ttfamily
  arXiv:0706.1131}}\relax
\mciteBstWouldAddEndPuncttrue
\mciteSetBstMidEndSepPunct{\mcitedefaultmidpunct}
{\mcitedefaultendpunct}{\mcitedefaultseppunct}\relax
\EndOfBibitem
\bibitem{Schwaller:2015gea}
P.~Schwaller, D.~Stolarski, and A.~Weiler,
  \ifthenelse{\boolean{articletitles}}{\emph{{Emerging Jets}},
  }{}\href{https://doi.org/10.1007/JHEP05(2015)059}{JHEP \textbf{05} (2015)
  059}, \href{http://arxiv.org/abs/1502.05409}{{\normalfont\ttfamily
  arXiv:1502.05409}}\relax
\mciteBstWouldAddEndPuncttrue
\mciteSetBstMidEndSepPunct{\mcitedefaultmidpunct}
{\mcitedefaultendpunct}{\mcitedefaultseppunct}\relax
\EndOfBibitem
\bibitem{Cohen:2015toa}
T.~Cohen, M.~Lisanti, and H.~K. Lou,
  \ifthenelse{\boolean{articletitles}}{\emph{{Semivisible Jets: Dark Matter
  Undercover at the LHC}},
  }{}\href{https://doi.org/10.1103/PhysRevLett.115.171804}{Phys.\ Rev.\ Lett.\
  \textbf{115} (2015) 171804},
  \href{http://arxiv.org/abs/1503.00009}{{\normalfont\ttfamily
  arXiv:1503.00009}}\relax
\mciteBstWouldAddEndPuncttrue
\mciteSetBstMidEndSepPunct{\mcitedefaultmidpunct}
{\mcitedefaultendpunct}{\mcitedefaultseppunct}\relax
\EndOfBibitem
\bibitem{Sirunyan:2018njd}
CMS collaboration, A.~M. Sirunyan {\em et~al.},
  \ifthenelse{\boolean{articletitles}}{\emph{{Search for new particles decaying
  to a jet and an emerging jet}},
  }{}\href{https://doi.org/10.1007/JHEP02(2019)179}{JHEP \textbf{02} (2019)
  179}, \href{http://arxiv.org/abs/1810.10069}{{\normalfont\ttfamily
  arXiv:1810.10069}}\relax
\mciteBstWouldAddEndPuncttrue
\mciteSetBstMidEndSepPunct{\mcitedefaultmidpunct}
{\mcitedefaultendpunct}{\mcitedefaultseppunct}\relax
\EndOfBibitem
\bibitem{Aad:2019tua}
ATLAS collaboration, G.~Aad {\em et~al.},
  \ifthenelse{\boolean{articletitles}}{\emph{{Search for light long-lived
  neutral particles produced in $pp$ collisions at $\sqrt{s} =$ 13 TeV and
  decaying into collimated leptons or light hadrons with the ATLAS detector}},
  }{}\href{https://doi.org/10.1140/epjc/s10052-020-7997-4}{Eur.\ Phys.\ J.\ C
  \textbf{80} (2020) 450},
  \href{http://arxiv.org/abs/1909.01246}{{\normalfont\ttfamily
  arXiv:1909.01246}}\relax
\mciteBstWouldAddEndPuncttrue
\mciteSetBstMidEndSepPunct{\mcitedefaultmidpunct}
{\mcitedefaultendpunct}{\mcitedefaultseppunct}\relax
\EndOfBibitem
\bibitem{Ilten:2016tkc}
P.~Ilten {\em et~al.}, \ifthenelse{\boolean{articletitles}}{\emph{{Proposed
  Inclusive Dark Photon Search at LHCb}},
  }{}\href{https://doi.org/10.1103/PhysRevLett.116.251803}{Phys.\ Rev.\ Lett.\
  \textbf{116} (2016) 251803},
  \href{http://arxiv.org/abs/1603.08926}{{\normalfont\ttfamily
  arXiv:1603.08926}}\relax
\mciteBstWouldAddEndPuncttrue
\mciteSetBstMidEndSepPunct{\mcitedefaultmidpunct}
{\mcitedefaultendpunct}{\mcitedefaultseppunct}\relax
\EndOfBibitem
\bibitem{Han:2007ae}
T.~Han, Z.~Si, K.~M. Zurek, and M.~J. Strassler,
  \ifthenelse{\boolean{articletitles}}{\emph{{Phenomenology of hidden valleys
  at hadron colliders}},
  }{}\href{https://doi.org/10.1088/1126-6708/2008/07/008}{JHEP \textbf{07}
  (2008) 008}, \href{http://arxiv.org/abs/0712.2041}{{\normalfont\ttfamily
  arXiv:0712.2041}}\relax
\mciteBstWouldAddEndPuncttrue
\mciteSetBstMidEndSepPunct{\mcitedefaultmidpunct}
{\mcitedefaultendpunct}{\mcitedefaultseppunct}\relax
\EndOfBibitem
\bibitem{Li:2019ulz}
L.~Li and Y.~Tsai, \ifthenelse{\boolean{articletitles}}{\emph{{Detector-size
  Upper Bounds on Dark Hadron Lifetime from Cosmology}},
  }{}\href{https://doi.org/10.1007/JHEP05(2019)072}{JHEP \textbf{05} (2019)
  072}, \href{http://arxiv.org/abs/1901.09936}{{\normalfont\ttfamily
  arXiv:1901.09936}}\relax
\mciteBstWouldAddEndPuncttrue
\mciteSetBstMidEndSepPunct{\mcitedefaultmidpunct}
{\mcitedefaultendpunct}{\mcitedefaultseppunct}\relax
\EndOfBibitem
\bibitem{Cheng:2015buv}
H.-C. Cheng, S.~Jung, E.~Salvioni, and Y.~Tsai,
  \ifthenelse{\boolean{articletitles}}{\emph{{Exotic Quarks in Twin Higgs
  Models}}, }{}\href{https://doi.org/10.1007/JHEP03(2016)074}{JHEP \textbf{03}
  (2016) 074}, \href{http://arxiv.org/abs/1512.02647}{{\normalfont\ttfamily
  arXiv:1512.02647}}\relax
\mciteBstWouldAddEndPuncttrue
\mciteSetBstMidEndSepPunct{\mcitedefaultmidpunct}
{\mcitedefaultendpunct}{\mcitedefaultseppunct}\relax
\EndOfBibitem
\bibitem{Knapen:2021eip}
S.~Knapen, J.~Shelton, and D.~Xu,
  \ifthenelse{\boolean{articletitles}}{\emph{{Perturbative benchmark models for
  a dark shower search program}},
  }{}\href{http://arxiv.org/abs/2103.01238}{{\normalfont\ttfamily
  arXiv:2103.01238}}\relax
\mciteBstWouldAddEndPuncttrue
\mciteSetBstMidEndSepPunct{\mcitedefaultmidpunct}
{\mcitedefaultendpunct}{\mcitedefaultseppunct}\relax
\EndOfBibitem
\bibitem{Craig:2015pha}
N.~Craig, A.~Katz, M.~Strassler, and R.~Sundrum,
  \ifthenelse{\boolean{articletitles}}{\emph{{Naturalness in the Dark at the
  LHC}}, }{}\href{https://doi.org/10.1007/JHEP07(2015)105}{JHEP \textbf{07}
  (2015) 105}, \href{http://arxiv.org/abs/1501.05310}{{\normalfont\ttfamily
  arXiv:1501.05310}}\relax
\mciteBstWouldAddEndPuncttrue
\mciteSetBstMidEndSepPunct{\mcitedefaultmidpunct}
{\mcitedefaultendpunct}{\mcitedefaultseppunct}\relax
\EndOfBibitem
\bibitem{Gripaios:2009pe}
B.~Gripaios, A.~Pomarol, F.~Riva, and J.~Serra,
  \ifthenelse{\boolean{articletitles}}{\emph{{Beyond the Minimal Composite
  Higgs Model}}, }{}\href{https://doi.org/10.1088/1126-6708/2009/04/070}{JHEP
  \textbf{04} (2009) 070},
  \href{http://arxiv.org/abs/0902.1483}{{\normalfont\ttfamily
  arXiv:0902.1483}}\relax
\mciteBstWouldAddEndPuncttrue
\mciteSetBstMidEndSepPunct{\mcitedefaultmidpunct}
{\mcitedefaultendpunct}{\mcitedefaultseppunct}\relax
\EndOfBibitem
\bibitem{Vecchi:2013bja}
L.~Vecchi, \ifthenelse{\boolean{articletitles}}{\emph{{The Natural Composite
  Higgs}}, }{}\href{http://arxiv.org/abs/1304.4579}{{\normalfont\ttfamily
  arXiv:1304.4579}}\relax
\mciteBstWouldAddEndPuncttrue
\mciteSetBstMidEndSepPunct{\mcitedefaultmidpunct}
{\mcitedefaultendpunct}{\mcitedefaultseppunct}\relax
\EndOfBibitem
\bibitem{Sanz:2015sua}
V.~Sanz and J.~Setford, \ifthenelse{\boolean{articletitles}}{\emph{{Composite
  Higgses with seesaw EWSB}},
  }{}\href{https://doi.org/10.1007/JHEP12(2015)154}{JHEP \textbf{12} (2015)
  154}, \href{http://arxiv.org/abs/1508.06133}{{\normalfont\ttfamily
  arXiv:1508.06133}}\relax
\mciteBstWouldAddEndPuncttrue
\mciteSetBstMidEndSepPunct{\mcitedefaultmidpunct}
{\mcitedefaultendpunct}{\mcitedefaultseppunct}\relax
\EndOfBibitem
\bibitem{Chala:2016ykx}
M.~Chala, G.~Nardini, and I.~Sobolev,
  \ifthenelse{\boolean{articletitles}}{\emph{{Unified explanation for dark
  matter and electroweak baryogenesis with direct detection and gravitational
  wave signatures}},
  }{}\href{https://doi.org/10.1103/PhysRevD.94.055006}{Phys.\ Rev.\
  \textbf{D94} (2016) 055006},
  \href{http://arxiv.org/abs/1605.08663}{{\normalfont\ttfamily
  arXiv:1605.08663}}\relax
\mciteBstWouldAddEndPuncttrue
\mciteSetBstMidEndSepPunct{\mcitedefaultmidpunct}
{\mcitedefaultendpunct}{\mcitedefaultseppunct}\relax
\EndOfBibitem
\bibitem{Balkin:2017aep}
R.~Balkin, M.~Ruhdorfer, E.~Salvioni, and A.~Weiler,
  \ifthenelse{\boolean{articletitles}}{\emph{{Charged Composite Scalar Dark
  Matter}}, }{}\href{https://doi.org/10.1007/JHEP11(2017)094}{JHEP \textbf{11}
  (2017) 094}, \href{http://arxiv.org/abs/1707.07685}{{\normalfont\ttfamily
  arXiv:1707.07685}}\relax
\mciteBstWouldAddEndPuncttrue
\mciteSetBstMidEndSepPunct{\mcitedefaultmidpunct}
{\mcitedefaultendpunct}{\mcitedefaultseppunct}\relax
\EndOfBibitem
\bibitem{DaRold:2019ccj}
L.~Da~Rold and A.~N. Rossia, \ifthenelse{\boolean{articletitles}}{\emph{{The
  Minimal Simple Composite Higgs Model}},
  }{}\href{http://arxiv.org/abs/1904.02560}{{\normalfont\ttfamily
  arXiv:1904.02560}}\relax
\mciteBstWouldAddEndPuncttrue
\mciteSetBstMidEndSepPunct{\mcitedefaultmidpunct}
{\mcitedefaultendpunct}{\mcitedefaultseppunct}\relax
\EndOfBibitem
\bibitem{Ramos:2019qqa}
M.~Ramos, \ifthenelse{\boolean{articletitles}}{\emph{{Composite dark matter
  phenomenology in the presence of lighter degrees of freedom}},
  }{}\href{https://doi.org/10.1007/JHEP07(2020)128}{JHEP \textbf{07} (2020)
  128}, \href{http://arxiv.org/abs/1912.11061}{{\normalfont\ttfamily
  arXiv:1912.11061}}\relax
\mciteBstWouldAddEndPuncttrue
\mciteSetBstMidEndSepPunct{\mcitedefaultmidpunct}
{\mcitedefaultendpunct}{\mcitedefaultseppunct}\relax
\EndOfBibitem
\bibitem{Niehoff:2015bfa}
C.~Niehoff, P.~Stangl, and D.~M. Straub,
  \ifthenelse{\boolean{articletitles}}{\emph{{Violation of lepton flavour
  universality in composite Higgs models}},
  }{}\href{https://doi.org/10.1016/j.physletb.2015.05.063}{Phys.\ Lett.\
  \textbf{B747} (2015) 182},
  \href{http://arxiv.org/abs/1503.03865}{{\normalfont\ttfamily
  arXiv:1503.03865}}\relax
\mciteBstWouldAddEndPuncttrue
\mciteSetBstMidEndSepPunct{\mcitedefaultmidpunct}
{\mcitedefaultendpunct}{\mcitedefaultseppunct}\relax
\EndOfBibitem
\bibitem{Niehoff:2015iaa}
C.~Niehoff, P.~Stangl, and D.~M. Straub,
  \ifthenelse{\boolean{articletitles}}{\emph{{Direct and indirect signals of
  natural composite Higgs models}},
  }{}\href{https://doi.org/10.1007/JHEP01(2016)119}{JHEP \textbf{01} (2016)
  119}, \href{http://arxiv.org/abs/1508.00569}{{\normalfont\ttfamily
  arXiv:1508.00569}}\relax
\mciteBstWouldAddEndPuncttrue
\mciteSetBstMidEndSepPunct{\mcitedefaultmidpunct}
{\mcitedefaultendpunct}{\mcitedefaultseppunct}\relax
\EndOfBibitem
\bibitem{Carmona:2015ena}
A.~Carmona and F.~Goertz, \ifthenelse{\boolean{articletitles}}{\emph{{Lepton
  Flavor and Nonuniversality from Minimal Composite Higgs Setups}},
  }{}\href{https://doi.org/10.1103/PhysRevLett.116.251801}{Phys.\ Rev.\ Lett.\
  \textbf{116} (2016) 251801},
  \href{http://arxiv.org/abs/1510.07658}{{\normalfont\ttfamily
  arXiv:1510.07658}}\relax
\mciteBstWouldAddEndPuncttrue
\mciteSetBstMidEndSepPunct{\mcitedefaultmidpunct}
{\mcitedefaultendpunct}{\mcitedefaultseppunct}\relax
\EndOfBibitem
\bibitem{Megias:2016bde}
E.~Megias, G.~Panico, O.~Pujolas, and M.~Quiros,
  \ifthenelse{\boolean{articletitles}}{\emph{{A Natural origin for the LHCb
  anomalies}}, }{}\href{https://doi.org/10.1007/JHEP09(2016)118}{JHEP
  \textbf{09} (2016) 118},
  \href{http://arxiv.org/abs/1608.02362}{{\normalfont\ttfamily
  arXiv:1608.02362}}\relax
\mciteBstWouldAddEndPuncttrue
\mciteSetBstMidEndSepPunct{\mcitedefaultmidpunct}
{\mcitedefaultendpunct}{\mcitedefaultseppunct}\relax
\EndOfBibitem
\bibitem{GarciaGarcia:2016nvr}
I.~Garcia~Garcia, \ifthenelse{\boolean{articletitles}}{\emph{{LHCb anomalies
  from a natural perspective}},
  }{}\href{https://doi.org/10.1007/JHEP03(2017)040}{JHEP \textbf{03} (2017)
  040}, \href{http://arxiv.org/abs/1611.03507}{{\normalfont\ttfamily
  arXiv:1611.03507}}\relax
\mciteBstWouldAddEndPuncttrue
\mciteSetBstMidEndSepPunct{\mcitedefaultmidpunct}
{\mcitedefaultendpunct}{\mcitedefaultseppunct}\relax
\EndOfBibitem
\bibitem{Sannino:2017utc}
F.~Sannino, P.~Stangl, D.~M. Straub, and A.~E. Thomsen,
  \ifthenelse{\boolean{articletitles}}{\emph{{Flavor Physics and Flavor
  Anomalies in Minimal Fundamental Partial Compositeness}},
  }{}\href{https://doi.org/10.1103/PhysRevD.97.115046}{Phys.\ Rev.\
  \textbf{D97} (2018) 115046},
  \href{http://arxiv.org/abs/1712.07646}{{\normalfont\ttfamily
  arXiv:1712.07646}}\relax
\mciteBstWouldAddEndPuncttrue
\mciteSetBstMidEndSepPunct{\mcitedefaultmidpunct}
{\mcitedefaultendpunct}{\mcitedefaultseppunct}\relax
\EndOfBibitem
\bibitem{Carmona:2017fsn}
A.~Carmona and F.~Goertz, \ifthenelse{\boolean{articletitles}}{\emph{{Recent
  $B$ physics anomalies: a first hint for compositeness?}},
  }{}\href{https://doi.org/10.1140/epjc/s10052-018-6437-1}{Eur.\ Phys.\ J.\
  \textbf{C78} (2018) 979},
  \href{http://arxiv.org/abs/1712.02536}{{\normalfont\ttfamily
  arXiv:1712.02536}}\relax
\mciteBstWouldAddEndPuncttrue
\mciteSetBstMidEndSepPunct{\mcitedefaultmidpunct}
{\mcitedefaultendpunct}{\mcitedefaultseppunct}\relax
\EndOfBibitem
\bibitem{Chala:2018igk}
M.~Chala and M.~Spannowsky,
  \ifthenelse{\boolean{articletitles}}{\emph{{Behavior of composite resonances
  breaking lepton flavor universality}},
  }{}\href{https://doi.org/10.1103/PhysRevD.98.035010}{Phys.\ Rev.\ D
  \textbf{98} (2018) 035010},
  \href{http://arxiv.org/abs/1803.02364}{{\normalfont\ttfamily
  arXiv:1803.02364}}\relax
\mciteBstWouldAddEndPuncttrue
\mciteSetBstMidEndSepPunct{\mcitedefaultmidpunct}
{\mcitedefaultendpunct}{\mcitedefaultseppunct}\relax
\EndOfBibitem
\bibitem{Aaij:2016kfs}
LHCb collaboration, R.~Aaij {\em et~al.},
  \ifthenelse{\boolean{articletitles}}{\emph{{Search for decays of neutral
  beauty mesons into four muons}},
  }{}\href{https://doi.org/10.1007/JHEP03(2017)001}{JHEP \textbf{03} (2017)
  001}, \href{http://arxiv.org/abs/1611.07704}{{\normalfont\ttfamily
  arXiv:1611.07704}}\relax
\mciteBstWouldAddEndPuncttrue
\mciteSetBstMidEndSepPunct{\mcitedefaultmidpunct}
{\mcitedefaultendpunct}{\mcitedefaultseppunct}\relax
\EndOfBibitem
\bibitem{Blance:2019ixw}
A.~Blance, M.~Chala, M.~Ramos, and M.~Spannowsky,
  \ifthenelse{\boolean{articletitles}}{\emph{{Novel $B$-decay signatures of
  light scalars at high energy facilities}},
  }{}\href{https://doi.org/10.1103/PhysRevD.100.115015}{Phys.\ Rev.\ D
  \textbf{100} (2019) 115015},
  \href{http://arxiv.org/abs/1907.13151}{{\normalfont\ttfamily
  arXiv:1907.13151}}\relax
\mciteBstWouldAddEndPuncttrue
\mciteSetBstMidEndSepPunct{\mcitedefaultmidpunct}
{\mcitedefaultendpunct}{\mcitedefaultseppunct}\relax
\EndOfBibitem
\bibitem{Chala:2019vzu}
M.~Chala, U.~Egede, and M.~Spannowsky,
  \ifthenelse{\boolean{articletitles}}{\emph{{Searching new physics in rare
  $B$-meson decays into multiple muons}},
  }{}\href{https://doi.org/10.1140/epjc/s10052-019-6946-6}{Eur.\ Phys.\ J.\
  \textbf{C79} (2019) 431},
  \href{http://arxiv.org/abs/1902.10156}{{\normalfont\ttfamily
  arXiv:1902.10156}}\relax
\mciteBstWouldAddEndPuncttrue
\mciteSetBstMidEndSepPunct{\mcitedefaultmidpunct}
{\mcitedefaultendpunct}{\mcitedefaultseppunct}\relax
\EndOfBibitem
\bibitem{DiLuzio:2017fdq}
L.~Di~Luzio, M.~Kirk, and A.~Lenz,
  \ifthenelse{\boolean{articletitles}}{\emph{{Updated $B_s$-mixing constraints
  on new physics models for $b\to s\ell^+\ell^-$ anomalies}},
  }{}\href{https://doi.org/10.1103/PhysRevD.97.095035}{Phys.\ Rev.\
  \textbf{D97} (2018) 095035},
  \href{http://arxiv.org/abs/1712.06572}{{\normalfont\ttfamily
  arXiv:1712.06572}}\relax
\mciteBstWouldAddEndPuncttrue
\mciteSetBstMidEndSepPunct{\mcitedefaultmidpunct}
{\mcitedefaultendpunct}{\mcitedefaultseppunct}\relax
\EndOfBibitem
\bibitem{Cacciapaglia:2020kgq}
G.~Cacciapaglia, C.~Pica, and F.~Sannino,
  \ifthenelse{\boolean{articletitles}}{\emph{{Fundamental Composite Dynamics: A
  Review}}, }{}\href{https://doi.org/10.1016/j.physrep.2020.07.002}{Phys.\
  Rept.\  \textbf{877} (2020) 1},
  \href{http://arxiv.org/abs/2002.04914}{{\normalfont\ttfamily
  arXiv:2002.04914}}\relax
\mciteBstWouldAddEndPuncttrue
\mciteSetBstMidEndSepPunct{\mcitedefaultmidpunct}
{\mcitedefaultendpunct}{\mcitedefaultseppunct}\relax
\EndOfBibitem
\bibitem{Gherghetta:2020ofz}
T.~Gherghetta and M.~D. Nguyen, \ifthenelse{\boolean{articletitles}}{\emph{{A
  Composite Higgs with a Heavy Composite Axion}},
  }{}\href{https://doi.org/10.1007/JHEP12(2020)094}{JHEP \textbf{12} (2020)
  094}, \href{http://arxiv.org/abs/2007.10875}{{\normalfont\ttfamily
  arXiv:2007.10875}}\relax
\mciteBstWouldAddEndPuncttrue
\mciteSetBstMidEndSepPunct{\mcitedefaultmidpunct}
{\mcitedefaultendpunct}{\mcitedefaultseppunct}\relax
\EndOfBibitem
\bibitem{Dugan:1984hq}
M.~J. Dugan, H.~Georgi, and D.~B. Kaplan,
  \ifthenelse{\boolean{articletitles}}{\emph{{Anatomy of a Composite Higgs
  Model}}, }{}\href{https://doi.org/10.1016/0550-3213(85)90221-4}{Nucl.\ Phys.\
  B \textbf{254} (1985) 299}\relax
\mciteBstWouldAddEndPuncttrue
\mciteSetBstMidEndSepPunct{\mcitedefaultmidpunct}
{\mcitedefaultendpunct}{\mcitedefaultseppunct}\relax
\EndOfBibitem
\bibitem{Ferretti:2013kya}
G.~Ferretti and D.~Karateev,
  \ifthenelse{\boolean{articletitles}}{\emph{{Fermionic UV completions of
  Composite Higgs models}},
  }{}\href{https://doi.org/10.1007/JHEP03(2014)077}{JHEP \textbf{03} (2014)
  077}, \href{http://arxiv.org/abs/1312.5330}{{\normalfont\ttfamily
  arXiv:1312.5330}}\relax
\mciteBstWouldAddEndPuncttrue
\mciteSetBstMidEndSepPunct{\mcitedefaultmidpunct}
{\mcitedefaultendpunct}{\mcitedefaultseppunct}\relax
\EndOfBibitem
\bibitem{Cacciapaglia:2019bqz}
G.~Cacciapaglia, G.~Ferretti, T.~Flacke, and H.~Ser\^odio,
  \ifthenelse{\boolean{articletitles}}{\emph{{Light scalars in composite Higgs
  models}}, }{}\href{https://doi.org/10.3389/fphy.2019.00022}{Front.\ in Phys.\
   \textbf{7} (2019) 22},
  \href{http://arxiv.org/abs/1902.06890}{{\normalfont\ttfamily
  arXiv:1902.06890}}\relax
\mciteBstWouldAddEndPuncttrue
\mciteSetBstMidEndSepPunct{\mcitedefaultmidpunct}
{\mcitedefaultendpunct}{\mcitedefaultseppunct}\relax
\EndOfBibitem
\bibitem{Cacciapaglia:2017iws}
G.~Cacciapaglia, G.~Ferretti, T.~Flacke, and H.~Serodio,
  \ifthenelse{\boolean{articletitles}}{\emph{{Revealing timid pseudo-scalars
  with taus at the LHC}},
  }{}\href{https://doi.org/10.1140/epjc/s10052-018-6183-4}{Eur.\ Phys.\ J.\ C
  \textbf{78} (2018) 724},
  \href{http://arxiv.org/abs/1710.11142}{{\normalfont\ttfamily
  arXiv:1710.11142}}\relax
\mciteBstWouldAddEndPuncttrue
\mciteSetBstMidEndSepPunct{\mcitedefaultmidpunct}
{\mcitedefaultendpunct}{\mcitedefaultseppunct}\relax
\EndOfBibitem
\bibitem{Belyaev:2016ftv}
A.~Belyaev {\em et~al.}, \ifthenelse{\boolean{articletitles}}{\emph{{Di-boson
  signatures as Standard Candles for Partial Compositeness}},
  }{}\href{https://doi.org/10.1007/JHEP01(2017)094}{JHEP \textbf{01} (2017)
  094}, \href{http://arxiv.org/abs/1610.06591}{{\normalfont\ttfamily
  arXiv:1610.06591}}, [Erratum: JHEP 12, 088 (2017)]\relax
\mciteBstWouldAddEndPuncttrue
\mciteSetBstMidEndSepPunct{\mcitedefaultmidpunct}
{\mcitedefaultendpunct}{\mcitedefaultseppunct}\relax
\EndOfBibitem
\bibitem{Haisch:2016hzu}
U.~Haisch and J.~F. Kamenik,
  \ifthenelse{\boolean{articletitles}}{\emph{{Searching for new spin-0
  resonances at LHCb}},
  }{}\href{https://doi.org/10.1103/PhysRevD.93.055047}{Phys.\ Rev.\ D
  \textbf{93} (2016) 055047},
  \href{http://arxiv.org/abs/1601.05110}{{\normalfont\ttfamily
  arXiv:1601.05110}}\relax
\mciteBstWouldAddEndPuncttrue
\mciteSetBstMidEndSepPunct{\mcitedefaultmidpunct}
{\mcitedefaultendpunct}{\mcitedefaultseppunct}\relax
\EndOfBibitem
\bibitem{Spira:1996if}
M.~Spira, \ifthenelse{\boolean{articletitles}}{\emph{{HIGLU and HDECAY:
  Programs for Higgs boson production at the LHC and Higgs boson decay
  widths}}, }{}\href{https://doi.org/10.1016/S0168-9002(97)00129-0}{Nucl.\
  Instrum.\ Meth.\ A \textbf{389} (1997) 357},
  \href{http://arxiv.org/abs/hep-ph/9610350}{{\normalfont\ttfamily
  arXiv:hep-ph/9610350}}\relax
\mciteBstWouldAddEndPuncttrue
\mciteSetBstMidEndSepPunct{\mcitedefaultmidpunct}
{\mcitedefaultendpunct}{\mcitedefaultseppunct}\relax
\EndOfBibitem
\bibitem{Bertone:2018dse}
V.~Bertone, R.~Gauld, and J.~Rojo,
  \ifthenelse{\boolean{articletitles}}{\emph{{Neutrino Telescopes as QCD
  Microscopes}}, }{}\href{https://doi.org/10.1007/JHEP01(2019)217}{JHEP
  \textbf{01} (2019) 217},
  \href{http://arxiv.org/abs/1808.02034}{{\normalfont\ttfamily
  arXiv:1808.02034}}\relax
\mciteBstWouldAddEndPuncttrue
\mciteSetBstMidEndSepPunct{\mcitedefaultmidpunct}
{\mcitedefaultendpunct}{\mcitedefaultseppunct}\relax
\EndOfBibitem
\bibitem{Carmona:2021seb}
A.~Carmona, C.~Scherb, and P.~Schwaller,
  \ifthenelse{\boolean{articletitles}}{\emph{{Charming ALPs}},
  }{}\href{http://arxiv.org/abs/2101.07803}{{\normalfont\ttfamily
  arXiv:2101.07803}}\relax
\mciteBstWouldAddEndPuncttrue
\mciteSetBstMidEndSepPunct{\mcitedefaultmidpunct}
{\mcitedefaultendpunct}{\mcitedefaultseppunct}\relax
\EndOfBibitem
\bibitem{Alwall:2011uj}
J.~Alwall {\em et~al.}, \ifthenelse{\boolean{articletitles}}{\emph{{MadGraph 5
  : Going Beyond}}, }{}\href{https://doi.org/10.1007/JHEP06(2011)128}{JHEP
  \textbf{06} (2011) 128},
  \href{http://arxiv.org/abs/1106.0522}{{\normalfont\ttfamily
  arXiv:1106.0522}}\relax
\mciteBstWouldAddEndPuncttrue
\mciteSetBstMidEndSepPunct{\mcitedefaultmidpunct}
{\mcitedefaultendpunct}{\mcitedefaultseppunct}\relax
\EndOfBibitem
\bibitem{Artoisenet:2013puc}
P.~Artoisenet {\em et~al.}, \ifthenelse{\boolean{articletitles}}{\emph{{A
  framework for Higgs characterisation}},
  }{}\href{https://doi.org/10.1007/JHEP11(2013)043}{JHEP \textbf{11} (2013)
  043}, \href{http://arxiv.org/abs/1306.6464}{{\normalfont\ttfamily
  arXiv:1306.6464}}\relax
\mciteBstWouldAddEndPuncttrue
\mciteSetBstMidEndSepPunct{\mcitedefaultmidpunct}
{\mcitedefaultendpunct}{\mcitedefaultseppunct}\relax
\EndOfBibitem
\bibitem{Sjostrand:2014zea}
T.~Sj\"ostrand {\em et~al.}, \ifthenelse{\boolean{articletitles}}{\emph{{An
  introduction to PYTHIA 8.2}},
  }{}\href{https://doi.org/10.1016/j.cpc.2015.01.024}{Comput.\ Phys.\ Commun.\
  \textbf{191} (2015) 159},
  \href{http://arxiv.org/abs/1410.3012}{{\normalfont\ttfamily
  arXiv:1410.3012}}\relax
\mciteBstWouldAddEndPuncttrue
\mciteSetBstMidEndSepPunct{\mcitedefaultmidpunct}
{\mcitedefaultendpunct}{\mcitedefaultseppunct}\relax
\EndOfBibitem
\bibitem{Aaij:2016yip}
LHCb collaboration, R.~Aaij {\em et~al.},
  \ifthenelse{\boolean{articletitles}}{\emph{{Measurement of $CP$ violation in
  $B^0 \!\rightarrow D^+ D^-$ decays}},
  }{}\href{https://doi.org/10.1103/PhysRevLett.117.261801}{Phys.\ Rev.\ Lett.\
  \textbf{117} (2016) 261801},
  \href{http://arxiv.org/abs/1608.06620}{{\normalfont\ttfamily
  arXiv:1608.06620}}\relax
\mciteBstWouldAddEndPuncttrue
\mciteSetBstMidEndSepPunct{\mcitedefaultmidpunct}
{\mcitedefaultendpunct}{\mcitedefaultseppunct}\relax
\EndOfBibitem
\bibitem{dEnterria:2016ids}
D.~d'Enterria and A.~M. Snigirev,
  \ifthenelse{\boolean{articletitles}}{\emph{{Triple parton scatterings in
  high-energy proton-proton collisions}},
  }{}\href{https://doi.org/10.1103/PhysRevLett.118.122001}{Phys.\ Rev.\ Lett.\
  \textbf{118} (2017) 122001},
  \href{http://arxiv.org/abs/1612.05582}{{\normalfont\ttfamily
  arXiv:1612.05582}}\relax
\mciteBstWouldAddEndPuncttrue
\mciteSetBstMidEndSepPunct{\mcitedefaultmidpunct}
{\mcitedefaultendpunct}{\mcitedefaultseppunct}\relax
\EndOfBibitem
\bibitem{Filimonova:2019tuy}
A.~Filimonova, R.~Sch\"afer, and S.~Westhoff,
  \ifthenelse{\boolean{articletitles}}{\emph{{Probing dark sectors with
  long-lived particles at BELLE II}},
  }{}\href{https://doi.org/10.1103/PhysRevD.101.095006}{Phys.\ Rev.\ D
  \textbf{101} (2020) 095006},
  \href{http://arxiv.org/abs/1911.03490}{{\normalfont\ttfamily
  arXiv:1911.03490}}\relax
\mciteBstWouldAddEndPuncttrue
\mciteSetBstMidEndSepPunct{\mcitedefaultmidpunct}
{\mcitedefaultendpunct}{\mcitedefaultseppunct}\relax
\EndOfBibitem
\bibitem{Winkler:2018qyg}
M.~W. Winkler, \ifthenelse{\boolean{articletitles}}{\emph{{Decay and detection
  of a light scalar boson mixing with the Higgs boson}},
  }{}\href{https://doi.org/10.1103/PhysRevD.99.015018}{Phys.\ Rev.\
  \textbf{D99} (2019) 015018},
  \href{http://arxiv.org/abs/1809.01876}{{\normalfont\ttfamily
  arXiv:1809.01876}}\relax
\mciteBstWouldAddEndPuncttrue
\mciteSetBstMidEndSepPunct{\mcitedefaultmidpunct}
{\mcitedefaultendpunct}{\mcitedefaultseppunct}\relax
\EndOfBibitem
\bibitem{Bondarenko:2019vrb}
K.~Bondarenko {\em et~al.}, \ifthenelse{\boolean{articletitles}}{\emph{{Direct
  detection and complementary constraints for sub-GeV dark matter}},
  }{}\href{https://doi.org/10.1007/JHEP03(2020)118}{JHEP \textbf{03} (2020)
  118}, \href{http://arxiv.org/abs/1909.08632}{{\normalfont\ttfamily
  arXiv:1909.08632}}\relax
\mciteBstWouldAddEndPuncttrue
\mciteSetBstMidEndSepPunct{\mcitedefaultmidpunct}
{\mcitedefaultendpunct}{\mcitedefaultseppunct}\relax
\EndOfBibitem
\bibitem{Lees:2015rxq}
BaBar collaboration, J.~P. Lees {\em et~al.},
  \ifthenelse{\boolean{articletitles}}{\emph{{Search for Long-Lived Particles
  in $e^+e^-$ Collisions}},
  }{}\href{https://doi.org/10.1103/PhysRevLett.114.171801}{Phys.\ Rev.\ Lett.\
  \textbf{114} (2015) 171801},
  \href{http://arxiv.org/abs/1502.02580}{{\normalfont\ttfamily
  arXiv:1502.02580}}\relax
\mciteBstWouldAddEndPuncttrue
\mciteSetBstMidEndSepPunct{\mcitedefaultmidpunct}
{\mcitedefaultendpunct}{\mcitedefaultseppunct}\relax
\EndOfBibitem
\bibitem{DAgnolo:2017dbv}
R.~T. D'Agnolo, D.~Pappadopulo, and J.~T. Ruderman,
  \ifthenelse{\boolean{articletitles}}{\emph{{Fourth Exception in the
  Calculation of Relic Abundances}},
  }{}\href{https://doi.org/10.1103/PhysRevLett.119.061102}{Phys.\ Rev.\ Lett.\
  \textbf{119} (2017) 061102},
  \href{http://arxiv.org/abs/1705.08450}{{\normalfont\ttfamily
  arXiv:1705.08450}}\relax
\mciteBstWouldAddEndPuncttrue
\mciteSetBstMidEndSepPunct{\mcitedefaultmidpunct}
{\mcitedefaultendpunct}{\mcitedefaultseppunct}\relax
\EndOfBibitem
\bibitem{Garny:2017rxs}
M.~Garny, J.~Heisig, B.~L{\"u}lf, and S.~Vogl,
  \ifthenelse{\boolean{articletitles}}{\emph{{Coannihilation without chemical
  equilibrium}}, }{}\href{https://doi.org/10.1103/PhysRevD.96.103521}{Phys.\
  Rev.\ D \textbf{96} (2017) 103521},
  \href{http://arxiv.org/abs/1705.09292}{{\normalfont\ttfamily
  arXiv:1705.09292}}\relax
\mciteBstWouldAddEndPuncttrue
\mciteSetBstMidEndSepPunct{\mcitedefaultmidpunct}
{\mcitedefaultendpunct}{\mcitedefaultseppunct}\relax
\EndOfBibitem
\bibitem{Bharucha:2018pfu}
A.~Bharucha, F.~Br{\"u}mmer, and N.~Desai,
  \ifthenelse{\boolean{articletitles}}{\emph{{Next-to-minimal dark matter at
  the LHC}}, }{}\href{https://doi.org/10.1007/JHEP11(2018)195}{JHEP \textbf{11}
  (2018) 195}, \href{http://arxiv.org/abs/1804.02357}{{\normalfont\ttfamily
  arXiv:1804.02357}}\relax
\mciteBstWouldAddEndPuncttrue
\mciteSetBstMidEndSepPunct{\mcitedefaultmidpunct}
{\mcitedefaultendpunct}{\mcitedefaultseppunct}\relax
\EndOfBibitem
\bibitem{Filimonova:2018qdc}
A.~Filimonova and S.~Westhoff, \ifthenelse{\boolean{articletitles}}{\emph{{Long
  live the Higgs portal!}},
  }{}\href{https://doi.org/10.1007/JHEP02(2019)140}{JHEP \textbf{02} (2019)
  140}, \href{http://arxiv.org/abs/1812.04628}{{\normalfont\ttfamily
  arXiv:1812.04628}}\relax
\mciteBstWouldAddEndPuncttrue
\mciteSetBstMidEndSepPunct{\mcitedefaultmidpunct}
{\mcitedefaultendpunct}{\mcitedefaultseppunct}\relax
\EndOfBibitem
\bibitem{ArkaniHamed:2006mb}
N.~Arkani-Hamed, A.~Delgado, and G.~F. Giudice,
  \ifthenelse{\boolean{articletitles}}{\emph{{The Well-tempered neutralino}},
  }{}\href{https://doi.org/10.1016/j.nuclphysb.2006.02.010}{Nucl.\ Phys.\ B
  \textbf{741} (2006) 108},
  \href{http://arxiv.org/abs/hep-ph/0601041}{{\normalfont\ttfamily
  arXiv:hep-ph/0601041}}\relax
\mciteBstWouldAddEndPuncttrue
\mciteSetBstMidEndSepPunct{\mcitedefaultmidpunct}
{\mcitedefaultendpunct}{\mcitedefaultseppunct}\relax
\EndOfBibitem
\bibitem{Nagata:2015pra}
N.~Nagata, H.~Otono, and S.~Shirai,
  \ifthenelse{\boolean{articletitles}}{\emph{{Probing Bino-Wino Coannihilation
  at the LHC}}, }{}\href{https://doi.org/10.1007/JHEP10(2015)086}{JHEP
  \textbf{10} (2015) 086},
  \href{http://arxiv.org/abs/1506.08206}{{\normalfont\ttfamily
  arXiv:1506.08206}}\relax
\mciteBstWouldAddEndPuncttrue
\mciteSetBstMidEndSepPunct{\mcitedefaultmidpunct}
{\mcitedefaultendpunct}{\mcitedefaultseppunct}\relax
\EndOfBibitem
\bibitem{Khachatryan:2014mea}
CMS collaboration, V.~Khachatryan {\em et~al.},
  \ifthenelse{\boolean{articletitles}}{\emph{{Search for Displaced
  Supersymmetry in events with an electron and a muon with large impact
  parameters}}, }{}\href{https://doi.org/10.1103/PhysRevLett.114.061801}{Phys.\
  Rev.\ Lett.\  \textbf{114} (2015) 061801},
  \href{http://arxiv.org/abs/1409.4789}{{\normalfont\ttfamily
  arXiv:1409.4789}}\relax
\mciteBstWouldAddEndPuncttrue
\mciteSetBstMidEndSepPunct{\mcitedefaultmidpunct}
{\mcitedefaultendpunct}{\mcitedefaultseppunct}\relax
\EndOfBibitem
\bibitem{Blekman:2020hwr}
F.~Blekman {\em et~al.}, \ifthenelse{\boolean{articletitles}}{\emph{{Soft
  displaced leptons at the LHC}},
  }{}\href{https://doi.org/10.1007/JHEP11(2020)112}{JHEP \textbf{20} (2020)
  112}, \href{http://arxiv.org/abs/2007.03708}{{\normalfont\ttfamily
  arXiv:2007.03708}}\relax
\mciteBstWouldAddEndPuncttrue
\mciteSetBstMidEndSepPunct{\mcitedefaultmidpunct}
{\mcitedefaultendpunct}{\mcitedefaultseppunct}\relax
\EndOfBibitem
\bibitem{Fox:2008kb}
P.~J. Fox and E.~Poppitz,
  \ifthenelse{\boolean{articletitles}}{\emph{{Leptophilic Dark Matter}},
  }{}\href{https://doi.org/10.1103/PhysRevD.79.083528}{Phys.\ Rev.\ D
  \textbf{79} (2009) 083528},
  \href{http://arxiv.org/abs/0811.0399}{{\normalfont\ttfamily
  arXiv:0811.0399}}\relax
\mciteBstWouldAddEndPuncttrue
\mciteSetBstMidEndSepPunct{\mcitedefaultmidpunct}
{\mcitedefaultendpunct}{\mcitedefaultseppunct}\relax
\EndOfBibitem
\bibitem{Freitas:2014jla}
A.~Freitas and S.~Westhoff,
  \ifthenelse{\boolean{articletitles}}{\emph{{Leptophilic Dark Matter in Lepton
  Interactions at LEP and ILC}},
  }{}\href{https://doi.org/10.1007/JHEP10(2014)116}{JHEP \textbf{10} (2014)
  116}, \href{http://arxiv.org/abs/1408.1959}{{\normalfont\ttfamily
  arXiv:1408.1959}}\relax
\mciteBstWouldAddEndPuncttrue
\mciteSetBstMidEndSepPunct{\mcitedefaultmidpunct}
{\mcitedefaultendpunct}{\mcitedefaultseppunct}\relax
\EndOfBibitem
\bibitem{Evans:2016zau}
J.~A. Evans and J.~Shelton,
  \ifthenelse{\boolean{articletitles}}{\emph{{Long-Lived Staus and Displaced
  Leptons at the LHC}}, }{}\href{https://doi.org/10.1007/JHEP04(2016)056}{JHEP
  \textbf{04} (2016) 056},
  \href{http://arxiv.org/abs/1601.01326}{{\normalfont\ttfamily
  arXiv:1601.01326}}\relax
\mciteBstWouldAddEndPuncttrue
\mciteSetBstMidEndSepPunct{\mcitedefaultmidpunct}
{\mcitedefaultendpunct}{\mcitedefaultseppunct}\relax
\EndOfBibitem
\bibitem{Barducci:2018esg}
D.~Barducci {\em et~al.},
  \ifthenelse{\boolean{articletitles}}{\emph{{Characterizing dark matter
  interacting with extra charged leptons}},
  }{}\href{https://doi.org/10.1103/PhysRevD.97.075006}{Phys.\ Rev.\ D
  \textbf{97} (2018) 075006},
  \href{http://arxiv.org/abs/1801.02707}{{\normalfont\ttfamily
  arXiv:1801.02707}}\relax
\mciteBstWouldAddEndPuncttrue
\mciteSetBstMidEndSepPunct{\mcitedefaultmidpunct}
{\mcitedefaultendpunct}{\mcitedefaultseppunct}\relax
\EndOfBibitem
\bibitem{Junius:2019dci}
S.~Junius, L.~Lopez-Honorez, and A.~Mariotti,
  \ifthenelse{\boolean{articletitles}}{\emph{{A feeble window on leptophilic
  dark matter}}, }{}\href{https://doi.org/10.1007/JHEP07(2019)136}{JHEP
  \textbf{07} (2019) 136},
  \href{http://arxiv.org/abs/1904.07513}{{\normalfont\ttfamily
  arXiv:1904.07513}}\relax
\mciteBstWouldAddEndPuncttrue
\mciteSetBstMidEndSepPunct{\mcitedefaultmidpunct}
{\mcitedefaultendpunct}{\mcitedefaultseppunct}\relax
\EndOfBibitem
\bibitem{Cai:2017mow}
Y.~Cai, T.~Han, T.~Li, and R.~Ruiz,
  \ifthenelse{\boolean{articletitles}}{\emph{{Lepton Number Violation: Seesaw
  Models and Their Collider Tests}},
  }{}\href{https://doi.org/10.3389/fphy.2018.00040}{Front.\ in Phys.\
  \textbf{6} (2018) 40},
  \href{http://arxiv.org/abs/1711.02180}{{\normalfont\ttfamily
  arXiv:1711.02180}}\relax
\mciteBstWouldAddEndPuncttrue
\mciteSetBstMidEndSepPunct{\mcitedefaultmidpunct}
{\mcitedefaultendpunct}{\mcitedefaultseppunct}\relax
\EndOfBibitem
\bibitem{Drewes:2019fou}
M.~Drewes and J.~Hajer, \ifthenelse{\boolean{articletitles}}{\emph{{Heavy
  Neutrinos in displaced vertex searches at the LHC and HL-LHC}},
  }{}\href{https://doi.org/10.1007/JHEP02(2020)070}{JHEP \textbf{02} (2020)
  070}, \href{http://arxiv.org/abs/1903.06100}{{\normalfont\ttfamily
  arXiv:1903.06100}}\relax
\mciteBstWouldAddEndPuncttrue
\mciteSetBstMidEndSepPunct{\mcitedefaultmidpunct}
{\mcitedefaultendpunct}{\mcitedefaultseppunct}\relax
\EndOfBibitem
\bibitem{Elor:2018twp}
G.~Elor, M.~Escudero, and A.~Nelson,
  \ifthenelse{\boolean{articletitles}}{\emph{{Baryogenesis and Dark Matter from
  $B$ Mesons}}, }{}\href{https://doi.org/10.1103/PhysRevD.99.035031}{Phys.\
  Rev.\  \textbf{D99} (2019) 035031},
  \href{http://arxiv.org/abs/1810.00880}{{\normalfont\ttfamily
  arXiv:1810.00880}}\relax
\mciteBstWouldAddEndPuncttrue
\mciteSetBstMidEndSepPunct{\mcitedefaultmidpunct}
{\mcitedefaultendpunct}{\mcitedefaultseppunct}\relax
\EndOfBibitem
\bibitem{Elor:2020tkc}
G.~Elor and R.~McGehee, \ifthenelse{\boolean{articletitles}}{\emph{{Making the
  Universe at 20 MeV}},
  }{}\href{http://arxiv.org/abs/2011.06115}{{\normalfont\ttfamily
  arXiv:2011.06115}}\relax
\mciteBstWouldAddEndPuncttrue
\mciteSetBstMidEndSepPunct{\mcitedefaultmidpunct}
{\mcitedefaultendpunct}{\mcitedefaultseppunct}\relax
\EndOfBibitem
\bibitem{Alonso-Alvarez:2020}
G.~Alonso-\'Alvarez, G.~Elor, and M.~Escudero,
  \ifthenelse{\boolean{articletitles}}{\emph{{Collider Signals of Baryogenesis
  and Dark Matter from $B$ Mesons: A Roadmap to Discovery}},
  }{}\href{http://arxiv.org/abs/2101.02706}{{\normalfont\ttfamily
  arXiv:2101.02706}}\relax
\mciteBstWouldAddEndPuncttrue
\mciteSetBstMidEndSepPunct{\mcitedefaultmidpunct}
{\mcitedefaultendpunct}{\mcitedefaultseppunct}\relax
\EndOfBibitem
\bibitem{Alonso-Alvarez:2019fym}
G.~Alonso-\'Alvarez, G.~Elor, A.~E. Nelson, and H.~Xiao,
  \ifthenelse{\boolean{articletitles}}{\emph{{A Supersymmetric Theory of
  Baryogenesis and Sterile Sneutrino Dark Matter from $B$ Mesons}},
  }{}\href{https://doi.org/10.1007/JHEP03(2020)046}{JHEP \textbf{03} (2020)
  046}, \href{http://arxiv.org/abs/1907.10612}{{\normalfont\ttfamily
  arXiv:1907.10612}}\relax
\mciteBstWouldAddEndPuncttrue
\mciteSetBstMidEndSepPunct{\mcitedefaultmidpunct}
{\mcitedefaultendpunct}{\mcitedefaultseppunct}\relax
\EndOfBibitem
\bibitem{Sirunyan:2019ctn}
CMS collaboration, A.~M. Sirunyan {\em et~al.},
  \ifthenelse{\boolean{articletitles}}{\emph{{Search for supersymmetry in
  proton-proton collisions at 13 TeV in final states with jets and missing
  transverse momentum}}, }{}\href{https://doi.org/10.1007/JHEP10(2019)244}{JHEP
  \textbf{10} (2019) 244},
  \href{http://arxiv.org/abs/1908.04722}{{\normalfont\ttfamily
  arXiv:1908.04722}}\relax
\mciteBstWouldAddEndPuncttrue
\mciteSetBstMidEndSepPunct{\mcitedefaultmidpunct}
{\mcitedefaultendpunct}{\mcitedefaultseppunct}\relax
\EndOfBibitem
\bibitem{Aad:2020aze}
ATLAS collaboration, G.~Aad {\em et~al.},
  \ifthenelse{\boolean{articletitles}}{\emph{{Search for squarks and gluinos in
  final states with jets and missing transverse momentum using 139 fb$^{-1}$ of
  $\sqrt{s}$ =13 TeV $pp$ collision data with the ATLAS detector}},
  }{}\href{http://arxiv.org/abs/2010.14293}{{\normalfont\ttfamily
  arXiv:2010.14293}}\relax
\mciteBstWouldAddEndPuncttrue
\mciteSetBstMidEndSepPunct{\mcitedefaultmidpunct}
{\mcitedefaultendpunct}{\mcitedefaultseppunct}\relax
\EndOfBibitem
\bibitem{Aitken:2017wie}
K.~Aitken, D.~McKeen, T.~Neder, and A.~E. Nelson,
  \ifthenelse{\boolean{articletitles}}{\emph{{Baryogenesis from Oscillations of
  Charmed or Beautiful Baryons}},
  }{}\href{https://doi.org/10.1103/PhysRevD.96.075009}{Phys.\ Rev.\
  \textbf{D96} (2017) 075009},
  \href{http://arxiv.org/abs/1708.01259}{{\normalfont\ttfamily
  arXiv:1708.01259}}\relax
\mciteBstWouldAddEndPuncttrue
\mciteSetBstMidEndSepPunct{\mcitedefaultmidpunct}
{\mcitedefaultendpunct}{\mcitedefaultseppunct}\relax
\EndOfBibitem
\bibitem{Nelson:2019fln}
A.~E. Nelson and H.~Xiao,
  \ifthenelse{\boolean{articletitles}}{\emph{{Baryogenesis from B Meson
  Oscillations}}, }{}\href{https://doi.org/10.1103/PhysRevD.100.075002}{Phys.\
  Rev.\  \textbf{D100} (2019) 075002},
  \href{http://arxiv.org/abs/1901.08141}{{\normalfont\ttfamily
  arXiv:1901.08141}}\relax
\mciteBstWouldAddEndPuncttrue
\mciteSetBstMidEndSepPunct{\mcitedefaultmidpunct}
{\mcitedefaultendpunct}{\mcitedefaultseppunct}\relax
\EndOfBibitem
\bibitem{Lenz:2019lvd}
A.~Lenz and G.~Tetlalmatzi-Xolocotzi,
  \ifthenelse{\boolean{articletitles}}{\emph{{Model-independent bounds on new
  physics effects in non-leptonic tree-level decays of B-mesons}},
  }{}\href{https://doi.org/10.1007/JHEP07(2020)177}{JHEP \textbf{07} (2020)
  177}, \href{http://arxiv.org/abs/1912.07621}{{\normalfont\ttfamily
  arXiv:1912.07621}}\relax
\mciteBstWouldAddEndPuncttrue
\mciteSetBstMidEndSepPunct{\mcitedefaultmidpunct}
{\mcitedefaultendpunct}{\mcitedefaultseppunct}\relax
\EndOfBibitem
\bibitem{Cerri:2018ypt}
A.~Cerri {\em et~al.}, \ifthenelse{\boolean{articletitles}}{\emph{{Report from
  Working Group 4}: {Opportunities in Flavour Physics at the HL-LHC and
  HE-LHC}}, }{}\href{https://doi.org/10.23731/CYRM-2019-007.867}{CERN Yellow
  Rep.\ Monogr.\  \textbf{7} (2019) 867},
  \href{http://arxiv.org/abs/1812.07638}{{\normalfont\ttfamily
  arXiv:1812.07638}}\relax
\mciteBstWouldAddEndPuncttrue
\mciteSetBstMidEndSepPunct{\mcitedefaultmidpunct}
{\mcitedefaultendpunct}{\mcitedefaultseppunct}\relax
\EndOfBibitem
\bibitem{Barel:2020jvf}
M.~Z. Barel, K.~De~Bruyn, R.~Fleischer, and E.~Malami,
  \ifthenelse{\boolean{articletitles}}{\emph{{In Pursuit of New Physics with
  $B_d^0\to J/\psi K^0$ and $B_s^0\to J/\psi\phi$ Decays at the High-Precision
  Frontier}}, }{}\href{http://arxiv.org/abs/2010.14423}{{\normalfont\ttfamily
  arXiv:2010.14423}}\relax
\mciteBstWouldAddEndPuncttrue
\mciteSetBstMidEndSepPunct{\mcitedefaultmidpunct}
{\mcitedefaultendpunct}{\mcitedefaultseppunct}\relax
\EndOfBibitem
\bibitem{pdg}
ParticleDataGroup, M.~Tanabashi {\em et~al.},
  \ifthenelse{\boolean{articletitles}}{\emph{{Review of Particle Physics}},
  }{}\href{https://doi.org/10.1103/PhysRevD.98.030001}{Phys.\ Rev.\
  \textbf{D98} (2018) 030001}\relax
\mciteBstWouldAddEndPuncttrue
\mciteSetBstMidEndSepPunct{\mcitedefaultmidpunct}
{\mcitedefaultendpunct}{\mcitedefaultseppunct}\relax
\EndOfBibitem
\bibitem{Lees:2013kla}
BaBar collaboration, J.~P. Lees {\em et~al.},
  \ifthenelse{\boolean{articletitles}}{\emph{{Search for $B \to K^{(*)} \nu
  \overline \nu$ and invisible quarkonium decays}},
  }{}\href{https://doi.org/10.1103/PhysRevD.87.112005}{Phys.\ Rev.\ D
  \textbf{87} (2013) 112005},
  \href{http://arxiv.org/abs/1303.7465}{{\normalfont\ttfamily
  arXiv:1303.7465}}\relax
\mciteBstWouldAddEndPuncttrue
\mciteSetBstMidEndSepPunct{\mcitedefaultmidpunct}
{\mcitedefaultendpunct}{\mcitedefaultseppunct}\relax
\EndOfBibitem
\bibitem{Lutz:2013ftz}
Belle collaboration, O.~Lutz {\em et~al.},
  \ifthenelse{\boolean{articletitles}}{\emph{{Search for $B \to h^{(*)} \nu
  \bar{\nu}$ with the full Belle $\Upsilon(4S)$ data sample}},
  }{}\href{https://doi.org/10.1103/PhysRevD.87.111103}{Phys.\ Rev.\ D
  \textbf{87} (2013) 111103},
  \href{http://arxiv.org/abs/1303.3719}{{\normalfont\ttfamily
  arXiv:1303.3719}}\relax
\mciteBstWouldAddEndPuncttrue
\mciteSetBstMidEndSepPunct{\mcitedefaultmidpunct}
{\mcitedefaultendpunct}{\mcitedefaultseppunct}\relax
\EndOfBibitem
\bibitem{Lees:2019lme}
BaBar collaboration, J.~P. Lees {\em et~al.},
  \ifthenelse{\boolean{articletitles}}{\emph{{Search for $B^- \to \Lambda \bar
  p \nu \bar\nu$ with the BaBar experiment}},
  }{}\href{https://doi.org/10.1103/PhysRevD.100.111101}{Phys.\ Rev.\ D
  \textbf{100} (2019) 111101},
  \href{http://arxiv.org/abs/1908.07425}{{\normalfont\ttfamily
  arXiv:1908.07425}}\relax
\mciteBstWouldAddEndPuncttrue
\mciteSetBstMidEndSepPunct{\mcitedefaultmidpunct}
{\mcitedefaultendpunct}{\mcitedefaultseppunct}\relax
\EndOfBibitem
\bibitem{Brea:2020}
A.~Brea~Rodr\'iguez, X.~Cid~Vidal, C.~V\'azquez~Sierra, and S.~L\'opez
  Soli\~no, { to appear}\relax
\mciteBstWouldAddEndPuncttrue
\mciteSetBstMidEndSepPunct{\mcitedefaultmidpunct}
{\mcitedefaultendpunct}{\mcitedefaultseppunct}\relax
\EndOfBibitem
\bibitem{Poluektov:2019trg}
A.~Poluektov and A.~Morris,
  \ifthenelse{\boolean{articletitles}}{\emph{{Oscillations of $B_s^0$ mesons as
  a probe of decays with unreconstructed particles}},
  }{}\href{https://doi.org/10.1007/JHEP02(2020)163}{JHEP \textbf{02} (2020)
  163}, \href{http://arxiv.org/abs/1911.12729}{{\normalfont\ttfamily
  arXiv:1911.12729}}\relax
\mciteBstWouldAddEndPuncttrue
\mciteSetBstMidEndSepPunct{\mcitedefaultmidpunct}
{\mcitedefaultendpunct}{\mcitedefaultseppunct}\relax
\EndOfBibitem
\bibitem{Stone:2014mza}
S.~Stone and L.~Zhang, \ifthenelse{\boolean{articletitles}}{\emph{{Method of
  Studying $\Lambda_b^0$ decays with one missing particle}},
  }{}\href{https://doi.org/10.1155/2014/931257}{Adv.\ High Energy Phys.\
  \textbf{2014} (2014) 931257},
  \href{http://arxiv.org/abs/1402.4205}{{\normalfont\ttfamily
  arXiv:1402.4205}}\relax
\mciteBstWouldAddEndPuncttrue
\mciteSetBstMidEndSepPunct{\mcitedefaultmidpunct}
{\mcitedefaultendpunct}{\mcitedefaultseppunct}\relax
\EndOfBibitem
\bibitem{Bona:2007vi}
UTfit collaboration, M.~Bona {\em et~al.},
  \ifthenelse{\boolean{articletitles}}{\emph{{Model-independent constraints on
  $\Delta F=2$ operators and the scale of new physics}},
  }{}\href{https://doi.org/10.1088/1126-6708/2008/03/049}{JHEP \textbf{03}
  (2008) 049}, \href{http://arxiv.org/abs/0707.0636}{{\normalfont\ttfamily
  arXiv:0707.0636}}\relax
\mciteBstWouldAddEndPuncttrue
\mciteSetBstMidEndSepPunct{\mcitedefaultmidpunct}
{\mcitedefaultendpunct}{\mcitedefaultseppunct}\relax
\EndOfBibitem
\bibitem{Aaij:2019vot}
LHCb collaboration, R.~Aaij {\em et~al.},
  \ifthenelse{\boolean{articletitles}}{\emph{{Updated measurement of
  time-dependent {\it CP}-violating observables in $B^{0}_{s}\to J/\psi K^+
  K^-$ decays}}, }{}\href{https://doi.org/10.1140/epjc/s10052-019-7159-8}{Eur.\
  Phys.\ J.\  \textbf{C79} (2019) 706},
  \href{http://arxiv.org/abs/1906.08356}{{\normalfont\ttfamily
  arXiv:1906.08356}}\relax
\mciteBstWouldAddEndPuncttrue
\mciteSetBstMidEndSepPunct{\mcitedefaultmidpunct}
{\mcitedefaultendpunct}{\mcitedefaultseppunct}\relax
\EndOfBibitem
\bibitem{Aad:2020jfw}
ATLAS collaboration, G.~Aad {\em et~al.},
  \ifthenelse{\boolean{articletitles}}{\emph{{Measurement of the $CP$-violating
  phase $\phi_s$ in $B^0_s \to J/\psi\phi$ decays in ATLAS at 13 TeV}},
  }{}\href{http://arxiv.org/abs/2001.07115}{{\normalfont\ttfamily
  arXiv:2001.07115}}\relax
\mciteBstWouldAddEndPuncttrue
\mciteSetBstMidEndSepPunct{\mcitedefaultmidpunct}
{\mcitedefaultendpunct}{\mcitedefaultseppunct}\relax
\EndOfBibitem
\bibitem{Sirunyan:2020vke}
CMS collaboration, A.~M. Sirunyan {\em et~al.},
  \ifthenelse{\boolean{articletitles}}{\emph{{Measurement of the $CP$-violating
  phase $\phi_\mathrm{s}$ in the B$^0_\mathrm{s}\to$ J$/\psi\, \phi$(1020) $\to
  \mu^+\mu^-$K$^+$K$^-$ channel in proton-proton collisions at $\sqrt{s} =$ 13
  TeV}}, }{}\href{http://arxiv.org/abs/2007.02434}{{\normalfont\ttfamily
  arXiv:2007.02434}}\relax
\mciteBstWouldAddEndPuncttrue
\mciteSetBstMidEndSepPunct{\mcitedefaultmidpunct}
{\mcitedefaultendpunct}{\mcitedefaultseppunct}\relax
\EndOfBibitem
\bibitem{Gligorov:2017nwh}
V.~V. Gligorov, S.~Knapen, M.~Papucci, and D.~J. Robinson,
  \ifthenelse{\boolean{articletitles}}{\emph{{Searching for Long-lived
  Particles: A Compact Detector for Exotics at LHCb}},
  }{}\href{https://doi.org/10.1103/PhysRevD.97.015023}{Phys.\ Rev.\ D
  \textbf{97} (2018) 015023},
  \href{http://arxiv.org/abs/1708.09395}{{\normalfont\ttfamily
  arXiv:1708.09395}}\relax
\mciteBstWouldAddEndPuncttrue
\mciteSetBstMidEndSepPunct{\mcitedefaultmidpunct}
{\mcitedefaultendpunct}{\mcitedefaultseppunct}\relax
\EndOfBibitem
\bibitem{Minkowski:1977sc}
P.~Minkowski, \ifthenelse{\boolean{articletitles}}{\emph{{$\mu \to e\gamma$ at
  a Rate of One Out of $10^{9}$ Muon Decays?}},
  }{}\href{https://doi.org/10.1016/0370-2693(77)90435-X}{Phys.\ Lett.\ B
  \textbf{67} (1977) 421}\relax
\mciteBstWouldAddEndPuncttrue
\mciteSetBstMidEndSepPunct{\mcitedefaultmidpunct}
{\mcitedefaultendpunct}{\mcitedefaultseppunct}\relax
\EndOfBibitem
\bibitem{Mohapatra:1979ia}
R.~N. Mohapatra and G.~Senjanovic,
  \ifthenelse{\boolean{articletitles}}{\emph{{Neutrino Mass and Spontaneous
  Parity Nonconservation}},
  }{}\href{https://doi.org/10.1103/PhysRevLett.44.912}{Phys.\ Rev.\ Lett.\
  \textbf{44} (1980) 912}\relax
\mciteBstWouldAddEndPuncttrue
\mciteSetBstMidEndSepPunct{\mcitedefaultmidpunct}
{\mcitedefaultendpunct}{\mcitedefaultseppunct}\relax
\EndOfBibitem
\bibitem{Asaka:2005pn}
T.~Asaka and M.~Shaposhnikov, \ifthenelse{\boolean{articletitles}}{\emph{{The
  $\nu$MSM, dark matter and baryon asymmetry of the universe}},
  }{}\href{https://doi.org/10.1016/j.physletb.2005.06.020}{Phys.\ Lett.\ B
  \textbf{620} (2005) 17},
  \href{http://arxiv.org/abs/hep-ph/0505013}{{\normalfont\ttfamily
  arXiv:hep-ph/0505013}}\relax
\mciteBstWouldAddEndPuncttrue
\mciteSetBstMidEndSepPunct{\mcitedefaultmidpunct}
{\mcitedefaultendpunct}{\mcitedefaultseppunct}\relax
\EndOfBibitem
\bibitem{Shaposhnikov:2006nn}
M.~Shaposhnikov, \ifthenelse{\boolean{articletitles}}{\emph{{A Possible
  symmetry of the nuMSM}},
  }{}\href{https://doi.org/10.1016/j.nuclphysb.2006.11.003}{Nucl.\ Phys.\ B
  \textbf{763} (2007) 49},
  \href{http://arxiv.org/abs/hep-ph/0605047}{{\normalfont\ttfamily
  arXiv:hep-ph/0605047}}\relax
\mciteBstWouldAddEndPuncttrue
\mciteSetBstMidEndSepPunct{\mcitedefaultmidpunct}
{\mcitedefaultendpunct}{\mcitedefaultseppunct}\relax
\EndOfBibitem
\bibitem{Drewes:2018gkc}
M.~Drewes, J.~Hajer, J.~Klaric, and G.~Lanfranchi,
  \ifthenelse{\boolean{articletitles}}{\emph{{NA62 sensitivity to heavy neutral
  leptons in the low scale seesaw model}},
  }{}\href{https://doi.org/10.1007/JHEP07(2018)105}{JHEP \textbf{07} (2018)
  105}, \href{http://arxiv.org/abs/1801.04207}{{\normalfont\ttfamily
  arXiv:1801.04207}}\relax
\mciteBstWouldAddEndPuncttrue
\mciteSetBstMidEndSepPunct{\mcitedefaultmidpunct}
{\mcitedefaultendpunct}{\mcitedefaultseppunct}\relax
\EndOfBibitem
\bibitem{ATL-DAQ-PUB-2017-001}
ATLAS collaboration,  \ifthenelse{\boolean{articletitles}}{\emph{{Trigger Menu
  in 2016}}}{},
  \href{https://cds.cern.ch/record/2242069}{ATL-DAQ-PUB-2017-001}, CERN,
  Geneva, 2017\relax
\mciteBstWouldAddEndPuncttrue
\mciteSetBstMidEndSepPunct{\mcitedefaultmidpunct}
{\mcitedefaultendpunct}{\mcitedefaultseppunct}\relax
\EndOfBibitem
\bibitem{Bobrovskyi:2011vx}
S.~Bobrovskyi, W.~Buchmuller, J.~Hajer, and J.~Schmidt,
  \ifthenelse{\boolean{articletitles}}{\emph{{Quasi-stable neutralinos at the
  LHC}}, }{}\href{https://doi.org/10.1007/JHEP09(2011)119}{JHEP \textbf{09}
  (2011) 119}, \href{http://arxiv.org/abs/1107.0926}{{\normalfont\ttfamily
  arXiv:1107.0926}}\relax
\mciteBstWouldAddEndPuncttrue
\mciteSetBstMidEndSepPunct{\mcitedefaultmidpunct}
{\mcitedefaultendpunct}{\mcitedefaultseppunct}\relax
\EndOfBibitem
\bibitem{Bobrovskyi:2012dc}
S.~Bobrovskyi, J.~Hajer, and S.~Rydbeck,
  \ifthenelse{\boolean{articletitles}}{\emph{{Long-lived higgsinos as probes of
  gravitino dark matter at the LHC}},
  }{}\href{https://doi.org/10.1007/JHEP02(2013)133}{JHEP \textbf{02} (2013)
  133}, \href{http://arxiv.org/abs/1211.5584}{{\normalfont\ttfamily
  arXiv:1211.5584}}\relax
\mciteBstWouldAddEndPuncttrue
\mciteSetBstMidEndSepPunct{\mcitedefaultmidpunct}
{\mcitedefaultendpunct}{\mcitedefaultseppunct}\relax
\EndOfBibitem
\bibitem{Antusch:2017hhu}
S.~Antusch, E.~Cazzato, and O.~Fischer,
  \ifthenelse{\boolean{articletitles}}{\emph{{Sterile Neutrino Searches via
  Displaced Vertices at Lhcb}},
  }{}\href{https://doi.org/10.1016/j.physletb.2017.09.057}{Phys.\ Lett.\
  \textbf{B774} (2017) 114},
  \href{http://arxiv.org/abs/1706.05990}{{\normalfont\ttfamily
  arXiv:1706.05990}}\relax
\mciteBstWouldAddEndPuncttrue
\mciteSetBstMidEndSepPunct{\mcitedefaultmidpunct}
{\mcitedefaultendpunct}{\mcitedefaultseppunct}\relax
\EndOfBibitem
\bibitem{Boiarska:2019jcw}
I.~Boiarska {\em et~al.}, \ifthenelse{\boolean{articletitles}}{\emph{{Probing
  Baryon Asymmetry of the Universe at Lhc and Ship}},
  }{}\href{http://arxiv.org/abs/1902.04535}{{\normalfont\ttfamily
  arXiv:1902.04535}}\relax
\mciteBstWouldAddEndPuncttrue
\mciteSetBstMidEndSepPunct{\mcitedefaultmidpunct}
{\mcitedefaultendpunct}{\mcitedefaultseppunct}\relax
\EndOfBibitem
\bibitem{Gago:2015vma}
A.~M. Gago {\em et~al.}, \ifthenelse{\boolean{articletitles}}{\emph{{Probing
  the Type I Seesaw Mechanism with Displaced Vertices at the LHC}},
  }{}\href{https://doi.org/10.1140/epjc/s10052-015-3693-1}{Eur.\ Phys.\ J.\
  \textbf{C75} (2015) 470},
  \href{http://arxiv.org/abs/1505.05880}{{\normalfont\ttfamily
  arXiv:1505.05880}}\relax
\mciteBstWouldAddEndPuncttrue
\mciteSetBstMidEndSepPunct{\mcitedefaultmidpunct}
{\mcitedefaultendpunct}{\mcitedefaultseppunct}\relax
\EndOfBibitem
\bibitem{Degrande:2016aje}
C.~Degrande, O.~Mattelaer, R.~Ruiz, and J.~Turner,
  \ifthenelse{\boolean{articletitles}}{\emph{{Fully-Automated Precision
  Predictions for Heavy Neutrino Production Mechanisms at Hadron Colliders}},
  }{}\href{https://doi.org/10.1103/PhysRevD.94.053002}{Phys.\ Rev.\
  \textbf{D94} (2016) 053002},
  \href{http://arxiv.org/abs/1602.06957}{{\normalfont\ttfamily
  arXiv:1602.06957}}\relax
\mciteBstWouldAddEndPuncttrue
\mciteSetBstMidEndSepPunct{\mcitedefaultmidpunct}
{\mcitedefaultendpunct}{\mcitedefaultseppunct}\relax
\EndOfBibitem
\bibitem{Das:2017rsu}
A.~Das, Y.~Gao, and T.~Kamon, \ifthenelse{\boolean{articletitles}}{\emph{{Heavy
  Neutrino Search via Semileptonic Higgs Decay at the Lhc}},
  }{}\href{https://doi.org/10.1140/epjc/s10052-019-6937-7}{Eur.\ Phys.\ J.\
  \textbf{C79} (2019) 424},
  \href{http://arxiv.org/abs/1704.00881}{{\normalfont\ttfamily
  arXiv:1704.00881}}\relax
\mciteBstWouldAddEndPuncttrue
\mciteSetBstMidEndSepPunct{\mcitedefaultmidpunct}
{\mcitedefaultendpunct}{\mcitedefaultseppunct}\relax
\EndOfBibitem
\bibitem{Ruiz:2017yyf}
R.~Ruiz, M.~Spannowsky, and P.~Waite,
  \ifthenelse{\boolean{articletitles}}{\emph{{Heavy Neutrinos from Gluon
  Fusion}}, }{}\href{https://doi.org/10.1103/PhysRevD.96.055042}{Phys.\ Rev.\
  \textbf{D96} (2017) 055042},
  \href{http://arxiv.org/abs/1706.02298}{{\normalfont\ttfamily
  arXiv:1706.02298}}\relax
\mciteBstWouldAddEndPuncttrue
\mciteSetBstMidEndSepPunct{\mcitedefaultmidpunct}
{\mcitedefaultendpunct}{\mcitedefaultseppunct}\relax
\EndOfBibitem
\bibitem{Abada:2018sfh}
A.~Abada, N.~Bernal, M.~Losada, and X.~Marcano,
  \ifthenelse{\boolean{articletitles}}{\emph{{Inclusive Displaced Vertex
  Searches for Heavy Neutral Leptons at the LHC}},
  }{}\href{https://doi.org/10.1007/JHEP01(2019)093}{JHEP \textbf{01} (2019)
  093}, \href{http://arxiv.org/abs/1807.10024}{{\normalfont\ttfamily
  arXiv:1807.10024}}\relax
\mciteBstWouldAddEndPuncttrue
\mciteSetBstMidEndSepPunct{\mcitedefaultmidpunct}
{\mcitedefaultendpunct}{\mcitedefaultseppunct}\relax
\EndOfBibitem
\bibitem{Das:2017zjc}
A.~Das, P.~S.~B. Dev, and C.~S. Kim,
  \ifthenelse{\boolean{articletitles}}{\emph{{Constraining Sterile Neutrinos
  from Precision Higgs Data}},
  }{}\href{https://doi.org/10.1103/PhysRevD.95.115013}{Phys.\ Rev.\
  \textbf{D95} (2017) 115013},
  \href{http://arxiv.org/abs/1704.00880}{{\normalfont\ttfamily
  arXiv:1704.00880}}\relax
\mciteBstWouldAddEndPuncttrue
\mciteSetBstMidEndSepPunct{\mcitedefaultmidpunct}
{\mcitedefaultendpunct}{\mcitedefaultseppunct}\relax
\EndOfBibitem
\bibitem{Das:2016hof}
A.~Das, P.~Konar, and S.~Majhi,
  \ifthenelse{\boolean{articletitles}}{\emph{{Production of Heavy Neutrino in
  Next-To-Leading Order QCD at the Lhc and Beyond}},
  }{}\href{https://doi.org/10.1007/JHEP06(2016)019}{JHEP \textbf{06} (2016)
  019}, \href{http://arxiv.org/abs/1604.00608}{{\normalfont\ttfamily
  arXiv:1604.00608}}\relax
\mciteBstWouldAddEndPuncttrue
\mciteSetBstMidEndSepPunct{\mcitedefaultmidpunct}
{\mcitedefaultendpunct}{\mcitedefaultseppunct}\relax
\EndOfBibitem
\bibitem{Cottin:2018nms}
G.~Cottin, J.~C. Helo, and M.~Hirsch,
  \ifthenelse{\boolean{articletitles}}{\emph{{Displaced Vertices as Probes of
  Sterile Neutrino Mixing at the Lhc}},
  }{}\href{https://doi.org/10.1103/PhysRevD.98.035012}{Phys.\ Rev.\
  \textbf{D98} (2018) 035012},
  \href{http://arxiv.org/abs/1806.05191}{{\normalfont\ttfamily
  arXiv:1806.05191}}\relax
\mciteBstWouldAddEndPuncttrue
\mciteSetBstMidEndSepPunct{\mcitedefaultmidpunct}
{\mcitedefaultendpunct}{\mcitedefaultseppunct}\relax
\EndOfBibitem
\bibitem{Aad:2016naf}
ATLAS collaboration, G.~Aad {\em et~al.},
  \ifthenelse{\boolean{articletitles}}{\emph{{Measurement of $W^{\pm}$ and
  $Z$-boson production cross sections in $pp$ collisions at $\sqrt{s}=13$ TeV
  with the ATLAS detector}},
  }{}\href{https://doi.org/10.1016/j.physletb.2016.06.023}{Phys.\ Lett.\
  \textbf{B759} (2016) 601},
  \href{http://arxiv.org/abs/1603.09222}{{\normalfont\ttfamily
  arXiv:1603.09222}}\relax
\mciteBstWouldAddEndPuncttrue
\mciteSetBstMidEndSepPunct{\mcitedefaultmidpunct}
{\mcitedefaultendpunct}{\mcitedefaultseppunct}\relax
\EndOfBibitem
\bibitem{LHCb-bbbar}
LHCb collaboration, \ifthenelse{\boolean{articletitles}}{\emph{{$\bar{b}b$
  production angle plots}}, }{}
  \url{https://lhcb.web.cern.ch/lhcb/speakersbureau/html/bb_ProductionAngles.html}\relax
\mciteBstWouldAddEndPuncttrue
\mciteSetBstMidEndSepPunct{\mcitedefaultmidpunct}
{\mcitedefaultendpunct}{\mcitedefaultseppunct}\relax
\EndOfBibitem
\bibitem{Cacciari:1998it}
M.~Cacciari, M.~Greco, and P.~Nason,
  \ifthenelse{\boolean{articletitles}}{\emph{{The P(T) Spectrum in Heavy Flavor
  Hadroproduction}},
  }{}\href{https://doi.org/10.1088/1126-6708/1998/05/007}{JHEP \textbf{05}
  (1998) 007}, \href{http://arxiv.org/abs/hep-ph/9803400}{{\normalfont\ttfamily
  arXiv:hep-ph/9803400}}\relax
\mciteBstWouldAddEndPuncttrue
\mciteSetBstMidEndSepPunct{\mcitedefaultmidpunct}
{\mcitedefaultendpunct}{\mcitedefaultseppunct}\relax
\EndOfBibitem
\bibitem{Bondarenko:2018ptm}
K.~Bondarenko, A.~Boyarsky, D.~Gorbunov, and O.~Ruchayskiy,
  \ifthenelse{\boolean{articletitles}}{\emph{{Phenomenology of Gev-Scale Heavy
  Neutral Leptons}}, }{}\href{https://doi.org/10.1007/JHEP11(2018)032}{JHEP
  \textbf{11} (2018) 032},
  \href{http://arxiv.org/abs/1805.08567}{{\normalfont\ttfamily
  arXiv:1805.08567}}\relax
\mciteBstWouldAddEndPuncttrue
\mciteSetBstMidEndSepPunct{\mcitedefaultmidpunct}
{\mcitedefaultendpunct}{\mcitedefaultseppunct}\relax
\EndOfBibitem
\bibitem{SHiP:2018xqw}
SHiP collaboration, C.~Ahdida {\em et~al.},
  \ifthenelse{\boolean{articletitles}}{\emph{{Sensitivity of the Ship
  Experiment to Heavy Neutral Leptons}},
  }{}\href{https://doi.org/10.1007/JHEP04(2019)077}{JHEP \textbf{04} (2019)
  077}, \href{http://arxiv.org/abs/1811.00930}{{\normalfont\ttfamily
  arXiv:1811.00930}}\relax
\mciteBstWouldAddEndPuncttrue
\mciteSetBstMidEndSepPunct{\mcitedefaultmidpunct}
{\mcitedefaultendpunct}{\mcitedefaultseppunct}\relax
\EndOfBibitem
\bibitem{Alekhin:2015byh}
S.~Alekhin {\em et~al.}, \ifthenelse{\boolean{articletitles}}{\emph{{A Facility
  to Search for Hidden Particles at the Cern Sps: the Ship Physics Case}},
  }{}\href{https://doi.org/10.1088/0034-4885/79/12/124201}{Rept.\ Prog.\ Phys.\
   \textbf{79} (2016) 124201},
  \href{http://arxiv.org/abs/1504.04855}{{\normalfont\ttfamily
  arXiv:1504.04855}}\relax
\mciteBstWouldAddEndPuncttrue
\mciteSetBstMidEndSepPunct{\mcitedefaultmidpunct}
{\mcitedefaultendpunct}{\mcitedefaultseppunct}\relax
\EndOfBibitem
\bibitem{ArkaniHamed:2008qn}
N.~Arkani-Hamed, D.~P. Finkbeiner, T.~R. Slatyer, and N.~Weiner,
  \ifthenelse{\boolean{articletitles}}{\emph{{A Theory of Dark Matter}},
  }{}\href{https://doi.org/10.1103/PhysRevD.79.015014}{Phys.\ Rev.\ D
  \textbf{79} (2009) 015014},
  \href{http://arxiv.org/abs/0810.0713}{{\normalfont\ttfamily
  arXiv:0810.0713}}\relax
\mciteBstWouldAddEndPuncttrue
\mciteSetBstMidEndSepPunct{\mcitedefaultmidpunct}
{\mcitedefaultendpunct}{\mcitedefaultseppunct}\relax
\EndOfBibitem
\bibitem{Cheung:2009qd}
C.~Cheung, J.~T. Ruderman, L.-T. Wang, and I.~Yavin,
  \ifthenelse{\boolean{articletitles}}{\emph{{Kinetic Mixing as the Origin of
  Light Dark Scales}},
  }{}\href{https://doi.org/10.1103/PhysRevD.80.035008}{Phys.\ Rev.\ D
  \textbf{80} (2009) 035008},
  \href{http://arxiv.org/abs/0902.3246}{{\normalfont\ttfamily
  arXiv:0902.3246}}\relax
\mciteBstWouldAddEndPuncttrue
\mciteSetBstMidEndSepPunct{\mcitedefaultmidpunct}
{\mcitedefaultendpunct}{\mcitedefaultseppunct}\relax
\EndOfBibitem
\bibitem{Pospelov:2008zw}
M.~Pospelov, \ifthenelse{\boolean{articletitles}}{\emph{{Secluded U(1) below
  the weak scale}}, }{}\href{https://doi.org/10.1103/PhysRevD.80.095002}{Phys.\
  Rev.\ D \textbf{80} (2009) 095002},
  \href{http://arxiv.org/abs/0811.1030}{{\normalfont\ttfamily
  arXiv:0811.1030}}\relax
\mciteBstWouldAddEndPuncttrue
\mciteSetBstMidEndSepPunct{\mcitedefaultmidpunct}
{\mcitedefaultendpunct}{\mcitedefaultseppunct}\relax
\EndOfBibitem
\bibitem{Okun:1982xi}
L.~B. Okun, \ifthenelse{\boolean{articletitles}}{\emph{{Limits of
  electrodynamics: paraphotons?}}, }{}Sov.\ Phys.\ JETP \textbf{56} (1982)
  502\relax
\mciteBstWouldAddEndPuncttrue
\mciteSetBstMidEndSepPunct{\mcitedefaultmidpunct}
{\mcitedefaultendpunct}{\mcitedefaultseppunct}\relax
\EndOfBibitem
\bibitem{Holdom:1985ag}
B.~Holdom, \ifthenelse{\boolean{articletitles}}{\emph{{Two U(1)'s and Epsilon
  Charge Shifts}},
  }{}\href{https://doi.org/10.1016/0370-2693(86)91377-8}{Phys.\ Lett.\ B
  \textbf{166} (1986) 196}\relax
\mciteBstWouldAddEndPuncttrue
\mciteSetBstMidEndSepPunct{\mcitedefaultmidpunct}
{\mcitedefaultendpunct}{\mcitedefaultseppunct}\relax
\EndOfBibitem
\bibitem{Ilten:2015hya}
P.~Ilten, J.~Thaler, M.~Williams, and W.~Xue,
  \ifthenelse{\boolean{articletitles}}{\emph{{Dark photons from charm mesons at
  LHCb}}, }{}\href{https://doi.org/10.1103/PhysRevD.92.115017}{Phys.\ Rev.\ D
  \textbf{92} (2015) 115017},
  \href{http://arxiv.org/abs/1509.06765}{{\normalfont\ttfamily
  arXiv:1509.06765}}\relax
\mciteBstWouldAddEndPuncttrue
\mciteSetBstMidEndSepPunct{\mcitedefaultmidpunct}
{\mcitedefaultendpunct}{\mcitedefaultseppunct}\relax
\EndOfBibitem
\bibitem{Bauer:2018onh}
M.~Bauer, P.~Foldenauer, and J.~Jaeckel,
  \ifthenelse{\boolean{articletitles}}{\emph{{Hunting All the Hidden Photons}},
  }{}\href{https://doi.org/10.1007/JHEP07(2018)094}{JHEP \textbf{18} (2020)
  094}, \href{http://arxiv.org/abs/1803.05466}{{\normalfont\ttfamily
  arXiv:1803.05466}}\relax
\mciteBstWouldAddEndPuncttrue
\mciteSetBstMidEndSepPunct{\mcitedefaultmidpunct}
{\mcitedefaultendpunct}{\mcitedefaultseppunct}\relax
\EndOfBibitem
\bibitem{Bauer:2020itv}
M.~Bauer, P.~Foldenauer, and M.~Mosny,
  \ifthenelse{\boolean{articletitles}}{\emph{{On the Flavour Structure of
  Anomaly-free Hidden Photon Models}},
  }{}\href{http://arxiv.org/abs/2011.12973}{{\normalfont\ttfamily
  arXiv:2011.12973}}\relax
\mciteBstWouldAddEndPuncttrue
\mciteSetBstMidEndSepPunct{\mcitedefaultmidpunct}
{\mcitedefaultendpunct}{\mcitedefaultseppunct}\relax
\EndOfBibitem
\bibitem{Sirunyan:2018nnz}
CMS collaboration, A.~M. Sirunyan {\em et~al.},
  \ifthenelse{\boolean{articletitles}}{\emph{{Search for an $L_{\mu}-L_{\tau}$
  gauge boson using Z$\to4\mu$ events in proton-proton collisions at $\sqrt{s}
  =$ 13 TeV}}, }{}\href{https://doi.org/10.1016/j.physletb.2019.01.072}{Phys.\
  Lett.\ B \textbf{792} (2019) 345},
  \href{http://arxiv.org/abs/1808.03684}{{\normalfont\ttfamily
  arXiv:1808.03684}}\relax
\mciteBstWouldAddEndPuncttrue
\mciteSetBstMidEndSepPunct{\mcitedefaultmidpunct}
{\mcitedefaultendpunct}{\mcitedefaultseppunct}\relax
\EndOfBibitem
\bibitem{Foldenauer:2019dai}
P.~Foldenauer, \ifthenelse{\boolean{articletitles}}{\emph{{Dark Sectors from
  the Hidden Photon Perspective}},
  }{}\href{https://doi.org/10.22323/1.350.0179}{PoS \textbf{LHCP2019} (2019)
  179}, \href{http://arxiv.org/abs/1907.10630}{{\normalfont\ttfamily
  arXiv:1907.10630}}\relax
\mciteBstWouldAddEndPuncttrue
\mciteSetBstMidEndSepPunct{\mcitedefaultmidpunct}
{\mcitedefaultendpunct}{\mcitedefaultseppunct}\relax
\EndOfBibitem
\bibitem{Lees:2014xha}
BaBar collaboration, J.~P. Lees {\em et~al.},
  \ifthenelse{\boolean{articletitles}}{\emph{{Search for a Dark Photon in
  $e^+e^-$ Collisions at BaBar}},
  }{}\href{https://doi.org/10.1103/PhysRevLett.113.201801}{Phys.\ Rev.\ Lett.\
  \textbf{113} (2014) 201801},
  \href{http://arxiv.org/abs/1406.2980}{{\normalfont\ttfamily
  arXiv:1406.2980}}\relax
\mciteBstWouldAddEndPuncttrue
\mciteSetBstMidEndSepPunct{\mcitedefaultmidpunct}
{\mcitedefaultendpunct}{\mcitedefaultseppunct}\relax
\EndOfBibitem
\bibitem{Lees:2017lec}
BaBar collaboration, J.~P. Lees {\em et~al.},
  \ifthenelse{\boolean{articletitles}}{\emph{{Search for Invisible Decays of a
  Dark Photon Produced in ${e}^{+}{e}^{-}$ Collisions at BaBar}},
  }{}\href{https://doi.org/10.1103/PhysRevLett.119.131804}{Phys.\ Rev.\ Lett.\
  \textbf{119} (2017) 131804},
  \href{http://arxiv.org/abs/1702.03327}{{\normalfont\ttfamily
  arXiv:1702.03327}}\relax
\mciteBstWouldAddEndPuncttrue
\mciteSetBstMidEndSepPunct{\mcitedefaultmidpunct}
{\mcitedefaultendpunct}{\mcitedefaultseppunct}\relax
\EndOfBibitem
\bibitem{Coloma:2020gfv}
P.~Coloma, M.~C. Gonzalez-Garcia, and M.~Maltoni,
  \ifthenelse{\boolean{articletitles}}{\emph{{Neutrino Oscillation Constraints
  on U(1)' Models: from Non-Standard Interactions to Long-Range Forces}},
  }{}\href{http://arxiv.org/abs/2009.14220}{{\normalfont\ttfamily
  arXiv:2009.14220}}\relax
\mciteBstWouldAddEndPuncttrue
\mciteSetBstMidEndSepPunct{\mcitedefaultmidpunct}
{\mcitedefaultendpunct}{\mcitedefaultseppunct}\relax
\EndOfBibitem
\bibitem{Alexander:2018png}
M.~Alexander {\em et~al.}, \ifthenelse{\boolean{articletitles}}{\emph{{Mapping
  the material in the LHCb vertex locator using secondary hadronic
  interactions}},
  }{}\href{https://doi.org/10.1088/1748-0221/13/06/P06008}{JINST \textbf{13}
  (2018) P06008}, \href{http://arxiv.org/abs/1803.07466}{{\normalfont\ttfamily
  arXiv:1803.07466}}\relax
\mciteBstWouldAddEndPuncttrue
\mciteSetBstMidEndSepPunct{\mcitedefaultmidpunct}
{\mcitedefaultendpunct}{\mcitedefaultseppunct}\relax
\EndOfBibitem
\bibitem{Kors:2004dx}
B.~Kors and P.~Nath, \ifthenelse{\boolean{articletitles}}{\emph{{A Stueckelberg
  extension of the standard model}},
  }{}\href{https://doi.org/10.1016/j.physletb.2004.02.051}{Phys.\ Lett.\ B
  \textbf{586} (2004) 366},
  \href{http://arxiv.org/abs/hep-ph/0402047}{{\normalfont\ttfamily
  arXiv:hep-ph/0402047}}\relax
\mciteBstWouldAddEndPuncttrue
\mciteSetBstMidEndSepPunct{\mcitedefaultmidpunct}
{\mcitedefaultendpunct}{\mcitedefaultseppunct}\relax
\EndOfBibitem
\bibitem{Feldman:2006ce}
D.~Feldman, Z.~Liu, and P.~Nath,
  \ifthenelse{\boolean{articletitles}}{\emph{{Probing a very narrow Z-prime
  boson with CDF and D0 data}},
  }{}\href{https://doi.org/10.1103/PhysRevLett.97.021801}{Phys.\ Rev.\ Lett.\
  \textbf{97} (2006) 021801},
  \href{http://arxiv.org/abs/hep-ph/0603039}{{\normalfont\ttfamily
  arXiv:hep-ph/0603039}}\relax
\mciteBstWouldAddEndPuncttrue
\mciteSetBstMidEndSepPunct{\mcitedefaultmidpunct}
{\mcitedefaultendpunct}{\mcitedefaultseppunct}\relax
\EndOfBibitem
\bibitem{Feldman:2007wj}
D.~Feldman, Z.~Liu, and P.~Nath,
  \ifthenelse{\boolean{articletitles}}{\emph{{The Stueckelberg Z-prime
  Extension with Kinetic Mixing and Milli-Charged Dark Matter From the Hidden
  Sector}}, }{}\href{https://doi.org/10.1103/PhysRevD.75.115001}{Phys.\ Rev.\ D
  \textbf{75} (2007) 115001},
  \href{http://arxiv.org/abs/hep-ph/0702123}{{\normalfont\ttfamily
  arXiv:hep-ph/0702123}}\relax
\mciteBstWouldAddEndPuncttrue
\mciteSetBstMidEndSepPunct{\mcitedefaultmidpunct}
{\mcitedefaultendpunct}{\mcitedefaultseppunct}\relax
\EndOfBibitem
\bibitem{Du:2019mlc}
M.~Du, Z.~Liu, and V.~Q. Tran,
  \ifthenelse{\boolean{articletitles}}{\emph{{Enhanced Long-Lived Dark Photon
  Signals at the LHC}}, }{}\href{https://doi.org/10.1007/JHEP05(2020)055}{JHEP
  \textbf{05} (2020) 055},
  \href{http://arxiv.org/abs/1912.00422}{{\normalfont\ttfamily
  arXiv:1912.00422}}\relax
\mciteBstWouldAddEndPuncttrue
\mciteSetBstMidEndSepPunct{\mcitedefaultmidpunct}
{\mcitedefaultendpunct}{\mcitedefaultseppunct}\relax
\EndOfBibitem
\bibitem{Alwall:2014hca}
J.~Alwall {\em et~al.}, \ifthenelse{\boolean{articletitles}}{\emph{{The
  automated computation of tree-level and next-to-leading order differential
  cross sections, and their matching to parton shower simulations}},
  }{}\href{https://doi.org/10.1007/JHEP07(2014)079}{JHEP \textbf{07} (2014)
  079}, \href{http://arxiv.org/abs/1405.0301}{{\normalfont\ttfamily
  arXiv:1405.0301}}\relax
\mciteBstWouldAddEndPuncttrue
\mciteSetBstMidEndSepPunct{\mcitedefaultmidpunct}
{\mcitedefaultendpunct}{\mcitedefaultseppunct}\relax
\EndOfBibitem
\bibitem{Carloni:2010tw}
L.~Carloni and T.~Sjostrand,
  \ifthenelse{\boolean{articletitles}}{\emph{{Visible Effects of Invisible
  Hidden Valley Radiation}},
  }{}\href{https://doi.org/10.1007/JHEP09(2010)105}{JHEP \textbf{09} (2010)
  105}, \href{http://arxiv.org/abs/1006.2911}{{\normalfont\ttfamily
  arXiv:1006.2911}}\relax
\mciteBstWouldAddEndPuncttrue
\mciteSetBstMidEndSepPunct{\mcitedefaultmidpunct}
{\mcitedefaultendpunct}{\mcitedefaultseppunct}\relax
\EndOfBibitem
\bibitem{Carloni:2011kk}
L.~Carloni, J.~Rathsman, and T.~Sjostrand,
  \ifthenelse{\boolean{articletitles}}{\emph{{Discerning Secluded Sector gauge
  structures}}, }{}\href{https://doi.org/10.1007/JHEP04(2011)091}{JHEP
  \textbf{04} (2011) 091},
  \href{http://arxiv.org/abs/1102.3795}{{\normalfont\ttfamily
  arXiv:1102.3795}}\relax
\mciteBstWouldAddEndPuncttrue
\mciteSetBstMidEndSepPunct{\mcitedefaultmidpunct}
{\mcitedefaultendpunct}{\mcitedefaultseppunct}\relax
\EndOfBibitem
\bibitem{CMStiming}
CMS collaboration, \ifthenelse{\boolean{articletitles}}{\emph{{Technical
  proposal for a MIP timing detector in the CMS experiment Phase 2 upgrade}},
  }{}Tech.\ Rep.\ CERN-LHCC-2017-027.\ LHCC-P-009, CERN, Geneva (2017)\relax
\mciteBstWouldAddEndPuncttrue
\mciteSetBstMidEndSepPunct{\mcitedefaultmidpunct}
{\mcitedefaultendpunct}{\mcitedefaultseppunct}\relax
\EndOfBibitem
\bibitem{Liu:2018wte}
J.~Liu, Z.~Liu, and L.-T. Wang,
  \ifthenelse{\boolean{articletitles}}{\emph{{Enhancing Long-Lived Particles
  Searches at the LHC with Precision Timing Information}},
  }{}\href{https://doi.org/10.1103/PhysRevLett.122.131801}{Phys.\ Rev.\ Lett.\
  \textbf{122} (2019) 131801},
  \href{http://arxiv.org/abs/1805.05957}{{\normalfont\ttfamily
  arXiv:1805.05957}}\relax
\mciteBstWouldAddEndPuncttrue
\mciteSetBstMidEndSepPunct{\mcitedefaultmidpunct}
{\mcitedefaultendpunct}{\mcitedefaultseppunct}\relax
\EndOfBibitem
\bibitem{Allaire:2018bof}
C.~Allaire {\em et~al.}, \ifthenelse{\boolean{articletitles}}{\emph{{Beam test
  measurements of Low Gain Avalanche Detector single pads and arrays for the
  ATLAS High Granularity Timing Detector}},
  }{}\href{https://doi.org/10.1088/1748-0221/13/06/P06017}{JINST \textbf{13}
  (2018) P06017}, \href{http://arxiv.org/abs/1804.00622}{{\normalfont\ttfamily
  arXiv:1804.00622}}\relax
\mciteBstWouldAddEndPuncttrue
\mciteSetBstMidEndSepPunct{\mcitedefaultmidpunct}
{\mcitedefaultendpunct}{\mcitedefaultseppunct}\relax
\EndOfBibitem
\bibitem{Allaire:2019ioj}
ATLAS LAr-HGTD Group, C.~Allaire, \ifthenelse{\boolean{articletitles}}{\emph{{A
  High-Granularity Timing Detector in ATLAS: Performance at the HL-LHC}},
  }{}\href{https://doi.org/10.1016/j.nima.2018.05.028}{Nucl.\ Instrum.\ Meth.\
  A \textbf{924} (2019) 355}\relax
\mciteBstWouldAddEndPuncttrue
\mciteSetBstMidEndSepPunct{\mcitedefaultmidpunct}
{\mcitedefaultendpunct}{\mcitedefaultseppunct}\relax
\EndOfBibitem
\bibitem{Beacham:2019nyx}
J.~Beacham {\em et~al.}, \ifthenelse{\boolean{articletitles}}{\emph{{Physics
  Beyond Colliders at CERN: Beyond the Standard Model Working Group Report}},
  }{}\href{https://doi.org/10.1088/1361-6471/ab4cd2}{J.\ Phys.\ G \textbf{47}
  (2020) 010501}, \href{http://arxiv.org/abs/1901.09966}{{\normalfont\ttfamily
  arXiv:1901.09966}}\relax
\mciteBstWouldAddEndPuncttrue
\mciteSetBstMidEndSepPunct{\mcitedefaultmidpunct}
{\mcitedefaultendpunct}{\mcitedefaultseppunct}\relax
\EndOfBibitem
\bibitem{Pospelov:2007mp}
M.~Pospelov, A.~Ritz, and M.~B. Voloshin,
  \ifthenelse{\boolean{articletitles}}{\emph{{Secluded WIMP Dark Matter}},
  }{}\href{https://doi.org/10.1016/j.physletb.2008.02.052}{Phys.\ Lett.\
  \textbf{B662} (2008) 53},
  \href{http://arxiv.org/abs/0711.4866}{{\normalfont\ttfamily
  arXiv:0711.4866}}\relax
\mciteBstWouldAddEndPuncttrue
\mciteSetBstMidEndSepPunct{\mcitedefaultmidpunct}
{\mcitedefaultendpunct}{\mcitedefaultseppunct}\relax
\EndOfBibitem
\bibitem{Krnjaic:2015mbs}
G.~Krnjaic, \ifthenelse{\boolean{articletitles}}{\emph{{Probing Light Thermal
  Dark-Matter With a Higgs Portal Mediator}},
  }{}\href{https://doi.org/10.1103/PhysRevD.94.073009}{Phys.\ Rev.\
  \textbf{D94} (2016) 073009},
  \href{http://arxiv.org/abs/1512.04119}{{\normalfont\ttfamily
  arXiv:1512.04119}}\relax
\mciteBstWouldAddEndPuncttrue
\mciteSetBstMidEndSepPunct{\mcitedefaultmidpunct}
{\mcitedefaultendpunct}{\mcitedefaultseppunct}\relax
\EndOfBibitem
\bibitem{Dev:2017dui}
P.~S.~B. Dev, R.~N. Mohapatra, and Y.~Zhang,
  \ifthenelse{\boolean{articletitles}}{\emph{{Long Lived Light Scalars as Probe
  of Low Scale Seesaw Models}},
  }{}\href{https://doi.org/10.1016/j.nuclphysb.2017.07.021}{Nucl.\ Phys.\
  \textbf{B923} (2017) 179},
  \href{http://arxiv.org/abs/1703.02471}{{\normalfont\ttfamily
  arXiv:1703.02471}}\relax
\mciteBstWouldAddEndPuncttrue
\mciteSetBstMidEndSepPunct{\mcitedefaultmidpunct}
{\mcitedefaultendpunct}{\mcitedefaultseppunct}\relax
\EndOfBibitem
\bibitem{Fayet:1974pd}
P.~Fayet, \ifthenelse{\boolean{articletitles}}{\emph{{Supergauge Invariant
  Extension of the Higgs Mechanism and a Model for the electron and Its
  Neutrino}}, }{}\href{https://doi.org/10.1016/0550-3213(75)90636-7}{Nucl.\
  Phys.\  \textbf{B90} (1975) 104}\relax
\mciteBstWouldAddEndPuncttrue
\mciteSetBstMidEndSepPunct{\mcitedefaultmidpunct}
{\mcitedefaultendpunct}{\mcitedefaultseppunct}\relax
\EndOfBibitem
\bibitem{Shaposhnikov:2006xi}
M.~Shaposhnikov and I.~Tkachev, \ifthenelse{\boolean{articletitles}}{\emph{{The
  nuMSM, inflation, and dark matter}},
  }{}\href{https://doi.org/10.1016/j.physletb.2006.06.063}{Phys.\ Lett.\
  \textbf{B639} (2006) 414},
  \href{http://arxiv.org/abs/hep-ph/0604236}{{\normalfont\ttfamily
  arXiv:hep-ph/0604236}}\relax
\mciteBstWouldAddEndPuncttrue
\mciteSetBstMidEndSepPunct{\mcitedefaultmidpunct}
{\mcitedefaultendpunct}{\mcitedefaultseppunct}\relax
\EndOfBibitem
\bibitem{Graham:2015cka}
P.~W. Graham, D.~E. Kaplan, and S.~Rajendran,
  \ifthenelse{\boolean{articletitles}}{\emph{{Cosmological Relaxation of the
  Electroweak Scale}},
  }{}\href{https://doi.org/10.1103/PhysRevLett.115.221801}{Phys.\ Rev.\ Lett.\
  \textbf{115} (2015) 221801},
  \href{http://arxiv.org/abs/1504.07551}{{\normalfont\ttfamily
  arXiv:1504.07551}}\relax
\mciteBstWouldAddEndPuncttrue
\mciteSetBstMidEndSepPunct{\mcitedefaultmidpunct}
{\mcitedefaultendpunct}{\mcitedefaultseppunct}\relax
\EndOfBibitem
\bibitem{Curtin:2013fra}
D.~Curtin {\em et~al.}, \ifthenelse{\boolean{articletitles}}{\emph{{Exotic
  decays of the 125 GeV Higgs boson}},
  }{}\href{https://doi.org/10.1103/PhysRevD.90.075004}{Phys.\ Rev.\ D
  \textbf{90} (2014) 075004},
  \href{http://arxiv.org/abs/1312.4992}{{\normalfont\ttfamily
  arXiv:1312.4992}}\relax
\mciteBstWouldAddEndPuncttrue
\mciteSetBstMidEndSepPunct{\mcitedefaultmidpunct}
{\mcitedefaultendpunct}{\mcitedefaultseppunct}\relax
\EndOfBibitem
\bibitem{Aaboud:2018aqj}
ATLAS collaboration, M.~Aaboud {\em et~al.},
  \ifthenelse{\boolean{articletitles}}{\emph{{Search for long-lived particles
  produced in $pp$ collisions at $\sqrt{s}=13$ TeV that decay into displaced
  hadronic jets in the ATLAS muon spectrometer}},
  }{}\href{https://doi.org/10.1103/PhysRevD.99.052005}{Phys.\ Rev.\ D
  \textbf{99} (2019) 052005},
  \href{http://arxiv.org/abs/1811.07370}{{\normalfont\ttfamily
  arXiv:1811.07370}}\relax
\mciteBstWouldAddEndPuncttrue
\mciteSetBstMidEndSepPunct{\mcitedefaultmidpunct}
{\mcitedefaultendpunct}{\mcitedefaultseppunct}\relax
\EndOfBibitem
\bibitem{Aaboud:2019opc}
ATLAS collaboration, M.~Aaboud {\em et~al.},
  \ifthenelse{\boolean{articletitles}}{\emph{{Search for long-lived neutral
  particles in $pp$ collisions at $\sqrt{s}$ = 13 TeV that decay into displaced
  hadronic jets in the ATLAS calorimeter}},
  }{}\href{https://doi.org/10.1140/epjc/s10052-019-6962-6}{Eur.\ Phys.\ J.\ C
  \textbf{79} (2019) 481},
  \href{http://arxiv.org/abs/1902.03094}{{\normalfont\ttfamily
  arXiv:1902.03094}}\relax
\mciteBstWouldAddEndPuncttrue
\mciteSetBstMidEndSepPunct{\mcitedefaultmidpunct}
{\mcitedefaultendpunct}{\mcitedefaultseppunct}\relax
\EndOfBibitem
\bibitem{Sirunyan:2018vlw}
CMS collaboration, A.~M. Sirunyan {\em et~al.},
  \ifthenelse{\boolean{articletitles}}{\emph{{Search for long-lived particles
  decaying into displaced jets in proton-proton collisions at $\sqrt{s}=$ 13
  TeV}}, }{}\href{https://doi.org/10.1103/PhysRevD.99.032011}{Phys.\ Rev.\ D
  \textbf{99} (2019) 032011},
  \href{http://arxiv.org/abs/1811.07991}{{\normalfont\ttfamily
  arXiv:1811.07991}}\relax
\mciteBstWouldAddEndPuncttrue
\mciteSetBstMidEndSepPunct{\mcitedefaultmidpunct}
{\mcitedefaultendpunct}{\mcitedefaultseppunct}\relax
\EndOfBibitem
\bibitem{Sirunyan:2018owy}
CMS collaboration, A.~M. Sirunyan {\em et~al.},
  \ifthenelse{\boolean{articletitles}}{\emph{{Search for invisible decays of a
  Higgs boson produced through vector boson fusion in proton-proton collisions
  at $\sqrt{s} =$ 13 TeV}},
  }{}\href{https://doi.org/10.1016/j.physletb.2019.04.025}{Phys.\ Lett.\ B
  \textbf{793} (2019) 520},
  \href{http://arxiv.org/abs/1809.05937}{{\normalfont\ttfamily
  arXiv:1809.05937}}\relax
\mciteBstWouldAddEndPuncttrue
\mciteSetBstMidEndSepPunct{\mcitedefaultmidpunct}
{\mcitedefaultendpunct}{\mcitedefaultseppunct}\relax
\EndOfBibitem
\bibitem{Aaboud:2019rtt}
ATLAS collaboration, M.~Aaboud {\em et~al.},
  \ifthenelse{\boolean{articletitles}}{\emph{{Combination of searches for
  invisible Higgs boson decays with the ATLAS experiment}},
  }{}\href{https://doi.org/10.1103/PhysRevLett.122.231801}{Phys.\ Rev.\ Lett.\
  \textbf{122} (2019) 231801},
  \href{http://arxiv.org/abs/1904.05105}{{\normalfont\ttfamily
  arXiv:1904.05105}}\relax
\mciteBstWouldAddEndPuncttrue
\mciteSetBstMidEndSepPunct{\mcitedefaultmidpunct}
{\mcitedefaultendpunct}{\mcitedefaultseppunct}\relax
\EndOfBibitem
\bibitem{ATLAS:2020cjb}
ATLAS collaboration, \ifthenelse{\boolean{articletitles}}{\emph{{Search for
  invisible Higgs boson decays with vector boson fusion signatures with the
  ATLAS detector using an integrated luminosity of 139 fb$^{-1}$}}, }{}\relax
\mciteBstWouldAddEndPuncttrue
\mciteSetBstMidEndSepPunct{\mcitedefaultmidpunct}
{\mcitedefaultendpunct}{\mcitedefaultseppunct}\relax
\EndOfBibitem
\bibitem{Anastasiou:2016cez}
C.~Anastasiou {\em et~al.}, \ifthenelse{\boolean{articletitles}}{\emph{{High
  precision determination of the gluon fusion Higgs boson cross-section at the
  LHC}}, }{}\href{https://doi.org/10.1007/JHEP05(2016)058}{JHEP \textbf{05}
  (2016) 058}, \href{http://arxiv.org/abs/1602.00695}{{\normalfont\ttfamily
  arXiv:1602.00695}}\relax
\mciteBstWouldAddEndPuncttrue
\mciteSetBstMidEndSepPunct{\mcitedefaultmidpunct}
{\mcitedefaultendpunct}{\mcitedefaultseppunct}\relax
\EndOfBibitem
\bibitem{Curtin:2018mvb}
D.~Curtin {\em et~al.}, \ifthenelse{\boolean{articletitles}}{\emph{{Long-Lived
  Particles at the Energy Frontier: The MATHUSLA Physics Case}},
  }{}\href{https://doi.org/10.1088/1361-6633/ab28d6}{Rept.\ Prog.\ Phys.\
  \textbf{82} (2019) 116201},
  \href{http://arxiv.org/abs/1806.07396}{{\normalfont\ttfamily
  arXiv:1806.07396}}\relax
\mciteBstWouldAddEndPuncttrue
\mciteSetBstMidEndSepPunct{\mcitedefaultmidpunct}
{\mcitedefaultendpunct}{\mcitedefaultseppunct}\relax
\EndOfBibitem
\bibitem{Feng:2017uoz}
J.~L. Feng, I.~Galon, F.~Kling, and S.~Trojanowski,
  \ifthenelse{\boolean{articletitles}}{\emph{{ForwArd Search ExpeRiment at the
  LHC}}, }{}\href{https://doi.org/10.1103/PhysRevD.97.035001}{Phys.\ Rev.\ D
  \textbf{97} (2018) 035001},
  \href{http://arxiv.org/abs/1708.09389}{{\normalfont\ttfamily
  arXiv:1708.09389}}\relax
\mciteBstWouldAddEndPuncttrue
\mciteSetBstMidEndSepPunct{\mcitedefaultmidpunct}
{\mcitedefaultendpunct}{\mcitedefaultseppunct}\relax
\EndOfBibitem
\bibitem{Sjostrand:2007gs}
T.~Sj\"{o}strand, S.~Mrenna, and P.~Skands,
  \ifthenelse{\boolean{articletitles}}{\emph{{A brief introduction to PYTHIA
  8.1}}, }{}\href{https://doi.org/10.1016/j.cpc.2008.01.036}{Comput.\ Phys.\
  Commun.\  \textbf{178} (2008) 852},
  \href{http://arxiv.org/abs/0710.3820}{{\normalfont\ttfamily
  arXiv:0710.3820}}\relax
\mciteBstWouldAddEndPuncttrue
\mciteSetBstMidEndSepPunct{\mcitedefaultmidpunct}
{\mcitedefaultendpunct}{\mcitedefaultseppunct}\relax
\EndOfBibitem
\bibitem{Batell:2018fqo}
B.~Batell, A.~Freitas, A.~Ismail, and D.~Mckeen,
  \ifthenelse{\boolean{articletitles}}{\emph{{Probing Light Dark Matter with a
  Hadrophilic Scalar Mediator}},
  }{}\href{https://doi.org/10.1103/PhysRevD.100.095020}{Phys.\ Rev.\ D
  \textbf{100} (2019) 095020},
  \href{http://arxiv.org/abs/1812.05103}{{\normalfont\ttfamily
  arXiv:1812.05103}}\relax
\mciteBstWouldAddEndPuncttrue
\mciteSetBstMidEndSepPunct{\mcitedefaultmidpunct}
{\mcitedefaultendpunct}{\mcitedefaultseppunct}\relax
\EndOfBibitem
\bibitem{Dobrescu:2000yn}
B.~A. Dobrescu and K.~T. Matchev,
  \ifthenelse{\boolean{articletitles}}{\emph{{Light axion within the
  next-to-minimal supersymmetric standard model}},
  }{}\href{https://doi.org/10.1088/1126-6708/2000/09/031}{JHEP \textbf{09}
  (2000) 031}, \href{http://arxiv.org/abs/hep-ph/0008192}{{\normalfont\ttfamily
  arXiv:hep-ph/0008192}}\relax
\mciteBstWouldAddEndPuncttrue
\mciteSetBstMidEndSepPunct{\mcitedefaultmidpunct}
{\mcitedefaultendpunct}{\mcitedefaultseppunct}\relax
\EndOfBibitem
\bibitem{Haisch:2018kqx}
U.~Haisch, J.~F. Kamenik, A.~Malinauskas, and M.~Spira,
  \ifthenelse{\boolean{articletitles}}{\emph{{Collider constraints on light
  pseudoscalars}}, }{}\href{https://doi.org/10.1007/JHEP03(2018)178}{JHEP
  \textbf{03} (2018) 178},
  \href{http://arxiv.org/abs/1802.02156}{{\normalfont\ttfamily
  arXiv:1802.02156}}\relax
\mciteBstWouldAddEndPuncttrue
\mciteSetBstMidEndSepPunct{\mcitedefaultmidpunct}
{\mcitedefaultendpunct}{\mcitedefaultseppunct}\relax
\EndOfBibitem
\bibitem{Gunion:1989we}
J.~F. Gunion, H.~E. Haber, G.~L. Kane, and S.~Dawson, {\em {The Higgs Hunter's
  Guide}}, vol.~80, 2000\relax
\mciteBstWouldAddEndPuncttrue
\mciteSetBstMidEndSepPunct{\mcitedefaultmidpunct}
{\mcitedefaultendpunct}{\mcitedefaultseppunct}\relax
\EndOfBibitem
\bibitem{Branco:2011iw}
G.~C. Branco {\em et~al.}, \ifthenelse{\boolean{articletitles}}{\emph{{Theory
  and phenomenology of two-Higgs-doublet models}},
  }{}\href{https://doi.org/10.1016/j.physrep.2012.02.002}{Phys.\ Rept.\
  \textbf{516} (2012) 1},
  \href{http://arxiv.org/abs/1106.0034}{{\normalfont\ttfamily
  arXiv:1106.0034}}\relax
\mciteBstWouldAddEndPuncttrue
\mciteSetBstMidEndSepPunct{\mcitedefaultmidpunct}
{\mcitedefaultendpunct}{\mcitedefaultseppunct}\relax
\EndOfBibitem
\bibitem{Aaij:2015awa}
LHCb collaboration, R.~Aaij {\em et~al.},
  \ifthenelse{\boolean{articletitles}}{\emph{{Forward production of $\Upsilon$
  mesons in $pp$ collisions at $\sqrt{s}=7$ and 8TeV}},
  }{}\href{https://doi.org/10.1007/JHEP11(2015)103}{JHEP \textbf{11} (2015)
  103}, \href{http://arxiv.org/abs/1509.02372}{{\normalfont\ttfamily
  arXiv:1509.02372}}\relax
\mciteBstWouldAddEndPuncttrue
\mciteSetBstMidEndSepPunct{\mcitedefaultmidpunct}
{\mcitedefaultendpunct}{\mcitedefaultseppunct}\relax
\EndOfBibitem
\bibitem{Chatrchyan:2012am}
CMS collaboration, S.~Chatrchyan {\em et~al.},
  \ifthenelse{\boolean{articletitles}}{\emph{{Search for a Light Pseudoscalar
  Higgs Boson in the Dimuon Decay Channel in $pp$ Collisions at $\sqrt{s}=7$
  TeV}}, }{}\href{https://doi.org/10.1103/PhysRevLett.109.121801}{Phys.\ Rev.\
  Lett.\  \textbf{109} (2012) 121801},
  \href{http://arxiv.org/abs/1206.6326}{{\normalfont\ttfamily
  arXiv:1206.6326}}\relax
\mciteBstWouldAddEndPuncttrue
\mciteSetBstMidEndSepPunct{\mcitedefaultmidpunct}
{\mcitedefaultendpunct}{\mcitedefaultseppunct}\relax
\EndOfBibitem
\bibitem{Lees:2012iw}
BaBar collaboration, J.~P. Lees {\em et~al.},
  \ifthenelse{\boolean{articletitles}}{\emph{{Search for di-muon decays of a
  low-mass Higgs boson in radiative decays of the \ensuremath{\Upsilon}(1S)}},
  }{}\href{https://doi.org/10.1103/PhysRevD.87.031102}{Phys.\ Rev.\ D
  \textbf{87} (2013) 031102},
  \href{http://arxiv.org/abs/1210.0287}{{\normalfont\ttfamily
  arXiv:1210.0287}}, [Erratum: Phys.Rev.D 87, 059903 (2013)]\relax
\mciteBstWouldAddEndPuncttrue
\mciteSetBstMidEndSepPunct{\mcitedefaultmidpunct}
{\mcitedefaultendpunct}{\mcitedefaultseppunct}\relax
\EndOfBibitem
\bibitem{Lees:2012te}
BaBar collaboration, J.~P. Lees {\em et~al.},
  \ifthenelse{\boolean{articletitles}}{\emph{{Search for a low-mass scalar
  Higgs boson decaying to a tau pair in single-photon decays of
  $\Upsilon(1S)$}}, }{}\href{https://doi.org/10.1103/PhysRevD.88.071102}{Phys.\
  Rev.\ D \textbf{88} (2013) 071102},
  \href{http://arxiv.org/abs/1210.5669}{{\normalfont\ttfamily
  arXiv:1210.5669}}\relax
\mciteBstWouldAddEndPuncttrue
\mciteSetBstMidEndSepPunct{\mcitedefaultmidpunct}
{\mcitedefaultendpunct}{\mcitedefaultseppunct}\relax
\EndOfBibitem
\bibitem{Khachatryan:2015baw}
CMS collaboration, V.~Khachatryan {\em et~al.},
  \ifthenelse{\boolean{articletitles}}{\emph{{Search for a Low-Mass
  Pseudoscalar Higgs Boson Produced in Association with a $b\bar{b}$ Pair in
  $pp$ Collisions at $\sqrt{s} =$ 8 TeV}},
  }{}\href{https://doi.org/10.1016/j.physletb.2016.05.003}{Phys.\ Lett.\ B
  \textbf{758} (2016) 296},
  \href{http://arxiv.org/abs/1511.03610}{{\normalfont\ttfamily
  arXiv:1511.03610}}\relax
\mciteBstWouldAddEndPuncttrue
\mciteSetBstMidEndSepPunct{\mcitedefaultmidpunct}
{\mcitedefaultendpunct}{\mcitedefaultseppunct}\relax
\EndOfBibitem
\bibitem{CMS:2017yta}
CMS collaboration, \ifthenelse{\boolean{articletitles}}{\emph{{Search for new
  resonances in the diphoton final state in the mass range between 70 and 110
  GeV in pp collisions at $\sqrt{s}=$ 8 and 13 TeV}}, }{}\relax
\mciteBstWouldAddEndPuncttrue
\mciteSetBstMidEndSepPunct{\mcitedefaultmidpunct}
{\mcitedefaultendpunct}{\mcitedefaultseppunct}\relax
\EndOfBibitem
\bibitem{CMS:2016rjp}
CMS collaboration, \ifthenelse{\boolean{articletitles}}{\emph{{Search for a
  neutral MSSM Higgs boson decaying into $\tau\tau$ with
  $12.9~\mathrm{fb}^{-1}$ of data at $\sqrt{s}=13~\mathrm{TeV}$}}, }{}\relax
\mciteBstWouldAddEndPuncttrue
\mciteSetBstMidEndSepPunct{\mcitedefaultmidpunct}
{\mcitedefaultendpunct}{\mcitedefaultseppunct}\relax
\EndOfBibitem
\bibitem{Drees:1989du}
M.~Drees and K.-i. Hikasa, \ifthenelse{\boolean{articletitles}}{\emph{{Heavy
  Quark Thresholds in Higgs Physics}},
  }{}\href{https://doi.org/10.1103/PhysRevD.41.1547}{Phys.\ Rev.\ D \textbf{41}
  (1990) 1547}\relax
\mciteBstWouldAddEndPuncttrue
\mciteSetBstMidEndSepPunct{\mcitedefaultmidpunct}
{\mcitedefaultendpunct}{\mcitedefaultseppunct}\relax
\EndOfBibitem
\bibitem{Baumgart:2012pj}
M.~Baumgart and A.~Katz,
  \ifthenelse{\boolean{articletitles}}{\emph{{Implications of a New Light
  Scalar Near the Bottomonium Regime}},
  }{}\href{https://doi.org/10.1007/JHEP08(2012)133}{JHEP \textbf{08} (2012)
  133}, \href{http://arxiv.org/abs/1204.6032}{{\normalfont\ttfamily
  arXiv:1204.6032}}\relax
\mciteBstWouldAddEndPuncttrue
\mciteSetBstMidEndSepPunct{\mcitedefaultmidpunct}
{\mcitedefaultendpunct}{\mcitedefaultseppunct}\relax
\EndOfBibitem
\bibitem{Peccei:1977hh}
R.~D. Peccei and H.~R. Quinn, \ifthenelse{\boolean{articletitles}}{\emph{{CP
  Conservation in the Presence of Instantons}},
  }{}\href{https://doi.org/10.1103/PhysRevLett.38.1440}{Phys.\ Rev.\ Lett.\
  \textbf{38} (1977) 1440}\relax
\mciteBstWouldAddEndPuncttrue
\mciteSetBstMidEndSepPunct{\mcitedefaultmidpunct}
{\mcitedefaultendpunct}{\mcitedefaultseppunct}\relax
\EndOfBibitem
\bibitem{Peccei:1977ur}
R.~D. Peccei and H.~R. Quinn,
  \ifthenelse{\boolean{articletitles}}{\emph{{Constraints Imposed by CP
  Conservation in the Presence of Instantons}},
  }{}\href{https://doi.org/10.1103/PhysRevD.16.1791}{Phys.\ Rev.\ D \textbf{16}
  (1977) 1791}\relax
\mciteBstWouldAddEndPuncttrue
\mciteSetBstMidEndSepPunct{\mcitedefaultmidpunct}
{\mcitedefaultendpunct}{\mcitedefaultseppunct}\relax
\EndOfBibitem
\bibitem{Weinberg:1977ma}
S.~Weinberg, \ifthenelse{\boolean{articletitles}}{\emph{{A New Light Boson?}},
  }{}\href{https://doi.org/10.1103/PhysRevLett.40.223}{Phys.\ Rev.\ Lett.\
  \textbf{40} (1978) 223}\relax
\mciteBstWouldAddEndPuncttrue
\mciteSetBstMidEndSepPunct{\mcitedefaultmidpunct}
{\mcitedefaultendpunct}{\mcitedefaultseppunct}\relax
\EndOfBibitem
\bibitem{Wilczek:1977pj}
F.~Wilczek, \ifthenelse{\boolean{articletitles}}{\emph{{Problem of Strong $P$
  and $T$ Invariance in the Presence of Instantons}},
  }{}\href{https://doi.org/10.1103/PhysRevLett.40.279}{Phys.\ Rev.\ Lett.\
  \textbf{40} (1978) 279}\relax
\mciteBstWouldAddEndPuncttrue
\mciteSetBstMidEndSepPunct{\mcitedefaultmidpunct}
{\mcitedefaultendpunct}{\mcitedefaultseppunct}\relax
\EndOfBibitem
\bibitem{Zhitnitsky:1980tq}
A.~R. Zhitnitsky, \ifthenelse{\boolean{articletitles}}{\emph{{On Possible
  Suppression of the Axion Hadron Interactions. (In Russian)}}, }{}Sov.\ J.\
  Nucl.\ Phys.\  \textbf{31} (1980) 260\relax
\mciteBstWouldAddEndPuncttrue
\mciteSetBstMidEndSepPunct{\mcitedefaultmidpunct}
{\mcitedefaultendpunct}{\mcitedefaultseppunct}\relax
\EndOfBibitem
\bibitem{Dine:1981rt}
M.~Dine, W.~Fischler, and M.~Srednicki,
  \ifthenelse{\boolean{articletitles}}{\emph{{A Simple Solution to the Strong
  CP Problem with a Harmless Axion}},
  }{}\href{https://doi.org/10.1016/0370-2693(81)90590-6}{Phys.\ Lett.\ B
  \textbf{104} (1981) 199}\relax
\mciteBstWouldAddEndPuncttrue
\mciteSetBstMidEndSepPunct{\mcitedefaultmidpunct}
{\mcitedefaultendpunct}{\mcitedefaultseppunct}\relax
\EndOfBibitem
\bibitem{Gorghetto:2018ocs}
M.~Gorghetto and G.~Villadoro,
  \ifthenelse{\boolean{articletitles}}{\emph{{Topological Susceptibility and
  QCD Axion Mass: QED and NNLO corrections}},
  }{}\href{https://doi.org/10.1007/JHEP03(2019)033}{JHEP \textbf{03} (2019)
  033}, \href{http://arxiv.org/abs/1812.01008}{{\normalfont\ttfamily
  arXiv:1812.01008}}\relax
\mciteBstWouldAddEndPuncttrue
\mciteSetBstMidEndSepPunct{\mcitedefaultmidpunct}
{\mcitedefaultendpunct}{\mcitedefaultseppunct}\relax
\EndOfBibitem
\bibitem{Georgi:1986df}
H.~Georgi, D.~B. Kaplan, and L.~Randall,
  \ifthenelse{\boolean{articletitles}}{\emph{{Manifesting the Invisible Axion
  at Low-energies}},
  }{}\href{https://doi.org/10.1016/0370-2693(86)90688-X}{Phys.\ Lett.\ B
  \textbf{169} (1986) 73}\relax
\mciteBstWouldAddEndPuncttrue
\mciteSetBstMidEndSepPunct{\mcitedefaultmidpunct}
{\mcitedefaultendpunct}{\mcitedefaultseppunct}\relax
\EndOfBibitem
\bibitem{Calibbi:2020jvd}
L.~Calibbi, D.~Redigolo, R.~Ziegler, and J.~Zupan,
  \ifthenelse{\boolean{articletitles}}{\emph{{Looking forward to
  Lepton-flavor-violating ALPs}},
  }{}\href{http://arxiv.org/abs/2006.04795}{{\normalfont\ttfamily
  arXiv:2006.04795}}\relax
\mciteBstWouldAddEndPuncttrue
\mciteSetBstMidEndSepPunct{\mcitedefaultmidpunct}
{\mcitedefaultendpunct}{\mcitedefaultseppunct}\relax
\EndOfBibitem
\bibitem{Abbott:1982af}
L.~F. Abbott and P.~Sikivie, \ifthenelse{\boolean{articletitles}}{\emph{{A
  Cosmological Bound on the Invisible Axion}},
  }{}\href{https://doi.org/10.1016/0370-2693(83)90638-X}{Phys.\ Lett.\ B
  \textbf{120} (1983) 133}\relax
\mciteBstWouldAddEndPuncttrue
\mciteSetBstMidEndSepPunct{\mcitedefaultmidpunct}
{\mcitedefaultendpunct}{\mcitedefaultseppunct}\relax
\EndOfBibitem
\bibitem{Preskill:1982cy}
J.~Preskill, M.~B. Wise, and F.~Wilczek,
  \ifthenelse{\boolean{articletitles}}{\emph{{Cosmology of the Invisible
  Axion}}, }{}\href{https://doi.org/10.1016/0370-2693(83)90637-8}{Phys.\ Lett.\
  B \textbf{120} (1983) 127}\relax
\mciteBstWouldAddEndPuncttrue
\mciteSetBstMidEndSepPunct{\mcitedefaultmidpunct}
{\mcitedefaultendpunct}{\mcitedefaultseppunct}\relax
\EndOfBibitem
\bibitem{Dine:1982ah}
M.~Dine and W.~Fischler, \ifthenelse{\boolean{articletitles}}{\emph{{The Not So
  Harmless Axion}},
  }{}\href{https://doi.org/10.1016/0370-2693(83)90639-1}{Phys.\ Lett.\ B
  \textbf{120} (1983) 137}\relax
\mciteBstWouldAddEndPuncttrue
\mciteSetBstMidEndSepPunct{\mcitedefaultmidpunct}
{\mcitedefaultendpunct}{\mcitedefaultseppunct}\relax
\EndOfBibitem
\bibitem{Nomura:2008ru}
Y.~Nomura and J.~Thaler, \ifthenelse{\boolean{articletitles}}{\emph{{Dark
  Matter through the Axion Portal}},
  }{}\href{https://doi.org/10.1103/PhysRevD.79.075008}{Phys.\ Rev.\ D
  \textbf{79} (2009) 075008},
  \href{http://arxiv.org/abs/0810.5397}{{\normalfont\ttfamily
  arXiv:0810.5397}}\relax
\mciteBstWouldAddEndPuncttrue
\mciteSetBstMidEndSepPunct{\mcitedefaultmidpunct}
{\mcitedefaultendpunct}{\mcitedefaultseppunct}\relax
\EndOfBibitem
\bibitem{Freytsis:2010ne}
M.~Freytsis and Z.~Ligeti, \ifthenelse{\boolean{articletitles}}{\emph{{On dark
  matter models with uniquely spin-dependent detection possibilities}},
  }{}\href{https://doi.org/10.1103/PhysRevD.83.115009}{Phys.\ Rev.\ D
  \textbf{83} (2011) 115009},
  \href{http://arxiv.org/abs/1012.5317}{{\normalfont\ttfamily
  arXiv:1012.5317}}\relax
\mciteBstWouldAddEndPuncttrue
\mciteSetBstMidEndSepPunct{\mcitedefaultmidpunct}
{\mcitedefaultendpunct}{\mcitedefaultseppunct}\relax
\EndOfBibitem
\bibitem{Dolan:2014ska}
M.~J. Dolan, F.~Kahlhoefer, C.~McCabe, and K.~Schmidt-Hoberg,
  \ifthenelse{\boolean{articletitles}}{\emph{{A taste of dark matter: Flavour
  constraints on pseudoscalar mediators}},
  }{}\href{https://doi.org/10.1007/JHEP03(2015)171}{JHEP \textbf{03} (2015)
  171}, \href{http://arxiv.org/abs/1412.5174}{{\normalfont\ttfamily
  arXiv:1412.5174}}, [Erratum: JHEP 07, 103 (2015)]\relax
\mciteBstWouldAddEndPuncttrue
\mciteSetBstMidEndSepPunct{\mcitedefaultmidpunct}
{\mcitedefaultendpunct}{\mcitedefaultseppunct}\relax
\EndOfBibitem
\bibitem{Hochberg:2018rjs}
Y.~Hochberg {\em et~al.}, \ifthenelse{\boolean{articletitles}}{\emph{{Strongly
  interacting massive particles through the axion portal}},
  }{}\href{https://doi.org/10.1103/PhysRevD.98.115031}{Phys.\ Rev.\ D
  \textbf{98} (2018) 115031},
  \href{http://arxiv.org/abs/1806.10139}{{\normalfont\ttfamily
  arXiv:1806.10139}}\relax
\mciteBstWouldAddEndPuncttrue
\mciteSetBstMidEndSepPunct{\mcitedefaultmidpunct}
{\mcitedefaultendpunct}{\mcitedefaultseppunct}\relax
\EndOfBibitem
\bibitem{Aloni:2018vki}
D.~Aloni, Y.~Soreq, and M.~Williams,
  \ifthenelse{\boolean{articletitles}}{\emph{{Coupling QCD-Scale Axionlike
  Particles to Gluons}},
  }{}\href{https://doi.org/10.1103/PhysRevLett.123.031803}{Phys.\ Rev.\ Lett.\
  \textbf{123} (2019) 031803},
  \href{http://arxiv.org/abs/1811.03474}{{\normalfont\ttfamily
  arXiv:1811.03474}}\relax
\mciteBstWouldAddEndPuncttrue
\mciteSetBstMidEndSepPunct{\mcitedefaultmidpunct}
{\mcitedefaultendpunct}{\mcitedefaultseppunct}\relax
\EndOfBibitem
\bibitem{Marciano:2016yhf}
W.~J. Marciano, A.~Masiero, P.~Paradisi, and M.~Passera,
  \ifthenelse{\boolean{articletitles}}{\emph{{Contributions of axionlike
  particles to lepton dipole moments}},
  }{}\href{https://doi.org/10.1103/PhysRevD.94.115033}{Phys.\ Rev.\
  \textbf{D94} (2016) 115033},
  \href{http://arxiv.org/abs/1607.01022}{{\normalfont\ttfamily
  arXiv:1607.01022}}\relax
\mciteBstWouldAddEndPuncttrue
\mciteSetBstMidEndSepPunct{\mcitedefaultmidpunct}
{\mcitedefaultendpunct}{\mcitedefaultseppunct}\relax
\EndOfBibitem
\bibitem{Jaeckel:2015jla}
J.~Jaeckel and M.~Spannowsky,
  \ifthenelse{\boolean{articletitles}}{\emph{{Probing MeV to 90 GeV axion-like
  particles with LEP and LHC}},
  }{}\href{https://doi.org/10.1016/j.physletb.2015.12.037}{Phys.\ Lett.\
  \textbf{B753} (2016) 482},
  \href{http://arxiv.org/abs/1509.00476}{{\normalfont\ttfamily
  arXiv:1509.00476}}\relax
\mciteBstWouldAddEndPuncttrue
\mciteSetBstMidEndSepPunct{\mcitedefaultmidpunct}
{\mcitedefaultendpunct}{\mcitedefaultseppunct}\relax
\EndOfBibitem
\bibitem{Dobrich:2015jyk}
B.~Döbrich {\em et~al.}, \ifthenelse{\boolean{articletitles}}{\emph{{ALPtraum:
  ALP production in proton beam dump experiments}},
  }{}\href{https://doi.org/10.1007/JHEP02(2016)018}{JHEP \textbf{02} (2016)
  018}, \href{http://arxiv.org/abs/1512.03069}{{\normalfont\ttfamily
  arXiv:1512.03069}}, [JHEP02,018(2016)]\relax
\mciteBstWouldAddEndPuncttrue
\mciteSetBstMidEndSepPunct{\mcitedefaultmidpunct}
{\mcitedefaultendpunct}{\mcitedefaultseppunct}\relax
\EndOfBibitem
\bibitem{Izaguirre:2016dfi}
E.~Izaguirre, T.~Lin, and B.~Shuve,
  \ifthenelse{\boolean{articletitles}}{\emph{{Searching for Axionlike Particles
  in Flavor-Changing Neutral Current Processes}},
  }{}\href{https://doi.org/10.1103/PhysRevLett.118.111802}{Phys.\ Rev.\ Lett.\
  \textbf{118} (2017) 111802},
  \href{http://arxiv.org/abs/1611.09355}{{\normalfont\ttfamily
  arXiv:1611.09355}}\relax
\mciteBstWouldAddEndPuncttrue
\mciteSetBstMidEndSepPunct{\mcitedefaultmidpunct}
{\mcitedefaultendpunct}{\mcitedefaultseppunct}\relax
\EndOfBibitem
\bibitem{Knapen:2016moh}
S.~Knapen, T.~Lin, H.~K. Lou, and T.~Melia,
  \ifthenelse{\boolean{articletitles}}{\emph{{Searching for Axionlike Particles
  with Ultraperipheral Heavy-Ion Collisions}},
  }{}\href{https://doi.org/10.1103/PhysRevLett.118.171801}{Phys.\ Rev.\ Lett.\
  \textbf{118} (2017) 171801},
  \href{http://arxiv.org/abs/1607.06083}{{\normalfont\ttfamily
  arXiv:1607.06083}}\relax
\mciteBstWouldAddEndPuncttrue
\mciteSetBstMidEndSepPunct{\mcitedefaultmidpunct}
{\mcitedefaultendpunct}{\mcitedefaultseppunct}\relax
\EndOfBibitem
\bibitem{Mariotti:2017vtv}
A.~Mariotti, D.~Redigolo, F.~Sala, and K.~Tobioka,
  \ifthenelse{\boolean{articletitles}}{\emph{{New LHC bound on low-mass
  diphoton resonances}},
  }{}\href{https://doi.org/10.1016/j.physletb.2018.06.039}{Phys.\ Lett.\
  \textbf{B783} (2018) 13},
  \href{http://arxiv.org/abs/1710.01743}{{\normalfont\ttfamily
  arXiv:1710.01743}}\relax
\mciteBstWouldAddEndPuncttrue
\mciteSetBstMidEndSepPunct{\mcitedefaultmidpunct}
{\mcitedefaultendpunct}{\mcitedefaultseppunct}\relax
\EndOfBibitem
\bibitem{Bauer:2017ris}
M.~Bauer, M.~Neubert, and A.~Thamm,
  \ifthenelse{\boolean{articletitles}}{\emph{{Collider Probes of Axion-Like
  Particles}}, }{}\href{https://doi.org/10.1007/JHEP12(2017)044}{JHEP
  \textbf{12} (2017) 044},
  \href{http://arxiv.org/abs/1708.00443}{{\normalfont\ttfamily
  arXiv:1708.00443}}\relax
\mciteBstWouldAddEndPuncttrue
\mciteSetBstMidEndSepPunct{\mcitedefaultmidpunct}
{\mcitedefaultendpunct}{\mcitedefaultseppunct}\relax
\EndOfBibitem
\bibitem{CidVidal:2018blh}
X.~Cid~Vidal {\em et~al.}, \ifthenelse{\boolean{articletitles}}{\emph{{New
  Axion Searches at Flavor Factories}},
  }{}\href{https://doi.org/10.1007/JHEP06(2020)141,
  10.1007/JHEP01(2019)113}{JHEP \textbf{01} (2019) 113},
  \href{http://arxiv.org/abs/1810.09452}{{\normalfont\ttfamily
  arXiv:1810.09452}}, [Erratum: JHEP06,141(2020)]\relax
\mciteBstWouldAddEndPuncttrue
\mciteSetBstMidEndSepPunct{\mcitedefaultmidpunct}
{\mcitedefaultendpunct}{\mcitedefaultseppunct}\relax
\EndOfBibitem
\bibitem{Bauer:2018uxu}
M.~Bauer, M.~Heiles, M.~Neubert, and A.~Thamm,
  \ifthenelse{\boolean{articletitles}}{\emph{{Axion-Like Particles at Future
  Colliders}}, }{}\href{https://doi.org/10.1140/epjc/s10052-019-6587-9}{Eur.\
  Phys.\ J.\  \textbf{C79} (2019) 74},
  \href{http://arxiv.org/abs/1808.10323}{{\normalfont\ttfamily
  arXiv:1808.10323}}\relax
\mciteBstWouldAddEndPuncttrue
\mciteSetBstMidEndSepPunct{\mcitedefaultmidpunct}
{\mcitedefaultendpunct}{\mcitedefaultseppunct}\relax
\EndOfBibitem
\bibitem{Harland-Lang:2019zur}
L.~Harland-Lang, J.~Jaeckel, and M.~Spannowsky,
  \ifthenelse{\boolean{articletitles}}{\emph{{A fresh look at ALP searches in
  fixed target experiments}},
  }{}\href{https://doi.org/10.1016/j.physletb.2019.04.045}{Phys.\ Lett.\
  \textbf{B793} (2019) 281},
  \href{http://arxiv.org/abs/1902.04878}{{\normalfont\ttfamily
  arXiv:1902.04878}}\relax
\mciteBstWouldAddEndPuncttrue
\mciteSetBstMidEndSepPunct{\mcitedefaultmidpunct}
{\mcitedefaultendpunct}{\mcitedefaultseppunct}\relax
\EndOfBibitem
\bibitem{Ebadi:2019gij}
J.~Ebadi, S.~Khatibi, and M.~Mohammadi~Najafabadi,
  \ifthenelse{\boolean{articletitles}}{\emph{{New probes for axionlike
  particles at hadron colliders}},
  }{}\href{https://doi.org/10.1103/PhysRevD.100.015016}{Phys.\ Rev.\
  \textbf{D100} (2019) 015016},
  \href{http://arxiv.org/abs/1901.03061}{{\normalfont\ttfamily
  arXiv:1901.03061}}\relax
\mciteBstWouldAddEndPuncttrue
\mciteSetBstMidEndSepPunct{\mcitedefaultmidpunct}
{\mcitedefaultendpunct}{\mcitedefaultseppunct}\relax
\EndOfBibitem
\bibitem{Mimasu:2014nea}
K.~Mimasu and V.~Sanz, \ifthenelse{\boolean{articletitles}}{\emph{{ALPs at
  Colliders}}, }{}\href{https://doi.org/10.1007/JHEP06(2015)173}{JHEP
  \textbf{06} (2015) 173},
  \href{http://arxiv.org/abs/1409.4792}{{\normalfont\ttfamily
  arXiv:1409.4792}}\relax
\mciteBstWouldAddEndPuncttrue
\mciteSetBstMidEndSepPunct{\mcitedefaultmidpunct}
{\mcitedefaultendpunct}{\mcitedefaultseppunct}\relax
\EndOfBibitem
\bibitem{Brivio:2017ije}
I.~Brivio {\em et~al.}, \ifthenelse{\boolean{articletitles}}{\emph{{ALPs
  Effective Field Theory and Collider Signatures}},
  }{}\href{https://doi.org/10.1140/epjc/s10052-017-5111-3}{Eur.\ Phys.\ J.\
  \textbf{C77} (2017) 572},
  \href{http://arxiv.org/abs/1701.05379}{{\normalfont\ttfamily
  arXiv:1701.05379}}\relax
\mciteBstWouldAddEndPuncttrue
\mciteSetBstMidEndSepPunct{\mcitedefaultmidpunct}
{\mcitedefaultendpunct}{\mcitedefaultseppunct}\relax
\EndOfBibitem
\bibitem{Aloni:2019ruo}
D.~Aloni, C.~Fanelli, Y.~Soreq, and M.~Williams,
  \ifthenelse{\boolean{articletitles}}{\emph{{Photoproduction of Axionlike
  Particles}}, }{}\href{https://doi.org/10.1103/PhysRevLett.123.071801}{Phys.\
  Rev.\ Lett.\  \textbf{123} (2019) 071801},
  \href{http://arxiv.org/abs/1903.03586}{{\normalfont\ttfamily
  arXiv:1903.03586}}\relax
\mciteBstWouldAddEndPuncttrue
\mciteSetBstMidEndSepPunct{\mcitedefaultmidpunct}
{\mcitedefaultendpunct}{\mcitedefaultseppunct}\relax
\EndOfBibitem
\bibitem{Batell:2009jf}
B.~Batell, M.~Pospelov, and A.~Ritz,
  \ifthenelse{\boolean{articletitles}}{\emph{{Multi-lepton Signatures of a
  Hidden Sector in Rare B Decays}},
  }{}\href{https://doi.org/10.1103/PhysRevD.83.054005}{Phys.\ Rev.\ D
  \textbf{83} (2011) 054005},
  \href{http://arxiv.org/abs/0911.4938}{{\normalfont\ttfamily
  arXiv:0911.4938}}\relax
\mciteBstWouldAddEndPuncttrue
\mciteSetBstMidEndSepPunct{\mcitedefaultmidpunct}
{\mcitedefaultendpunct}{\mcitedefaultseppunct}\relax
\EndOfBibitem
\bibitem{Chakraborty:2021wda}
S.~Chakraborty {\em et~al.}, \ifthenelse{\boolean{articletitles}}{\emph{{Heavy
  QCD Axion in $b\to s$ transition: Enhanced Limits and Projections}},
  }{}\href{http://arxiv.org/abs/2102.04474}{{\normalfont\ttfamily
  arXiv:2102.04474}}\relax
\mciteBstWouldAddEndPuncttrue
\mciteSetBstMidEndSepPunct{\mcitedefaultmidpunct}
{\mcitedefaultendpunct}{\mcitedefaultseppunct}\relax
\EndOfBibitem
\bibitem{MartinCamalich:2020dfe}
J.~Martin~Camalich {\em et~al.},
  \ifthenelse{\boolean{articletitles}}{\emph{{Quark Flavor Phenomenology of the
  QCD Axion}}, }{}\href{https://doi.org/10.1103/PhysRevD.102.015023}{Phys.\
  Rev.\ D \textbf{102} (2020) 015023},
  \href{http://arxiv.org/abs/2002.04623}{{\normalfont\ttfamily
  arXiv:2002.04623}}\relax
\mciteBstWouldAddEndPuncttrue
\mciteSetBstMidEndSepPunct{\mcitedefaultmidpunct}
{\mcitedefaultendpunct}{\mcitedefaultseppunct}\relax
\EndOfBibitem
\bibitem{Calibbi:2016hwq}
L.~Calibbi {\em et~al.}, \ifthenelse{\boolean{articletitles}}{\emph{{Minimal
  axion model from flavor}},
  }{}\href{https://doi.org/10.1103/PhysRevD.95.095009}{Phys.\ Rev.\ D
  \textbf{95} (2017) 095009},
  \href{http://arxiv.org/abs/1612.08040}{{\normalfont\ttfamily
  arXiv:1612.08040}}\relax
\mciteBstWouldAddEndPuncttrue
\mciteSetBstMidEndSepPunct{\mcitedefaultmidpunct}
{\mcitedefaultendpunct}{\mcitedefaultseppunct}\relax
\EndOfBibitem
\bibitem{Bauer:2019gfk}
M.~Bauer {\em et~al.}, \ifthenelse{\boolean{articletitles}}{\emph{{Axionlike
  Particles, Lepton-Flavor Violation, and a New Explanation of $a_\mu$ and
  $a_e$}}, }{}\href{https://doi.org/10.1103/PhysRevLett.124.211803}{Phys.\
  Rev.\ Lett.\  \textbf{124} (2020) 211803},
  \href{http://arxiv.org/abs/1908.00008}{{\normalfont\ttfamily
  arXiv:1908.00008}}\relax
\mciteBstWouldAddEndPuncttrue
\mciteSetBstMidEndSepPunct{\mcitedefaultmidpunct}
{\mcitedefaultendpunct}{\mcitedefaultseppunct}\relax
\EndOfBibitem
\bibitem{Cornella:2019uxs}
C.~Cornella, P.~Paradisi, and O.~Sumensari,
  \ifthenelse{\boolean{articletitles}}{\emph{{Hunting for ALPs with Lepton
  Flavor Violation}}, }{}\href{https://doi.org/10.1007/JHEP01(2020)158}{JHEP
  \textbf{01} (2020) 158},
  \href{http://arxiv.org/abs/1911.06279}{{\normalfont\ttfamily
  arXiv:1911.06279}}\relax
\mciteBstWouldAddEndPuncttrue
\mciteSetBstMidEndSepPunct{\mcitedefaultmidpunct}
{\mcitedefaultendpunct}{\mcitedefaultseppunct}\relax
\EndOfBibitem
\bibitem{Albrecht:2019zul}
J.~Albrecht, E.~Stamou, R.~Ziegler, and R.~Zwicky,
  \ifthenelse{\boolean{articletitles}}{\emph{{Probing flavoured Axions in the
  Tail of $B_q \to \mu^+\mu^-$}},
  }{}\href{http://arxiv.org/abs/1911.05018}{{\normalfont\ttfamily
  arXiv:1911.05018}}\relax
\mciteBstWouldAddEndPuncttrue
\mciteSetBstMidEndSepPunct{\mcitedefaultmidpunct}
{\mcitedefaultendpunct}{\mcitedefaultseppunct}\relax
\EndOfBibitem
\bibitem{Davoudiasl:2021haa}
H.~Davoudiasl, R.~Marcarelli, N.~Miesch, and E.~T. Neil,
  \ifthenelse{\boolean{articletitles}}{\emph{{Searching for Flavor-Violating
  ALPs in Higgs Decays}},
  }{}\href{http://arxiv.org/abs/2105.05866}{{\normalfont\ttfamily
  arXiv:2105.05866}}\relax
\mciteBstWouldAddEndPuncttrue
\mciteSetBstMidEndSepPunct{\mcitedefaultmidpunct}
{\mcitedefaultendpunct}{\mcitedefaultseppunct}\relax
\EndOfBibitem
\bibitem{Hughes:1971}
V.~W. Hughes and B.~MaglicBull.\ Am.\ Phys.\ Soc.\  \textbf{16} (1971) 65\relax
\mciteBstWouldAddEndPuncttrue
\mciteSetBstMidEndSepPunct{\mcitedefaultmidpunct}
{\mcitedefaultendpunct}{\mcitedefaultseppunct}\relax
\EndOfBibitem
\bibitem{CidVidal:2019qub}
X.~Cid~Vidal {\em et~al.},
  \ifthenelse{\boolean{articletitles}}{\emph{{Discovering True Muonium at
  LHCb}}, }{}\href{https://doi.org/10.1103/PhysRevD.100.053003}{Phys.\ Rev.\ D
  \textbf{100} (2019) 053003},
  \href{http://arxiv.org/abs/1904.08458}{{\normalfont\ttfamily
  arXiv:1904.08458}}\relax
\mciteBstWouldAddEndPuncttrue
\mciteSetBstMidEndSepPunct{\mcitedefaultmidpunct}
{\mcitedefaultendpunct}{\mcitedefaultseppunct}\relax
\EndOfBibitem
\bibitem{Essig:2013lka}
R.~Essig {\em et~al.}, \ifthenelse{\boolean{articletitles}}{\emph{{Working
  Group Report: New Light Weakly Coupled Particles}}, }{} in {\em {Community
  Summer Study 2013}: {Snowmass on the Mississippi}}, 2013,
  \href{http://arxiv.org/abs/1311.0029}{{\normalfont\ttfamily
  arXiv:1311.0029}}\relax
\mciteBstWouldAddEndPuncttrue
\mciteSetBstMidEndSepPunct{\mcitedefaultmidpunct}
{\mcitedefaultendpunct}{\mcitedefaultseppunct}\relax
\EndOfBibitem
\bibitem{Alexander:2016aln}
J.~Alexander {\em et~al.}, \ifthenelse{\boolean{articletitles}}{\emph{{Dark
  Sectors 2016 Workshop: Community Report}}, }{} 2016,
  \href{http://arxiv.org/abs/1608.08632}{{\normalfont\ttfamily
  arXiv:1608.08632}}\relax
\mciteBstWouldAddEndPuncttrue
\mciteSetBstMidEndSepPunct{\mcitedefaultmidpunct}
{\mcitedefaultendpunct}{\mcitedefaultseppunct}\relax
\EndOfBibitem
\bibitem{Strassler:2006ri}
M.~J. Strassler and K.~M. Zurek,
  \ifthenelse{\boolean{articletitles}}{\emph{{Discovering the Higgs through
  highly-displaced vertices}},
  }{}\href{https://doi.org/10.1016/j.physletb.2008.02.008}{Phys.\ Lett.\
  \textbf{B661} (2008) 263},
  \href{http://arxiv.org/abs/hep-ph/0605193}{{\normalfont\ttfamily
  arXiv:hep-ph/0605193}}\relax
\mciteBstWouldAddEndPuncttrue
\mciteSetBstMidEndSepPunct{\mcitedefaultmidpunct}
{\mcitedefaultendpunct}{\mcitedefaultseppunct}\relax
\EndOfBibitem
\bibitem{Strassler:2008bv}
M.~J. Strassler, \ifthenelse{\boolean{articletitles}}{\emph{{Why Unparticle
  Models with Mass Gaps are Examples of Hidden Valleys}},
  }{}\href{http://arxiv.org/abs/0801.0629}{{\normalfont\ttfamily
  arXiv:0801.0629}}\relax
\mciteBstWouldAddEndPuncttrue
\mciteSetBstMidEndSepPunct{\mcitedefaultmidpunct}
{\mcitedefaultendpunct}{\mcitedefaultseppunct}\relax
\EndOfBibitem
\bibitem{Cesarotti:2020ngq}
C.~Cesarotti, M.~Reece, and M.~J. Strassler,
  \ifthenelse{\boolean{articletitles}}{\emph{{The Efficacy of Event Isotropy as
  an Event Shape Observable}},
  }{}\href{http://arxiv.org/abs/2011.06599}{{\normalfont\ttfamily
  arXiv:2011.06599}}\relax
\mciteBstWouldAddEndPuncttrue
\mciteSetBstMidEndSepPunct{\mcitedefaultmidpunct}
{\mcitedefaultendpunct}{\mcitedefaultseppunct}\relax
\EndOfBibitem
\bibitem{Polchinski:2002jw}
J.~Polchinski and M.~J. Strassler,
  \ifthenelse{\boolean{articletitles}}{\emph{{Deep inelastic scattering and
  gauge / string duality}},
  }{}\href{https://doi.org/10.1088/1126-6708/2003/05/012}{JHEP \textbf{05}
  (2003) 012}, \href{http://arxiv.org/abs/hep-th/0209211}{{\normalfont\ttfamily
  arXiv:hep-th/0209211}}\relax
\mciteBstWouldAddEndPuncttrue
\mciteSetBstMidEndSepPunct{\mcitedefaultmidpunct}
{\mcitedefaultendpunct}{\mcitedefaultseppunct}\relax
\EndOfBibitem
\bibitem{Hofman:2008ar}
D.~M. Hofman and J.~Maldacena,
  \ifthenelse{\boolean{articletitles}}{\emph{{Conformal collider physics:
  Energy and charge correlations}},
  }{}\href{https://doi.org/10.1088/1126-6708/2008/05/012}{JHEP \textbf{05}
  (2008) 012}, \href{http://arxiv.org/abs/0803.1467}{{\normalfont\ttfamily
  arXiv:0803.1467}}\relax
\mciteBstWouldAddEndPuncttrue
\mciteSetBstMidEndSepPunct{\mcitedefaultmidpunct}
{\mcitedefaultendpunct}{\mcitedefaultseppunct}\relax
\EndOfBibitem
\bibitem{Lin:2007fa}
S.~Lin and E.~Shuryak, \ifthenelse{\boolean{articletitles}}{\emph{{Toward the
  AdS/CFT gravity dual for high energy collisions. 2. The stress tensor on the
  boundary}}, }{}\href{https://doi.org/10.1103/PhysRevD.77.085014}{Phys.\ Rev.\
   \textbf{D77} (2008) 085014},
  \href{http://arxiv.org/abs/0711.0736}{{\normalfont\ttfamily
  arXiv:0711.0736}}\relax
\mciteBstWouldAddEndPuncttrue
\mciteSetBstMidEndSepPunct{\mcitedefaultmidpunct}
{\mcitedefaultendpunct}{\mcitedefaultseppunct}\relax
\EndOfBibitem
\bibitem{Hagedorn:1965st}
R.~Hagedorn, \ifthenelse{\boolean{articletitles}}{\emph{{Statistical
  thermodynamics of strong interactions at high-energies}}, }{}Nuovo Cim.\
  Suppl.\  \textbf{3} (1965) 147\relax
\mciteBstWouldAddEndPuncttrue
\mciteSetBstMidEndSepPunct{\mcitedefaultmidpunct}
{\mcitedefaultendpunct}{\mcitedefaultseppunct}\relax
\EndOfBibitem
\bibitem{Hatta:2008qx}
Y.~Hatta and T.~Matsuo, \ifthenelse{\boolean{articletitles}}{\emph{{Thermal
  hadron spectrum in e+e- annihilation from gauge/string duality}},
  }{}\href{https://doi.org/10.1103/PhysRevLett.102.062001}{Phys.\ Rev.\ Lett.\
  \textbf{102} (2009) 062001},
  \href{http://arxiv.org/abs/0807.0098}{{\normalfont\ttfamily
  arXiv:0807.0098}}\relax
\mciteBstWouldAddEndPuncttrue
\mciteSetBstMidEndSepPunct{\mcitedefaultmidpunct}
{\mcitedefaultendpunct}{\mcitedefaultseppunct}\relax
\EndOfBibitem
\bibitem{Cesarotti:2020uod}
C.~Cesarotti, M.~Reece, and M.~J. Strassler,
  \ifthenelse{\boolean{articletitles}}{\emph{{Spheres To Jets: Tuning Event
  Shapes with 5d Simplified Models}},
  }{}\href{http://arxiv.org/abs/2009.08981}{{\normalfont\ttfamily
  arXiv:2009.08981}}\relax
\mciteBstWouldAddEndPuncttrue
\mciteSetBstMidEndSepPunct{\mcitedefaultmidpunct}
{\mcitedefaultendpunct}{\mcitedefaultseppunct}\relax
\EndOfBibitem
\bibitem{Kang:2008ea}
J.~Kang and M.~A. Luty, \ifthenelse{\boolean{articletitles}}{\emph{{Macroscopic
  Strings and 'Quirks' at Colliders}},
  }{}\href{https://doi.org/10.1088/1126-6708/2009/11/065}{JHEP \textbf{11}
  (2009) 065}, \href{http://arxiv.org/abs/0805.4642}{{\normalfont\ttfamily
  arXiv:0805.4642}}\relax
\mciteBstWouldAddEndPuncttrue
\mciteSetBstMidEndSepPunct{\mcitedefaultmidpunct}
{\mcitedefaultendpunct}{\mcitedefaultseppunct}\relax
\EndOfBibitem
\bibitem{Farina:2017cts}
M.~Farina and M.~Low, \ifthenelse{\boolean{articletitles}}{\emph{{Constraining
  Quirky Tracks with Conventional Searches}},
  }{}\href{https://doi.org/10.1103/PhysRevLett.119.111801}{Phys.\ Rev.\ Lett.\
  \textbf{119} (2017) 111801},
  \href{http://arxiv.org/abs/1703.00912}{{\normalfont\ttfamily
  arXiv:1703.00912}}\relax
\mciteBstWouldAddEndPuncttrue
\mciteSetBstMidEndSepPunct{\mcitedefaultmidpunct}
{\mcitedefaultendpunct}{\mcitedefaultseppunct}\relax
\EndOfBibitem
\bibitem{Knapen:2017kly}
S.~Knapen, H.~K. Lou, M.~Papucci, and J.~Setford,
  \ifthenelse{\boolean{articletitles}}{\emph{{Tracking down Quirks at the Large
  Hadron Collider}},
  }{}\href{https://doi.org/10.1103/PhysRevD.96.115015}{Phys.\ Rev.\ D
  \textbf{96} (2017) 115015},
  \href{http://arxiv.org/abs/1708.02243}{{\normalfont\ttfamily
  arXiv:1708.02243}}\relax
\mciteBstWouldAddEndPuncttrue
\mciteSetBstMidEndSepPunct{\mcitedefaultmidpunct}
{\mcitedefaultendpunct}{\mcitedefaultseppunct}\relax
\EndOfBibitem
\bibitem{stealth_workshop}
\ifthenelse{\boolean{articletitles}}{\emph{{STEALTH physics at LHCb: unleashing
  the full power of LHCb to probe new physics}}, }{}
  \url{https://indico.cern.ch/event/849862/}, 2020\relax
\mciteBstWouldAddEndPuncttrue
\mciteSetBstMidEndSepPunct{\mcitedefaultmidpunct}
{\mcitedefaultendpunct}{\mcitedefaultseppunct}\relax
\EndOfBibitem
\end{mcitethebibliography}

\newpage

\end{document}